\documentclass[reqno,a4paper,11pt]{book}
\usepackage{amssymb,amstext,amsthm,amsfonts,mathrsfs,amsbsy,amsmath,array,bbm,multirow}
\usepackage{slashed,fancyhdr,hojadeestilotesisdoctoral,bbding}
\usepackage{a4wide} 
\usepackage[english,greek]{babel}
\usepackage[usenames,dvipsnames,svgnames,table]{xcolor}
\usepackage[pdftex]{graphicx}
\usepackage{xparse}
\usepackage{tikz}

\NewDocumentCommand\mycite{mgggg}{\IfNoValueTF{#5}{\IfNoValueTF{#3}{\IfNoValueTF{#2}{\singlecite{#1}}{\singlecitedetail{#1}{#2}}}{\multicite{#1}{#2}{#3}}}{\multimulticite{#1}{#2}{#3}{#4}{#5}}} 
\usepackage[authoryear,round,square,colon,comma]{natbib}
\bibpunct{}{}{,}{n}{}{,}

\usepackage[pdftex,unicode,implicit]{hyperref}
\hypersetup{%
  pdftitle    = {Killing spinors - Beyond Supergravity},
  pdfkeywords = {Supergravity,SUGRA,fakeSupergravity,fSUGRA,Killing spinors,Einstein-Weyl}, 
  pdfsubject  = {hep-th, gr-qc, math.dg},
  pdfauthor   = {A.~Palomo-Lozano},
  pdfcreator  = {pdf\LaTeXe\ with package \flqq hyperref\frqq},
  pdffitwindow= false,  
  bookmarksopen   = false,
  bookmarksopenlevel = 1,
  bookmarksnumbered = true,
  unicode     = true,
  plainpages  = false,
  citecolor   = black,
  urlcolor    = black,
  linkcolor   = black,
}

\begin{document}
\setcounter{tocdepth}{1}
\selectlanguage{english}
\thispagestyle{empty}
\phantom{a}
\begin{center}
{
\small{
\renewcommand{\tabcolsep}{2em}
\begin{tabular}{cc}  
\textbf{Universidad}& \textbf{Consejo Superior de} \\ 
\textbf{Aut\'onoma de Madrid} & \textbf{Investigaciones Cient\'ificas} \\  
\includegraphics[scale=0.2]{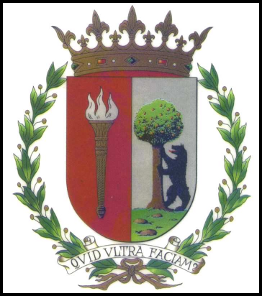} & \includegraphics[scale=0.2]{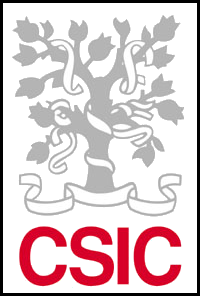} \\
Departamento de F\'isica Te\'orica & Instituto de F\'isica Te\'orica\\\
Facultad de Ciencias\
\end{tabular}}} 
\end{center}

\vspace{\stretch{4}}
\begin{center}

\setlength{\baselineskip}{3\baselineskip}
\textbf{
{\sizea {\sizeA K}ILLING {\sizeA S}PINORS -}\\
{\sizea {\sizeA B}EYOND {\sizeA S}UPERGRAVITY}}

\vspace{\stretch{4}}
{Alberto R. Palomo Lozano}
\vspace{\stretch{1}}
\end{center}
\clearpage

\thispagestyle{empty}
\phantom{1}

\clearpage

\thispagestyle{empty}
\phantom{a}
\begin{center}
{
\renewcommand{\tabcolsep}{2em}
\begin{tabular}{cc}  
\textbf{Universidad}& \textbf{Consejo Superior de} \\ 
\textbf{Aut\'onoma de Madrid} & \textbf{Investigaciones Cient\'ificas} \\  
\includegraphics[scale=0.3]{escudo_uam.png} & \includegraphics[scale=0.3]{logo_csic.png} \\
Departamento de F\'isica Te\'orica & Instituto de F\'isica Te\'orica\\
Facultad de Ciencias\\
\end{tabular}} 
\end{center}
\vspace{2cm}

\vspace{\stretch{5}}
\begin{center}
\setlength{\baselineskip}{2\baselineskip}
\textbf{
{\sizea {\sizeA E}SPINORES DE {\sizeA K}ILLING -}\\
{\sizea {\sizeA M}\'AS {\sizeA A}LL\'A DE LA {\sizeA S}UPERGRAVEDAD}}
	
\vspace{\stretch{1}}
\setlength{\baselineskip}{0.5\baselineskip}
\vspace{\stretch{6}}

\begin{minipage}{16cm}
\begin{center}
Memoria de Tesis Doctoral presentada ante el Departamento de F\'isica Te\'orica\\
de la Universidad Aut\'onoma de Madrid para la obtenci\'on del t\'itulo de Doctor en Ciencias
\end{center}
\end{minipage}
	
\vspace{\stretch{4}}
Tesis Doctoral dirigida por:
\\[1ex]
\textbf{Dr. D. Patrick A. A. Meessen}\\
Investigador \emph{Ram\'on y Cajal}, Universidad de Oviedo
\\[3ex]
y tutelada por:
\\[1ex]
\textbf{Dr. D. Tom\'as Ort\'in Miguel}\\
Cient\'ifico Titular, Consejo Superior de Investigaciones Cient\'ificas
\\[5ex]
2012
\end{center}

\newpage

\thispagestyle{empty}
\phantom{1}

\clearpage

\thispagestyle{empty}
\hspace{\stretch{3}}
\begin{minipage}{25em}
\begin{flushright}
{\em ``For the things we have to learn\\
before we can do,\\
we learn by doing"}\\
\vspace{2ex}
\textgreek{`Aristot'elhc}\phantom{"}
\end{flushright}
\end{minipage}
\hspace{\stretch{1}}
\vspace{8ex}

El trabajo de un f\'isico te\'orico es probablemente uno de los trabajos m\'as solitarios que existen. La elaboraci\'on de esta memoria es el resultado de numerosas horas en soledad dedicadas primero a entender/ reproducir literatura y c\'alculos previos, frutos del esfuerzo de otras personas, y sobre los que se cimentan los trabajos que forman el n\'ucleo de mis a\~nos como doctorando, que han sido finalmente plasmados en estas p\'aginas. Dicho esto, hubiera sido imposible llegar hasta aqu\'i de haber trabajado exclusivamente por mi cuenta, sin haber contado con la ayuda de varias personas. Quisiera por tanto agradecerles su apoyo durante estos a\~nos.

Un lugar privilegiado ocupan mis supervisores de tesis, Tom\'as y Patrick, a quienes quiero agradecer primero el haberme ofrecido la oportunidad de cumplir uno de mis sue\~nos, y tambi\'en el haberme dedicado un tiempo y esfuerzo sin el que no podr\'ia haber culminado esta tesis. Patrick, con qui\'en he trabajado m\'as directamente, ha sido mucho m\'as que un director acad\'emico; me ha instruido en aspectos t\'ecnicos relacionados con las construcciones matem\'aticas que figuran en estas p\'aginas, pero tambi\'en hemos hablado de la vida en la Academia, y las implicaciones personales que conlleva esta profesi\'on. Es una de las personas acad\'emicamente m\'as capacitadas que jam\'as he conocido\footnote{Hace algunos a\~nos dije de organizar una fiesta cuando descubriera un fallo significativo en alguno de los articulos en que hemos trabajado juntos. Aunque logr\'e encontrar uno, seguramente era de transcripcci\'on, y no fundamental como yo hubiese querido.}, y tiene adem\'as un sentido del humor y una capacidad de auto-desaprobaci\'on que encuentro muy refrescante, y de lo que espero haberme empapado un poco. De Tom\'as he aprendido, o al menos as\'i creo que lo ha intentado \'el, a ser m\'as pr\'actico, y no andarme por las nubes. 

Este trabajo representa el final de mi andadura como estudiante de posgrado, pero no quisiera olvidarme de tres buenos amigos que me orientaron en los comienzos en Inglaterra, y a quienes debo en gran parte la que ser\'ia mi futura vinculaci\'on con la Universidad Aut\'onoma de Madrid, Daniel Cremades, Manuel Donaire y Javier Fdz. Alcazar. 

As\'i mismo, debo dar las gracias a mis colaboradores, con quienes he compartido ideas y e-mails estos a\~nos, y de los que he aprendido otra forma de trabajar. Estos son: Jai Grover, Jan Gutowski, Carlos Herdeiro y  Wafic Sabra. No puedo tampoco olvidarme de Ulf Gran, bajo cuya direcci\'on e instrucci\'on pas\'e seis fant\'asticos meses en la \emph{Chalmers Tekniska H\"ogskola} de Gotemburgo, Suecia, y que me instruy\'o en el \emph{m\'etodo espinorial} de atacar la clasificaci\'on de soluciones a teor\'ias de Supergravedad.

Aunque he hecho poca \emph{vida social} en el IFT durante estos a\~nos, he llegado a entablar unos v\'inculos que sobrepasaban lo puramente acad\'emico con mis compa\~neros de doctorado, a los que quisiera dar las gracias por los buenos momentos pasados. Tambi\'en a Paco, de \emph{Limcamar}, con quien he compartido reflexiones sobre f\'utbol y el devenir de la Liga 2011-2012. As\'i mismo, no podr\'ia haber llevado a buen culmen toda la parte log\'istica sin la ayuda de un fant\'astico equipo administrativo, encabezado por Isabel P\'erez. !`Muchas gracias a todos! Quisiera tambi\'en agradecer al director del IFT, Alberto Casas, el haber sido siempre tan coagente con mis peticiones relacionadas con el nuevo edificio, y a Carlos Pena lo f\'acil que resulta hablar con \'el, que aprend\'i cuando estaba organizando el minicurso sobre branas. Es un placer tambi\'en mencionar la ayuda inform\'atica recibida por parte de Giovanni Ram\'irez, Andr\'es D\'iaz-Gil y Charo Villamariz, as\'i como de la familia Carton en los \'ultimos d\'ias antes de la entrega de esta tesis, porque prefiero tirarme por la ventana antes de que falle el ordenador. En este sentido, me gustar\'ia hacer menci\'on de las comunidades de \emph{open source} GNU/ Linux/ \LaTeX, que se encargan de producir y mantener software de muy alta calidad, gracias al cual me ha resultado mucho m\'as f\'acil desarrollar mis tareas. Si todav\'ia usas un sistema operativo propietario, realmente vale la pena hacer la transici\'on a una distribuci\'on Linux. Juan Jover me ayud\'o cuando yo la hice a \emph{Ubuntu}, all\'a por la release 7.10 \emph{Gutsy Gibbon}.

\thispagestyle{empty}
Debo tambi\'en agradecer el apoyo econ\'omico recibido durante estos a\~nos por el \emph{Consejo Superior de Investigaciones Cient\'ificas}, en el contexto de las ayudas para la investigaci\'on `Junta para la Ampliaci\'on de Estudios' JAE 2007.

Probablemente la persona m\'as feliz por este trabajo sea mi padre, que lleva cuatro a\~nos pregunt\'andome si todav\'ia no hab\'ia acabado la tesis. ``!`Tesis, tesis, tesis!", ha repetido en numerosas ocasiones. Es pues que quiero dedicarle estas p\'aginas, al igual que a mi madre y hermana, por proveeer el balance adecuado, y por soportar mi tendencia a estar despierto a altas horas de la noche, as\'i como a mi t\'ia Luz, por \emph{estar siempre ah\'i}, y a mi re-t\'io Mat\'ias, que ahora D.E.P.~junto a su Se\~nor. 

Por \'ultimo, quisiera terminar recordando una peque\~na frase que se atribuye a Arist\'oteles, pero que a mi me transmiti\'o el que fuera mi profesor de \first{M\'etodos num\'ericos}, el Dr. Ventzeslav Valev, que dec\'ia que \emph{``aprendemos haciendo, no viendo"}. Es una m\'axima que en su momento no llegu\'e a comprender del todo, pero que durante el proceso de este doctorado he llegado a asimilar y reconocer. \'Este ha sido para mi un viaje que me ha instruido m\'as all\'a de lo puramente acad\'emico, tambi\'en acerca de mis propias capacidades, mis l\'imites, e incluso mis mecanismos epistemol\'ogicos, que espero me sirvan en el futuro.\\

\vspace{1cm}
\hspace{\stretch{1}}Madrid, junio 2012.
\clearpage

\addtocontents{toc}{\protect\thispagestyle{empty}}
\tableofcontents
\thispagestyle{empty}
\clearpage

\renewcommand{\leftmark}{\MakeUppercase{Introduction}}
\chapter{Introduction}
\label{intro}
These pages are the result of the research undertaken in the \first{Grupo de \mbox{Gravitaci\'on} y Cuerdas}, led by Prof. Tom\'as Ort\'in 
at the \href{http://www.ift.uam-csic.es/en}{Instituto de F\'isica Te\'orica UAM/ CSIC}, during the tenure of my doctoral degree. They are intended to give an account of the kind of problems attacked during this time, as well as an illustration of why they can prove of interest, and the methods considered to solve them. In particular, it is a study of various constructions which arise in the context of supergravity theories, and along those lines we have made use of methods developed during the programme of classification of supersymmetric solutions to said theories, in order to obtain our results.

In a very concise manner, Supergravity (SUGRA, for short) is a mathematical construction made from adding the algebra of Supersymmetry (which gives the \emph{Super} in SUGRA) to the framework of General Relativity. This gives a field theory, which is interesting because one can recognise in it a spin-2 massless particle, identified with the graviton (the postulated particle mediating the gravitational force). In this way the `Physics' behind the mathematical construction appears quite naturally in the construction. For the interested reader, there are several available books and review articles which give a detailed account of these topics.

The outline of this work is the following. We start in section \ref{Killingspinors} with a quick introduction to Killing spinors and their use in the study of Supergravity. In particular, we comment on the interest for supersymmetric solutions to supergravity theories, from where we transition to the core topic of this thesis, which is the application of techniques arising in the search for the latter to other contexts. We begin by discussing fakeSupergravity in detail. This theory can be understood as a spin-off from genuine (regular) SUGRA, which arises by introducing a change of sign on the potential of the latter, and can be used as a scaffolding from where to obtain backgrounds with a positive cosmological constant. 
At last, we comment on possibilities of using these supersymmetric techniques to tackle problems of interest in Mathematics.

Chapter \ref{classificationSUGRAs} summarises the programme of classification of supersymmetric solutions to supergravity theories. This was initiated in 1983 by Paul Tod, and was pushed to within its calculational limits in the last decade, where many interesting characterisations were produced. The chapter gives details of the spinor bilinear method pioneered by Gauntlett \emph{et al.} in \mycite{Gauntlett:2002nw}, as well as the spinorial geometry approach of Gillard \emph{et al.} \mycite{Gillard:2004xq}, including the r\^ole that the mathematical concept of G-structures plays. Chapter \ref{4d} gives the full classification of solutions of $\N=2$ $d=4$ gauged fakeSUGRA, by considering bilinears formed out of spinors. The solution includes both the case where the vector bilinear's absolute value is larger than zero, dubbed the \emph{timelike} case, and that when it is zero, \emph{i.e.}~the \emph{null} case. Chapter \ref{5dminimal} considers the problem of characterising the solutions to five-dimensional minimal fSUGRA, also by means of the spinor-bilinear formalism. Chapter \ref{5dgauged} generalises this work, by presenting the classification of $\N=1$ $d=5$ fSUGRA coupled to Abelian vector multiplets, but now using the spinorial geometry approach. Chapter \ref{EWspaces} is an application of the techniques considered in previous sections to the mathematical investigation of Einstein-Weyl spaces. The problem of classifying this kind of spaces has largely been ignored, and only few certain examples have been proposed as of present. We add to the spectrum of solutions by considering a modified connection and using the techniques that were useful in classifying fSUGRA solutions, only this time we obtain a characterisation of Einstein-Weyl (EW) spaces.

Annexes \ref{resumen}-\ref{conclusiones} give a summary and conclusions of the work presented, in Spanish language. The last part of the thesis has the appendices, which contain information used during the main chapters of the work. Appendix \ref{conventions} has the preferred conventions. It includes the definitions for the tensorial calculus used extensively in both four and five dimensions, given in appendix \ref{tensorconventions}. Appendix \ref{spinorialstructures} provides the conventions for the spinorial geometry structures. Appendix \ref{app:bilinearsandfierzs} gives account of the spinor bilinears and Fierzs identities, which are relevant to the classification method of chapters \ref{4d} and \ref{5dminimal}. Appendix \ref{scalargeometries} defines the geometry of the scalar manifolds which appear naturally in the theories considered. We discuss K\"ahler-Hodge, special K\"ahler, quaternionic-K\"ahler and real special K\"ahler geometries, which appear in the context of pure SUGRA theories with vector and hyperscalar multiplets in both four and five dimensions. Appendix \ref{someusefulgeometry} has some geometric information useful in the characterisation of the null cases in both $d=4$ and minimal $d=5$ fSUGRA, as well as a little note on Kundt waves, which recurrently show up as a solution to this class. Appendix \ref{sec:Weylgeometry} contains a scholium on Weyl geometry, including annotations on Einstein-Weyl spaces and their subclass of Gauduchon-Tod, which appear naturally as the base-space geometry in the timelike case in $d=4$ and in the null case of $d=5$. Appendix \ref{appsec:Sim} includes some basic information on the Similitude group, which appears when studying null case scenarios. Finally, appendix \ref{appsec:lorentz} is a short introduction to the Lorentz and the Spin groups, and the 2-1 relationship existent between them.  

\section{Killing spinors}
\label{Killingspinors}
Spinorial fields parallelised by some connection have been studied in both Mathematics and Physics for some time now. In Mathematics, this idea is succinctly expressed in the language of Killing spinors (see e.g.~\mycite{Baum:1991xx} for a review). These are spinors $\epsilon$ such that
\begin{equation}
\nabla_X \epsilon=\lambda X\cdot \epsilon\ ,
\end{equation}
for a vector field $X$, where $\lambda\in\mathbb{C}$ is usually called the \emph{Killing number}, and $\cdot$ is the Clifford product. Moreover, we say that a spinor is covariantly constant (or parallel) if it is Killing with Killing number $\lambda=0$, so that 
\begin{equation}
\nabla \epsilon=0\ .
\end{equation}
A very remarkable result is the correspondence between Riemannian manifolds admitting connections with special holonomy\footnote{See appendix \ref{app:simhol} for some information on the holonomy group and special holonomy manifolds.} and parallel spinors \mycite{Hitchin:1974}. In hindsight, this is clear since if the spin manifold $M^d$ carries a Killing spinor, then $M$ is Einstein and has Ricci scalar
\begin{equation}
R=4d(d-1)\lambda^2\ .
\end{equation}
One can then easily see why special holonomy manifolds are Ricci-flat. 


More so, Killing spinors have a lot of relevance in Physics. They were for example used to corroborate \mycite{Witten:1981mf} the positive energy theorem of classical General Relativity (stability of Minkowski space as a ground state) \mycite{Schon:1979rg,Schon:1981vd}. Moreover, they naturally appear in the context of Supergravity (see \emph{e.g.}~\mycite{Duff:1986hr}). This is because they serve to describe spinorial fields parallel w.r.t.~a supercovariant derivative which encodes the vanishing of the supersymmetric variation of the fermionic superpartners. In other words, solutions having unbroken SUSY admit Killing spinors. 

Supersymmetric solutions to SUGRA theories are prescriptions for the bosonic fields of the latter such that they solve the equations of motion arising from its action. In principle, a solution to a SUGRA theory, \emph{i.e.}~a theory whose action is invariant under the SUSY transformations
\begin{equation}
\begin{array}{l}
\delta_\epsilon e^a_\mu=\frac{1}{2}\bar{\epsilon}\, \gamma^a\psi_\mu\ ,\quad \delta_\epsilon \psi_\mu=D_\mu \epsilon+\dots\\
\\
\quad\dots
\end{array}
\end{equation}
need not be supersymmetric itself. A supersymmetric solution\footnote{It is also customary to refer to those solutions preserving some residual SUSY as BPS solutions, since they saturate the Bogomol'nyi-Prasad-Sommerfeld bound.} is defined as one which preserves a certain amount (all, or a fraction) of the original supersymmetry. This means that if we act on the solution fields with the (corresponding) elements of the SUSY group, these remain invariant. Nothing prevents us from trying to find non-BPS solutions to a SUGRA theory, but solving the equations of motion will in general not prove an easy task. Looking for solutions which preserve at least some fraction of the original SUSY will be more managable, and it is in this sense that supersymmetric solutions to supergravity theories have been of interest in the last couple of decades.

For one, if Nature were to hold a symmetry between fermions and bosons at some high energy level, it is presumed that its vacuum would be described by some kind of supersymmetric solution. Furthermore, SUGRA has been notably relevant to string theorists ever since it was interpreted as a low-energy limit of Superstring Theory, providing with the framework of an effective field theory where one has the ability to calculate, and whose solutions are bone fide backgrounds which capture many of the gravitational aspects present in the full picture. In this sense, supersymmetric solutions offer a very convenient scenario where to test and develop ideas for the latter. For instance, the study of the quantum regimes where string theories purportedly live is by no means an easy task; supersymmetric solutions, on the other hand, have further stability properties granted by the presence of Supersymmetry, which constrains the equations of motion that need to be solved, and this can be used in certain cases to prove their non-renormalisability under quantum corrections (see \emph{e.g.}~\mycite{Kallosh:1993wx}{Banks:1998nr,Kallosh:1998qs,Coley:2007yx}{Meessen:2007ef}). 

An example of this is the microscopic interpretation of black hole (BH) states, that led to the renowned matching of macroscopic and microscopic black hole entropy \mycite{Strominger:1996sh}. This was achieved by considering an asymptotically-flat five-dimensional extremal black hole, where the non-renormalisability of the mass-charge relation for supersymmetric bound states allows for the counting of microstates. Moreover, the analysis makes use of the so-called \emph{attractor mechanism} of \mycite{Ferrara:1995ih}{Strominger:1996kf}{Ferrara:1996dd}, which was discovered in the study of supergravity models, and says that the matter fields on the event horizon of a supersymmetric black hole depend only on the electric and magnetic charges of the solution, and not on the asymptotic values of the fields. 

Another instance of foremost importance is that which led to the very famous AdS/CFT conjecture, that relates a string theory/ gravity theory to a quantum field theory (without gravity) on its boundary. In \mycite{Maldacena:1997re}, Maldacena described the correspondence between an extremal black hole composed of $N$ D3-branes, whose near-horizon geometry is given by the maximally-symmetric and maximally-supersymmetric $AdS_5 \times S^5$, and the $\N=4$ super Yang-Mills theory on $\mathbb{R}\times S^3$, to which the string theory drifts as the string scale goes to zero. This $AdS_5$ geometry is actually the only maximally-supersymmetric ground state for five-dimensional minimal gauged SUGRA, and it is presumed that other less-supersymmetric solutions would be of interest in further studies of the conjecture, as well as in other brane-world constructions of a similar nature. In particular, these BPS solutions with smaller residual SUSY may correspond to states of the conformal field theory expanded around operators which have a non-zero vacuum expectation value. 

In view of this, it is clear that the connection between the study of supersymmetric solutions and our understanding of String Theory has proven very fruitful. However, it should be noted that most of the advances in the topic have been concerned with the study of asymptotically-flat or anti-De Sitter solutions. Very few advances have been produced in the area of De Sitter solutions, and little is known about them at present. We shall return to this topic in the following section.

\section{Beyond Supergravity}
\label{beyondSUGRA}
In this section we introduce the core topic of this thesis, which is the application of techniques arising from the classification of supersymmetric solutions to supergravity theories to other contexts. We start with a description and motivation to the theory of fakeSupergravity, which is considered in chapters \ref{4d}, \ref{5dminimal} and \ref{5dgauged}. Later on we will see that the arguments can be refined to also classify geometries.

As commented above, not much is known about the behaviour of De Sitter solutions. From a Supergravity perspective, this is because there are few models allowing for a positive cosmological constant, and the ones that do are very complicated, which hinders our ability to find non-trivial solutions. Even though some solutions do exist, they are not general enough, and thus far the achievements obtained with the likes of the attractor mechanism, or the AdS/CFT correspondence, are unmatched. It is in this context that the study of solutions to fakeSupergravity is motivated.

Roughly speaking, fakeSupergravity (fSUGRA) arises by performing a Wick rotation on the Fayet-Iliopoulos (FI) term \mycite{Fayet:1974jb} of a standard SUGRA, which modifies the gauge group of the theory. Its connection with cosmological solutions is given by the works of Kastor and Traschen (KT). In \mycite{Kastor:1992nn}, they created an asymptotically-De Sitter charged multi-black hole solution by observing that the extreme Reissner-Nordstr\"om-De Sitter black hole solution written in spherical coordinates could be transformed to the time-dependent conforma-static form
\begin{equation}
\label{eq:KT1}
ds^{2} \; =\; \Omega^{-2}d\tau^{2} \ -\ \Omega^{2}\ d\vec{x}^{2}_{(3)} \hspace{.6cm}\mbox{with}\hspace{.6cm} \Omega \ =\ H\tau \ +\ \frac{m}{r} \; ,
\end{equation}
where $3H^{2}$ is the cosmological constant. 
As the $r$-dependent part of $\Omega$ is a spherically-symmetric harmonic function, the multi-BH solutions can be 
created by changing it to a more general harmonic function.

Seeing the similarity of the above solution and the supersymmetric solutions to minimal $\N=2$ $d=4$ Supergravity \mycite{Tod:1983pm}, whose bosonic part is just E-M theory, Kastor and Traschen proceeded to show in \mycite{Kastor:1993mj} that their multi-BH solution solved the spinorial equations 
\begin{equation}
  \label{eq:FM6a}
  \nabla_{a}\epsilon_{I} \, =\, -\textstyle{\frac{iH}{2}}\gamma_{a}\ \varepsilon_{IJ}\epsilon^{J}
                             \ +\ H\ A_{a}\epsilon_{I}
                             \ +\ iF^{+}_{ab}\gamma^{b}\varepsilon_{IJ}\epsilon^{J}\; .
\end{equation}
This fermionic rule can be derived from the supersymmetry variations of minimal gauged $\N=2$ $d=4$ Supergravity, which has
an anti-De Sitter type cosmological constant $\Lambda = -3g^{2}$, by Wick-rotating $g\rightarrow iH$. Eq.~(\ref{eq:FM6a}) looks a lot like a Killing spinor equation, however, unlike a proper KSE, this one does not arise from SUSY considerations. Therefore it is common to refer to an equation like this as a fake-Killing spinor equation (fKSE) \mycite{Freedman:2003ax}. 

In this direction, further works were produced by London in \mycite{London:1995ib}, where he generalised the KT solutions to produce higher-dimensional multi-black holes, by showing that his solutions solved too a suitable fKSE. Also, Shiromizu included spinning solutions in a stringy theory in \mycite{Shiromizu:1999xj}. Similarly, the first solutions to five-dimensional E-M-dS theory were given in \mycite{Klemm:2000vn,Klemm:2000gh}, which were based on proposed Ans\"atze. A little later, Behrndt and Cveti\v{c} generalised in \mycite{Behrndt:2003cx} the KT backgrounds to solutions to four- and five-dimensional SUGRA theories coupled to vector multiplets, by noting that the difference between the cosmological solutions of \mycite{Kastor:1992nn} and the usual supersymmetric solutions (see \emph{e.g.}~eq.~(2.6) in \mycite{Tod:1983pm}) is given by the linear $\tau$-dependence in $\Omega$. They thus proposed a substitution rule, in which they called for adding a piece (linear in the time-coordinate) to the harmonic functions which recurrently appear in the study of supersymmetric solutions to SUGRA theories.

Furthermore, they showed that their solutions solved fKSEs that could be obtained from the KSEs of gauged supergravity with vectors, by Wick-rotating the coupling constant. They pointed out that this is equivalent to considering an $\mathbb{R}$-gauged symmetry. This can be seen explicitly \emph{e.g.}~in the construction of gauged $\N=2$ $d=4$ Supergravity coupled to vector multiplets, which calls for the inclusion of a $U(1)$ Fayet-Iliopoulos term. In terms of the KSE of theory,
this FI term is gauged, proportional to the coupling constant (see \emph{e.g.}~\mycite{Andrianopoli:1996cm}). By Wick-rotating the coupling constant, this is equivalent to performing a Wick-rotating on the gauge group, which now becomes $\mathbb{R}$. 

In summary, the work of Kastor and Traschen opened up a whole new window of research, where new backgrounds can be obtained from the study of theories allowing for parallel spinors w.r.t.~a new supercovariant derivative, different from the one prescribed by SUSY. In this sense, Grover \emph{et al.}~pioneered in \mycite{Grover:2008jr} the programme of systematic classification of solutions to theories admitting fKSEs. They used the spinorial geometry techniques of \mycite{Gillard:2004xq} to characterise the timelike solutions of minimal $\N=1$ $d=5$ fakeSupergravity. One of the novel results they obtained is that the geometry contains a base-space which is a hyper-K\"ahler manifold with torsion (HKT), whereas it is of hyper-K\"ahler type in ungauged $\N=1$ $d=5$ SUGRA \mycite{Gauntlett:2002nw}, and K\"ahler in the gauged version \mycite{Gauntlett:2003fk}. Moreover, ungauged and gauged Supergravities permit the embedding of gravitational models with a vanishing or negative cosmological constant, respectively. Since the procedure involved in fSUGRAs generates a change of sign in the potential of the theory, the resulting cosmological constant is positive, and fSUGRA is often referred to as De Sitter Supergravity.

However, a word of caution should be issued here: it has long been known that although actions with local dS Supersymmetry exist, they violate the positive-defineteness of the Hilbert space, so the corresponding supergravity theory contains ghosts \mycite{Pilch:1984aw,Lukierski:1984it}. This of course poses a problem with unitarity, and thus one has to be particularly careful with the interpretation of physical results coming from studies of such a setting. In this sense, the idea of using spinors that are not parallel under the standard covariant derivative prescribed by SUSY was (to the best of our knowledge) first introduced in \mycite{Freedman:2003ax,Celi:2004st}. These fields were later employed to establish a duality between supersymmetric domain-wall solutions of a supergravity theory and supersymmetric cosmologies  \mycite{Skenderis:2007sm}, in the context of the Domain Wall/ Cosmology correspondence (cf.~\mycite{Cvetic:1994ya}{Skenderis:2006jq,Skenderis:2006rr,Skenderis:2006fb}{Bergshoeff:2007cg}). In their construction, Skenderis, Townsend and Van Proeyen require the closure of the SUSY algebra, on top of the rotation of the coupling constant. This they do by introducing a further reality condition on the spinorial and matter fields of the theory, which produces the expected ghost fields. They have indeed a SUGRA action invariant under the De Sitter group. 

The situation for our course of study is different, in that fSUGRA is considered as a solution-generating technique. We only perform a Wick-rotation on the coupling constant (FI term) of the corresponding SUGRA theory, and solve the condition of parallelity of the spinorial field under the (resulting) supercovariant connection. There is no demand for the SUSY algebra to hold; this does away with the ghosts, and hence the solutions to fSUGRA are faithful physical backgrounds. Furthermore, starting from a fakeSupergravity theory and taking the limit of vanishing FI term, one recovers an ordinary Supergravity theory with vanishing coupling constant (Minkowski SUGRA).\vspace{\baselineskip}


We have just seen some instances of the link between Killing spinors and certain classes of geometrical spaces. This suggests that the techniques used in the classification of fSUGRA solutions can be extrapolated to also classify geometries. A case study of this is given in chapter \ref{EWspaces}, where we have considered a non-vanishing spinorial field fulfilling a particular condition of parallelity. The condition has been chosen so that its integrability condition gives rise to the equation defining Einstein-Weyl spaces, thus offering a way to classify these. By way of the bilinear method, we come to characterise all possible EW manifolds which arise in this SUSY-like manner.


\cleardoublepage

\renewcommand{\leftmark}{\MakeUppercase{Chapter \thechapter. Classification of SUGRA solutions}}
\chapter{\texorpdfstring{Classification of supersymmetric SUGRA solutions}{Classification of SUGRA solutions}}
\label{classificationSUGRAs}
In the previous chapter, we have commented on the interest of finding BPS solutions to supergravity theories. A lot of efforts have been devoted to this end throughout the last decades. The classical approach is to take some motivated Ansatz for the bosonic fields, and seek examples that admit the vanishing of the fermionic superpartners, see \emph{e.g.}~\mycite{Gibbons:1981ja}{Kallosh:1992ii, Kallosh:1993yg, Gibbons:1993xt, Ferrara:1995ih, Tod:1995jf, Bergshoeff:1996gg, Behrndt:1997ny,Sabra:1997yd}{LopesCardoso:2000qm}. 
While this approach has given some results, it is of course useful to obtain a more systematic method for finding supersymmetric solutions. In particular, given a supergravity theory, it is natural to ask whether one can obtain all supersymmetric solutions of that theory. This was done in 1983 by Tod for the case of minimal $\N=2$ $d=4$ Supergravity \mycite{Tod:1983pm}. He obtained the most-general background which would saturate the positivity bound for the ADM mass of an asymptotically-flat spacetime that satisfies the dominant-energy condition \mycite{Hawking:1970eq}. Tod's work was a completion of an analysis that Gibbons and Hull had started in \mycite{Gibbons:1982fy}, where they noticed that a solution saturating the bound would imply that a Dirac spinor $\epsilon^i$ (for $i=1,2$) would serve as a supersymmetry transformation. Many years later, Gauntlett and collaborators took the method one step further by classifying all the supersymmetric solutions of minimal five-dimensional SUGRA \mycite{Gauntlett:2002nw}. One then naturally asks why did it take so long, from the original article by Tod, to the feverish activity that started in 2002. The reason is that the original article by Tod employed the Newman-Penrose formalism for General Relativity \mycite{Newman:1961qr}, which is inherently limited to four dimensions. Gauntlett \emph{et al.}~surpassed that limitation by considering a novel techique in the characterisation of solutions. They asummed the existence of a Killing spinor, and constructed bilinears out of them. These bilinears satisfy a number of algebraic and differential equations, and their analysis gives a characterisation of the supersymmetric configurations\footnote{When we refer to a configuration, we mean the prescription for the fields of the theory arising from the existence of the Killing spinor. It will become a solution once the field equations of the theory are also satisfied.}.


Roughly, the process of characterisation is the following: we consider the vanishing of the variations (with respect to a SUSY parameter) of the supersymmetric partners of the bosonic fields of the theory, which gives rules for these fields. We shall generically refer to these rules as the Killing spinor equations (KSEs). They are prescribed in terms of the relevant bosonic supergravity fields, as well as the assumed Killing spinor\footnote{Notice that because supersymmetric partners have spin either $1/2$ or $3/2$, the variation also has to be half-spin-valued, which a combination of integer-valued bosonic fields and a spinor will respect.}. Among them, there is one that can be interpreted as a rule for the parallel propagation of the spinor, in the Mathematical sense. One then uses these rules to obtain the most general form of the bosonic fields, of course compatible with the KSEs and the existence of a preserved spinor. They form a supersymmetric configuration, since they are consequence of assuming the existence of a non-vanishing spinorial field, which is interpreted as the supersymmetric transformation.    


Moreover, the configurations thus obtained need to also solve the dynamical field equations. Since the rules obtained above are linear in derivatives, and the equations of motion (EOMs) are second order, one cannot hope to obtain a recipe for solutions to the EOMs straight away. Instead, one uses the supersymmetric configurations as Ans\"atze from which to obtain supersymmetric solutions, by solving the field equations. In this sense, an observation made by Gauntlett \emph{et al.}~in \mycite{Gauntlett:2002nw} (originating from the work in \mycite{deWit:1986xg}, and further formalised in \mycite{Bellorin:2005hy}) reduces the amount of work necessary to find the conditions that a supersymmetric field configuration needs to fulfill. The key idea is that the existence of Killing spinors (\emph{i.e.}~the solutions preserving some supersymmetry) implies relations between the EOMs.


This was already seen in \mycite{Kallosh:1993wx}, where they called these relations Killing spinor identities (KSIs), although at that time they used them to prove invariance under quantum corrections of certain supersymmetric black holes solutions. In particular, these KSIs relate equations for fields of spin differing by $1/2$. In turn, this implies that one only needs to check explicitly a certain number of the components of the field equations, since others are automatically fulfilled by means of these relations between them. 
The KSI is based on the fact that the invariance of an action under a (super)symmetry implies the following gauge identity
\begin{equation}
\label{eq:KSI3}
\delta\mathcal{S} \; =\; \int\; \delta\Phi^{A}\,\frac{\delta\left( \sqrt{g}\mathcal{S}\right)}{\sqrt{g}\delta\Phi^{A}}\; =\; \int\ \delta\Phi^{A}\ \mathcal{E}_{A} \quad \longrightarrow\quad 0 \; =\; \delta\Phi^{A}\, \mathcal{E}_{A}(\Phi ) \ ,
\end{equation}
where we are using a superset of fields $\Phi^{A}=\{ B^{a},F^{\alpha}\}$), and we have introduced the notation in which the equation of motion for a field $\Phi^{A}$ is written as $\mathcal{E}_{A}(\Phi )=0$. If one then considers the functional derivative of the last equation in (\ref{eq:KSI3}) w.r.t.~some fermion field, and evaluate the resulting identity for purely bosonic configurations that solve the Killing spinor equations, 
{\em i.e.\/} $F^{\alpha}\ =\ \delta_{\epsilon}F^{\alpha}\mid_{F=0}\ =\ 0$, ones sees that
\begin{equation}
\label{eq:KSI8}
0 \; =\; \left. \frac{\delta}{\delta F^{\beta}}\left[ \delta_{\epsilon}B^{a}\right]\right|_{F=0}\; \mathcal{E}_{a} \ .  
\end{equation}
This equation is the \emph{Killing Spinor Identity}, and must hold for any supersymmetric system. Equivalently, the KSIs
can be seen as a subset of the integrability conditions of the KSEs.

We now proceed to describe the bilinear formalism of \mycite{Gauntlett:2002nw}. Spinorial geometry techniques, first introduced in \mycite{Gillard:2004xq}, are discussed in section \ref{spinorialgeometry}. 

\section{Bilinear formalism}
\label{bilinearformalism}
The bilinear formalism is a method for characterising supersymmetric solutions. Thus, it starts with assuming at least one Killing spinor; this is an $\epsilon$ such that
\begin{equation}
\mathfrak{D}_\mu\epsilon=0\ ,
\end{equation}
where $\mathfrak{D}$ is the covariant derivative that results from demanding the vanishing of the gravitino variation (w.r.t.~$\epsilon$) for the theory in question, \emph{i.e.}~$\delta_\epsilon \psi^i_\mu=0$. To be specific, lets focus on the original pure $\N=1$ $d=5$ theory of \mycite{Cremmer:1981xx}, where we shall repeat the calculations presented in \mycite{Gauntlett:2002nw}. The bosonic field content of this theory is given by a F\"unfbein (the graviton) ${e^a}_\mu$ and the graviphoton $A_\mu$; the fermionic content by the gravitinos $\psi^i_\mu$, which are taken to be sympletic-Majorana. There are two of them ($i=1,2$), accordingly with taking $Sp(1)$ as the group defining the condition on the spinors.

The action of the theory is 
\begin{equation}
\label{eq:ciiactionoftheory}
S=\frac{1}{4\pi G}\int \left( -\frac{1}{4} R\, dvol -\frac{1}{2} F \wedge \star F - \frac{2}{3\sqrt{3}} F\wedge F\wedge A\right)\ ,
\end{equation}
and the field equations are
\begin{eqnarray}
\label{eq:ciiEini}
0&=&\displaystyle R_{\mu\nu}+2\,(F_{\mu\rho}{F_\nu}^\rho-\frac{1}{6}g_{\mu\nu}F_{\tau\sigma}F^{\tau\sigma})\ ,\\
\label{eq:ciiMaxwell}
0&=&\displaystyle d \star F+\frac{2}{\sqrt{3}}F\wedge F\ .
\end{eqnarray}
The supersymmetric transformation rules for the bosonic fields are given by \mycite{Cremmer:1981xx}
\begin{eqnarray}
\delta_\epsilon {e^a}_\mu&=&\frac{i}{2}\,\bar{\epsilon}_i\gamma^a\psi^i_\mu\\
\delta_\epsilon A_\mu&=&i\sqrt{3}\,\bar{\epsilon}_i \psi^i_\mu\ .
\end{eqnarray}
and they shall be used in the context of the KSIs. The gravitino KSE is
\begin{equation}
\label{eq:ciiKSE}
\delta_\epsilon \psi^i_\mu= \mathfrak{D}_\mu \epsilon^i=\left(\nabla_\mu+\frac{1}{4\sqrt{3}}(\gamma_{\mu\nu\rho}-4\eta_{\mu\nu}\gamma_\rho)F^{\nu\rho}\right)\epsilon^i=0\ ,
\end{equation}
where $\epsilon^i$ are again sympletic-Majorana spinors, and, as the KSE is linear in them, we can take them to be classical commuting spinors.

Continuing with the formalism, we proceed to construct bilinears out of the two spinors (see also appendix \ref{sec:bilinears}),
\begin{eqnarray}
\f		& = & i\,\bar{\epsilon}_i \epsilon^i\\
V^a 		& = & i\,\bar{\epsilon}_i \gamma^a \epsilon^i\\
\Phi^{r\ ab} 	& = & {{(\sigma^r)}_i}^j\, \bar{\epsilon}_j \gamma^{ab} \epsilon^i\ .
\end{eqnarray}
The key step in the method is relating them to the fields in the theory through the KSE (\ref{eq:ciiKSE}). In order to do this, we now summon the very important Fierz identities, cf. appendix \ref{app-idFierz}. These allow us to obtain several algebraic identities which shall play a major part in the analysis. The five-dimensional identities are given in eqs.~(\ref{fsquared})-(\ref{phiepsilon}), and in particular (\ref{fsquared}) says
\begin{equation}
\label{eq:sexertest}
V^{a}V_{a}  = \f^{2}\, ,\\
 \end{equation}

\noindent This implies that $V^a$ is either timelike, null or vanishing\footnote{The possibility of having a zero $V$ is excluded in regions where the spinor is non-vanishing, and we shall only differentiate among timelike or null case.}. The reasoning needed in these two cases are different, and thus we shall treat them sequentially. Its solutions, just as in Tod's analysis, will hence vary depending on the case. 

We now differentiate the bilinears and use the KSE to obtain the following differential equations
\begin{eqnarray}
\label{eq:ciidf}
d\f & = & -\frac{2}{\sqrt{3}} \iota_V F\\
\label{eq:ciidV}
\nabla_{(\mu} V_{\nu)} & = & 0\\
\label{eq:ciidVt}
dV^\flat & = & \frac{4\f}{\sqrt{3}}F+\frac{2}{\sqrt{3}}\star(F\wedge V^\flat)\\
\label{eq:ciidphi}
\nabla_\mu (\Phi^r)_{\nu\rho} & = & \frac{2}{\sqrt{3}}F^{\sigma\tau}\left(g_{\sigma[\nu}(\star\Phi^r)_{\rho]\mu\tau} -g_{\mu[\sigma}(\star\Phi^r)_{\tau]\nu\rho}-\frac{1}{2}g_{\mu[\nu}(\star\Phi^r)_{\rho]\sigma\tau}\right)\ ,
\end{eqnarray}
where $\iota_V F$ is the interior product of $F$ and $V$ defined as $\iota_V F=\iota_{V^\mu \partial_\mu} (\frac{1}{2}F_{\nu\rho}dx^\nu \wedge dx^\rho)=V^\mu F_{\mu\nu} dx^\nu$, and $V^\flat$ is the one-form obtained from $V$ through the musical isomorphism $\flat:TM\rightarrow T^* M$. In components, this is the action of raising/ lowering indices with the metric. Notice eq.~(\ref{eq:ciidV}) is saying that $V$ is a Killing vector\footnote{This will not be the case in the study of fakeSUGRA solutions.}. Furthermore, taking the exterior derivative of eq.~(\ref{eq:ciidf}), one gets\
\begin{equation*}
0=d(\iota_V F)\ ,
\end{equation*}
which, by the Bianchi identity of the field strength and the definition of the Lie derivative, implies that 
\be
\mathcal{L}_VF=0\ .
\ee
This result, plus eq.~(\ref{eq:ciidV}), means that $V$ is a symmetry of the putative solution $(g,F)$. Also, by totally antisymmetrising eq.~(\ref{eq:ciidphi}) over the free indices we see that
\begin{equation}
\label{eq:ciiclosedtwoforms}
d\Phi^r=0\ ,
\end{equation}
hence the $\Phi^r$ are closed 2-forms.

Before splitting the analysis in the two possible cases, let us consider a subset of the integrability condition on the supersymmetry transformation of the gravitino. This will give an equation which Kallosh and Ort\'in first called the Killing spinor identity. It is   
\begin{eqnarray}
0&=&4 \gamma^\nu\nabla_{[\mu}\delta_\epsilon \psi^i_{\nu]}\nonumber \\
\label{eq:intcond}
 &=&\left\{ ({\mathcal{E}_\mu}^\sigma-\frac{1}{3}{g_\mu}^\sigma {\mathcal{E}_\rho}^\rho)\gamma_\sigma+\frac{\gamma_\mu}{\sqrt{3}}(\mathcal{M}_\sigma \gamma^\sigma+\frac{2}{3}\mathcal{B}_\sigma \gamma^\sigma)-\sqrt{3}(\mathcal{M}_\mu+\frac{2}{3}\mathcal{B}_\mu) \right\} \epsilon^i\ ,
\end{eqnarray}
where $\mathcal{E}_{\mu\nu}$ is, when equal to zero, the Einstein equation, $\mathcal{M}_\mu$ the Maxwell equation and $\mathcal{B}_{\mu\nu\sigma}$ the Bianchi identity, such that
\begin{equation}
{\mathcal{E}_\mu}^\nu={e_\mu}^a {\mathcal{E}_a}^\nu\ ,\quad 
{\mathcal{E}_a}^\nu=\displaystyle-\frac{1}{2\sqrt{g}}\frac{\delta S}{\delta {e^a}_\nu}\ ,\quad
\mathcal{M}^{\mu}=\displaystyle \frac{1}{\sqrt{g}}\frac{\delta S}{\delta A_\mu}\ ,\quad
\displaystyle \mathcal{B}_{\mu\nu\sigma}=(dF)_{\mu\nu\sigma}\ .
\end{equation}
We can get an identity for ${\mathcal{E}_\rho}^\rho$ by acting on (\ref{eq:intcond}) with $\gamma^\mu$ (from the left); plugging it back into (\ref{eq:intcond}) one gets the KSI relating Einstein, Maxwell and Bianchi equations
\begin{equation}
\label{eq:KSI-EM}
0=\left\{ \left({\mathcal{E}_\mu}^\sigma-\sqrt{3}\,{(\star\mathcal{B})_\mu}^\sigma\right)\gamma_\sigma-\frac{\sqrt{3}}{4}\mathcal{M}_\mu\right\}\epsilon^i\ .
\end{equation}
Acting on it with $i\bar{\epsilon}_i \gamma_\nu$, one gets again the structure of the bilinears, which upon recalling the Fierz identities gives
\begin{equation}
\label{eq:KSI-EMc1}
0=\f \left(\mathcal{E}_{\mu\nu}-\sqrt{3}\, (\star\mathcal{B})_{\mu\nu} \right)-\frac{\sqrt{3}}{4}\mathcal{M}_\mu V_\nu\ .
\end{equation}
Acting instead with $i\bar{\epsilon}_i$, one arrives at
\begin{equation}
\label{eq:KSI-EMc2}
0=\left(\mathcal{E}_{\mu\nu}-\sqrt{3}\, (\star\mathcal{B})_{\mu\nu} \right)V^\nu-\frac{\sqrt{3}}{4} \f\mathcal{M}_\mu\ ,
\end{equation}
which in the timelike $\f\ne 0$ case is equivalent to eq.~(\ref{eq:KSI-EMc1}), by means of the Fierz identity (\ref{fsquared}).\\
One can also act on eq.~(\ref{eq:KSI-EM}) with $\bar{\epsilon}_j{(\sigma^r)_i}^j\gamma^\nu$ \mycite{Bellorin:2007yp}, which gives
\begin{equation}
\label{eq:KSI-EMc3}
0=\left(\mathcal{E}_{\mu\sigma}-\sqrt{3}\, (\star\mathcal{B})_{\mu\sigma} \right) \Phi^{r\,\nu\sigma}\ .
\end{equation}

In the timelike case we can separate eq.~(\ref{eq:KSI-EMc1}) in symmetric and antisymmetric parts, which read
\begin{eqnarray}
\label{eq:KSIs11}
\mathcal{E}_{\mu\nu}&=&\displaystyle\frac{\sqrt{3}}{4\f}\,\mathcal{M}_{(\mu} V_{\nu)}\\
\label{eq:KSIs12}
(\star B)_{\mu\nu}&=&\displaystyle-\frac{1}{4\f}\,\mathcal{M}_{[\mu}V_{\nu]}\ .
\end{eqnarray}
Notice that this says that if the Maxwell equantion is satisfied, \emph{i.e.}~$\mathcal{M}_\mu=0\quad \forall\, \mu$, then both the Einstein equation and the Bianchi identity are also automatically satisfied. These equations will be used in section \ref{ciitimelikecase}, when we analyse the equations of motion of the Ansatz for the timelike supersymmetric configurations.\vspace{\baselineskip}

\noindent In the null $(\f=0)$ case, the tensorial equation (\ref{eq:KSI-EMc1}) can be expressed as
\begin{eqnarray}
\label{eq:KSIs21}
\mathcal{M}_\mu V_\nu &=& 0\ ,
\end{eqnarray}
which implies that the Maxwell equation is identically satisfied. 

To proceed, we shall distinguish between the two possible cases.

\subsection{Null case}
\label{ciinullcase}
In this case $V^2=\f^2=0$ and hence $\f=0$. Eq.~(\ref{eq:ciidVt}) expresses that $V^\flat\wedge dV^\flat=0$, \emph{i.e.}~$V^\flat$ is hypersurface-orthogonal. The Frobenius theorem of differential geometry then implies that $V^\flat$ can be written as $V^\flat=fdu$, where $f$ and $u$ are functions. Furthermore, since
\begin{equation}
\iota_V dV^\flat=0\ ,
\end{equation}
$V$ is tangent to surfaces of constant $u$; this allows us to introduce coordinates $(u,v,y^m)\ (m=1,2,3)$ such that the surfaces tangent to $V$ are parametrised by $v$. Hence $V=\partial_v$. The metric then has the form
\begin{equation}
\label{eq:metricadeKundtnull}
ds^2=2fdu(dv+Hdu+w)-f^{-2}h_{mn}dx^m dx^n\ ,
\end{equation}
where the function $f$, the function $H$ and the three-dimensional metric $h_{mn}$ do not depend on $v$, because by eq.~(\ref{eq:ciidV}) $V$ is Killing. This is a Kundt metric, as given in eq.~(\ref{eq:KundtMetric}). Using identities (\ref{VPhi}) and (\ref{VStarPhi}) one easily arrives at
\begin{equation}
\Phi^r=\Phi^r_{um}\,du \wedge dx^m\ .
\end{equation}
Since the $\Phi^r$ forms were previously determined to be closed, cf. eq.~(\ref{eq:ciiclosedtwoforms}), this implies that $(d\Phi^r)_{uvn}=0$, whence $\Phi^r$ is $v$-independent, and also that $(d\Phi^r)_{umn}=0$, which means that by Poincar\'e's lemma we can locally write $\Phi^r$ as an exact form, \emph{i.e.}
\begin{equation}
\Phi^r_{um}=\partial_{m}y^r\ ,
\end{equation}
for some function $y^r=y^r(u,x^m)$, as $\Phi^r$ should not depend on $v$. Precisely because of the coordinate dependence of the new functions $y^r$, one can perform a coordinate transformation that respects the metric (\ref{eq:metricadeKundtnull}), and rewrite $\Phi^r$ as\footnote{Observe that the coordinate transformation sends the $x^i$ (in the metric and in $\Phi^r$) to $y^i$, which we relabel again to $x^i$ for convenience.} 
\begin{equation}
\Phi^r=du \wedge dx^r\ .
\end{equation}
Moreover, we can use the Fierz (\ref{quaternions}) to find
\begin{equation}
h^{mn}{\Phi^r}_{um}{\Phi^s}_{un}=h^{rs}=\delta^{rs}\ ,
\end{equation}
so the base-space metric $h_{mn}$ is Euclidean, \emph{i.e.}~flat.

We now proceed to obtain the form of $F$. To do so, it is easier to work with an orthogonal frame where the full metric is flat; a convenient one is given by
\begin{equation}
\begin{array}{l@{\qquad}l}
    e^{+}  = fdu\ , & \theta_{+}  = f^{-1}(\partial_{u} \ -\ H \partial_{v}) \ , \\
    e^{-}  = dv + Hdu + w \ ,& \theta_{-} = \partial_{v} \ ,\\
    e^{i}  = f^{-1}{e^{i}}_{m}\ dx^{m}\ ,& \theta_{i} = f{e^{m}}_{i} \left( \partial_{m} \ -\ w_{m}\partial_{v}\right) \ ,
\end{array}
\end{equation}
where ${e^i}_m=\delta^i_m$, ${e^m}_i=\delta^m_i$ because of the flatness of the base-space; the metric is consequently given by 
\begin{equation}
ds^2=e^+ \otimes e^- + e^- \otimes e^+ - e^i \otimes e^i\ ,
\end{equation}
where $i=1,2,3$. An inmediate thing to see is that due to eq.~(\ref{eq:ciidf}), $\iota_V F=0$, and just as in the timelike case we use it to symplify the field strength to
\begin{equation}
F=F_{+i}\:e^+ \wedge e^i+\frac{1}{2}F_{ij}\:e^i \wedge e^j\ .
\end{equation} 
Likewise, using eqs.~(\ref{eq:ciidV}) and (\ref{eq:ciidphi}) one obtains the full form, which reads 
\begin{equation}
\label{eq:formadeFnula}
F=\frac{2}{\sqrt{3}}f^2 du \wedge \star_{(3)}\tilde{d}w+ 2\sqrt{3}f^{-2} \star_{(3)} \tilde{d}f \ ,
\end{equation}
where $\tilde{d}$ is the derivate w.r.t.~to the base-space coordinates $x^i$, \emph{i.e.}~$\tilde{d}=dx^i\,\theta_i$. The Fierz identity eq.~(\ref{vepsilon}) in the tangent-space base reads
\begin{equation}
\label{eq:ciiotracondiciondequiralidad}
\gamma^+ \epsilon=0\ .
\end{equation}
It follows from the analysis of the KSE that this implies having a constant spinor. Since eq.~(\ref{eq:ciiotracondiciondequiralidad}) is the only restriction applied, and $\gamma^+$ has rank 2, this entails that at least half of the supersymmetry is preserved.

We have so far obtained supersymmetric configurations to the theory in question, but we would now like to obtain solutions. For this we need to impose the Einstein field equation, as well as Maxwell's equation and the Bianchi identity on the gauge field strength. The analysis of these will also give us the final pieces in the characterisation, the descriptions of $f$ and $H$ above. Recall that, because of the KSIs, Maxwell's equation (\ref{eq:ciiMaxwell}) is identically satisfied. The Bianchi identity, $dF=0$, gives the following constraints
\begin{eqnarray}
\label{ciiharmonicity}
0 & = & \nabla^2 f^{-1}\\
\label{ciiBianchiid}
0 & = & \tilde{d} (f^2 \star_{(3)} \tilde{d}w)+ 3 \star_{(3)} \partial_u \star_{(3)} \tilde{d} (f^{-1})\ .
\end{eqnarray}
Eq.~(\ref{ciiharmonicity}) implies that $f^{-1}=f^{-1}(u,x^i)$ is a harmonic function, and eq.~(\ref{ciiBianchiid}) can be recast as
\begin{equation}
-4\epsilon_{ijk}\nabla_j w_k=f^{-2} \nabla_i \phi\ ,
\end{equation}
for an undetermined function $\phi=\phi(u,x^i)$. Its integrability condition implies that it is generically solved in terms of another harmonic function $L=L(u,x^i)$.

We now analyse the gravity field equation. In view of eq.~(\ref{eq:KSI-EMc3}), it is easily seen that the components
\begin{equation}
\mathcal{E}_{+-}\ ,\quad \mathcal{E}_{+i}\ , \quad\mathcal{E}_{--}\ ,\quad \mathcal{E}_{-i}\ ,\quad \mathcal{E}_{ij}
\end{equation}
all vanish, thus only $\mathcal{E}_{++}$ needs to be explicitly solved. The explicit form of this equation of 
motion is not too enlightening and we shall content ourselves with the knowledge that it can be solved
by a function $H_{0}=H_{0}(u,x^{i})$ that is harmonic w.r.t.~the $x^i$ coordinates.

Observe that we still have remaining gauge freedoms, \emph{e.g.}~shifts to the coordinate $v\rightarrow v+g(u,x^i)$, which can lead to simplification of the results. We will, however, refrain from doing so, since the purpose of this section is solely to introduce the method. Interested readers can of course refer to \mycite{Gauntlett:2002nw} for details, as well as the solution to the problem by means of spinorial geometry techniques, in section \ref{sec:nullcasespinorialgeometry}. The end result is that the general supersymmetric null solution to $d=5$ SUGRA is a geometry given by the metric (\ref{eq:metricadeKundtnull}) and the field strength (\ref{eq:formadeFnula}). The functions $f$ and $H$ are harmonic, and fulfill the eq.~$\mathcal{E}_{++}=0$, which is solved by expressing $H$ in terms of a harmonic function $H_0$. The 1-form $w$ is given by eq.~(\ref{ciiBianchiid}), also solved by means of another harmonic function, in this case $L$. Moreover, the three-dimensional base-space is flat.

\subsection{Timelike case}
\label{ciitimelikecase}
In the timelike case, since $V^2$ is larger than zero, eq.~(\ref{eq:sexertest}) implies that $\f^2>0$. The two possible cases are positive and negative $\f$. We focus solely on the positive one, for illustrative purposes; the other case is obtained analogously, and grants the same solution modulo some minus signs and replaces the self-duality conditions by antiself-duality, and vice versa.

Because $V$ is timelike, we introduce a time coordinate adapted to its flow, \emph{i.e.}~$V=\partial_t$. The metric can then be decomposed in \emph{conforma-stationary} form
\begin{equation}
\label{ciiconforma}
ds^2=f^2(dt+w)^2-f^{-1}h_{mn}dx^m dx^n\ ,
\end{equation}
where $h_{mn}$ is the metric on the four-dimensional base-space $M_{(4)}$, which is four-dimensional ($m,n=1,...,4$), and $w=w_m\,dx^m$ is a 1-form. Because $V$ is a Killing vector field, the function $f$, the metric $h_{mn}$ and the 1-form $w$ are time-independent.
Notice that the presence of the $f^{-1}$ factor multiplying the base-space metric guarantees that the Laplacian operator acting on time-independent fields can be expressed solely in terms of $h_{mn}$. 

We choose the following Vielbein 
\begin{equation}
\begin{array}{ll}
    e^{0} = f(dt+w)\ , &\qquad \theta_{0} = f^{-1}\partial_t\ , \\
    e^{i} = f^{-1/2}{e^i}_m\ dx^{m}\ ,&\qquad\theta_{i} = -f^{1/2} w_i \partial_t + f^{1/2} {e^m}_i \partial_m\ ,
\end{array}
\end{equation}
where $e^a$ and $\theta_a$ are canonically-dual, \emph{i.e.}~$e^a(\theta_b)=\delta^a_b$, and we have defined ${e^i}_m {e^i}_n\equiv h_{mn}$, $w_i\equiv w_m {e^m}_i$ and ${e^i}_m {e^m}_j\equiv\delta^i_j$, for $i,j=1,\ldots,4$. The metric is then given by 
\begin{equation}
ds^2 = e^0 \otimes e^0 - e^i \otimes e^i\ .
\end{equation}

The Fierz identity (\ref{VPhi}) shows that the two-forms $\Phi^r$ live in the base-space, and by (\ref{VStarPhi}) they are also antiself-dual. (\ref{quaternions}) can then be written as
\begin{equation}
\label{ciiquat}
{(\Phi^r)_m}^n{(\Phi^s)_n}^p=-\delta^{rs}\delta_m^p+\varepsilon^{rst} {(\Phi^t)_m}^p\ .
\end{equation}
This is the algebra of imaginary unit quaternions (see appendix \ref{appsec:complexstructures}). By eq.~(\ref{eq:ciidphi}) we can see that $\Phi^r$ is covariantly constant w.r.t.~the Levi-Civit\`a connection on the base-space, induced from the five-dimensional one. Since $\Phi^r$ is also compatible with the metric (they are 2-forms and thus antisymmetric in their indices), we can say that the base-space has an integrable quaternionic structure with three closed K\"ahler forms (given by $(\Phi^r)_{mn}$), and whence $(M_{(4)},h_{mn})$ is a hyper-K\"ahler manifold.

Since, as commented above, $\omega$ is time-independent, its exterior derivative $d\omega$ is a 2-form that lives on the base-space $M_{(4)}$. We can therefore decompose
\begin{equation}
\label{eq:formasdeGdualestimelikebilinears}
fdw= G^+ + G^-\ ,
\end{equation}
where $G^\pm$ are (anti)self-dual 2-forms on the basespace, \emph{i.e.}~$\star_{(4)}G^{\pm}=\pm G^{\pm}$. We can use this decomposition to study the field strength. This can be generically be written as
\begin{equation}
F=F_{0i}\,e^0 \wedge e^i+\frac{1}{2}F_{ij}\,e^i \wedge e^j\ .
\end{equation}
Then after some calculations with eqs.~(\ref{eq:ciidf}) and (\ref{eq:ciidVt}) one obtains (respectively) the form of $F_{0i}$ and $F_{ij}$, such that 
\begin{equation}
\label{eq:ciiformadeF}
F=\frac{1}{2\sqrt{3}}\left( -3f^{-2} V^\flat \wedge df + G^{+} + 3G^{-}\right)\ .
\end{equation}

We now show that the necessary conditions obtained so far are also sufficient to satisfy the KSE for a non-vanishing $\epsilon^i$, and hence the configuration has some unbroken SUSY. Eq.~(\ref{vepsilon}) in a Vielbein basis implies that \begin{equation}
\label{eq:ciicondiciondequiralidad}
\gamma^0\epsilon^i=\epsilon^i\ .
\end{equation}
Notice this equation also says that $\epsilon^i$ is chiral with respect to the associated spin structure defined on the base-space, \emph{i.e.} 
\begin{equation}
\epsilon^i=\gamma^{01234}\epsilon^i=\gamma^{1234}\gamma^0\epsilon^i=\gamma^{1234}\epsilon^i\ ,
\end{equation}
having used eq.~(\ref{eq:identidaddegammas}), and chosen the orientation $\epsilon^{01234}=1$. Invoking again (\ref{eq:identidaddegammas}), one obtains that the antisymmetric product of two gamma matrices acting on the spinor is antiself-dual w.r.t.~the base-space metric, \emph{i.e.} 
\begin{equation}
(\star_{(4)}\gamma)^{jk}\epsilon^i=-\gamma^{jk}\epsilon^i\ ;
\end{equation}
whence the product
\begin{equation}
G^+_{ij} \gamma^{ij}\epsilon=0
\end{equation}
identically. This is useful for analysing the time-component of the KSE (\ref{eq:ciiKSE}), which results in the statement that the spinor is time-independent, \emph{i.e.}~$\epsilon=\epsilon(x)$. The spatial components can be generically solved by taking an epsilon of the form
\begin{equation}
\epsilon(t,x^m)=\epsilon(x^m)=f^{1/2}\,\eta(x^m)\ ,
\end{equation}
where $\eta$ is such that $\nabla_m \eta=0$. In other words, $(M_{(4)},h_{mn})$ admits a parallel spinor. Furthermore, as the base-space is hyper-K\"ahler, we are assured that such a spinor exists \mycite{Hitchin:1974}. 
The projector $\frac{1}{2}(\gamma^0-\mathbbm{1})$ has rank 2, and consequently the configurations presented preserve at least $1/2$ SUSY, as in the null case.

To finish the analysis, we proceed to impose the equations of motion: demanding the fulfillment of the Maxwell equation implies the following equation 
\begin{equation}
\label{eq:ciicondiciondecerradura}
\nabla^m \nabla_m f^{-1}=\frac{2}{9}(G^+)_{mn} {(G^+)}^{mn} \ .
\end{equation}
The Bianchi identity $dF=0$, in turn, says that $G^+$ is closed, \emph{i.e.} 
\begin{equation}
\label{eq:ciicondiciondecerradurados}
dG^+=0\ .
\end{equation}

The remaining EOMs do not impose any further conditions, as by the KSIs all of the components of the Einstein field equation are automatically satisfied. To summarise, we have that supersymmetric timelike solutions to minimal $d=5$ SUGRA are given by the metric (\ref{ciiconforma}), where the base-space is time-independent and described by a hyper-K\"ahler geometry $(M_{(4)},h_{mn})$ and antiself-dual K\"ahler 2-forms. Also, there is a globally-defined time-independent function $f$, and a time-independent 1-form $w$ locally-defined on $M_{(4)}$, such that $fdw=G^+ + G^-$. Moreover, the field strength is given by
\begin{equation}
F=\frac{1}{2\sqrt{3}}\left( -3 (dt+w) \wedge df + 3fdw - 2G^{+}\right)\ ,
\end{equation}
and eqs.~(\ref{eq:ciicondiciondecerradura}) and (\ref{eq:ciicondiciondecerradurados}) ought to be satisfied.

\subsection{G-structures}
\label{Gstructures}
So far in this chapter we have described the bilinear method for characterising supersymmetric solutions to SUGRA theories. One of its foremost ingredients is the construction of forms out of the supersymmetric parameters (spinors) of the theory. By means of the KSE, these forms are analysed and prescribe the resulting geometry. While we did not comment on it above, historically the construction of these forms was strongly suggested by the mathematical concept of $G$-structures, and in fact one can reinterpret the method from their point of view. We proceed to give a brief description of these structures, and their relevance to our study.

Pure gravity supersymmetric solutions (where the only turned-on field is the vielbein) are given by Ricci-flat metrics with special holonomy. By this we mean metrics on a manifold $M^d$ whose torsion-free, metric connection has holonomy group smaller than the most-general $Gl(d)$ group\footnote{Having chosen a metric-compatible connection, $SO(d)$ is in fact the largest possible holonomy group allowed.}, and such that it paralellises a spinor, \emph{i.e.}
\begin{equation}
\nabla \epsilon=0\ .
\end{equation}
The classification of such possible metrics, in Euclidean signature, was famously given by Berger in 1955 (see appendix \ref{app:simhol} for some more detail). Alas, here we are interested in a Lorentzian signature, which is much less studied. Furthermore, we are dealing with theories which also have matter, and hence the spinor will in general no longer be parallel w.r.t.~the Levi-Civit\`a connection. It is in this sense that $G$-structures come to our aid, as they provide us with a framework with which to generalise the concept of special holomomy.

We define $G$-structures in terms of the frame bundle $F(M)$ on our manifold $M^d$ \mycite{Joyce:2000,Gauntlett:2002sc}. Generically, any element of $GL(d)$ will take one choice of frame into another. However, as soon as there is any additional structure defined on $M$, the group $G$ preserving such structure will be strictly smaller than GL($d$). This smaller group will serve as the structure group for a frame sub-bundle, and it defines a $G$-structure. This is similar to $SO(1,d)$ being the largest possible holonomy group on a manifold with a connection that respects lenghts. Thus, the existence of a $G$-structure implies that the possible change of frames of $F(M)$ is given by $G$.

There is an equivalent definition of $G$-structures using tensors (see \emph{e.g.}~\mycite{Gauntlett:2002sc}). In particular, any tensor can be decomposed into representations of the group $G$. We want to consider those tensors that are non-vanishing, globally-defined and invariant under $G$. Precisely because of the global aspect, if there is such an invariant tensor, it means that the structure group of $F(M)$ is no longer $GL(d)$, but rather $G$ (or a subgroup of it). So we can establish a correspondence between the existence of a $G$-invariant tensor, which is something that can easily be calculated, and having a $G$-structure.

Additionally, there is a relation between $G$-invariant tensors and torsionful connections \mycite{Salamon:1989}, which is what is needed for our desired generalisation. We consider the standard covariant derivative $\nabla$ of a $G$-invariant form $\Omega$; it can be decomposed into irreducible $G$-modules $\mathcal{W}_i$
\begin{equation}
\label{eq:ciiGmodules}
\nabla\Omega \rightarrow  \oplus\,\mathcal{W}_i \simeq \Lambda^1 \otimes g^\bot\ ,
\end{equation}
where $g^\bot$ are the elements in $\mathfrak{so}(1,d)$ which are not in the Lie algebra $\mathfrak{g}$ of $G$. The right hand side is so because $\Omega$ is $G$-invariant, and hence $\mathfrak{g} \circ \Omega=0$. Equation (\ref{eq:ciiGmodules}) can be explicitly expanded as
\begin{equation}
\nabla_m \Omega_{n_1\ldots n_r}=-{K_{mn_1}}^p\: \Omega_{pn_2\ldots n_r}-{K_{mn_2}}^p\: \Omega_{n_1 p\ldots n_r}-\ldots -{K_{mn_r}}^p\: \Omega_{n_1\ldots n_{r-1}p}\ ,
\end{equation}
where the ${K_{mn}}^p \in \oplus\, \mathcal{W}_i$ label the modules. The decomposition allows us to define a new covariant derivative $\nabla'\equiv\nabla+K$ with torsion $K$. Since $\nabla' \Omega=0$ by construction, the holonomy group of this new connection is inside of $G$. Note that this serves as the desired generalisation, since when all the modules $\mathcal{W}_i$ vanish, it is equivalent to $\nabla\Omega=0$, \emph{i.e.}~having special holonomy inside of $G$. Having the modules turned on implies deviating away from special holonomy, which is what we want for classifying supersymmetric solutions to SUGRA theories with matter.

In this sense, the bilinears made out of spinors which we had in the previous two sections are $G$-invariant tensors, and they encode the existence of a $G$-structure, where the ${K_{mn}}^p$ are given by the matter content of the theories studied. This $G$-structure is defined in terms of the supersymmetry parameters $\epsilon^i$, which are solutions to the KSE of (\ref{eq:ciiKSE}). The bilinears constructed in the timelike case determine an $SU(2)$-structure, while in the null case its an $\mathbb{R}^3$-structure.



\section{Spinorial geometry techniques}
\label{spinorialgeometry}
In this section we introduce the \emph{Spinorial Geometry} technique for classifying supersymmetric solutions to supergravity theories. In \mycite{Gillard:2004xq} Gillard \emph{et al.}~characterised the solutions of $d=11$ SUGRA for $\N=1,2,3,4$ SUSYs, by recasting the supersymmetric parameters (spinors) in the language of differential forms, and using the gauge symmetry of the theory $Spin(1,10)$ to greatly simplify the calculations. This approach stands on the shoulders of G-structures, and proves very adequate to treat problems involving a high number of dimensions and several preserved supersymmetries, where the bilinears method presented in the previous section becomes increasingly unmanageable. In this sense, this method has been used to extensively classify solutions in many scenarios (see \emph{e.g.}~\mycite{Gran:2005wn}{Gran:2005wu,Gran:2005ct,Gran:2005wf,Gran:2006dq,Gran:2007fu,Cacciatori:2007vn,Gran:2008vx,Elvin:2009xx}{Akyol:2012cq}). The study of chapter \ref{5dgauged} has been produced considering these techniques\footnote{The results can also be obtained by employing the bilinear method.}, which we now review.

The starting idea behind the formalism is to consider the spinors as forms \mycite{Wang:1989xx}{Harvey:1990xx}{Lawson:1998yr}. This gives us the means to introduce an explicit basis of spinors, which (and this is where the effectiveness of the method lies) due to the gauge symmetry of the theory, we can reduce to a simple expression/s. These canonical spinors are then introduced into the Killing spinor equation for an explicit evaluation, which translates into a linear system of algebraic and differential equations. These equations no longer have Gamma matrices in them, and can be solved to obtain the form of the resulting geometry.

From an abstract point of view, the spinorial fields for different theories come in different irreducible representations of the Spin group, depending on the particular algebraic structure associated to the signature of the theory in question, which allows or not for certain restrictions on the spinors. A summary of this is given \emph{e.g.}~in \mycite{FigueroaO'Farrill:xxxx} or \mycite{Ortin:2004ms}{appendix B}. These spinorial representations each have a number of $\mathbb{R}$ independent components, with the minimal number corresponding to the smaller representation in a given dimension $d$.

Generically, the space of Dirac spinors has $2^{\lfloor d/2 \rfloor}$ (complex) components, and one can recast them in terms of the complexified space of forms on $\mathbb{R}^{\lfloor d/2\rfloor}$, which is denoted mathematically by $\Delta=\text{Span}_{\mathbb{C}}(1,e_1,\ldots,e_{\lfloor d/2 \rfloor})=\Lambda^* (\mathbb{R}^{\lfloor d/2 \rfloor} \otimes \mathbb{C})$. The canonical basis for $\Delta$ is then given by
\begin{equation}
\xi_i=\{1,e_{i_1},e_{i_1 i_2},e_{i_1 i_2 i_3},\ldots,e_{i_1\ldots i_{\lfloor d/2 \rfloor}}\}\ ,
\end{equation}
where each index $i$ takes values in $\{1,\ldots, \lfloor d/2 \rfloor\}$ and $e_{i_1 i_2\ldots i_p}\equiv e_{i_1} \wedge e_{i_2} \wedge \ldots e_{i_p}$. The dimensionality of the space of forms is thus given by the sum of the number of $p$-forms, where $p=0,\ldots,\lfloor d/2 \rfloor$. This can be shown to equal
\begin{equation}
\sum_{d=0}^{\lfloor d/2 \rfloor} \left(
\begin{array}{c}
\lfloor d/2 \rfloor\\ d
\end{array}
\right)=2^{\lfloor d/2 \rfloor}\ ,
\end{equation}
thus agreeing with the (complex) dimensionality of the space of Dirac spinors. A general Dirac spinor can then be written as
\begin{equation}
\label{eq:ciiespinorgenericoenterminodeformas}
\epsilon=\mu^{(0)}\,1+\mu^{(1)}_{i_1}\, e_{i_1}+\mu^{(2)}_{i_1 i_2}\, e_{i_1 i_2}+\ldots+\mu^{(\lfloor d/2 \rfloor)}_{i_1\ldots i_{\lfloor d/2 \rfloor}}\, e_{i_1\ldots i_{\lfloor d/2 \rfloor}}\ ,
\end{equation}
where all the $\mu^{(n)}$s are complex-valued functions, antisymmetric in all their $n$-indices.

The given isomorphism also needs to carry a recipe for the action of the Lorentz algebra on the spinors. In this sense, the representation chosen for the action of the Gamma matrices on $\Delta$ will depend on the theory in question, since these need to respect the Clifford algebra defining-relation 
\begin{equation}
\{\gamma_a,\gamma_b\}=2\eta_{ab}\,\mathbbm{1}\ ,
\end{equation}
where $\eta$ in our case is diagonal mostly-minus. For odd-dimensional $M^{(1,d)}$ theories\footnote{Interested readers can consult \mycite{Gran:2005wn} for an explicit representation suitable for type IIB theory. In $(1,d^\text{odd})$, one can essentially use the equations in (\ref{eq:formadelaaccion}) to represent $d-1$ of the gamma matrices, and cook-up one more for the $d^\text{th}$ one.}, one can \emph{e.g.}~choose 
\begin{equation}
\label{eq:formadelaaccion}
\gamma_i\,\epsilon= i (e_i\wedge \epsilon+ \iota_{e_i} \epsilon)\ ,\qquad \gamma_{i+(d/2)}\,\epsilon=-e_i\wedge \epsilon + \iota_{e_i}\epsilon\ ,
\end{equation}
where $i=1,\ldots, (d/2)$ and $\epsilon$ is a Killing spinor that can be expressed in term of forms, as in eq.~(\ref{eq:ciiespinorgenericoenterminodeformas}). Observe that the action of the $\gamma_i$\,s sends the space of even forms to the space of odd form, and vice versa
. The additional gamma matrix is given by 
\begin{equation}
\gamma_0=k\,\gamma_1\ldots\gamma_{d}\ ,
\end{equation}
where $k=k(\eta,d)$ is a constant that equals either $i$ or $1$, depending on the metric's signature and the dimensionality of the theory, so that $\gamma_0$ squares to $\mathbbm{1}$, in accordance with our choice of metric.

To study smaller representations of the Spin group, we need a definition for the inner products on the space of forms. We do so by using the charge conjugation matrix
\begin{equation}
\mathcal{C}=\gamma_{(d/2)+1}\ldots\gamma_{d}\ ,
\end{equation}
and an extension of the Euclidean inner product on $\mathbb{R}^{d/2}$ to $\Delta$
\begin{equation}
\langle z^i\,\xi_i,w^j\,\xi_j \rangle=-\sum_{i=1}^{2^{\lfloor d/2\rfloor}}(z^i)^* w^i\ ,
\end{equation}
where $(z^i)^*$ is the complex conjugate of $z^i$, and only identical forms (up to reordering of indices) have non-vanishing product. This inner product is not invariant under the Spin group
, but we can define the \emph{Dirac} inner product on $\Delta$ as
\begin{equation}
D(\eta_1,\eta_2)=\langle \gamma_0\,\eta_1, \eta_2\rangle= \bar{\eta}_1\,\eta_2\ ,
\end{equation}
which is so by construction. 
Notice that $(\bar{\eta}_1)_\alpha=(\eta_1^\dagger\gamma_0)_\alpha={({\eta_1}^\beta)}^*(\gamma_0)_{\beta\alpha}$ is the Dirac conjugate of $\eta_1^{\,\alpha}$. We also define the Majorana inner product as\footnote{Note that on even-dimensional theories there are actually two Spin-invariant Majorana inner products: one defined as  $\tilde{B}(\eta_1,\eta_2)=\langle \gamma_{0}\gamma_{(d/2)+1}\ldots\gamma_{(d-1)}\,\eta_1^*,\eta_2\rangle$, and one defined as $\tilde{A}(\eta_1,\eta_2)=\langle \gamma_{1}\ldots\gamma_{(d/2)}\,\eta_1^*,\eta_2\rangle$ \mycite{Gran:2005wn}.
}
\begin{equation}
\label{eq:ciiproductointeriorMajorana}
B(\eta_1,\eta_2)=\langle \mathcal{C}\,\eta^*_1,\eta_2 \rangle= \eta_1^{\,c}\,\eta_2\ ,
\end{equation}
where $B(-,-)$ is antisymmetric by virtue of the form of the gamma matrices (\ref{eq:formadelaaccion}). The Majorana conjugate of $\eta_1^{\,\alpha}$ is given by ${\eta_1^{\,c}}_\alpha=-(\eta_1^T \mathcal{C})_\alpha=-\eta_1^\beta \mathcal{C}_{\beta\alpha}$. The Majorana condition on $\eta$ is then given by demanding that the Dirac and Majorana conjugation are equal, \emph{i.e.}
\begin{equation}
\label{eq:realityconditiononspinors}
\eta^{*}=\gamma_0\,\mathcal{C}\,\eta\ .
\end{equation}
Spinors satisfying this condition are called Majorana spinors, and are used \emph{e.g.}~in the study of M-theory backgrounds \mycite{Gillard:2004xq,Gran:2005wu}.

For the purpose of establishing a connection between the geometrical structures of the tangent and curved spaces, note that one can also construct $k$-forms from the spinors (where $I,J$ go from $1$ to $N$) in the following way \mycite{Gillard:2004xq}
\begin{equation}
\label{eq:formasdeespinores}
\alpha^{IJ}\equiv\alpha(\eta^I,\eta^J)=\frac{1}{k!}\,B(\eta^I,\gamma_{a_1\,\ldots\,a_k}\eta^J)\,e^{a_1}\wedge \ldots \wedge e^{a_k}\ .
\end{equation}
Observe that by using Hodge duality one can find the higher order forms $\Omega^{(d/2)+1},\ldots,\Omega^{d}$ to be dual or antiself-dual to the lower order ones $\Omega^{1},\ldots,\Omega^{(d/2)}$. Also due to the symmetry property of $B\scriptstyle(-,-)$, $\alpha^{JI}=-\alpha^{IJ}$ for 0, 3 and 4-forms and $\alpha^{JI}=\alpha^{IJ}$ for 1,2 and 5-forms, so one only needs to compute $\alpha^{IJ}$ for pairs of spinors $(\eta^I,\eta^J)$, where $I\leq J$. This and other analogue constructions of forms give a method to obtain Killing vector fields, along whose directions to introduce adapted curved space coordinates. 

This completes the description of spinors in terms of forms. The identification allows us to explicitly construct a basis of spinors for any given theory, and furthermore cast them in canonical form. Moreover, it will permit us to find the stability subgroups of the spinors inside $Spin(1,n)$, which will in general be related to the G-structure of the solution backgrounds. The stability groups define classes of spinors (given by their orbits under such groups), so we can use a representative of each orbit to substitute into the KSE and operate on. This will generate a system of equations for the geometry and fluxes of the theory, all of which are locally expressed in terms of functions that are originally undetermined. The solution to these equations will give the resulting geometry, which is shaped by supersymmetry, hence serving to fully classify the theory's vacua. As an example, we repeat the analysis done on the supersymmetric solutions to minimal $\N=1$ $d=5$ SUGRA, this time using these techniques.

\subsection{\texorpdfstring{Null solutions to minimal $d=5$ SUGRA}{Null solutions to minimal d=5 SUGRA}}
\label{sec:nullcasespinorialgeometry}
In five dimensions, the minimal spinor used to construct the theory is a symplectic-Majorana (SM) spinor\footnote{Throughout this thesis we will be employing the notation that a symplectic-Majorana spinor is defined as a pair of Dirac spinors on which one imposes a reality condition. We say we have a single spinor since the spin structure group is $Sp(1)$, in our notation.}. The isomorphism outlined above is then to the complexified space of forms on $\mathbb{R}^2$, which we again denote by $\Delta=\Lambda^* (\mathbb{R}^2 \otimes \mathbb{C})$. A basis of $\Delta$ is given by 
\begin{equation}
\xi_i=\{1,e_1,e_2,e_{12}\}\ ,
\end{equation}
where $e_{12}\equiv e_{1}\wedge e_2$. The action of the Gamma matrices on $\Delta$ is represented by
\begin{equation}
\gamma_i\,\epsilon= i (e_i\wedge \epsilon+ \iota_{e_i} \epsilon)\ ,\qquad \gamma_{i+2}\,\epsilon=-e_i\wedge \epsilon + \iota_{e_i}\epsilon\ ,
\end{equation}
where $i=1,2$. The zeroth element of the Clifford algebra is given by $\gamma_0=\gamma_{1234}$; its action on $1$ and $e_{12}$ is given by the identity operator and on $e_1$ and $e_2$ by minus the identity
\begin{equation}
\begin{array}{ll}
\gamma_0\,1=1\ ,&\gamma_0\,e_1=-e_1\ ,\\
\gamma_0\,e_{12}=e_{12}\ ,\phantom{aaa}&\gamma_0\,e_2=-e_2\ , 
\end{array}
\end{equation}
and thus it squares to $\mathbbm{1}$.

A generic Dirac spinor can then be written as
\begin{equation}
\label{eq:genericDiracspinor}
\epsilon=\lambda\, 1+\mu_1 e_1+\mu_2 e_2+\sigma e_{12}\ ,
\end{equation}
where $\lambda$, $\mu_1$, $\mu_2$ and $\sigma$ are complex functions, representing the eight real degrees of freedom of an unconstrained complex spinor in five dimensions. One can see that imposing the reality condition (\ref{eq:realityconditiononspinors}) on $\epsilon$ does not work, which was to be expected since there are no Majorana spinors in $d=5$. Instead, we are lead to consider the symplectic-Majorana representation on a pair of Dirac spinors $\epsilon_1$ and $\epsilon_2$. This is given by 
\begin{equation}
{\epsilon_i}^*=\varepsilon^{ij}\mathcal{B}\epsilon_j\ ,
\end{equation}
where $\mathcal{B}=-\gamma_{0}\mathcal{C}=-\gamma_{034}$ (cf. appendix \ref{app-spinors}). Thus, we can generically write $\epsilon_2=-\mathcal{B}{\epsilon_1}^*$, where the $\text{dim}_{\mathbb{C}}(\epsilon_1)=2^{\lfloor 5/2\rfloor}=4$.

We now consider the null case, leaving the timelike one for section \ref{sec:timelikecasespinorialgeometry}. 
Precisely, being in the null class implies that we need to fulfill that (see eq.~(\ref{eq:defoff}))
which can be recast in terms of the following nullity condition \mycite{Grover:2008ih}
\begin{equation}
B(\epsilon_1,\epsilon_2)=B(\epsilon_1,\gamma_{034}{\epsilon_1}^*)\; =\; 0\ ,
\end{equation}
where the inner product is defined in (\ref{eq:ciiproductointeriorMajorana}). Since $\epsilon_1$ is a generic Dirac spinor, given by eq.~(\ref{eq:genericDiracspinor}), the condition then reads
\begin{equation}
\label{eq:ciinullitycondition}
|\lambda|^2-|\mu_1|^2-|\mu_2|^2+|\sigma|^2=0\ .
\end{equation}
We thus have seven real degrees of freedom, that make up the general null supersymmetric parameter. Furthermore, we can reduce the form of $\epsilon$ by using the $Spin(1,4)$ gauge freedom of the theory. This will make the calculations for solving the KSE much simpler. In particular, one can show that starting from any spinor of the form eq.~(\ref{eq:genericDiracspinor}) which solves (\ref{eq:ciinullitycondition}), by acting on it with $Spin(1,4)$, one can generate the seven DOFs. We can then take, without loss of generality, $\epsilon=1+e_1$.

Alternatively, one can see eq.~(\ref{eq:ciinullitycondition}) as saying that there is a null complex 4-vector living on $M^{(2,2)}_{\mathbb{C}}$, whose inner product is invariant under $U(2,2)$. In this sense, any element in $SU(2,2)$ will respect the length of the vector, \emph{i.e.}~the nullity condition. We can use this in our search for a simpler form of a null $\epsilon$ with the same generality as that of eq.~(\ref{eq:genericDiracspinor}). From an algebraic point of view, 
$SU(2,2)\simeq Spin(2,4) \supset Spin(1,4)$. The $4_\text{v}$ in the vector representation of $SU(2,2)$ goes to a $4_\text{s}$ in $Spin(2,4)$, which amounts to a chiral spinor. This also corresponds to a $4_\text{s}$ irrep in $Spin(1,4)$ \mycite{Slansky:1981yr}, the minimal spinor in $d=5$ being symplectic-Majorana. This implies that there is only one orbit of $Spin(1,4)$ on $\epsilon$, and hence starting from any element of the form (\ref{eq:genericDiracspinor}), satisfying eq.~(\ref{eq:ciinullitycondition}), one can generate the most-general seven DOFs. 

Moreover, the solutions we are about to find preserve at least half of the original supersymmetry. We show this by using the operator $\gamma_0\,C\ast$, which is invariant w.r.t.~the supercovariant derivative eq.~(\ref{eq:ciiKSE}), \emph{i.e.}
\begin{equation}
\gamma_{0}\,C\ast\gamma_\mu=\gamma_\mu\gamma_0\, C\, \ast\ ,
\end{equation}
for $\mu=0,i$, and where $\ast$ is the complex conjugation operator, \emph{i.e.}~$\ast \eta= \eta^*\ \ast$.  We then see that if $\epsilon=1+e_1$ is a Killing spinor, then so is $\gamma_0\,C\ast \epsilon=-e_2+e_{12}$.\vspace{\baselineskip}

In order to perform the calculation of the KSE, it is useful to cast the Gamma matrices in terms of a light-cone basis $\{\Gamma_a\}$
\begin{equation}
\label{eq:nullbasegammas}
\begin{array}{c}
\Gamma_+=\displaystyle\frac{1}{\sqrt{2}}(\gamma_0-\gamma_3)\ ,\quad \Gamma_-=\frac{1}{\sqrt{2}}(\gamma_0+\gamma_3)\\[4mm]
\Gamma_1=\displaystyle\frac{1}{\sqrt{2}}(\gamma_2-i\gamma_4)=\sqrt{2} i\: e_2 \wedge\ ,\quad \Gamma_{\bar{1}}=\frac{1}{\sqrt{2}}(\gamma_2+i\gamma_4)=\sqrt{2} i\: \iota_{e_2} \ ,\quad \Gamma_2=\gamma_1 \ .
\end{array}
\end{equation}
In terms of these coordinates, the metric is thus given by
\begin{equation}
ds^2=2e^+\otimes e^{-}-2e^1\otimes e^{\bar{1}}-e^2\otimes e^2\ .
\end{equation}
Imposing the KSE gives a linear system of equations, that can be solved to give a relation between the geometry (spin connection) and the field strength, \emph{i.e.}
\begin{equation}
\label{eq:fieldcomponents}
\begin{array}{c}
F_{+-}=F_{+1}=F_{+2}=0\ ,\\
\\
F_{-1}=\displaystyle\frac{i}{\sqrt{3}}\w_{-,12}\ ,\quad F_{-2}=\frac{-i}{\sqrt{3}}\w_{-,1\bar{1}}\ ,\quad F_{1\bar{1}}=-i\sqrt{3}\w_{-,+2}\ ,\quad F_{12}=-i\sqrt{3}\w_{1,+-}\ ,
\end{array}
\end{equation}
as well as the following restrictions on the spin connection components
\begin{eqnarray}
\label{eq:spincomponents1}
&\w_{+,+-}=\w_{+,+1}=\w_{+,+2}=\w_{+,1\bar{1}}=\w_{+,12}=\w_{-,+-}=\w_{1,+1}=&\nonumber\\
\nonumber\\
&=\w_{1,+\bar{1}}=\w_{1,+2}=\w_{1,12}=\w_{2,+1}=\w_{2,+2}=\w_{2,1\bar{1}}=0\ ,&\\
\nonumber\\
\label{eq:spincomponents2}
&\w_{1,\bar{1}2}=2\w_{-,+2}=-2\w_{2,+-}=\w_{\bar{1},12}\ ,&\\
\nonumber\\
\label{eq:spincomponents3}
&\w_{1,1\bar{1}}=-2\w_{-,+1}=2\w_{1,+-}=\w_{2,12}\ ,&
\end{eqnarray} 
and those related by complex conjugation. The 2-form is thus given by
\begin{equation}
F=\frac{iq}{2\sqrt{3}}\w_{-,jk}\,{\epsilon_i}^{jk}\,e^-\wedge e^i-\frac{i\sqrt{3}\,q}{2}{\epsilon_{ij}}^{k}\,\w_{k,+-}\,e^i\wedge e^j\ ,
\end{equation}
where\footnote{The constant $q$ is used to tune the field strenghts in the two approaches.} $q\,\epsilon_{1\bar{1}2}=1$. 

We now introduce the basis of vector fields $\{\theta_a\}$ dual to that of 1-forms. In particular, we perform the following calculations considering the vector field $\theta_+\equiv N=\partial_v$, which is normalised as $\iota_N (e^a)=e^a(\theta_+)=\delta^a_+$, and defines the direction of $v$, a coordinate on the curved manifold. Consider now the following Lie derivatives of the Funfbein along $N$
\begin{equation}
\begin{array}{l}
\mathcal{L}_N e^+=(\w_{a,+-}+\w_{+,-a})\, e^a\ ,\\
\mathcal{L}_N e^-=\w_{+,+a}\, e^a\ ,\\
\mathcal{L}_N e^i=-(\w_{a,+\bar{i}}+\w_{+,\bar{i}a})\, e^a\ ,
\end{array}
\end{equation}
for $a=\{+,-,i\}$, $i=\{1,\bar{1},2\}$ and $\bar{2}=2$, which imply
\begin{equation}
\label{eq:Liederofviels}
\begin{array}{lll}
(\mathcal{L}_N e^a)_+=0\ ,\qquad\qquad&(\mathcal{L}_N e^+)_-=0\ ,& (\mathcal{L}_N e^+)_i=\w_{i,+-}+\w_{+,-i}\ ,\\
&(\mathcal{L}_N e^-)_-=0\ ,& (\mathcal{L}_N e^-)_i=0\ ,\\
&(\mathcal{L}_N e^i)_-=\w_{+,-\bar{i}}-\w_{-,+\bar{i}}\ ,& (\mathcal{L}_N e^i)_j=0\ .
\end{array}
\end{equation}
Furthermore, one can use the residual gauge freedom (those spinorial transformations respecting the chosen representative for the spinor $\epsilon=1+e_1$) to simplify some of these results. In particular (see \emph{e.g.}~\mycite{Grover:2009ms}{appendix B}), one can set
\begin{equation}
\label{eq:choiceofLie}
(\mathcal{L}_N e^i)_-=0\ ,
\end{equation}
This, along with eqs.~(\ref{eq:spincomponents2}) and (\ref{eq:spincomponents3}), gives
\begin{equation}
\label{eq:condicionqueunespincomponents}
\w_{+,-i}=\w_{-,+i}=-\w_{i,+-}\ ,
\end{equation}
which implies that $\mathcal{L}_N e^+=0$, and allows us to express $de^a$ as
\begin{equation}
\label{eq:ciiVielbeinarray}
\begin{array}{lll}
de^+&=&\w_{-,-i}\,e^-\wedge e^i+\displaystyle\frac{1}{2}(\w_{i,-j}-\w_{j,-i})\,e^i \wedge e^j\ ,\\
\\
de^-&=&2\w_{-,+i}\,e^- \wedge e^i\ ,\\
\\
de^i&=&-(\w_{-,\bar{i}j}+\w_{j,-\bar{i}})e^- \wedge e^j+\displaystyle \frac{1}{2}(\w_{j,k\bar{i}}-\w_{k,j\bar{i}})\,e^j \wedge e^k\ ,\\
\end{array}
\end{equation}
We now focus on $de^-$, as it is hypersurface orthogonal (\emph{i.e.}~$de^-\wedge e^-=0$), which we shall now use in the analysis of the problem. The Frobenius theorem allows us to recast $e^-$ as
\begin{equation}
e^-=fdu\ ,
\end{equation}
where $f$ is a generic real function, which by the system of eqs.~(\ref{eq:Liederofviels}) we know it is independent of $v$, and where $u$ is another (curved) coordinate. 

To continue with the analysis, we integrate $\mathcal{L}_N e^+=0$ to obtain
\begin{equation}
e^+=dv+\alpha\ ,
\end{equation}
where $\alpha=H du + w_m dx^m$ is a 1-form independent of $v$, and the normalisation condition $e^+(\theta_+)=1$ has also been considered. The three remaining curved coordinates are given as
\begin{equation}
e^i=f^{-1}\,e^i_m dx^m\ ,
\end{equation}
where the $du$ terms have been eliminated by using again the residual gauge freedom of the theory\footnote{The induced effect on $e^+$ can be eliminated by considering a redefined $\alpha$ function.}, and $dv$-terms are prohibited by the condition $e^i(\theta_+)=0$. Furthermore, $\mathcal{L}_N e^i=0$ implies that the $e^i_m$ are $v$-independent, \emph{i.e.}~$e^i_m=e^i_m(u,x^n)$. 

Consider now other Lie derivatives on the Vielbein in terms of curved space coordinates. These establish relations that allow us to describe the field strength and its field equation. In particular, the analysis of $\mathcal{L}_{\theta_-} e^a$ and $\mathcal{L}_{\theta_i} e^a$ implies that 
%
%
\begin{eqnarray}
%
\label{eq:Vielbeinderivadasanalysisone}
\partial_u w_m&=&\w_{-,-i}\,e^i_m+\partial_m H\ ,\\
%
\label{eq:Vielbeinderivadasanalysistwo}
\partial_m f&=&2\,\w_{i,+-}\,e^i_m\ ,\\
%
\label{eq:Vielbeinderivadasanalysisthree}
f^{-1}\,e^i_m\,\partial_u f&=&\partial_u e^i_m+fe^j_m(\w_{-,\bar{i}j}+\w_{j,-\bar{i}})\ ,\\
%
\label{eq:Vielbeinderivadasanalysisfour}
(\tilde{d}w)_{mn}&=&2f^{-2}\,e^i_{[m}\,e^j_{n]}\,\w_{i,-j}\ ,\\
%
%
\label{eq:Vielbeinderivadasanalysisfive}
\partial_{[m} e^i_{n]}&=&0\ ,\\
&& \nonumber
\end{eqnarray}
%
where for the last equality we have used eqs.~(\ref{eq:spincomponents2}), (\ref{eq:spincomponents3}). By using the defining antisymmetry of the spin connection, one can then show that the base-space is flat, in accordance with the bilinears' analysis above. Thus, the solutions belonging the null case have a metric given by
\begin{equation}
ds^2=2fdu\,(dv+H du + w_m dx^m)-f^{-2}h_{mn}dx^m\,dx^n\ ,
\end{equation}
where $h_{mn}=e^i_m\, e^i_n=e^i_m\, e_{in}=\delta_{mn}$. Furthermore, by making the gauge choice
\begin{equation}
\w_{-,ij}=-\w_{i,-j}
\end{equation}
and using $e^j(\theta_i)=\delta^j_i=e^j_m\,e^m_i$, as well as eqs.~(\ref{eq:Vielbeinderivadasanalysistwo}) and (\ref{eq:Vielbeinderivadasanalysisfour}), we have a field strength given by
\begin{equation}
\begin{array}{lll}
F&=-\displaystyle\frac{iq}{2\sqrt{3}}\,f^2 du \wedge \star_{(3)}\tilde{d}w-\frac{i\sqrt{3}\,q}{2}f^{-2}\star_{\text{(3)}}\tilde{d}f\ ,
\end{array}
\end{equation}
Choosing $\epsilon_{123}=1/4$, we arrive at the same result for the field strength as in section \ref{ciinullcase}.


\subsection{\texorpdfstring{Timelike solutions to minimal $d=5$ SUGRA}{Timelike solutions to minimal d=5 SUGRA}}
\label{sec:timelikecasespinorialgeometry}
We now focus in the class of timelike solutions. Because we are still in the minimal five-dimensional case, the first two paragraphs of section \ref{sec:nullcasespinorialgeometry} still hold, which describe the spinorial structure of the theory. For this analysis, however, it will prove useful to follow \mycite{Grover:2008jr} and cast the Gamma matrices in terms of a different basis $\{\Gamma_0,\Gamma_\alpha,\Gamma_{\bar{\alpha}}\}$, where the $0$-direction will be associated with physical time, and $\alpha=\{1,2\},\ \bar{\alpha}=\{\bar{1},\bar{2}\}$
\begin{equation}
\label{eq:timelikebasegammas}
\begin{array}{c}
\Gamma_0=\gamma_0\ ,\\[4mm]
\Gamma_\alpha=\displaystyle\sqrt{2}i\, e_\alpha \wedge \ ,\quad \Gamma_{\bar{\alpha}}=\sqrt{2}i\, \iota_{e_{\alpha}}\ .
\end{array}
\end{equation}
The flat metric is thus given by
\begin{equation}
ds^2=e^0\otimes e^{0}-e^\alpha\otimes e^{\bar{\alpha}}-e^{\bar{\alpha}}\otimes e^\alpha\ .
\end{equation}
As in the previous analysis, we want to use the $Spin(1,4)$ symmetry of the theory to find a simple canonical form for the spinor. One can show that it can be reduced to $\epsilon=f\,1$. The analysis of the KSE (\ref{eq:ciiKSE}) with this spinor then gives the following set of equations
\begin{eqnarray}
\frac{\theta_0 f}{f}-\frac{1}{2}{\w_{0,\mu}}^\mu+\frac{1}{2\sqrt{3}}{F_\mu}^\mu&=&0\ ,\\
\w_{0,0\bar{\mu}}-\frac{2}{\sqrt{3}}F_{0\bar{\mu}}&=&0\ ,\\
\left(-\w_{0,\bar{\alpha}\bar{\beta}}+\frac{1}{\sqrt{3}}F_{\bar{\alpha}\bar{\beta}}\right)\epsilon^{\bar{\alpha}\bar{\beta}}&=&0\ ,
\end{eqnarray}
\begin{eqnarray}
\frac{\theta_\alpha f}{f}-\frac{1}{2}{\w_{\alpha,\mu}}^\mu+\frac{\sqrt{3}}{2}F_{0\alpha}&=&0\ ,\\
-\w_{\alpha,0\bar{\beta}}+\sqrt{3}F_{\alpha\bar{\beta}}+\frac{1}{\sqrt{3}}{F_\mu}^\mu \delta_\alpha^\beta&=&0\ ,\\
-\w_{\alpha,\bar{\mu}\bar{\nu}}\epsilon^{\bar{\mu}\bar{\nu}}-\frac{2}{\sqrt{3}}F^{0\mu}\epsilon_{\alpha\mu}&=&0\ ,
\end{eqnarray}
\begin{eqnarray}
\frac{\theta_{\bar{\alpha}}f}{f}-\frac{1}{2}{\w_{\bar{\alpha},\mu}}^\mu+\frac{1}{2\sqrt{3}}F_{0\bar{\alpha}}&=&0\ ,\\
\w_{\bar{\alpha},0\bar{\beta}}-\frac{1}{\sqrt{3}}F_{\bar{\alpha}\bar{\beta}}&=&0\ ,\\
\w_{\bar{\alpha},\bar{\mu}\bar{\nu}}\,\epsilon^{\bar{\mu}\bar{\nu}}&=&0\ .
\end{eqnarray}
Their analysis gives
\begin{eqnarray}
\label{eq:fieldcomponentstimelike}
F_{0\alpha}&=&\frac{\sqrt{3}}{2}\w_{0,0\alpha}\ ,\\
F_{\alpha\beta}&=&\sqrt{3}\,\w_{0,\alpha\beta}\ ,\\
F_{\alpha\bar{\beta}}&=&\frac{1}{\sqrt{3}}\left(\w_{\alpha,0\bar{\beta}}+{\w_{\mu,0}}^\mu\delta_{\alpha\bar{\beta}}\right)\ ,
\end{eqnarray}
as well as the following restrictions on the spin connection components
\begin{eqnarray}
\label{eq:spincomponentstimeone}
\w_{\alpha,0\beta}&=&\w_{0,\alpha\beta}\ ,\phantom{\frac{1}{2}}\\
\label{eq:spincomponentstimeoneextra}
\w_{\alpha,0\bar{\beta}}&=&-\w_{\bar{\beta},0\alpha }\ ,\phantom{\frac{1}{2}}\\
\label{eq:spincomponentstimetwo}
{\w_{\mu,0}}^\mu&=&{\w_{0,\mu}}^\mu\ ,\phantom{\frac{1}{2}}\\
\label{eq:spincomponentstimethree}
(\theta_0 f)/{f}&=&0\ ,\phantom{\frac{1}{2}}\\
\label{eq:spincomponentstimefour}
\w_{\bar{\mu},\mu\alpha}&=&-\w_{\alpha,\mu\bar{\mu}}\:=\:\frac{1}{2}\w_{0,0\alpha}\ ,\\
\label{eq:spincomponentstimefive}
\w_{\alpha,\beta\gamma}&=&0\ ,
\end{eqnarray} 
along with their complex conjugates. Furthermore,
\begin{equation}
\label{eq:ciicondicionqueobstruyeKilling}
\w_{0,0\alpha}=-2\frac{\theta_\alpha f}{f}
\end{equation}
As before, we use the conditions just found to analyse the geometry of the problem. We introduce curved-space coordinates $\{t,x^m\}$, where $m=(1,2,3,4)$, and express the F\"unfbein in terms of them 
\begin{equation}
e^0=g(dt+w_m dx^m)\ ,\quad e^\alpha=g^{-1/2}\,e^\alpha_m dx^m\ ,\quad e^{\bar{\alpha}}=g^{-1/2}\,e^{\bar{\alpha}}_m dx^m\ ,
\end{equation}
where $g$ is a generic function. The canonically-dual vector fields are given by 
\begin{equation}
\theta_0=g^{-1}\,\partial_t\ ,\quad \theta_\alpha=-g^{1/2}\,e^m_\alpha\,w_m \partial_t+g^{1/2}\,e^m_\alpha \partial_m\ ,\quad \theta_{\bar{\alpha}}=-g^{1/2}\,e^m_{\bar{\alpha}}\,w_m \partial_t+g^{1/2}\,e^m_{\bar{\alpha}} \partial_m\ ,
\end{equation}
and consequently $e^\alpha_m\,e^m_\beta=\delta^\alpha_\beta$, $e^\alpha_m\,e^m_{\bar{\beta}}=0$, $e^{\bar{\alpha}}_m\,e^m_{\bar{\beta}}=\delta^{\bar{\alpha}}_{\bar{\beta}}$. 
As expressed in \mycite{Gibbons:1984kp}, supersymmetric solutions to SUGRA theories will always have a Killing vector field, which we can use to simplify the problem. This was for example implicitly used in section \ref{ciitimelikecase} above, when the coordinate $t$ was chosen along the flow of Killing vector $V$, which guarantees that the fields contained in the metric are all $t$-independent. Similarly, we can consider the vector $V=f^2\,\theta_0=f^2\,g^{-1}\,\partial_t$. Because of eqs.~(\ref{eq:spincomponentstimethree}) and (\ref{eq:ciicondicionqueobstruyeKilling}), $V$ is Killing; we then choose $g=f^2$ so that the function $g$, the 1-form $w=w_m\,dx^m$ and the four-dimensional base-space ($B$) metric
\begin{equation}
ds_B^2=\tilde{e}^\alpha\otimes \tilde{e}^{\bar{\alpha}}+\tilde{e}^{\bar{\alpha}} \otimes \tilde{e}^\alpha=h_{mn}\,dx^m\otimes dx^n\ ,
\end{equation}
where $\tilde{e}^\alpha=g^{1/2}\,e^\alpha$ and $h_{mn}=e^\alpha_m e^{\bar{\alpha}}_n+e^{\bar{\alpha}}_m e^\alpha_n=e^\alpha_m e_{\alpha n} + e^{\bar{\alpha}}_m e_{\bar{\alpha}n}$, are all time-independent.\vspace{\baselineskip}

Consider now the Lie derivative of the F\"unfbein along $\theta_0$
\begin{equation}
\label{eq:derivadasdeLiedelfunfbeintime}
\mathcal{L}_{\theta_0} e^0=\,\w_{0,0a}\,e^a\ ,\quad  \mathcal{L}_{\theta_0} e^\alpha=-(\w_{0,\bar{\alpha}a}-\w_{a,\bar{\alpha}0})\,e^a\ ,\quad  \mathcal{L}_{\theta_0} e^{\bar{\alpha}}=-(\w_{0,\alpha a}-\w_{a,\alpha 0})\,e^a\ ,
\end{equation}
for $a=(0,\alpha,\bar{\alpha})$. The first implies that
\begin{equation}
\partial_t w_m=g^{-\frac{1}{2}}(\w_{0,0\alpha}\,{e^\alpha}_m+\w_{0,0\bar{\alpha}}\,{e^{\bar{\alpha}}}_m)+g^{-1}\partial_m g\ .
\end{equation}
The two last read
\begin{equation}
\begin{array}{l@{\qquad}l}
(\mathcal{L}_{\theta_0} e^\alpha)_0=0\ ,& (\mathcal{L}_{\theta_0} e^{\bar{\alpha}})_0= 0\ ,\\
(\mathcal{L}_{\theta_0} e^\alpha)_\beta=-(\w_{0,\bar{\alpha}\beta}-\w_{\beta,\bar{\alpha}0})\ ,&(\mathcal{L}_{\theta_0} e^{\bar{\alpha}})_\beta=-(\w_{0,\alpha\beta}-\w_{\beta,\alpha 0})=0\ ,\\
(\mathcal{L}_{\theta_0} e^\alpha)_{\bar{\beta}}=-(\w_{0,\bar{\alpha}\bar{\beta}}-\w_{\bar{\beta},\bar{\alpha}0})=0\ ,
& (\mathcal{L}_{\theta_0} e^{\bar{\alpha}})_{\bar{\beta}}=-(\w_{0,\alpha\bar{\beta}}-\w_{\bar{\beta},\alpha 0})\ ,
\end{array}
\end{equation}
where we have used the eq.~(\ref{eq:spincomponentstimeone}). Furthermore, time-independence of $g$ and $h_{mn}$ imply that
\begin{equation}
\label{eq:thirdelementinKillingequation}
\w_{\alpha,0\bar{\beta}}=\w_{0,\alpha\bar{\beta}}\ .
\end{equation}
Also, as in the null case analysis above, study the remaining Lie derivatives w.r.t.~curved coordinates. Among others, one obtains the following results
%
%
%
%
%
\begin{eqnarray}
\label{eq:arrayderivadasdeLietimelike1}
dg&=&-g(\w_{0,0\alpha}\,e^\alpha+\w_{0,0\bar{\alpha}}\,e^{\bar{\alpha}})\ ,\\
\label{eq:arrayderivadasdeLietimelike2}
g\,dw&=&\w_{0,\alpha\beta}\, e^\alpha \wedge e^\beta + \w_{0,\alpha\bar{\beta}}\, e^\alpha \wedge e^{\bar{\beta}} +\w_{0,\bar{\alpha}\beta}\, e^{\bar{\alpha}} \wedge e^\beta+\w_{0,\bar{\alpha}\bar{\beta}}\, e^{\bar{\alpha}} \wedge e^{\bar{\beta}}\ .
\end{eqnarray}
Notice that, by construction, eq.~(\ref{eq:arrayderivadasdeLietimelike1}) and the choice of $g=f^2$ is consistent with eq.~(\ref{eq:ciicondicionqueobstruyeKilling}).\vspace{\baselineskip}

Construct now three complex structures $J^i$ ($i=1,2,3$) on $B$, which give rise to the following antiself-dual\footnote{W.r.t.~the base-space $B$, where we have adopted the convention that $\epsilon_{1\bar{1}2\bar{2}}=1$.} K\"ahler 2-forms,
\begin{eqnarray}
K^1&=&{\tilde{e}}^1\wedge {\tilde{e}}^2+{\tilde{e}}^{\bar{1}}\wedge {\tilde{e}}^{\bar{2}}\ ,\\
K^2&=&-i\left({\tilde{e}}^1\wedge {\tilde{e}}^{\bar{1}}+{\tilde{e}}^2\wedge {\tilde{e}}^{\bar{2}}\right)\ ,\\
K^3&=&-i\left({\tilde{e}}^1\wedge {\tilde{e}}^2-{\tilde{e}}^{\bar{1}}\wedge {\tilde{e}}^{\bar{2}}\right)\ .
\end{eqnarray}  
These $J^i$ fulfill the algebra of imaginary unit quaternions, and the important thing to notice is that, by means of eqs.~(\ref{eq:spincomponentstimefour})-(\ref{eq:ciicondicionqueobstruyeKilling}), one can show that
\begin{eqnarray}
dK^i=0\ .
\end{eqnarray}
As commented in appendix \ref{appsec:complexstructures}, this implies that $B$ is a hyper-K\"ahler manifold.

We are left with the analysis of the field equations. Due to the KSIs, we only need to demand the fulfilling of the Maxwell equations, as the rest are automatically satisfied. In particular, the field strength is given by
\begin{eqnarray}
F&=&\frac{\sqrt{3}}{2}\,\w_{0,0\alpha}\,e^0 \wedge e^\alpha+\frac{\sqrt{3}}{2}\,\w_{0,0\bar{\alpha}}\,e^0 \wedge e^{\bar{\alpha}}+\frac{\sqrt{3}}{2}\,\w_{0,\alpha\beta}\,e^\alpha\wedge e^\beta+\frac{\sqrt{3}}{2}\,\w_{0,\bar{\alpha}\bar{\beta}}\,e^{\bar{\alpha}}\wedge e^{\bar{\beta}}\nonumber\\
&&+\frac{1}{2\sqrt{3}}\left(\w_{\alpha,0\bar{\beta}}+{\w_{\mu,0}}^\mu \delta_{\alpha\bar{\beta}}\right) e^\alpha\wedge e^{\bar{\beta}}+\frac{1}{2\sqrt{3}}\left(\w_{\bar{\alpha},0\beta}+{\w_{\bar{\mu},0}}^{\bar{\mu}} \delta_{\bar{\alpha}\beta}\right) e^{\bar{\alpha}}\wedge e^\beta\ .
\end{eqnarray}
We now use eqs.~(\ref{eq:thirdelementinKillingequation}), (\ref{eq:arrayderivadasdeLietimelike1}) and (\ref{eq:arrayderivadasdeLietimelike2}) to simplify this expression into
\begin{equation}
F=-\frac{\sqrt{3}}{2}(dt+w)\wedge dg +\frac{\sqrt{3}}{2}g\,dw-\frac{1}{\sqrt{3}}G^+\ ,
\end{equation}
where we have defined a $B$-self-dual 2-form $G^+$ (cf. eq.~(\ref{eq:formasdeGdualestimelikebilinears})) as
\begin{equation}
G^+=\w_{0,\alpha\bar{\beta}}\,e^\alpha\wedge e^{\bar{\beta}}+\w_{0,\bar{\alpha}\beta}\,e^{\bar{\alpha}}\wedge e^\beta+{\w_{0,\mu}}^\mu\, e^\alpha \wedge e^{\bar{\alpha}}\ ,
\end{equation}
\emph{i.e.}
\begin{equation}
G^+_{\alpha\bar{\beta}}=\w_{0,\alpha\bar{\beta}}+\frac{1}{2}{\epsilon_{\alpha\bar{\beta}}}^{\sigma\bar{\tau}}\w_{0,\sigma\bar{\tau}}\ .
\end{equation}
This is the same field strength we found in section \ref{ciitimelikecase}, and thus the same field equations apply.

Having detailed the two approaches to classifying solutions to Killing Spinor equations, and how to use them in order to obtain supersymmetric solutions to supergravity theories, we are ready to consider fakeSupergravity.

\cleardoublepage

\renewcommand{\leftmark}{\MakeUppercase{Chapter \thechapter. $N=2$ $d=4$ gauged fakeSUGRA}}
\chapter{\texorpdfstring{$\N=2$ $d=4$ gauged fakeSUGRA coupled to non-Abelian vectors}{N=2 d=4 gauged fakeSUGRA}}
\label{4d}
This chapter is a recapitulation of \mycite{Meessen:2009ma}, where we presented the classification of solutions to $\N=2$ $d=4$ fakeSupergravity coupled to non-Abelian vector multiplets, otherwise also referred to as Wick-rotated $\N=2$ \mbox{$d=4$} Supergravity, having allowed for gaugings of the isometries of the scalar manifold\footnote{For gauging in $\N=2$ SUGRA, see \emph{e.g.}~appendix \ref{appsec:SpecGeom}, or \mycite{Andrianopoli:1996cm} for a complete review.}. 
This is because, as explained in section \ref{beyondSUGRA}, this theory can be obtained from gauged $\N=2$ $d=4$ SUGRA coupled to non-Abelian vector multiplets by means of a Wick rotation. In this sense, since we are allowing for non-Abelian couplings, it is not the coupling constant that is rotated, but rather the Fayet-Iliopoulos term responsible for gauging the $\mathbb{R}$-symmetry. 

The outline of this chapter is the following: in section \ref{sec:DSVector} we set up the fake-Killing spinors equations that we are going to solve. We see that, as we are Wick rotating the FI term, the relations between the equations of motion one can derive from the integrability equation are similar to the ones obtained in the usual supersymmetric case, and hence the implications as far as the checking of equations of motion are identical. This was to be expected since we are not changing the characteristics of the Killing spinors. Some information about Special Geometry and the gauging of isometries in special geometries, needed to understand the set-up, is given in appendix (\ref{appsec:SpecGeom}).

As in the classification of supersymmetric solutions, we split the problem into two different cases, depending on the norm of the vector one constructs as a bilinear of the fake-Killing spinors; the timelike case, {\em i.e.\/} when the norm does not vanish, shall be treated in section \ref{sec:VectBil}. 
In section \ref{sec:Null} we will have a go at the null case, {\em i.e.\/} when the norm of the vector vanishes identically. In that section, we will be ignoring the possible non-Abelian couplings and hence we shall not obtain a complete characterisation; instead we will find that the solutions have holonomy group contained in $\Sim(2)$, and we shall discuss the general features of such a solution. This will be illustrated by two examples, namely the Nariai cosmos in the minimal theory in section \ref{sec:NullSols}, and a general class of solutions with holomorphic scalars that can be seen as a back-reacted intersection of a cosmic string with a Nariai/ Robinson-Bertotti solution, in section \ref{sec:Holomorphic}.

The reader might feel that the generic theories that can be treated in our setting are rather cryptic, as their connection with supergravity theories or Einstein-Yang-Mills-$\Lambda$ theories can be considered weak. However, it is well known that there are choices for the FI terms in gauged $\N=2$ $d=4$ SUGRA for which this equals the bosonic part of an ungauged supergravity \mycite{Cremmer:1984hj}. This means that for such choices, our fake-supersymmetric solutions are nothing more than non-BPS solutions to an ordinary ungauged supergravity. The easiest model in which one can see this happening is the model which can be obtained by dimensionally-reducing minimal $\N=1$ $d=5$ Supergravity, and we shall discuss some simple solutions to this model in section \ref{sec:PotIsNul}, as well as their uplift to five dimensions. Finally, in section \ref{sec:Concl}, we shall give our conclusions and a small outlook for related work in higher dimensions. The tensorial conventions are presented in appendix \ref{sec-d4conventions}, and the spinorial ones in appendix \ref{appsec:spinors4d}. Furthermore, the interested reader will find information about the normalisation of the bilinears and the curvatures for the null case in appendices \ref{sec:Bil} and \ref{sec:NullCurv}, respectively. 

\section{\texorpdfstring{Fake $\N=2$ Einstein-Yang-Mills}{Fake N=2 Einstein-Yang-Mills}}
\label{sec:DSVector}
The bosonic field content of $N=2$ $d=4$ Supergravity coupled to $n_{v}$ vector multiplets consists of the graviton, $g_{\mu\nu}$, $\bar{n}=n_{v}+1$ vector fields $A^{\Lambda}$ (where $\Lambda =0,\ldots,n_{v}$) and $n_{v}$ complex scalar fields $Z^{i}$ (for $i=1,\ldots ,n_{v}$). The self-interaction of the scalars and their interaction with the vector fields can be derived from a geometric structure called Special Geometry, of which more is given in appendix \ref{appsec:SpecGeom}.

As commented above, the scenario we want to consider can be obtained from ordinary $\N=2$ $d=4$ gauged SUGRA coupled to vector multiples (but not to hyper-multiplets) by Wick rotating the Fayet-Iliopoulos term\footnote{See appendix (\ref{sec:FIinfSUGRA}) for more information about FI terms and their Wick rotation.}. In other words, we Wick rotate the constant tri-holomorphic map $\mathsf{P}^{x}_{\Lambda}\rightarrow i\mathtt{C}_{\Lambda}\delta^{x}_{2}$, where $\mathtt{C}_{\Lambda}$ are real and constant. In usual supersymmetric studies the FI term gauges a $U(1)$ in the hyper-multiplets' $SU(2)$, and the effect of the Wick rotation is that we are gauging instead an $\mathbb{R}$-symmetry through the effective connection $\mathtt{C}_{\Lambda}A^{\Lambda}$ \mycite{Behrndt:2003cx}.

The presence of a FI term is compatible with the gauging of non-Abelian isometries of the scalar manifold, so long as the action of the gauge group commutes with the FI term (see {\em e.g.\/} \mycite{Andrianopoli:1996cm}). Taking the gauge algebra to have structure constants $f_{\Lambda\Sigma}{}^{\Gamma}$, this then implies that we must impose the constraint $f_{\Lambda\Sigma}{}^{\Omega}\mathtt{C}_{\Omega}=0$. One result of the introduction of the $\mathtt{C}_{\Lambda}$ constants is that the dimension of the possible non-Abelian gauge algebra is not $\bar{n}=n_v+1$, but rather $n_v$, as 1 vector field is already used as the connection for the $\mathbb{R}$-symmetry.

The gauging of isometries implies that the field strengths of the physical fields are given by
\begin{equation}
  \label{eq:Deriv1}
  \mathtt{D}Z^{i}\; \equiv\; dZ^{i}\ +\ gA^{\Lambda}\ \mathtt{K}^{i}_{\Lambda}\ ,\qquad F^{\Lambda} \; \equiv\; dA^{\Lambda} 
        \ +\ \textstyle{\frac{g}{2}}\ f_{\Sigma\Gamma}{}^{\Lambda}\ A^{\Sigma}\wedge A^{\Gamma} \ , 
\end{equation}
where $\mathtt{K}_{\Lambda}^{i}$ is the holomorphic part of the Killing vector $\mathtt{K}_{\Lambda}$ (see appendix \ref{sec:SGisom} for the minimal information needed, or \mycite{Andrianopoli:1996cm, Huebscher:2008yz} for a fuller account). One implication of the above definition is that $\mathtt{C}_{\Lambda}F^{\Lambda} = d\left[ \mathtt{C}_{\Lambda}A^{\Lambda}\right]$, so that the linear combination $\mathtt{C}_{\Lambda}A^{\Lambda}$ is indeed an Abelian vector field.   

As mentioned, we are introducing an $\mathbb{R}$-connection on top of the existent K\"ahler/ $U(1)$-symmetry due to the vector coupling. This means that we should define the covariant derivative on the fakeKilling spinors as\footnote{In the notation that we will follow throughout this chapter, $\mathbb{D}$ will be the total connection, whereas we will reserve $\mathtt{D}$ for the connection without the $\mathbb{R}$-part and $\mathfrak{D}$ for the K\"ahler-connection, {\em i.e.\/} the connection appearing in ungauged supergravity.}
\begin{eqnarray}
  \label{eq:16j}
  \mathbb{D}_{a}\epsilon_{I} 
       & =& \nabla_{a}\epsilon_{I}
       \, +\, \textstyle{\frac{i}{2}}\mathcal{Q}_{a}\epsilon_{I}
       \, +\, \frac{ig}{2}\ A_{a}^{\Lambda}\ \left[\mathtt{P}_{\Lambda}\ +\ i \mathtt{C}_{\Lambda}\right]\epsilon_{I} 
       \nonumber \\ 
    & \equiv& \mathtt{D}_{a}\epsilon_{I} \, -\, \textstyle{\frac{g}{2}}\ \mathtt{C}_{\Lambda}A^{\Lambda}_{a}\epsilon_{I}\; ,
\end{eqnarray}
where $\mathtt{P}_{\Lambda}$ is the momentum map corresponding to an isometry $\mathtt{K}_{\Lambda}$ of the special geometry.

Using the above definitions we can write the fake Killing spinor equations as
\begin{eqnarray}
  \label{eq:20j}
  \mathbb{D}_{a}\epsilon_{I} & =& -\varepsilon_{IJ}\ \mathcal{T}_{ab}^{+}\gamma^{b}\ \epsilon^{J}
          \; -\; \textstyle{\frac{ig}{4}}\ \mathtt{C}_{\Lambda}\mathcal{L}^{\Lambda}\, \gamma_{a}\ \varepsilon_{IJ}\epsilon^{J} \; ,\\
  \label{eq:20bj}
  \mathbb{D}_{a}\epsilon^{I} & =& \varepsilon^{IJ}\ \overline{\mathcal{T}^{+}}_{ab}\gamma^{b}\ \epsilon^{J}
          \; -\; \textstyle{\frac{ig}{4}}\ \mathtt{C}_{\Lambda}\overline{\mathcal{L}}^{\Lambda}\, \gamma_{a}\ \varepsilon^{IJ}\epsilon_{J} \; ,\\
  & & \nonumber \\
  \label{eq:20cj}
  i\slashed{\mathtt{D}}Z^{i}\ \epsilon^{I} & =& 
     -\varepsilon^{IJ}\ \slashed{G}^{i+}\epsilon_{J} 
     \; -\; \mathtt{W}^{i}\ \varepsilon^{IJ}\epsilon_{J} \; ,\\
  \label{eq:20dj}
  i\slashed{\mathtt{D}}\overline{Z}^{\bar{\imath}}\ \epsilon_{I} & =& 
     -\varepsilon_{IJ}\ \slashed{\overline{G}}^{\bar{\imath}-}\epsilon^{J} 
     \; -\; \overline{\mathtt{W}}^{\bar{\imath}}\ \varepsilon_{IJ}\epsilon^{J} \; ,   
\end{eqnarray}
where for clarity we have given also the rules for $\mathbb{D}_{a}\epsilon^{I}$ and ${\slashed{\mathtt{D}}\overline{Z}^{\bar{\imath}}}$ $\!\!\epsilon_{I}$ even though they can be obtained by complex conjugation from the other two rules. Furthermore, we have introduced the abbreviation
\begin{equation}
  \label{eq:Vect1}
  \mathtt{W}^{i} \; =\; -\textstyle{\frac{ig}{2}}\ \bar{f}^{i\Lambda}\ 
          \left[ \mathtt{P}_{\Lambda}\ +\ i\mathtt{C}_{\Lambda}\right]\ ,\quad
  \overline{\mathtt{W}^{\bar{\imath}}} \; =\; \overline{\mathtt{W}^{i}} \ ,
\end{equation}
and we have used the standard $\N=2$ $d=4$ SUGRA definitions \mycite{Andrianopoli:1996cm}
\begin{equation}
  \label{eq:DefFStrength}
  \mathcal{T}^{+}\; \equiv\; 2i\mathcal{L}_{\Lambda}\ F^{\Lambda +} \ ,\quad G^{i+} \; \equiv\; -\bar{f}^{i}_{\Lambda}\ F^{\Lambda\ +} \ .
\end{equation}

The integrability conditions for the above system of equations can easily be calculated and give rise to
\begin{equation}
  \label{eq:VectInt1}
  \mathcal{E}_{ab}\ \gamma^{b}\epsilon_{I} \; =\; -2i\ \mathcal{L}^{\Lambda}
           \left[
              \slashed{\mathcal{B}}_{\Lambda} \ -\ \mathcal{N}_{\Lambda\Sigma}\slashed{\mathcal{B}}^{\Sigma}
           \right]\ \varepsilon_{IJ}\gamma_{a}\epsilon^{J}\; ,
\end{equation}
where we defined not only the Bianchi identity as $\star \mathcal{B}^{\Lambda} = \mathtt{D}F^{\Lambda}\,(=0)$, but also the following equations, where $\mathcal{E}_{ab}=0$ is the Einstein equation, $\mathcal{B}_\Lambda=0$ the Maxwell equation and $\mathtt{V}$ the potential of the theory
\begin{eqnarray}
  \label{eq:VectEOM1}
  \mathcal{E}_{ab} & =& R_{ab}
     \ +\ 2\mathcal{G}_{i\bar{\jmath}}\mathtt{D}_{(a}Z^{i}\mathtt{D}_{b)}\overline{Z}^{\bar{\jmath}}
     \ +\ 4\mathrm{Im}\left(\mathcal{N}\right)_{\Lambda\Sigma}\left[
               F^{\Lambda}_{ac}F^{\Sigma}_{b}{}^{c} 
               -\textstyle{\frac{1}{4}}\eta_{ab}F^{\Lambda}_{cd}F^{\Sigma cd}
           \right]
     \ -\ \textstyle{\frac{1}{2}}\eta_{ab}\ \mathtt{V}\ ,\qquad\phantom{a}\\
   & & \nonumber \\
   \label{eq:VectEOM2}
  \star \mathcal{B}_{\Lambda} & =& \mathtt{D}\left[\ \mathcal{N}_{\Lambda\Sigma}\ F^{\Sigma -}
                  + \overline{\mathcal{N}}_{\Lambda\Sigma}\ F^{\Sigma +}
                 \right] 
                 - \textstyle{\frac{g}{2}}\mathrm{Re}\left(
                        \mathtt{K}_{\Lambda \bar{\imath}} \star\mathtt{D}\overline{Z}^{\bar{\imath}}
                     \right) 
           \equiv
             \mathtt{D}F_{\Lambda} - \textstyle{\frac{g}{2}}\mathrm{Re}\left(
                        \mathtt{K}_{\Lambda \bar{\imath}} \star\mathtt{D}\overline{Z}^{\bar{\imath}}
                     \right)\ ,\\
  & & \nonumber \\
  \label{eq:VectPot}
  \mathtt{V} & =& \textstyle{\frac{g^2}{2}}\left[
                       3\mathtt{C}_{\Lambda}\mathtt{C}_{\Sigma}\mathcal{L}^{\Lambda}\overline{\mathcal{L}}^{\Sigma}
                       \, +\, f_{i}^{\Lambda}\ \bar{f}^{i\Sigma}
                           \left(\mathtt{P}+i\mathtt{C}\right)_{\Lambda}
                            \left(\mathtt{P}+i\mathtt{C}\right)_{\Sigma}
                  \right] \ .
\end{eqnarray}
This potential is not real, and imposing it to be so implies that we must satisfy the constraint
\begin{equation}
  \label{eq:GaugeConstr}
  0\; =\; \mathrm{Im}\left(\mathcal{N}\right)^{-1|\Lambda\Sigma}\ \mathtt{P}_{\Lambda}\ \mathtt{C}_{\Sigma} \ ,
\end{equation}
which is a gauge-invariant statement. However, for our choice of non-Abelian gaugings compatible with the FI term, this constraint is satisfied identically; one can see that indeed contracting the last equation in eq.~(\ref{eq:SGK17}) with $f_{i}^{\Sigma}$ and using identities (\ref{eq:SGImpId}) and (\ref{eq:SGK10}) one obtains the identity 
\begin{equation}
\label{eq:GaugeConstr2}
\mathrm{Im}\left(\mathcal{N}\right)^{-1|\Lambda\Sigma}\ \mathtt{P}_{\Sigma} \; =\; 4i\ \mathcal{L}^{\Sigma}\overline{\mathcal{L}}^{\Omega}\ f_{\Sigma\Omega}{}^{\Lambda}\; ,
\end{equation}
which upon contracting with $\mathtt{C}_{\Lambda}$ 
gives the desired result. Therefore the potential $\mathtt{V}$ reads
\begin{equation}
\label{eq:Potential}
\begin{array}{lll}
\mathtt{V}&=& \textstyle{\frac{g^2}{2}}\left[
                       3\left|\mathtt{C}_{\Lambda}\mathcal{L}^{\Lambda}\right|^{2}
                       \; +\; f_{i}^{\Lambda}\bar{f}^{i\Sigma}\left(\
                              \mathtt{P}_{\Lambda}\mathtt{P}_{\Sigma}\ -\ \mathtt{C}_{\Lambda}\mathtt{C}_{\Sigma}
                        \right)\
                   \right]\ , \\
& & \\
& =& \textstyle{\frac{g^2}{2}}\left[
                       4\left|\mathtt{C}_{\Lambda}\mathcal{L}^{\Lambda}\right|^{2}
                       \; +\; \textstyle{\frac{1}{2}}\mathrm{Im}\left(\mathcal{N}\right)^{-1|\Lambda\Sigma}
                        \left(\
                              \mathtt{C}_{\Lambda}\mathtt{C}_{\Sigma}\ -\ \mathtt{P}_{\Lambda}\mathtt{P}_{\Sigma}
                        \right)\
                   \right]\ ,
\end{array}
\end{equation}
which is similar to the supersymmetric result in \mycite{Andrianopoli:1996cm}, upon Wick rotating the Fayet-Iliopoulos term.
Likewise, the above equations of motion can be obtained from the action
\begin{equation}
  \label{eq:VectAct}
  \int_{4}\sqrt{g}\left[
     R 
    + 2\mathcal{G}_{i\bar{\jmath}}\mathtt{D}_{a}Z^{i}\mathtt{D}^{a}\overline{Z}^{\bar{\jmath}}
    + 2\mathrm{Im}\left(\mathcal{N}\right)_{\Lambda\Sigma}F^{\Lambda}_{ab}F^{\Sigma ab}
    - 2\mathrm{Re}\left(\mathcal{N}\right)_{\Lambda\Sigma}F^{\Lambda}_{ab}\star F^{\Sigma ab}
    - \mathtt{V}
  \right] \ ,
\end{equation}
which as stated in the introduction has correctly normalised kinetic terms, and $\mathrm{Im}(\mathcal{N})$ is a negative definite matrix.

In SUGRA the integrability condition for the scalars relates the scalar equation of motion (EOM) with the Maxwell EOM, and the same happens here. A straightforward calculation results in 
\begin{equation}
  \label{eq:VectInt2}
  \mathcal{B}^{i}\epsilon_{I} \; =\; -2i\ \bar{f}^{i\Lambda}\left[
                        \slashed{\mathcal{B}}_{\Lambda}
                        \ -\ \mathcal{N}_{\Lambda\Sigma}\slashed{\mathcal{B}}^{\Sigma}
                  \right]\ \varepsilon_{IJ}\epsilon^{J} \; ,
\end{equation}
where we have introduced the equation of motion for the scalars $Z^{i}$ as
\begin{equation}
  \label{eq:VectEOM3}
  \mathcal{B}^{i} \, =\,
      \Box Z^{i} 
     -i\partial^{i}\overline{\mathcal{N}}_{\Lambda\Sigma}F^{\Lambda +}_{ab}F^{\Sigma +\ ab}
     +i\partial^{i}\mathcal{N}_{\Lambda\Sigma}F^{\Lambda -}_{ab}F^{\Sigma -\ ab}
     +\textstyle{\frac{1}{2}}\partial^{i}\mathtt{V} \; .
\end{equation}
It is thus clear that the integrability conditions for equations (\ref{eq:20j}\,-\,\ref{eq:20dj}) give relations between the EOMs. These, modulo the changes in the form of the tensors, are exactly the same as found in regular Supersymmetry, which was to be expected. The implication of relations (\ref{eq:VectInt1}) and (\ref{eq:VectInt2}) is then also the same as in SUSY \mycite{Gauntlett:2002nw,Bellorin:2005hy}, namely that the independent number of equations of motion one has to check in order to be sure that a given solution to eqs.~(\ref{eq:20j}\,-\,\ref{eq:20dj}) is also a solution to the equations of motion is greatly reduced\footnote{As we are using the same conventions as \mycite{Meessen:2006tu}, we can copy the arguments there as they stand.}. The minimal set of equations of motion one has to check depends on the norm of the vector bilinear $V_{a}=i\overline{\epsilon}^{I}\gamma_{a}\epsilon_{I}$.

In the timelike case we only need to solve the timelike direction of the Bianchi identity, {\em i.e.\/} $\imath_{V}\star\mathcal{B}^{\Lambda}=0$, and the Maxwell (Yang-Mills) equations. This case will be considered in section \ref{sec:VectBil}. If the norm of the bilinear is null, {\em i.e.\/} $V_{a}V^{a}=0$, then a convenient set of EOMs is given by $N^{a}N^{b}\mathcal{E}_{ab}=0\,$, $N^{a}\mathcal{B}_{\Lambda a}=0$ and $N^{a}\mathcal{B}_{a}^{\Lambda}=0\,$, where $N$ is a vector normalised by $V^{a}N_{a}=1\,$. This case will be considered in section \ref{sec:Null}.

\section{Analysis of the timelike case}
\label{sec:VectBil}
In this section we consider the timelike case and the strategy is the usual one employed in the characterisation of supersymmetric solutions. We analyse the differential constraints (coming from the fKSEs) on the bilinears constructed from the spinors $\epsilon_{I}$, which are defined in appendix \ref{sec:Bil}. We try try to solve these constraints generically in as little unknowns as possible. After they have been solved, we shall, following the comments made above, impose the Bianchi identity and the gauge field EOMs, to study the conditions that they generate. We shall then be left with a minimal set of functions, structures and conditions they have to satisfy in order to obtain fake-supersymmetric solutions; the resulting algorithm will be outlined in section \ref{sec:CosmMon}.

Let us start by discussing the differential constraints on the bilinears. Using eq.~(\ref{eq:20j}) and the definitions of the bilinears in appendix (\ref{sec:Bil}), we can calculate
\begin{eqnarray}
  \label{eq:VId1}
  \mathbb{D}X & =&  \textstyle{\frac{g}{4}}\ \mathtt{C}_{\Lambda}\mathcal{L}^{\Lambda}\ V
           \; +\; i\ \imath_{V}\mathcal{T}^{+}\; ,\\
    & & \nonumber\\
  \label{eq:VId2}
  \mathbb{D}_{a}V_{b} & =&  g|X|^{2}\ \mathtt{C}_{\Lambda}\mathcal{R}^{\Lambda}\ \eta_{ab}
      \ +\ 4\mathrm{Im}\left(
                \overline{X}\ \mathcal{T}^{+}_{ab}
            \right)\; ,\\
   & & \nonumber \\
  \label{eq:VId3}
  \mathbb{D}V^{x} & =& \textstyle{\frac{g}{2}}\mathtt{C}_{\Lambda}\mathcal{R}^{\Lambda}\ V\wedge V^{x}
     \ +\ \textstyle{\frac{g}{2}}\mathtt{C}_{\Lambda}\mathcal{I}^{\Lambda}\
            \star\left[ V\wedge V^{x}\right] \; ,
\end{eqnarray}
where, following \mycite{Meessen:2006tu}, we have introduced the real symplectic sections of K\"ahler weight zero,
\begin{equation}
  \label{eq:DefRandI}
  \mathcal{R} \; =\; \mathrm{Re}\left(\mathcal{V}/X\right) \ ,\quad \mathcal{I} \; =\; \mathrm{Im}\left(\mathcal{V}/X\right) \quad\longrightarrow\quad \frac{1}{2|X|^{2}}\ =\ \langle\mathcal{R}\,|\,\mathcal{I}\rangle \ ,
\end{equation}
where $\mathcal{V}$ and the symplectic inner product $\langle -| -\rangle$ are explained in appendix \ref{appsec:SpecGeom}. As in the ungauged theory, the $2\bar{n}$ real functions $\mathcal{I}$ play a fundamental r\^ole in the construction of BPS solutions and the $2\bar{n}$ real functions $\mathcal{R}$ depend on $\mathcal{I}$. Given a Special Geometric model, finding the explicit $\mathcal{I}$-dependence of $\mathcal{R}$ is known as the {\em stabilisation equation}, and solutions are known for different models.

A difference of this analysis w.r.t.~the usual supersymmetric case lies in the character of the bilinear $V$. In such a case it is always a Killing vector, whereas this is not true here, as can be seen from eq.~(\ref{eq:VId2}). We can of course still use it to introduce a timelike coordinate $\tau$ by choosing  an adapted coordinate system through $V=V^{a}\partial_{a}=\sqrt{2}\partial_{\tau}$, but now the components of the metric will depend explicitly on $\tau$, as was to be expected for instance from \mycite{Kastor:1993mj}.

As the $V^{x}$ contain information about the metric on the base-space, it is important to deduce its behaviour under translations along $V$; we calculate
\begin{equation}
  \label{eq:LieV}
  \pounds_{V}V^{x} \; =\; \imath_{V}dV^{x} +d\left(\imath_{V}V^{x}\right)
               \; =\; g \mathtt{C}_{\Lambda}\imath_{V}A^{\Lambda}\ V^{x} 
                      \ +\ 2g|X|^{2}\mathtt{C}_{\Lambda}\mathcal{R}^{\Lambda}\ V^{x} \; .
\end{equation}
This implies that by choosing the gauge-fixing
\begin{equation}
  \label{eq:VId4}
  \imath_{V}A^{\Lambda} \; =\; -2|X|^{2}\ \mathcal{R}^{\Lambda} \; ,
\end{equation}
we find that $\pounds_{V}V^{x}=0$. As a matter of fact, the above gauge-fixing is the actual result one obtains when considering timelike supersymmetric solutions in $\N=2$ $d=4$ Supergravity theories \mycite{Meessen:2006tu,Huebscher:2008yz}.

The above result has some interesting implications, the first of which is derived by contracting eq.~(\ref{eq:VId2})
with $V^{a}V^{b}$, namely
\begin{equation}\label{eq:Res1}
  \langle\nabla_{V}\mathcal{R}\,|\,\mathcal{I}\rangle + \langle\mathcal{R}\,|\,\nabla_{V}\mathcal{I}\rangle \; =\; \nabla_{V}\frac{1}{2|X|^{2}}  \; =\; g\,\mathtt{C}_{\Lambda}\,\mathcal{R}^{\Lambda} \ .
\end{equation}
We can rewrite the above equation to a simpler form by observing that 
\begin{eqnarray}
\langle \mathcal{V}/X\,|\, d\left(\mathcal{V}/X\right)\rangle & =& X^{-2}\ \langle\mathcal{V}\,|\,\mathfrak{D}\mathcal{V}\rangle\; -\; X^{-3}\mathfrak{D}X\; \langle\mathcal{V}\,|\,\mathcal{V}\rangle \; =\; 0\nonumber \\
& =& \langle\mathcal{R}\,|\,d\mathcal{R}\rangle \ -\ \langle\mathcal{I}\,|\,d\mathcal{I}\rangle \; +\; i\langle\mathcal{R}\,|\,d\mathcal{I}\rangle\; +\; i\langle\mathcal{I}\,|\,d\mathcal{R}\rangle \ ,
\end{eqnarray}
which 
implies\footnote{These expressions were derived in \mycite{Bellorin:2006xr} starting from a prepotential and using the homogeneity of the symplectic section $\mathcal{R}$. The derivation presented here is far less involved, and also holds in situations where no prepotential exists.} that
\begin{eqnarray}
  \label{eq:StabId1}
  \langle d\mathcal{R}\,|\,\mathcal{I}\rangle & =& \langle\mathcal{R}\,|\,d\mathcal{I}\rangle \ ,\\
  \label{eq:StabId2}
  \langle\mathcal{R}\,|\,d\mathcal{R}\rangle & =& \langle\mathcal{I}\,|\,d\mathcal{I}\rangle \ .
\end{eqnarray}
If we then introduce the real symplectic section $\mathtt{C}^{T}=(0,\mathtt{C}_{\Lambda})$, we can rewrite
eq.~(\ref{eq:Res1}) in the simple and suggestive form
\begin{equation}
  \label{eq:VId5a}
  0\; =\; \langle\mathcal{R}\,|\,\nabla_{V}\mathcal{I}\;+\; \textstyle{\frac{g}{2}}\mathtt{C}\,\rangle \ .
\end{equation}
The above equation could also have been obtained from the contraction of eq.~(\ref{eq:VId1}) with $V$, {\em i.e.\/}
\begin{equation}
  \label{eq:VId5b}
  {\frac{1}{\overline{X}}}\ \mathtt{D}_{V} {\frac{1}{X}}\; =\; -g\langle\mathcal{R}|\mathtt{C}\rangle
            \, +\, ig\langle\mathcal{I}|\mathtt{C}\rangle \ ,
\end{equation}
and taking its real part. By taking the imaginary part and using the identity
\begin{equation}
  \mathrm{Im}\left({\frac{1}{\overline{X}}}\,\mathtt{D}{\frac{1}{X}}\right) \; =\; -2\langle\mathcal{I}\,|\,\mathtt{D}\mathcal{I}\rangle \ ,
\end{equation}
we also find that we must have
\begin{equation}
  \label{eq:VId5c}
  0\; =\; \langle\mathcal{I}\,|\,\nabla_{V}\mathcal{I}\; +\; \textstyle{\frac{g}{2}}\mathtt{C}\rangle \ .
\end{equation}

These equations suggest that the derivative of the symplectic section $\mathcal{I}$ in the direction $V$ is constant. We can confirm this by considering eqs.~(\ref{eq:20cj}) and (\ref{eq:20dj}). The contraction of eq.~(\ref{eq:20cj}) with $\bar{\epsilon}^{K}\gamma_{a}\varepsilon_{KI}$ reads
\begin{equation}
\label{eq:VId6}
2\overline{X}\ \mathtt{D}Z^{i} \; =\; 4\ \imath_{V}G^{i+} \; -\; \mathtt{W}^{i}\ V\ ,
\end{equation}
which upon contraction with $V$ leads to 
\begin{equation}
\label{eq:VId7}
\mathtt{D}_{V}Z^{i} \; =\; -2\,X\,\mathtt{W}^{i} \ .
\end{equation}
Using the gauge-fixing condition (\ref{eq:VId4}), the identity $\bar{f}^{\Lambda i}\mathtt{P}_{\Lambda}=i\overline{\mathcal{L}}^{\Lambda}\mathtt{K}_{\Lambda}^{i}$ and the fact that for our choice of possible non-Abelian gauge groups we have 
$\mathcal{L}^{\Lambda}\mathtt{K}_{\Lambda}^{i}=0$ (cf. eq.~(\ref{eq:SGK16})), the above equation gets converted to 
\begin{equation}
\label{eq:VId7a}
\nabla_{V}Z^{i} \; =\; -g\, X\, \bar{f}^{\Lambda i}\, \mathtt{C}_{\Lambda} \ .
\end{equation}
Using the Special Geometry identity $\langle\mathcal{U}_{i}\,|\,\overline{\mathcal{U}}_{\bar{\jmath}}\rangle = i\,\mathcal{G}_{i\bar{\jmath}}\,$, we can rewrite the above equation to
\begin{equation}
  \langle \nabla_{V}\mathcal{I}+g\,\mathtt{C}\,|\, \overline{\mathcal{U}}_{\bar{\jmath}}\rangle
    \; =\; i\langle\nabla_{V}\mathcal{R}\,|\,\overline{\mathcal{U}}_{\bar{\jmath}}\rangle \ ,
\end{equation}
which can be manipulated by using Special Geometry properties and a renewed call to eq.~(\ref{eq:VId7}), giving
\begin{equation}
  \label{eq:Vad6}
  \langle\nabla_{V}\mathcal{I}\; +\; \textstyle{\frac{g}{2}}\mathtt{C}\, |\, \overline{\mathcal{U}}_{\bar{\jmath}}\rangle
   \; =\; 0 \ .
\end{equation}
The above equation, plus eqs.~(\ref{eq:VId5a}) and (\ref{eq:VId5c}), together with the completeness relation from Special Geometry eq.~(\ref{eq:SGSymplProj}), implies then 
\begin{equation}
\label{eq:import}
\nabla_{V}\mathcal{I} \; =\; -\textstyle{\frac{g}{2}}\ \mathtt{C} \ ,
\end{equation}
which implies that the $\tau$-dependence of the functions $\mathcal{I}$ is actually \emph{at most} linear; in fact it is so only for the $\mathcal{I}_{\Lambda}$, as the $\mathcal{I}^{\Lambda}$ are $\tau$-independent. 

At this point it is necessary to introduce a complete coordinate system ($\tau, y^{m}$), which we shall take to be adapted to $V$ and compatible with the Fierz identities in appendix \ref{sec:Bil}
\begin{equation}
\label{eq:Coord}
\begin{array}{llllll}
V^{a}\partial_{a}&=&\sqrt{2}\partial_{\tau}\ , & V&=& 2\sqrt{2}|X|^{2}\left( d\tau \ +\ \omega\right)\ , \\
& & & & & \\
V^{xa}\partial_{a}&=& -2\sqrt{2}|X|^{2}V^{xm}\,\left(\partial_{m}-\omega_{m}\partial_{\tau}\right)\ ,& V^{x} & =& \sqrt{2}\ V^{x}_{m}\ dy^{m} \ ,
\end{array}
\end{equation}
where $\omega =\omega_{m}dy^{m}$ is a (possibly $\tau$-dependent) 1-form and we have introduced $V^{xm}$ by $V^{xm}V^{y}_{m}=\delta^{xy}$; as the $V^{x}_{m}$ act as a Dreibein on a Riemannian space, the $x$-indices can be raised and lowered
with $\delta^{xy}$, so that we shall not be distinguishing between co- and contravariant $x$-indices.

Putting the Vierbein together with the Fierz identity (\ref{eq:Bil4}), we find that the metric takes on the
conforma-stationary form
\begin{equation}
\label{eq:37j}
ds^{2} \; =\; 2|X|^{2}\left( d\tau \ +\ \omega\right)^{2}\; -\; \textstyle{\frac{1}{2|X|^2}}\ h_{mn} dy^{m}dy^{n} \ ,
\end{equation}
where $h_{mn}=V^{x}_{m}V^{x}_{n}$ is the metric on the three-dimensional base-space.

W.r.t.~our choice of coordinates we have that $\pounds_{V}V^{x}=0$ implies $\partial_{\tau}V^{x}_{m}=0$. The $V^{x}$ are of course also constrained by eq.~(\ref{eq:VId3}), which in the chosen coordinate system and using the decomposition
\begin{equation}
\label{eq:Dec1}
A^{\Lambda} \; =\; -\textstyle{\frac{1}{2}}\mathcal{R}^{\Lambda}\; V \; +\; \tilde{A}^{\Lambda}_{m}\ dy^{m}
\;\ \longrightarrow\;\ F^{\Lambda} \; =\; -\textstyle{\frac{1}{2}}\mathtt{D}\left[\mathcal{R}^{\Lambda}V\right]\; +\; \tilde{F}^{\Lambda} \ ,
\end{equation}
reads
\begin{equation}
  \label{eq:ResVx}
  dV^{x} \; =\; g\,\mathtt{C}_{\Lambda}\;\tilde{A}^{\Lambda} \wedge V^{x}
         \; +\;\textstyle{\frac{g}{4}}\;\mathtt{C}_{\Lambda}\mathcal{I}^{\Lambda}\,\varepsilon^{xyz}\;V^{y}\wedge V^{z}\ .
\end{equation}
For consistency we must have $\mathtt{C}_{\Lambda}\partial_{\tau}\tilde{A}^{\Lambda}=0$. Furthermore, we can use the residual gauge freedom $\mathtt{C}_{\Lambda}\tilde{A}^{\Lambda}\rightarrow \mathtt{C}_{\Lambda}\tilde{A}^{\Lambda} + d\phi (y)\,$, $V^{x}\rightarrow e^{g\phi}V^{x}$ to take $\mathtt{C}_{\Lambda}\mathcal{I}^{\Lambda}$ to be constant, a possibility we will however not use. At last, the integrability condition $d^{2}V^{x}=0$ implies
\begin{equation}
  \label{eq:ResVxInt}
  0\; =\; \textstyle{\frac{g}{4}} \left[\varepsilon^{xyz}\; \mathtt{C}_{\Lambda}\tilde{F}^{\Lambda}_{yz} 
              \; +\; \sqrt{2}\;V_{x}^{m} \tilde{\mathbb{D}}_{m}\; \mathtt{C}_{\Lambda}\mathcal{I}^{\Lambda}\right] \ ,
\end{equation}
where we have introduced $\tilde{F}^{\Lambda}_{xy}\equiv V_{x}^{m}V_{y}^{n}\tilde{F}^{\Lambda}_{mn}$ and 
\begin{equation}\label{eq:DefTildeD}
    \tilde{\mathbb{D}}_{m}\mathcal{I}\; =\; \partial_{m} \mathcal{I}\; +\; g\mathtt{C}_{\Lambda}\tilde{A}^{\Lambda}_{m}\ \mathcal{I}\; +\; g\tilde{A}^{\Lambda}_{m}\; S_{\Lambda}\mathcal{I}\ , \qquad \tilde{\mathbb{D}}_{x} \;\equiv\ V_{x}^{m}\ \tilde{\mathbb{D}}_{m} \ .
\end{equation}

The system leading to (\ref{eq:ResVx}) was analysed by Gauduchon and Tod in \mycite{Gauduchon:1998}, in the study of four-dimensional hyper-hermitian Riemannian metrics admitting a tri-holomorphic conformal Killing vector. They observed that the geometry of the base-space belongs to a subclass of three-dimensional Einstein-Weyl spaces, called hyper-CR or (subsequently) Gauduchon-Tod spaces\footnote{See appendix \ref{sec:EWspaces} for some technical information about these geometries.}. The additional constraint demanded on the EW spaces is nothing more than the integrability condition (\ref{eq:ResVxInt}), which is called the {\em generalised Abelian monopole equation}. We shall see below that the equation determining the seed function $\mathcal{I}^{\Lambda}$ will be a generalised non-Abelian monopole equation, \emph{i.e.}~the straightforward generalisation of the standard Bogomol'nyi equation on $\mathbb{R}^{3}$ to GT spaces. 
\vspace{\baselineskip}

In \mycite{Behrndt:2003cx}, Behrndt and Cveti\v{c} realised that their five-dimensional cosmological solutions could be dimensionally reduced to four-dimensional ones, which raises the question of whether/ which of the solutions found by Grover {\em et al.\/}~in \mycite{Grover:2008jr} can be reduced to solutions that we are classifying. In such a case, we would be deadling with a map between the five-dimensional timelike case and the four-dimensional timelike case, and thus the dimensional reduction has to be over the four-dimensional base-space, which was found to be hyperK\"ahler-torsion \mycite{Grover:2008jr}. The key to identifying the subclass of five-dimensional solutions that can be reduced to ours then lies in a further result of Gauduchon and Tod (see \mycite{Gauduchon:1998}{remark 2}), which states that solutions to eqs.~(\ref{eq:ResVx}) and (\ref{eq:ResVxInt}) are obtained by reduction of a conformal hyper-K\"ahler space along a tri-holomorphic Killing vector. As discussed in \mycite{Grover:2008jr}{sec.~3.2}, this kind of spaces are actually particular instances of HKT spaces, which thus allows for the reduction. In fact, this inheritance of geometrical structures also ocurrs in ordinary supergravity theories in six, five and four dimensions \mycite{Meessen:2007te}, and it is reasonable to suppose that this also holds for fake (Wick-rotated) supergravities. As a final comment, let us mention that the three-dimensional Killing spinor equation on a GT manifold allows non-trivial solutions \mycite{Buchholz:2000,Buchholz:2000b}.

Before turning to the equations of motion, we deduce the following equation for $\omega$ from the antisymmetrised version of eq.~(\ref{eq:VId2}) and the explicit coordinate expression in (\ref{eq:Coord}). As can be seen, this calculation needs the explicit form for the 2-form $\mathcal{T}^{+}$, which can be obtained from eq.~(\ref{eq:VId1}) and the rule that a general imaginary self-dual 2-form $B^{+}$ is determined by its contraction with $V$ by means of (see \mycite{Caldarelli:2003pb} for more detail)
\begin{equation}
  \label{eq:Tplus}
  B^{+} \; =\; \frac{1}{4|X|^{2}}\left(V\wedge\ \imath_{V}B^{+} \; +\; i\star\left[V\wedge\ \imath_{V}B^{+}   
                       \right]\right)\ .
\end{equation}
The result thus reads
\begin{equation}
  \label{eq:Rot}
  d\omega \; +\; g\mathtt{C}_{\Lambda}\tilde{A}^{\Lambda}\wedge (d\tau +\omega ) \; =\; 
  \sqrt{2}\;\star\left[V\wedge\; \langle\mathcal{I}\,|\,\mathtt{D}\,\mathcal{I}\rangle\right] \ .
\end{equation}
Contracting the above equation with $V$ we find that 
\begin{equation}
  \label{eq:OmegaV}
   \pounds_{V}\omega \; =\; g\sqrt{2}\mathtt{C}_{\Lambda}\tilde{A}^{\Lambda} \quad\longrightarrow\quad \omega \; =\; g\mathtt{C}_{\Lambda}\tilde{A}^{\Lambda}\ \tau \; +\; \varpi \ ,
\end{equation}
where $\varpi = \varpi_{m}dy^{m}$ is $\tau$-independent. Substituting the above result into eq.~(\ref{eq:Rot}) and evaluating its R.H.S.~we obtain
\begin{equation}
\label{eq:Rot2}
d\varpi \; +\; g\mathtt{C}_{\Lambda}\tilde{A}^{\Lambda}\wedge \varpi +g\mathtt{C}_{\Lambda}\tilde{F}^{\Lambda}\ \tau\; =\; \textstyle{\frac{1}{2}}\ \langle\mathcal{I}\,|\,\tilde{\mathtt{D}}_{m}\,\mathcal{I}\;-\; \omega_{m}\partial_{\tau}\mathcal{I}\rangle\; V^{xm}\varepsilon^{xyz}\; V^{y}\wedge V^{z}\ .
\end{equation}
If now take the $\tau$-derivative of this equation, we shall find again eq.~(\ref{eq:ResVxInt}). The equation determining $\varpi$ is then found by splitting up the $\tau$-dependent part; it reads
\begin{equation}
\label{eq:Rot3}
\tilde{\mathbb{D}}\,\varpi \; =\; \textstyle{\frac{1}{2}}\; \varepsilon^{xyz}\; \langle \tilde{\mathcal{I}}\,|\,\tilde{\mathtt{D}}_{x}\tilde{\mathcal{I}} - \varpi_{x}\partial_{\tau}\mathcal{I}\rangle\, V^{y}\wedge V^{z} \; ,
\end{equation}
where we have introduced $\tilde{\mathcal{I}}=\mathcal{I}\,(\text{at } \tau =0)$.

The symplectic field strength $F^{T}=(F^{\Lambda},F_{\Lambda})$ can then easily be deduced to give the standard
supersymmetric result
\begin{eqnarray}
\label{eq:Fsympl}
F &=& -\textstyle{\frac{1}{2}}\; \mathtt{D}\left( \mathcal{R}\ V\right)\; -\; \textstyle{\frac{1}{2}}\ \star\left[V\wedge\ \mathbb{D}\mathcal{I}\right] \nonumber \\ 
& =& -\textstyle{\frac{1}{2}}\ \mathtt{D}\left( \mathcal{R}\ V\right)\; -\; \frac{\sqrt{2}}{8}\ \varepsilon^{xyz}\ V_{x}^{m}\left[\tilde{\mathbb{D}}_{m}\mathcal{I}\; -\; \omega_{m}\partial_{\tau}\mathcal{I}\right]\ V^{y}\wedge V^{z} \ ,
\end{eqnarray}
which agrees completely with the imposed gauge-fixing (\ref{eq:VId4}).

At this point we shall treat the Bianchi identity $\mathtt{D}F^{\Lambda}=0$ as in \mycite{Huebscher:2008yz}, namely as leading to a Bogomol'nyi equation determining the pair $(\tilde{A}^{\Lambda},\mathcal{I}^{\Lambda})$. Since we were given the gauge potential in eq.~(\ref{eq:Dec1}), the Bianchi identity is solved identically, and thus does not imply further contraints. Nevertheless one needs to make sure that such potential leads to a field strength with the form prescribed by fakeSupersymmetry in eq.~(\ref{eq:Fsympl}). If we impose that, we obtain 
\begin{equation}
\label{eq:Bogo}
\tilde{F}^{\Lambda}_{xy} \; =\; -\frac{1}{\sqrt{2}}\; \varepsilon^{xyz}\, \tilde{\mathbb{D}}_{z}\mathcal{I}^{\Lambda} \ ,
\end{equation}
which due to eq.~(\ref{eq:import}) it is manifestly $\tau$-independent. 
This is the equation which we referred to above as the generalisation of the standard Bogomol'nyi equation on $\mathbb{R}^{3}$ to a three-dimensional Gauduchon-Tod space. One can see that, upon contraction with $\mathtt{C}_{\Lambda}$, the above equation implies the constraint (\ref{eq:ResVxInt}).\vspace{\baselineskip}

We would now like to show that the timelike solutions (to the fKSEs) we have just characterised are indeed solutions to the EOMs. For this we need to impose the Yang-Mills equations, eq.~(\ref{eq:VectEOM2}). This equation consists of two parts, namely one in the time-direction, {\em i.e.\/} $\mathcal{B}_{\Lambda}^{t}$, and one in the space-like directions, $\mathcal{B}_{\Lambda}^{x}$. A tedious but straightforward calculation shows that $\mathcal{B}_{\Lambda}^{t}=0$ identically, in full concordance with the discussion in section \ref{sec:DSVector}. The EOMs in the $x$-direction, however, do not vanish identically. Instead, they impose the condition
\begin{equation}
\label{eq:MYM1}
\left(\tilde{\mathbb{D}}_{x}\ -\ \omega_{x}\partial_{\tau}\right)^{2}\mathcal{I}_{\Lambda} \; =\; \textstyle{\frac{g^{2}}{2}}\ f_{\Lambda (\Omega}{}^{\Gamma}f_{\Delta )\Gamma}{}^{\Sigma}\mathcal{I}^{\Omega}\mathcal{I}^{\Delta}\ \mathcal{I}_{\Sigma}
     \; -\; \textstyle{\frac{g^{2}}{2}}\ f_{\Lambda\Omega}{}^{\Sigma}\mathcal{I}^{\Omega}\mathcal{I}_{\Sigma}\,\mathtt{C}_{\Gamma}\mathcal{I}^{\Gamma}\ ,
\end{equation}
which in the limit $\mathtt{C}\rightarrow 0$ (\emph{i.e.}~fSUSY $\rightarrow$ SUSY) coincides with the result obtained in \mycite{Huebscher:2008yz}. A simplification on the above equation can be performed by employing eqs.~(\ref{eq:DefTildeD}) and (\ref{eq:OmegaV}), which implies that 
\begin{equation}
\partial_{\tau}\left(\tilde{\mathbb{D}}_{m}\mathcal{I}_{\Lambda}\ -\ \omega_{m}\partial_{\tau}\mathcal{I}_{\Lambda}\right)\; =\; \partial_{\tau}\partial_{m}\mathcal{I}_{\Lambda} \; =\; 0\ .
\end{equation}
Using this and the fact that $\mathcal{I}_{\Lambda}$ is linear in $\tau$, we can thus rewrite eq.~(\ref{eq:MYM1}) as
\begin{equation}
\label{eq:MYM2}
\tilde{\mathbb{D}}_{x}^{2}\ \tilde{\mathcal{I}}_{\Lambda}\;-\; \left(\tilde{\mathbb{D}}_{x}\varpi_{x}\right)\ \partial_{\tau}\mathcal{I}_{\Lambda} \; =\; \textstyle{\frac{g^{2}}{2}}\ f_{\Lambda (\Omega}{}^{\Gamma}f_{\Delta )\Gamma}{}^{\Sigma}\mathcal{I}^{\Omega}\mathcal{I}^{\Delta}\ \tilde{\mathcal{I}}_{\Sigma}\; -\; \textstyle{\frac{g^{2}}{2}}\ f_{\Lambda\Omega}{}^{\Sigma}\mathcal{I}^{\Omega}\tilde{\mathcal{I}}_{\Sigma}\;\mathtt{C}_{\Gamma}\mathcal{I}^{\Gamma}\ ,
\end{equation}
which is a $\tau$-independent equation. 


\subsection{Summary and further comments}
\label{sec:CosmMon}
Before making some general comments on the behaviour of the solutions, we describe how to construct solutions using the results obtained in the foregoing section. The first step is to decide which model to consider, {\em i.e.\/} one has to specify what special geometric manifold to consider, what non-Abelian groups to gauge, and furthermore the constants $\mathtt{C}_{\Lambda}$. Given this model, one must then decide what GT space will be describing the geometry of the three-dimensional base-space; this is equivalent to finding the triplet $(V^{x}, \mathtt{C}_{\Lambda}\tilde{A}^{\Lambda},\mathtt{C}_{\Lambda}\mathcal{I}^{\Lambda})$ that solves eq.~(\ref{eq:ResVx}). This allows in principle to solve the Bogomol'nyi equation (\ref{eq:Bogo}), giving $(\tilde{A}^{\Lambda},\mathcal{I}^{\Lambda})$.

The next step is to determine the $\tau$-independent part of the seed functions $\mathcal{I}_{\Lambda}$ \footnote{The $\tau$-dependence is fixed by eq.~(\ref{eq:import}).}, using equation (\ref{eq:MYM2}). As this equation contains not only the $\mathcal{I}_{\Lambda}$ but also $\varpi$, one is forced to determine both objects, and to make sure that eq.~(\ref{eq:Rot3}) is satisfied. After this, the only steps left include to determine the field strengths by means of eq.~(\ref{eq:Fsympl}), and writing down the physical scalars $Z^{i}=\mathcal{L}^{i}/\mathcal{L}^{0}$ and the metric by determining the stationarity 1-form $\omega$ through eq.~(\ref{eq:OmegaV}) and the metrical factor $|X|^{2}$ through eq.~(\ref{eq:DefRandI}). As usual, the explicit construction of the last fields goes through solving the {\em stabilisation equation}, which determines the symplectic section $\mathcal{R}$ in terms of the seed functions $\mathcal{I}$. For many models, solutions to this are known. 

However, the Bogomol'nyi equation eq.~(\ref{eq:Bogo}) is not easy to solve. Since we only know explicit solutions to the 
Bogomol'nyi equation on $\mathbb{R}^{3}$ \footnote{Observe that this is purely a non-Abelian restriction, as GT metrics (solutions to the generalised Abelian monopole equation) are known, see {\em e.g.\/} \mycite{Dunajski:1999qs, Calderbank:1999ad}.}, this means that for the moment the only non-trivial non-Abelian solution backgrounds we can build are the ones that follow from the supersymmetric ones, satisfying $\mathtt{C}_{\Lambda}\mathcal{I}^{\Lambda}=0$, by using the substitution rule $\mathcal{I}_{\Lambda}\rightarrow \mathcal{I}_{\Lambda}-g\,\mathtt{C}_{\Lambda}\,\tau /(2\sqrt{2})$. This  implies that $\mathtt{C}_{\Lambda}\tilde{A}^{\Lambda}$ is gauge trivial, and thus the base-space is still $\mathbb{R}^{3}$. This being so, the equations determining the $\tau$-independent part of the $\mathcal{I}$, \emph{i.e.}~eqs.~(\ref{eq:Bogo}) and (\ref{eq:MYM2}), reduce to the ones for $\N=2$ E-YM deduced in \mycite{Huebscher:2008yz}; indeed the only difference lies in the divergence of $\varpi$ occurring in eq.~(\ref{eq:MYM2}), and in the $\mathbb{R}^{3}$-case there is no obstruction to choosing it to vanish from the onset.\vspace{\baselineskip}

The construction of fake-supersymmetric solutions then boils down to the substitution principle put forward by Behrndt and Cveti\v{c} in \mycite{Behrndt:2003cx}: given a supersymmetric solution to $\N=2$ $d=4$ E-YM Supergravity, Abelian \mycite{Behrndt:1997ny} or non-Abelian \mycite{Huebscher:2008yz,Meessen:2008kb}, substitute $\mathcal{I}_{\Lambda}\rightarrow \mathcal{I}_{\Lambda}-g\mathtt{C}_{\Lambda}\,\tau/(2\sqrt{2})$ and impose the restriction $\mathtt{C}_{\Lambda}\mathcal{I}^{\Lambda}=0$. As explained before, when dealing with non-Abelian gauge groups not all choices for $\mathtt{C}_{\Lambda}$ are possible, and one must respect the constraint $f_{\Lambda\Sigma}{}^{\Gamma}\mathtt{C}_{\Gamma}=0$.

The first observation is that generically the asymptotic form of the solution is not De Sitter but rather Kasner, {\em i.e.\/} the $\tau$-expansion of the base-space is power-like, making the definition of asymptotic mass even more cumbersome than in the De Sitter case\footnote{As a side note, note that the resulting Kasner spaces have a timelike conformal isometry of the kind used in \mycite{Kastor:2002fu} to define a conformal energy.}. The second observation is that the metric has a curvature singularity at those events/ points for which $|X|^{-2}=0$, 
which may be located outside our chosen coordinate system. This raises the question of horizons, or, in other words, how to decide in a practical manner when does our solution describe a black hole. Observe that in the original solution for one single black hole of Kastor and Traschen, this question is readily resolved by changing coordinates to obtain the time-independent spherically symmetric extreme RNdS black hole, for which the criteria to have an horizon is known: the existence of a black hole in the original coordinate system can be expressed in terms of the existence of a Killing horizon for a timelike Killing vector, which covers the singularity. It is interesting to note that generically there is no such timelike Killing vector in this case. 

To see it, consider for instance the $\overline{\mathbb{CP}}^{1}$-model. This model only has one complex scalar field $Z$ living on the coset space $SL(2;\mathbb{R})/SO(2)$ and an associated K\"ahler potential $e^{\mathcal{K}}=1-|Z|^{2}$,
so that we have the constraint $0\leq |Z|^{2}< 1$. Making the choice $\mathtt{C}_{\Lambda}=(-2,0)$, the potential can be readily calculated to be
\begin{equation}
\label{eq:PotCPn}
\mathtt{V} \; =\; 2g^{2}\; \left[1\; +\; 2\ e^{\mathcal{K}}\right]\ ,
\end{equation}
which is manifestly positive. Imposing $\mathcal{I}^{0}=0$ in order to have $\mathbb{R}^{3}$ as the base-space, and $\mathcal{I}_{1}=0$ in order to have a static solution, {\em i.e.\/} $\omega =0$, the EOMs imply that a simple solution is given by
\begin{equation}
  \label{eq:PotCPn1}
  \mathcal{I}_{0} \; =\; \frac{g\tau}{\sqrt{2}} \ ,\quad \mathcal{I}^{1} \; =\; \sqrt{2}\ g\lambda \quad\longrightarrow\quad  \frac{1}{2|X|^{2}} \;=\; g^{2}\left[ \tau^{2} \; -\; \lambda^{2}\right] \ , 
\end{equation}
where $\lambda$ is a real constant. If $\lambda =0$ the above solution leads to $dS_{4}$, whereas if $\lambda\neq 0$ we can introduce a new coordinate $t$ through $\tau =\lambda\cosh\left( gt\right)$, such that the solution is given by
\begin{eqnarray}
  \label{eq:PotCPn2}
  ds^{2} & =& dt^{2}\; -\; \sinh^{2}\left( gt\right)\; d\vec{x}^{2}_{(3)} \ ,\\
  Z     & =& -i\, \cosh^{-1}\left( gt\right) \ .
\end{eqnarray}
This says that at late times the metric is $dS_{4}$, but it is singular when $t=0$; at that point in time also the scalar becomes problematic, as $|Z\,(t=0)|^{2}=1$, violating the bound, which in its turn implies that the contribution of the scalars to the energy-momentum tensor blows up. It is paramount that in this case no timelike Killing vector exists. Had we on the other hand taken $\mathcal{I}^{1}= \sqrt{2}gpr^{-1}$, for which a timelike Killing vector does exist, the metric can be transformed to the static form
\begin{equation}
  \label{eq:PotCPn3}
  ds^{2} \; =\; \textstyle{p^{2} + R^{2}-g^{2}R^{4}\over R^{2}}\ dt^{2}\; -\; \textstyle{ \frac{R^{4}}{(R^{2}+p^{2})(p^{2} + R^{2}-g^{2}R^{4}/4)}}\ dR^{2}\; -\; R^{2}dS^{2} \ .
\end{equation}
This metric has one Killing horizon, identified with the cosmological horizon for $R>0$, and has therefore a naked singularity located at $R=0$. In the static coordinates the scalar field reads $Z=-ip\ {\textstyle (p^{2}+R^{2})^{-1/2}}$, which explicitly breaks the bound $0\leq |Z|^{2}< 1$ at $R=0$, showing again the link between the regularity of the metric and that of the scalars.

In view of all this, it would be very desirable to have at hand a manageable prescription for deciding when a solution describes a black hole. In this respect, we would like to mention the isolated horizon formalism (see {\em e.g.~}\mycite{Ashtekar:2004cn}), which attempts to give a local definition of horizons, without a reference to the existence of timelike Killing vectors. This formalism was applied in the context of SUGRA in \mycite{Liko:2007mu,Booth:2008ru}, and similar work for fake SUGRAs might prove of interest.

\section{Analysis of the null case}
\label{sec:Null}
In this section we characterise the fake-supersymmetric solutions in the null case, \emph{i.e.}~when $V^{2}=0$. For simplicity we shall restrict ourselves to theories with no YM-type couplings; a full analysis is possible \mycite{Huebscher:2008yz}, but likely to not be very rewarding. As in the timelike case, the difference with the supersymmetric case lies in the fact that the vector-bilinear is not a Killing vector. Furthermore, introducing an adapted coordinate $v$ through $L=L^{a}\partial_{a}=\partial_{v}$, one can see that the metric will be explicitly $v$-dependent, unlike in the supersymmetric case. The aim of this section is then to determine this $v$-dependence, and to give two minimal and simple (albeit generic) solutions, that illustrate the changes generated by the $\mathbb{R}$-gauging.

In the null case the norm of the vector $V$ vanishes, whence $X=0$. This means that the two spinors $\epsilon_{I}$
are parallel, and following \mycite{Meessen:2006tu,Huebscher:2008yz} we shall put $\epsilon_{I}=\phi_{I}\,\epsilon$, for some functions $\phi_{I}$ and the independent spinor $\epsilon$. The decomposition of $\epsilon^{I}$ follows from its definition as $\epsilon^{I}=(\epsilon_{I})^{*}$, which then implies that $\epsilon^{I}=\phi^{I}\epsilon^{*}$, where we have defined $\phi^{I}=\overline{\phi_{I}}$. Furthermore, we can without loss of generality normalise the functions $\phi$ such that $\phi_{I}\phi^{I}=1$. Having taken into account this normalisation, one can write down the following completeness relation for the $I$-indices
\begin{equation}
  \label{eq:21j}
  \Delta_{I}{}^{J} \; =\; \phi_{I}\phi^{J} \; +\; \varepsilon_{IK}\Phi^{K}\ \varepsilon^{JL}\Phi_{L} \ ,
\end{equation}
which is such that $\Delta_{I}{}^{J}\phi_{J}=\phi_{I}\,$, $\Delta_{I}{}^{J} \varepsilon_{JK}\phi^{K}=\varepsilon_{IK}\phi^{K}$. Moreover, one can see that $\overline{\Delta_{I}{}^{J}}=\Delta_{J}{}^{I}$.

Projecting the fKSEs (\ref{eq:20j})-(\ref{eq:20dj}) onto the functions we obtain
\begin{eqnarray}
  \label{eq:22j}
  0 & =& \mathbb{D}_{a}\epsilon \; +\; \phi^{I}\nabla_{a}\phi_{I}\, \epsilon \ , \\
  \label{eq:22bj}
  0 & =& \left( \mathcal{T}_{ab}^{+}\ +\ \textstyle{ig\over 4}\mathtt{C}_{\Lambda}\mathcal{L}^{\Lambda}\ \eta_{ab}\right)\ \gamma^{b}\epsilon^{*} \; -\; \varepsilon^{IJ}\phi_{I}\nabla_{a}\phi_{J}\ \epsilon \ ,\\
  \label{eq:22cj}
  0 & =& i\slashed{\partial}Z^{i}\ \epsilon^{*} \ , \\
  \label{eq:22dj}
  0 & =& \left[ \slashed{G}^{i+} \; +\: \mathtt{W}^{i}\right]\ \epsilon \ .
\end{eqnarray}
We shall now introduce an auxiliary spinor $\eta$, normalised by $\overline{\epsilon}\eta = \textstyle{1\over \sqrt{2}} = -\overline{\eta}\epsilon$. This spinorial field allows us to introduce four new null vectors
\begin{equation}
  \label{eq:23j}
  \begin{array}{lcllcl}
    L_{a} & =& i\overline{\epsilon}\gamma_{a}\epsilon^{*}\ ,&\qquad N_{a} & =& i\overline{\eta}\gamma_{a}\eta^{*} \ ,\\
    M_{a} & =& i\overline{\eta}\gamma_{a}\epsilon^{*}\ ,& \qquad\overline{M}_{a} & =& i\overline{\epsilon}\gamma_{a}\eta^{*} \ ,
  \end{array}
\end{equation}
where $L$ and $N$ are real vectors and by construction $M^{*}=\overline{M}$, whence the notation. Observe that eq.~(\ref{eq:Bil2a}) implies that the vector $L$ is nothing but $V$, but where it has been denoted by $L$(ightlike), as we are now in the null case. Given the above definitions, it is a tedious yet straightforward calculation to show that they form an ordinary normalised null tetrad,  {\em i.e.\/} the only non-vanishing contractions are
\begin{equation}
  \label{eq:1j}
  L^{a}\, N_{a} \; =\; 1 \; =\; -M^{a}\, \overline{M}_{a}\ ,
\end{equation}
which implies that
\begin{equation}
\eta_{ab} \; =\; 2\, L_{(a}N_{b)} \; -\; 2\, M_{(a}\overline{M}_{b)} \ .
\end{equation}

Apart from these vectors, one can also define imaginary self-dual 2-forms analogous to the ones defined in eq.~(\ref{eq:Bil3}), by
\begin{equation}
  \label{eq:2j}
  \begin{array}{lcllcl}
    \Phi^{1}_{ab} & \equiv& \overline{\epsilon}\gamma_{ab}\epsilon\ ,&\qquad \Phi^{1} & =& \sqrt{2}\ L\wedge \overline{M} \ , \\
     \Phi^{2}_{ab} & \equiv& \overline{\epsilon}\gamma_{ab}\eta\ ,& \qquad\Phi^{2} & =& \textstyle{1\over \sqrt{2}}\left[ L\wedge N \ +\ M\wedge \overline{M}\right] \ , \\
     \Phi^{3}_{ab} & \equiv& \overline{\eta}\gamma_{ab}\eta\ , &\qquad\Phi^{3} & =& -\sqrt{2}\ N\wedge M \ , \\
  \end{array}
\end{equation}
where the identification on the RHS follows from the Fierz identities.

The introduction of the above auxiliary spinorial field is not unique, and one still has the freedom to rotate $\epsilon$ and $\eta$ by $\epsilon\rightarrow e^{i\theta}\epsilon$ and $\eta\rightarrow e^{-i\theta}\eta$. This does not affect $L$ nor $N$, but it does produce $M\rightarrow e^{-2i\theta}M$ and $\overline{M}\rightarrow e^{2i\theta}\overline{M}$, which we shall use it to get rid of a phase factor when introducing a coordinate expression for the tetrad. A second freedom is the shift $\eta\rightarrow \eta +\delta\,\epsilon$, with $\delta$ a complex function. This shift does not affect the normalisation condition, but on the vectors it generates
\begin{equation}
  \label{eq:3j}
  L\rightarrow L \ ,\qquad M\rightarrow M \; +\; \delta\, L \ ,\qquad N\rightarrow N \; +\; |\delta|^{2}\, L \;+\; \delta\, \overline{M} \; +\; \bar{\delta}\, M \ , 
\end{equation}
and can also be used to restrict the coordinate expressions of the tetrad.\vspace{\baselineskip}

The first step is introducing a coordinate $v$ through $L^{\flat}\equiv L^{a}\partial_{a}=\partial_{v}$, and using eq.~(\ref{eq:22j}) to derive
\begin{equation}
 \label{eq:5j}
   \nabla_{a}L_{b} \; =\; g\mathtt{C}_{\Lambda}A^{\Lambda}_{a}\, L_{b} \ .
\end{equation}
This equation says that $L$ is a recurrent null vector. Having such a vector field is the defining property of a space with holonomy $\mathrm{Sim}(d-2)$ (see \mycite{Gibbons:2007zu} for more information) and the combination $g\mathtt{C}_{\Lambda}A^{\Lambda}$ is called the recurrence 1-form. Antisymmetrising this expression we see that $dL= g\mathtt{C}_{\Lambda}A^{\Lambda}\wedge L$, which implies not only $\mathtt{C}_{\Lambda}F^{\Lambda}\wedge L =0\,$, but also $L\wedge dL =0$. This last result states that the vector $L$ is hyper-surface orthogonal, which implies the local existence of functions $Y$ and $u$ such that $L = Ydu$. Seeing that $L$ is charged under the $\mathbb{R}$-symmetry, we can always gauge-transform the function $Y$ away, arriving at
\begin{equation}
L=du\ ,
\end{equation}
whence also that $\mathtt{C}_{\Lambda}A^{\Lambda} = \Upsilon\, L\,$, for some function $\Upsilon$. We can write eq.~(\ref{eq:5j}) as 
\begin{equation}
  \label{eq:6j}
  \nabla_{a}\, L_{b} \; =\; g\Upsilon\, L_{a}L_{b} \ ,\quad\mbox{which immediately implies}\qquad \nabla_{L}L \; =\; 0 \ ,
\end{equation}
so that $L$ is a geodesic null vector. Given this information and the normalisation of the tetrad, we can choose coordinates $u$, $v$, $z$ and $\bar{z}$ such that
\begin{equation}
  \label{eq:8j}
  \begin{array}{lclclcl}
    L & =& du \ ,& &L^{\flat} & =& \partial_{v} \ ,   \\
    N & =& dv \ +\ Hdu \ +\varpi dz\ + \overline{\varpi}d\bar{z}\ ,&\quad& N^{\flat} & =& \partial_{u} \ -\ H\partial_{v} \ ,\\
    M & =& e^{U}dz  \ ,& & M^{\flat} & =& -e^{-U}\left( \partial_{\bar{z}} \ -\ \overline{\varpi}\partial_{v}\right) \ ,\\
    \overline{M} & =& e^{U}d\bar{z} \ ,&  &\overline{M}^{\flat} & =& -e^{-U}\left( \partial_{z} \ -\ \varpi\partial_{v}\right) \ ,
  \end{array}
\end{equation}
where we have used the $U(1)$ rotation $M\rightarrow e^{-2i\theta}M$ to get rid of a possible phase in the expression of $M$ and $\overline{M}$. The spin connection and the curvatures for this tetrad are given in appendix \ref{sec:NullCurv}. A last implication of the Fierz identities is that 
\begin{equation}
  \label{eq:4j}
  \varepsilon_{(4)} \;\equiv\; \textstyle{1\over 4!}\ \varepsilon_{abcd}\ e^{a}\wedge e^{b}\wedge e^{c}\wedge e^{d}
                   \; =\; i\ L\wedge N\wedge M\wedge \overline{M} \; =\; i\ e^{+}\wedge e^{-}\wedge e^{\bullet}\wedge e^{\bar{\bullet}} \ ,
\end{equation}
whence $\varepsilon^{+-\bullet\bar{\bullet}}=i$. Furthermore, one finds from (\ref{eq:6j})
\begin{equation}
  \label{eq:9j}
  \partial_{v}H \; =\; g\Upsilon \ ,\quad\mbox{and}\quad 0\ =\ \partial_{v}U \; =\; \partial_{v}\varpi\; =\; \partial_{v}\overline{\varpi} \ ,
\end{equation}
whence the only $v$-dependence of the metric resides in $H$. The resulting form of the metric 
\begin{equation}
\label{eq:Walkerfourdmetric}
ds^2=2du(dv+Hdu+\varpi dz+\overline{\varpi} d\bar{z})-2e^{2U}dzd\bar{z}
\end{equation}
is a Kundt wave, that is, it admits a non-expanding, shear- and twist-free geodesic null vector (cf. appendix \ref{appsec:Kundt} for more details). To see this take \emph{e.g.}~$L^\flat=\partial_v$. Moreover, it is in the Walker form, which implies that the space has holonomy contained in $\Sim(2)$ \mycite{Walker:1950}.

To determine $\Upsilon$ we shall be using the identity $\mathtt{C}_{\Lambda}F^{\Lambda} = d\left( \mathtt{C}_{\Lambda}A^{\Lambda}\right) = d\Upsilon\wedge L$, which presupposes knowing $F^{\Lambda}$. The generic form of $F^{\Lambda}$ can be derived from the fKSEs (\ref{eq:22bj}) and (\ref{eq:22dj}). Contraction of the first with $i\overline{\epsilon}$ and $i\overline{\eta}$ leads to
\begin{eqnarray}
  \label{eq:10j}
  \imath_{L}\mathcal{T}^{+} & =& \textstyle{ig\over 4}\ \mathtt{C}_{\Lambda}\mathcal{L}^{\Lambda}\ L \ ,\\
  \label{eq:10aj}
  \imath_{M}\mathcal{T}^{+} & =& \textstyle{ig\over 4}\ \mathtt{C}_{\Lambda}\mathcal{L}^{\Lambda}\ M \; +\; \textstyle{i\over \sqrt{2}} \phi_{I}\varepsilon^{IJ}\ d\phi_{J} \ .
\end{eqnarray}
Using these and the fact that, as $\mathcal{T}^{+}$ is an imaginary-self-dual 2-form it must be expressible in terms of the $\Phi$s defined in eq.~(\ref{eq:2j}), one obtains 
\begin{equation}
  \label{eq:11j}
  \mathcal{T}^{+} \; =\; \aleph\ L\wedge\overline{M} \; -\; \textstyle{ig\over 4}\ \mathtt{C}_{\Lambda}\mathcal{L}^{\Lambda}\, \left[ L\wedge N + M\wedge \overline{M}\right]\ , 
\end{equation}
with $\sqrt{2}\aleph \; =\;  i\phi_{I}\varepsilon^{IJ}\nabla_{N}\phi_{J}$, and moreover
\begin{equation}
  \label{eq:13j}
  \sqrt{2}\phi_{I}\varepsilon^{IJ}\nabla_{\overline{M}}\phi_{J} \; =\; g\mathtt{C}_{\Lambda}\mathcal{L}^{\Lambda}
  \ ,\qquad 0 \; =\; \phi_{I}\varepsilon^{IJ}\nabla_{L}\phi_{J} \; =\; \phi_{I}\varepsilon^{IJ}\nabla_{M}\phi_{J} \ .
\end{equation}
Giving eq.~(\ref{eq:22dj}) a similar treatment leads to 
\begin{equation}
  \label{eq:14j}
  G^{i+} \; =\; \aleph^{i}\ L\wedge\overline{M} \; -\; \textstyle{1\over 4}\ \mathtt{W}^{i} \left[ L\wedge N \; +\; M\wedge\overline{M}\right] \ ,
\end{equation}
where $\aleph^{i}$ are still undetermined functions. Using the rule $F^{\Lambda +}=i\overline{\mathcal{L}}^{\Lambda}\mathcal{T}^{+}+2f_{i}^{\Lambda}G^{i+}$ we find that 
\begin{equation}
\label{eq:15j}
F^{\Lambda +} \ =\ \varphi^{\Lambda}\ L\wedge\overline{M} \; +\; V^{\Lambda}\ \left[ L\wedge N + M\wedge\overline{M}\right] \ ,
\end{equation}
where we have introduced
\begin{equation}
  \label{eq:DefVLambda}
  V^{\Lambda} = \textstyle{g\over 8} \left(4\overline{\mathcal{L}}^{\Lambda}\mathcal{L}^{\Sigma} + \mathrm{Im}(\mathcal{N})^{-1|\Lambda\Sigma}\right)\mathtt{C}_{\Sigma}
\end{equation}
and 
\begin{equation}
\label{eq:17j}
\aleph=2i\,\mathcal{L}_{\Lambda}\ \varphi^{\Lambda}\ ,\quad \aleph^{i} \; =\; -\bar{f}_{\Lambda}^{i}\,\varphi^{\Lambda} \quad \longleftrightarrow \quad \varphi^{\Lambda} \; =\; i\aleph\ \overline{\mathcal{L}}^{\Lambda} + 2\aleph^{i}\,f_{i}^\Lambda \ .
\end{equation}
Using $F^{\Lambda} = F^{\Lambda +}+F^{\Lambda -}= 2\mathrm{Re}\left( F^{\Lambda +}\right)$ and using $d\Upsilon\wedge L=\mathtt{C}_{\Lambda}F^{\Lambda}$, we obtain
\begin{eqnarray}
  \label{eq:18j}
  \nabla_{L}\Upsilon & =& -\mathtt{C}_{\Lambda}\left[ V \ +\ \overline{V}\right]^{\Lambda} \ ,\\
  \label{eq:18bj}
  \nabla_{M}\Upsilon & =& \mathtt{C}_{\Lambda}\ \varphi^{\Lambda} \ ,\\
  \label{eq:18cj}
  \nabla_{\overline{M}}\Upsilon & =& \mathtt{C}_{\Lambda}\, \overline{\varphi}^{\Lambda} \ .
\end{eqnarray}
Eq.~(\ref{eq:18j}) is the key to the possible $v$-dependence. We want to integrate it to obtain $H$ through eq.~(\ref{eq:9j}); for this we need to know the coordinate dependence of the scalars $Z$. This can be obtained by contracting eq.~(\ref{eq:22cj}) with the $i\overline{\epsilon}$ and $i\overline{\eta}$. The result is that
\begin{equation}
  \label{eq:19j}
  0\, =\, \nabla_{L} Z^{i} \, =\, \partial_{v}Z^{i}\ ,\qquad 0\, =\, \nabla_{M} Z^{i} \, =\, e^{-U}\ \partial_{\bar{z}}Z^{i} \ ,
\end{equation}
so that the $Z^{i}$ depend only on $u$ and $z$. Likewise, the $\overline{Z}^{\bar{\imath}}$ depend only on $u$ and $\bar{z}$.


Using the fact that the scalars are $v$-independent, integration of eq.~(\ref{eq:18j}) is straightforward and leads to
\begin{eqnarray}
  \label{eq:7j}
  \Upsilon & =& -\textstyle{g\over 4}\left[4\left|\mathtt{C}_{\Lambda}\mathcal{L}^{\Lambda}\right|^{2}
                      \; +\; \mathrm{Im}\left(\mathcal{N}\right)^{-1|\Lambda\Sigma}\mathtt{C}_{\Lambda}\mathtt{C}_{\Sigma} 
                  \right]\, v \; +\; \Upsilon_{1}(u,z,\bar{z}) \ , \\
  \label{eq:7aj}
  H & =& -\textstyle{g^{2}\over 8}\left[4\left|\mathtt{C}_{\Lambda}\mathcal{L}^{\Lambda}\right|^{2}
                      \; +\; \mathrm{Im}\left(\mathcal{N}\right)^{-1|\Lambda\Sigma}\mathtt{C}_{\Lambda}\mathtt{C}_{\Sigma} 
                  \right]\; v^{2} \; +\; \Upsilon_{1}\ v \ +\ \Upsilon_{0}(u,z,\bar{z}) \ .
\end{eqnarray}
We could take $\Upsilon_{1}=0$ by doing a coordinate transformation $v\rightarrow v + f(u,z,\bar{z})$, but we shall be ignoring this possibility for the moment. $H$ can be written in terms of the potential $\mathtt{V}$ in eq.~(\ref{eq:Potential}) with $\mathtt{P}_{\Lambda}=0$, since we are ignoring possible non-Abelian couplings, as
\begin{equation}
  \label{eq:7bj}
  H \; =\; \textstyle{1\over 2}\left[ g^{2}\left|\mathtt{C}_{\Lambda}\mathcal{L}^{\Lambda}\right|^{2} \; -\; \mathtt{V}
         \right]\, v^{2} \; +\; \Upsilon_{1}v \; +\; \Upsilon_{0} \ ,
\end{equation}
which is calculationally advantageous when $\mathtt{V}$ is known.

At this point we have nearly completely specified the $v$-dependence of the solution, the only field missing being $A^{\Lambda}$; with this in mind it is worthwhile to impose the gauge-fixing $\imath_{L}A^{\Lambda}=0$, which is always possible and furthermore it is consistent with the earlier result $\mathtt{C}_{\Lambda}A^{\Lambda}=\Upsilon\ L$. As a result of this gauge fixing we have that 
\begin{equation}
  \label{eq:7cj}
  \partial_{v}A^{\Lambda} \; =\; \pounds_{L}A^{\Lambda} \; =\; d\left(\imath_{L}F^{\Lambda}\right)\; =\; -\left(V\; +\; \overline{V}\,\right)^{\Lambda}\, L \ ,
\end{equation}
so that 
\begin{equation}
  \label{eq:7dj}
  A^{\Lambda} \; =\; -\left(V\; +\; \overline{V}\; \right)^{\Lambda}\, v\, L \; +\; \tilde{A}^{\Lambda} 
                   \; =\; \textstyle{g\over 4}\; \mathsf{F}^{-1|\Lambda\Sigma}\,\mathtt{C}_{\Sigma}\, v\, L \; +\; \tilde{A}^{\Lambda} \ ,
\end{equation}
where $\tilde{A}^{\Lambda}$ is a $v$-independent 1-form satisfying $\imath_{L}\tilde{A}^{\Lambda}=0$, and $\mathsf{F}$ is the imaginary part of the prepotential's Hessian, cf. eq.~(\ref{eq:PrePotImNinv}). Given this expression, the Bianchi identity is automatically satisfied, but just as in the timelike case this does not (necessarily) mean that any $\tilde{A}^{\Lambda}$ leads to a field strength of the desired form. Calculating the comparison one finds that
\begin{equation}
\label{eq:7ej}
\begin{array}{ccl}
d\tilde{A}^{\Lambda} & =& \left( V-\overline{V}\right)^{\Lambda}\ M\wedge\overline{M}  \\
&& +\left( \phi^{\Lambda}+\theta_{M} (v\left( V+\overline{V}\right)^{\Lambda})\right)\; L\wedge\overline{M} \; +\; \left( \overline{\phi}^{\Lambda}+\theta_{\overline{M}}( v\left( V+\overline{V}\right)^{\Lambda})\right) L\wedge M \ .
\end{array}
\end{equation}

Let us at this point return to the fKSEs, and evaluate eq.~(\ref{eq:20cj}) using eqs.~(\ref{eq:14j}) and (\ref{eq:19j}). This results in 
\begin{equation}
  \label{eq:26j}
  i\theta_{+}Z^{i}\, \gamma^{+}\epsilon^{I} \; +\; i\theta_{\bullet}Z^{i}\, \gamma^{\bullet}\epsilon^{I} \; =\;
    -\varepsilon^{IJ}\left[ \mathtt{W}^{i}\gamma^{-}\; -\; 2\alpha^{i}\gamma^{\bar{\bullet}}\right]\ \gamma^{+}\epsilon_{J} \ .
\end{equation}
The above equation is readily seen to be solved by observing that the constraint
\begin{equation}
\gamma^{+}\epsilon_{J}=0
\end{equation}
leads to $\gamma^{+}\epsilon^{I}=0$ under complex conjugation, as well as to $\gamma^{\bar{\bullet}}\epsilon_{I}=0$ and
$\gamma^{\bullet}\epsilon^{I}=0$, due to having chiral spinors and the normalisation prescribed in eq.~(\ref{eq:4j}). A similar analysis on the fKSE (\ref{eq:20j}) in the $v$-direction shows that the spinor $\epsilon_{I}$ is $v$-independent, and thus also $\epsilon^{I}$. The other equations become
\begin{eqnarray}
  \label{eq:27j}
  \mathbb{D}_{\bar{\bullet}}\epsilon_{I} & =& 0\ ,\\
  \label{eq:27aj}
  \mathbb{D}_{\bullet}\epsilon_{I} & =& \textstyle{ig\over 2}\, \mathtt{C}_{\Lambda}\mathcal{\Lambda}\ \varepsilon_{IJ}\gamma^{\bar{\bullet}}\epsilon^{J} \ , \\
  \label{eq:27cj}
  \mathbb{D}_{+}\epsilon_{I} & =& -\aleph\, \varepsilon_{IJ}\gamma^{\bar{\bullet}}\epsilon^{J} \ .
\end{eqnarray}

Using the definition (\ref{eq:16j}) and the spin connection in eq.~(\ref{eq:NCspincon}), we can expand eqs.~(\ref{eq:27j}) and (\ref{eq:27aj}) as
\begin{eqnarray}
  \label{eq:28j}
  0 & =& \theta_{\bar{\bullet}}\epsilon_{I} \; -\; \textstyle{1\over 2}\theta_{\bar{\bullet}}\left( U +\textstyle{1\over 2}\mathcal{K}\right)\ \epsilon_{I} \ , \\
  \label{eq:28aj}
  0 & =& \theta_{\bullet}\epsilon_{I} \; +\; \textstyle{1\over 2}\theta_{\bullet}\left( U +\textstyle{1\over 2}\mathcal{K}\right)\ \epsilon_{I} \; -\; \textstyle{ig\over 2}\mathtt{C}_{\Lambda}\mathcal{L}^{\Lambda}\ \varepsilon_{IJ}\gamma^{\bar{\bullet}}\epsilon^{J} \ ,
\end{eqnarray}
The first is easily integrated by putting
\begin{equation}
\label{eq:29j}
\epsilon_{I} \; =\; \exp\left( \textstyle{1\over 2}\, S\right)\, \chi_{I} (u,z), \qquad\mbox{with}\qquad S \, \equiv\, U + \textstyle{1\over 2}\mathcal{K} \ ,
\end{equation}
which upon substitution into eq.~(\ref{eq:28aj}) leads to
\begin{equation}
  \label{eq:30j}
  \partial_{z}\chi_{I} \; +\; \left(\partial_{z}S\right)\chi_{I} \; =\; \textstyle{ig\over 2}\, \mathtt{C}_{\Lambda}\mathcal{X}^{\Lambda}\,\varepsilon_{IJ}\gamma^{\bar{\bullet}}\ e^{S}\chi^{J} \ .
\end{equation}
This last equation is potentially dangerous, as it has a residual $\bar{z}$-dependence (even though $\eta$ and $\mathcal{X}^{\Lambda}$ are $\bar{z}$-independent). We use avoiding this inconsistency as a mean to fix $S$; deriving eq.~(\ref{eq:30j}) w.r.t.~$\bar{z}$ and using the complex conjugated version of eq.~(\ref{eq:30j}) to get rid of $\eta^{I}$ in the resulting equations, the result is that $S$ has to satisfy
\begin{equation}
  \label{eq:31j}
  \partial_{z}\partial_{\bar{z}}S \; =\; -\textstyle{g^{2}\over 2}\, e^{2S}\, \left|\mathtt{C}_{\Lambda}\mathcal{X}^{\Lambda}\right|^{2} \quad\longrightarrow\quad e^{-2S} \, =\, \textstyle{g^{2}\over 2}\, \left|\mathtt{C}_{\Lambda}\mathcal{X}^{\Lambda}\right|^{2}\, \left( 1+|z|^{2}\right)^{2} \ .
\end{equation}
This unique choice for $S$ is a necessary condition for eqs.~(\ref{eq:28j}) and (\ref{eq:28aj}) admitting a solution, but it may not be sufficient. In the next subsection we shall discuss the simplest null case solution to the minimal theory, and show that the system can indeed be solved completely. The lesson to be learnt is that, once we introduce $S$, the system (\ref{eq:28j}, \ref{eq:28aj}) corresponds to an equation determining spinors on a 2-sphere, and it has solutions. 
In fact, the $\mathtt{C}_{\Lambda}\mathcal{X}^{\Lambda}$ factor can be absorbed by redefining $\chi_{I}=\sqrt{\mathtt{C}_{\Lambda}\mathcal{X}^{\Lambda}}\, \eta_{I}\,$, which converts eq.~(\ref{eq:31j}) into
\begin{equation}
\label{eq:31aj}
\partial_{z}\eta_{I} \ +\ \left(\partial_{z}\tilde{S}\right)\eta_{I} \, =\, \textstyle{ig\over 2}\,\varepsilon_{IJ}\gamma^{\bar{\bullet}}\ e^{\tilde{S}}\eta^{J}\ ,\qquad\mbox{with}\quad e^{-\tilde{S}} \ =\ \textstyle{g\over\sqrt{2}}\left( 1+|z|^{2}\right) \; , 
\end{equation}
which is just the spinor equation on $S^{2}$ in stereographic coordinates.

\subsection{The electrically-charged Nariai cosmos}
\label{sec:NullSols}
The minimal theory is obtained by selecting $\mathcal{V}^{T}=(1,-i/2)$, which leads to the monodromy matrix $\mathcal{N}=-i/2\,$, so that $\mathrm{Re}(\mathcal{N})=0$. If we further fix $\mathtt{C}_{0}=2$, the minimal De Sitter theory action
is given by
\begin{equation}
  \label{eq:MinDSaction}
  \int_{4}\sqrt{g}\left( R - F^{2} - 6g^{2}\right) \ .
\end{equation}
Using the general results obtained earlier, we can write down the following solution
\begin{eqnarray}
  \label{eq:NullNariai}
  ds^{2} & =& 2du\left( dv -g^{2}v^{2}du\right) - \frac{dzd\bar{z}}{g^{2}(1+|z|^{2})^{2}} \ ,\nonumber \\
  A     & =& -gv\,du\; .
\end{eqnarray}
This metric is nothing more than $dS_{2}\times S^{2}$, albeit in a non-standard coordinate system, and the solutions is known to the literature as the \emph{electrically-charged Nariai solution} \mycite{Nariai:1950, Nariai:1951}. Note that the local holonomy of the Nariai solution is not the full $\mathfrak{sim}(2)$, but rather $\mathfrak{so}(1,1)\oplus\mathfrak{so}(2)\subset \mathfrak{sim}(2)$ \mycite{Gibbons:2007zu}.

We now proceed to discuss the preserved fake-supersymmetries. For this, it is easier to write the metric as
\begin{equation}
\label{eq:NullNariai1}
ds^{2} \; =\; 2du\left( dv -g^{2}v^{2}du\right)-\frac{1}{4g^{2}}\left( d\theta^{2}+ \sin^{2}(\theta )d\varphi^{2}\right) \ ,
\end{equation}
and consider the fakeSupergravity equations in terms of a 2-component vector of Majorana spinors, also denoted by $\epsilon$,
\begin{equation}
\label{eq:NullNariai2}
\nabla_{a}\epsilon \ -\ gA_{a}\epsilon \; =\; -\textstyle{1\over 4}\slashed{F}\gamma_{a}\sigma^{2}\epsilon-\textstyle{g\over 2}\gamma_{a}\sigma^{2}\epsilon \ .
\end{equation}
The solution to the above equation is seen to be 
\begin{equation}
\label{eq:NullNariai3}
\epsilon \; =\; \exp\left( \textstyle{\theta\over 2}\gamma^{3}\sigma^{2}\right)\,\exp\left( -\textstyle{\varphi\over 2}\gamma^{34}\right)\, \epsilon_{0}\ , \qquad\mbox{with}\qquad \gamma^{+}\epsilon_{0}=0\ ,
\end{equation}
where $\epsilon_{0}$ is a 2-vector of constant spinors.

We finish this subsection with a couple of comments. In SUGRA one can associate a Lie superalgebra to a given supersymmetric solution \mycite{Gauntlett:1998kc, FigueroaO'Farrill:1999va}, and for the $AdS_{2}\times S^{2}$ maximally supersymmetric solutions in minimal $\N=2$ $d=4$, this algebra is $\mathfrak{su}(1,1|2)$. In the fake-supersymmetric case, however, one cannot assign a Lie superalgebra to the solution, as the vector bilinears which would represent the supertranslation part do not lead to Killing vectors; this fact is illustrated by eq.~(\ref{eq:5j}). 

In this sense, a perhaps worrisome point is the action of the Killing vectors 
on the preserved fakeSupersymmetry, especially since the Killing spinors are $u$- and $v$-independent. Since the spinors are gauge-dependent objects, one should consider an $\mathbb{R}$-covariant Lie derivative on them \mycite{Kosmann:1972,Ortin:2002qb}. This derivative is defined for Killing vectors $X$ and $Y$ as
\begin{equation}
\label{eq:Kosmann}
\mathbb{L}_{X}\epsilon \; =\; \nabla_{X}\epsilon \; +\; \textstyle{1\over 4}\left(\partial_{a}X_{b}\right)\, \gamma^{ab}\epsilon\; -\; g\xi_{X}\, \epsilon\ , \qquad\mbox{with}\quad \left\{
\begin{array}{rcl}
d\xi_{X} & =& \pounds_{X}A\ , \\
\xi_{[X,Y]} & =& \pounds_{X}\xi_{Y}-\pounds_{Y}\xi_{X}\ .
\end{array} \right.
\end{equation}
Using this Lie derivative, one can see that $\mathbb{L}_{X}\epsilon =0$ for any $X\in\mathrm{Isom}(dS_{2})$.

\subsection{Solutions with holomorphic scalars and deformations of the Nariai}
\label{sec:Holomorphic}
In the usual supersymmetric case there are two generic classes of null solutions whose Supersymmetry is straightforward to see: the first are the pp-waves which are characterised by the fact that the scalars depend only on $u$, and the {\em cosmic strings} which are characterised by vanishing vector potentials $A^{\Lambda}$, vanishing Sagnac connection $\varpi =0$, and a holomorphic spacetime dependence of the scalars, {\em i.e.\/} $Z^{i}=Z^{i}(z)$ \mycite{Tod:1995jf,Meessen:2006tu}. In this subsection we shall be considering the fSUGRA analogue of the latter case, and impose $\varpi =0$ and that $Z^{i}$ is a function of $z$ only. However, because of eq.~(\ref{eq:7dj}) the vector potentials cannot vanish, and we shall be looking for the minimal expression of $\tilde{A}^{\Lambda}$ for which the Bianchi identity, eq.~(\ref{eq:7ej}), is solved: minimality implies that $\phi^{\Lambda} = v e^{-U}\partial_{\bar{z}}\left( V^{\Lambda}+{\overline{V}}^\Lambda\right)$ and the Bianchi identity reduces to
\begin{equation}
  \label{eq:Hol1j}
  d\tilde{A}^{\Lambda} \; =\; 2i\ \mathrm{Im}\left(\frac{\mathcal{X}^{\Lambda}}{g\mathtt{C}_{\Sigma}\mathcal{X}^{\Sigma}}\right) \, \frac{dz\wedge d\bar{z}}{(1+|z|^{2})^{2}} \ ,
\end{equation}
a solution to which exists locally, and determines $\tilde{A}_{u}=0$ and $\tilde{A}_{z}^{\Lambda}$ and $\tilde{A}^{\Lambda}_{\bar{z}}$ as functions of $z$ and $\bar{z}$.

Given the above identifications, we can use eq.~(\ref{eq:15j}) to calculate the constraints imposed by the Maxwell equations, {\em i.e.\/} $\mathcal{B}_{\Lambda}=0$ in eq.~(\ref{eq:VectEOM2}), which leads to
\begin{eqnarray}
  \label{eq:Hol2a}
  \mathcal{N}_{\Lambda\Sigma}\, \partial_{z}\left( V +\overline{V}\right)^{\Sigma} & =&\partial_{z}\left[ \overline{\mathcal{N}}_{\Lambda\Sigma}V^{\Sigma} \; +\; \mathcal{N}_{\Lambda\Sigma}\overline{V}^{\Sigma}\right] \ ,\\
  \label{eq:Hol2b}
  \overline{\mathcal{N}}_{\Lambda\Sigma}\, \partial_{\bar{z}}\left( V +\overline{V}\right)^{\Sigma} & =&\partial_{\bar{z}}\left[ \overline{\mathcal{N}}_{\Lambda\Sigma}V^{\Sigma}\; +\; \mathcal{N}_{\Lambda\Sigma}\overline{V}^{\Sigma}\right]\ , \\
  \label{eq:Hol2c}
  \partial_{z}\left[\overline{\mathcal{N}}_{\Lambda\Sigma}\, \partial_{\bar{z}}\left( V +\overline{V}\right)^{\Sigma}
              \right]
      & =& 
  \partial_{z}\left[\mathcal{N}_{\Lambda\Sigma}\, \partial_{z}\left( V +\overline{V}\right)^{\Sigma}
              \right] \ ,
\end{eqnarray}
where the contribution due to $\tilde{A}^{\Lambda}$ has been dropped out identically. As eq.~(\ref{eq:Hol2b}) is the complex conjugated version of (\ref{eq:Hol2a}), and eq.~(\ref{eq:Hol2c}) is the integrability condition for (\ref{eq:Hol2a}) and (\ref{eq:Hol2b}), we only need to demand eq.~(\ref{eq:Hol2a}) to hold. Using the holomorphicity of the scalars to write $\partial_{z}=\partial_{z}Z^{i}\ \partial_{i}$, one can express it as an equation in Special Geometry, namely
\begin{equation}
\begin{array}{ccl}
\label{eq:Hol3a}
\partial_{i}\overline{\mathcal{N}}_{\Lambda\Sigma}\, V^{\Sigma}+\partial_{i}\mathcal{N}_{\Lambda\Sigma}\, \overline{V}^{\Sigma} & =& 2i\mathrm{Im}\left(\mathcal{N}\right)_{\Lambda\Sigma}\, \partial_{i}V^{\Sigma}\\
& =& gi\overline{\mathcal{L}}_{\Lambda}\,\mathtt{C}_{\Gamma}f_{i}^{\Gamma}-\textstyle{gi\over 4}\partial_{i}\mathrm{Im}\left(\mathcal{N}\right)_{\Lambda\Sigma}\,\mathrm{Im}\left(\mathcal{N}\right)^{-1|\Sigma\Gamma}\mathtt{C}_{\Gamma}\ .
\end{array}
\end{equation}
Some straightforward algebra using the expressions (\ref{eq:SGdifiN}) and (\ref{eq:SGdifbiN}) shows that the above equation holds, and thus the Maxwell equations are solved for arbitrary scalar functions $Z^{i}(z)$.

Had we obtained the most general solution above, we would be certain that the fields solve the fKSEs, 
and thus the KSIs would have told us that we only need to further verify $\mathcal{E}_{++}=0$ to make sure that the proposed configuration solves all the equations of motion. We did however check that all of the EOMs are indeed satisfied, which, as was to be expected from the discussion above, they all reduced to Special Geometry calculations.

In conclusion, given an expression for $Z^{i}=Z^{i}(z)$, we need to find the local expression for $\tilde{A}^{\Lambda}$ from eq.~(\ref{eq:Hol1j}), so the solution is given by
\begin{eqnarray}
  \label{eq:NarGen1}
  ds^{2} & =& 2du\left( dv - \textstyle{1\over 2}H_{0}v^{2}\ du\right) \; -\; \frac{4}{g^{2}\left|\mathtt{C}_{\Lambda}\mathcal{L}^{\Lambda}\right|^{2}}\frac{dzd\bar{z}}{(1+|z|^{2})^{2}} \ , \\
  & & \nonumber \\
  \label{eq:NarGen2}
  A^{\Lambda} & =& \textstyle{g\over 4}\mathsf{F}^{-1|\Lambda\Sigma}\mathtt{C}_{\Sigma}\, v\,du\; +\; \tilde{A}^{\Lambda} \ ,
\end{eqnarray}
where
\begin{equation}
  \label{eq:NarGen3}
  H_{0} \; =\; \mathtt{V} \; -\; g^{2}\left| \mathtt{C}_{\Lambda}\mathcal{L}^{\Lambda}\right|^{2} \ .
\end{equation}
Furthermore, one can obtain deformations of the Nariai cosmos by taking the scalars $Z^{i}$ to be constants, in which case the $z\bar{z}$-part of the metric describes a 2-sphere of radius $g|\mathtt{C}_{\Lambda}\mathcal{L}|$. Depending on $H_{0}$, the $uv$-part of the metric describes $dS_{2}$ ($H_{0}>0$), $\mathbb{R}^{1,1}$ ($H_{0}=0$) or $AdS_{2}$ ($H_{0}<0$). As before, these spaces have local holonomy contained in $\mathfrak{sim}(2)$. The solution for generic $Z^{i}(z)$ however has proper $\mathfrak{sim}(2)$ holonomy. 

\section{\texorpdfstring{Non-BPS solutions to $\N=2$ SUGRA}{Non-BPS solutions to N=2 SUGRA}}
\label{sec:PotIsNul}
As is well known, there are models in $\N=2$ $d=4$ SUGRA coupled to vector multiplets for which one can choose the 
Fayet-Iliopoulos terms such that the hyper-multiplet contribution to the potential vanishes (see {\em e.g.\/} \mycite{Cremmer:1984hj} or \mycite{Andrianopoli:1996cm}{sec.~9} for a discussion of this point). Since the construction we are considering in this chapter is a Wick-rotated version of the general supersymmetric set-up, with no hyperscalars, this implies that there are fake-supersymmetric models in which the only contribution to the potential comes from the gauging of the isometries, as the FI contributions to the latter vanish. In that case, the bosonic action (\ref{eq:VectAct}) coincides with that of an ordinary YMH-type of supergravity theory, and for those specific models the solutions we obtained are in fact non-BPS solutions of a regular supergravity theory. 
Let us illustrate this with the dimensional reduction of minimal $d=5$ SUGRA.

The dimensional reduction of minimal five-dimensional SUGRA leads to a specific $\N=2$ $d=4$ SUGRA, namely minimal SUGRA coupled to one vector multiplet, with a prepotential given by
\begin{equation}
\label{eq:24j}
\mathcal{F}\left( \mathcal{X}\right) \, =\, -{\frac{1}{8}}\,\frac{\left(\mathcal{X}^{1}\right)^{3}}{\mathcal{X}^{0}} \  .
\end{equation}
With the usual choice $Z=\mathcal{X}^{1}/\mathcal{X}^{0}$, one finds that the scalar manifold is $\mathrm{SL}(2;\mathbb{R})/\mathrm{U}(1)$ with the corresponding K\"ahler potential $e^{\mathcal{K}}\ =\ \mathrm{Im}^{3}\left( Z\right)$ (note that this implies the constraint $\mathrm{Im}\left( Z\right)>0)$. Ignoring the possibility of gauging isometries of the scalar manifold, so that $\mathtt{P}=0$, we can calculate the potential in eq.~(\ref{eq:Potential}) to find
\begin{equation}
\label{eq:25j}
\mathtt{V} \; =\; \textstyle{\frac{2g^2}{3}}\left[\mathtt{C}_{1}^{2}\; \mathrm{Im}^{-1}(Z) \; +\;6\mathtt{C}_{0}\mathtt{C}_{1}\; \mathrm{Re}(Z)\,\mathrm{Im}^{-3}(Z)\right] \ .
\end{equation}
There are two interesting subclasses we can consider. The first one is $\mathtt{C}_{\Lambda}=(0,\mathtt{C}_{1})$, for which the potential is of the correct form to correspond to the dimensionally-reduced version of the theory considered in \mycite{Grover:2008jr}.

The second case is $\mathtt{C}_{\Lambda}=(\mathtt{C}_{0},0)$, which means that the potential vanishes. By construction this not only means that we can construct non-BPS solutions to the four-dimensional supergravity theory, but also that it can be oxidised to minimal five-dimensional SUGRA. 
A simple timelike static solution for this latter case can be found by putting $\mathcal{I}^{0}=0\,$, so that the base-space is $\mathbb{R}^{3}$, and $\mathcal{I}_{1}=0$ as to ensure staticity, {\em i.e.\/} $\omega = 0\,$. The regularity of the solution to the stabilisation equations, or equivalently the consistency of the metrical factor $|X|^{2}$, imposes the constraint $\mathcal{I}_{0}\left(\mathcal{I}^{1}\right)^{3}<0$. With this information, the solution is determined by
\begin{equation}
\label{eq:32j}
\frac{1}{2|X|^{2}} \; =\; \sqrt{2\, \left| \mathcal{I}_{0}\left(\mathcal{I}^{1}\right)^{3}\right|}\ , \qquad Z \; =\; 2i\sqrt{\left|\frac{\mathcal{I}_{0}}{\mathcal{I}^{1}}\right|} \ ,
\end{equation}
so that the solution is asymptotically Kasner. As the effective radius of the compactified fifth direction is proportional to $\mathrm{Im}(Z)$, which grows linear in $\tau$, this solution is asymptotically decompactifying. The resulting five-dimensional metric is found to be (shifting $\mathcal{I}^{1}\rightarrow \sqrt{2}\ H$)
\begin{equation}
\label{eq:33j}
ds_{(5)}^{2} \; =\; 2H^{-1}\, dy\left(d\tau \; -\; 2\sqrt{2}\, \left|\mathcal{I}_{0}\right|\, dy\right) \; -\; H^{2}\,d\vec{x}^{2}\ .
\end{equation}
which can be transformed to a Walker metric for a space of holonomy $\mathrm{Sim}(3)$ \mycite{Walker:1950}. Observe that the relation between $(d+1)$-dimensional spaces of holonomy in $\mathrm{Sim}(d-1)$ and time-dependent black holes, of which the foregoing is one example, was first introduced in \mycite{Gibbons:2007zu}.\vspace{\baselineskip}

The generic solution in subsection \ref{sec:Holomorphic} can readily be adapted to the model at hand, and reads
\begin{equation}
\label{eq:34j}
ds^{2} \;  =\;  2du\left( dv \; +\; \lambda^{2}\, v^{2}\, \mathcal{Z}^{-3}\, du\right) \; -\; {\frac{2}{\lambda^2}}\ \mathcal{Z}^{3}\ \displaystyle\frac{dzd\bar{z}}{(1+|z|^{2})^{2}} \ , 
\end{equation}
where we have introduced the abbreviations $\sqrt{2}\lambda = g\mathtt{C}_{0}$ and $\mathcal{Z}=\mathrm{Im}(Z)$. The vector fields are given by the expression (\ref{eq:7dj}) with $\tilde{A}^{0}=0$. $\tilde{A}^{1}$ needs to satisfy
\begin{equation}
  \label{eq:35j}
  d\tilde{A}^{1} \; =\; {\sqrt{2}i}{\lambda}\, \mathcal{Z}\, \frac{dz\wedge d\bar{z}}{(1+|z|^{2})^{2}} \ , 
\end{equation}
which presupposes knowing the explicit dependence of $Z$ on $z$. Lifting this solution up to five dimensions we obtain, after the coordinate transformations $v\rightarrow e^{\sqrt{2}\lambda y}\,w$ where $y$ is the fifth direction, the following solution 
\begin{eqnarray}
\label{eq:36j}
ds^{2}_{(5)} & =& 2\mathcal{Z}^{-1}\ e^{\sqrt{2}\lambda y}\, du\,dw \; -\; \mathcal{Z}^{2}\left[dy^{2} \; +\; {\frac{2}{\lambda^2}}\, \frac{dzd\bar{z}}{(1+|z|^{2})^{2}} \right] \ , \\
\hat{A} & =& \sqrt{3}\ \mathrm{Re}\left( Z\right)\ \left[dy \, +\, 2\sqrt{2}\lambda\ \mathcal{Z}^{-3}\ vdu\right] \, -\, \sqrt{3}\ \tilde{A}^{1} \ ,
\end{eqnarray}
where $\tilde{A}^{1}$ is determined by the condition (\ref{eq:35j}). This solution is a deformation of the maximally-supersymmetric $AdS_{3}\times S^{2}$ solution, and deformations of the other maximally-supersymmetric five-dimensional solutions can be obtained by using the $Sp(2;\mathbb{R})$-duality transformations before oxidation, in a similar way to how the four- and five-dimensional 
vacua are related (see {\em e.g.\/} \mycite{LozanoTellechea:2002pn}).

Let us end this section by pointing out that there are more models for which the FI contribution to the potential vanishes \mycite{Andrianopoli:1996cm}. One of them is the $\mathcal{ST}[2,m]$ model, which in ungauged supergravity allows for the embedding of monopoles and the construction of non-Abelian black holes \mycite{Huebscher:2008yz}. 
A convenient parameterisation is given by the symplectic section
\begin{equation}
\label{eq:38j}
\mathcal{V} \, =\, \left(\begin{array}{c} \mathrm{L}^{\Lambda} \\ \eta_{\Lambda\Sigma}\,\mathrm{S}\, \mathrm{L}^{\Sigma}\end{array}\right)\ , \qquad\mbox{with}\quad
\left\{\begin{array}{lcl}
\eta & =& \mathrm{diag}([+]^{2},[-]^{m})\ , \\
     & & \\
  0  & =& \eta_{\Lambda\Sigma} \mathrm{L}^{\Lambda}\mathrm{L}^{\Sigma}\ .
\end{array}\right. 
\end{equation}
The FI part of the potential is easily calculated to give \mycite{Cremmer:1984hj,Andrianopoli:1996cm}
\begin{equation}
\label{eq:39j}
\mathtt{V}_{FI} \; =\; -\textstyle{\frac{g^2}{4}}\, \mathrm{Im}^{-1}\left(\mathrm{S}\right)\, \mathtt{C}_{\Lambda}\,\eta^{\Lambda\Sigma}\,\mathtt{C}_{\Sigma} \ ,
\end{equation}
so that $\mathtt{V}_{FI}=0$ whenever $\mathtt{C}$ is a null vector w.r.t.~$\eta$.

Taking $\mathcal{ST}[2,4]$ as the model to work with and $\mathtt{C}$ to be a null vector, we can gauge an $SU(2)$ group, and by further taking $\mathtt{C}_{\Lambda}\mathcal{I}^{\Lambda}=0$ (which implies that the base-space is $\mathbb{R}^{3}$) we can generalise the solutions found in \mycite{Meessen:2008kb} to cosmological solutions. For that, take the indices $\Lambda$ to run over $(0,+,-, i)$ \footnote{Where $0$ is a timelike direction, $\pm$ are null directions and $i=1,2,3$.} and let $\mathtt{C}_{+}$ be the only non-vanishing element in $\mathtt{C}$. We find a static solution, {\em i.e.\/} $\omega =0$, by selecting $\mathcal{I}^{\pm}=\mathcal{I}_{0}=\mathcal{I}_{i}=0\,$; this allows for the embedding of a 't Hooft-Polyakov monopole in the $\mathcal{I}^{i}$. If we then further take $\tilde{\mathcal{I}}_{+}=0$, where $\tilde{\mathcal{I}}$ is the part of the function $\mathcal{I}$ independent of $\tau$, and normalise the metric on constant-$\tau$ slices to be asymptotically $\mathbb{R}^{3}$ (which is equivalent to taking $\tilde{\mathcal{I}}_{-}$ and $\mathcal{I}^{0}$ to be suitable constants) the metric is determined through eq.~(\ref{eq:37j}) and 
\begin{equation}
\label{eq:40j}
\frac{1}{2|X|^{2}}\; =\; \sqrt{\tau}\ \sqrt{ 1\, +\, \textstyle{\frac{\mu^2}{g^2}}\left[ 1-\overline{H}^{2}\right] } \ , 
\end{equation}
where $\overline{H}$ is a completely regular function of $r\in\mathbb{R}$ coming from the 't Hooft-Polyakov monopole, which reads
\begin{equation}
\label{eq:41j}
\overline{H} \; =\; \coth \left(\mu r\right) \; -\; \frac{1}{\mu r} \ ,
\end{equation}
and is monotonic with $\overline{H}\,(r=0)=0$ and asymptoting to $\overline{H}\,(r\rightarrow\infty )=1$. This means that the constant-$\tau$ slices are geodesically-complete. The full metric however suffers from an initial singularity at $\tau =0$, and also from Kasner expansion (the base-space has a time-dependence which behaves power-like).

More general backgrounds can be constructed by considering the hairy or coloured (general Abelian) solutions of \mycite{Behrndt:1997ny}, and (non-Abelian) of \mycite{Meessen:2008kb,Huebscher:2008yz}. Comments made in section (\ref{sec:CosmMon}) apply to these solutions.

\section{Summary of the chapter}
\label{sec:Concl}
In this chapter we have studied the fake-supersymmetric solutions that can be obtained from $\N=2$ $d=4$ gauged Supergravity coupled to (non-Abelian) vector multiplets, by Wick rotating the FI term needed to obtain a gauged supergravity. As is usual in the classification of supersymmetric backgrounds, the solutions are divided into two classes, denoted the timelike and the null case, which are distinguished by the norm of the vector built out of the preserved Killing spinor. 

In the timelike case we have found that the metric is of the standard conforma-stationary form, also appearing naturally in the supersymmetric timelike solutions, with the difference that the metric is to have a specific time dependence. This is such that there is a natural substitution principle, as first pointed out by Behrndt and Cveti\v{c} \mycite{Behrndt:2003cx}, for creating solutions from the known supersymmetric backgrounds to $\N=2$ $d=4$ SUGRA coupled to (non-Abelian) vector multiplets. Apart from this time-dependence, we find that the base-space must be a subclass of three-dimensional Einstein-Weyl spaces known as hyper-CR or Gauduchon-Tod spaces \mycite{Gauduchon:1998}, and that half of the seed functions, namely the $\mathcal{I}^{\Lambda}$, must obey the Bogomol'nyi equation generalised to GT spaces.

In the null case we have found that the solutions must have a holonomy contained in $\mathrm{Sim}(2)$, which arguably can be considered to be a minor detail. It was however shown in \mycite{Coley:2008th} that the purely gravitational solutions of this kind have rather special properties with respect to quantum corrections, and it is not unconceivable that this holds for the more general class of solutions with $\mathrm{Sim}(2)$ holonomy in supergravity theories, such as the one presented in section \ref{sec:PotIsNul}.

We did not develop a full-fledged characterisation of the solutions in the null case, but instead focussed on the new characteristics induced by the interplay between $\mathrm{Sim}(2)$ holonomy and Special Geometry. The general solution to the null case is the Nariai solution in eq.~(\ref{eq:NullNariai}) with the substitution $g_{uu}=-2g^{2}v^{2}\rightarrow -2g^{2}v^{2} +2\Upsilon_{0}(z,\bar{z})$, where $\partial_{z}\partial_{\bar{z}}\Upsilon_{0}=0$. This can be seen from \mycite{Gutowski:2009vb}, that gives the characterisation for the minimal case. The end result is what can be considered to be a back-reacted solution describing the intersection of a Nariai/ Robinson-Bertotti 
space with a generic (stringy) cosmic string \mycite{Meessen:2006tu}.

The fact that the holonomy is contained in $\mathrm{Sim}(2)$ is due to having gauged an $\mathbb{R}$-symmetry, whence one can deduce that the null vector (constructed as a bilinear of the preserved Killing spinor) is a gauge-covariantly constant null vector. In other words, it is a recurrent null vector, and thus why the four-dimensional space has holonomy $\Sim(2)$ \mycite{Gibbons:2007zu}. As the Wick rotation needed to create  fake supergravities from ordinary gauged supergravities will always introduce an $\mathbb{R}$-gauging, one might be inclined to think that null fake-supersymmetric solutions will always have infinitesimal holonomy in $\mathfrak{sim}(d-2)$. This is however only partially true; consider for instance the theory studied by Grover {\em et al.\/}~\mycite{Grover:2008jr}. In that case one can see that the recurrency condition (\ref{eq:5j}) still holds {\em but} with the Levi-Civit\`a connection replaced with a metric compatible, torsionful connection, where the torsion is completely antisymmetric and proportional to the Hodge dual of the graviphoton field strength. As the connection is metric, the link between the recurrency relation and $\Sim$ holonomy going through {\em mutatis mutandis}, we see that in fake $\N=1$ $d=5$ gauged Supergravity theories, there is a $\mathrm{Sim}(3)$ holonomy even though in general this is not associated to the Levi-Civit\`a connection.

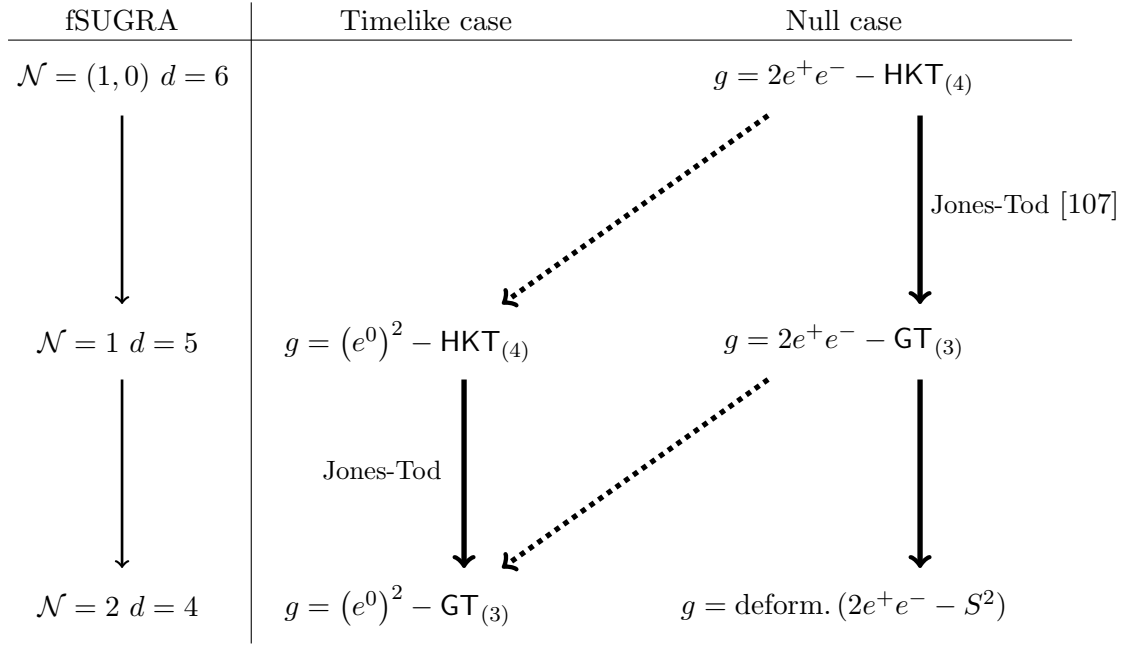
\begin{figure}[h!]
\centering
\begin{tikzpicture}[scale=1]
  \draw[color=black] (0,8) -- (1.5,8) node[color=black,above] {fSUGRA} 
          -- (5.5,8) node[color=black,above] {Timelike case} 
          -- (11,8) node[color=black,above] {Null case} -- (14,8);
  \draw[color=black] (3.2,0) -- (3.2,8.5);
  \draw[color=white] (0,7.5) node[color=black,right] {$\N=(1,0)$ $d=6$};
  \draw[color=white] (11,7.5) node[color=black] {$g= 2e^{+}e^{-}-\mathsf{HKT}_{(4)}$};
  \draw[color=black, line width=2, ->] (12,7) -- (12,4.5);
  \draw[color=white] (12,5.8) node[color=black,right] {{\small Jones-Tod} \mycite{Tod:1985}};
  \draw[color=black, line width=2, dotted, ->] (10,7) -- (6.5,4.5);
  \draw[color=black, line width=1, ->] (1.5,7) -- (1.5 ,4.5);
  \draw[color=white] (0,4) node[color=black,right] {~~$\N=1$ $d=5$};
  \draw[color=white] (3.5,4) node[color=black,right] {$g= \left( e^{0}\right)^{2} - {\mathsf{HKT}}_{(4)}$};
  \draw[color=white] (11,4) node[color=black] {$g= 2e^{+}e^{-}-{\mathsf{GT}^{}}_{(3)}$};
  \draw[color=black, line width=1, ->] (1.5,3.5) -- (1.5,1);
  \draw[color=black, line width=2, ->] (12,3.5) -- (12,1);
  \draw[color=white] (4,2.3) node[color=black,right] {{\small Jones-Tod}};
  \draw[color=black, line width=2, ->] (6,3.5) -- (6,1);
  \draw[color=black, line width=2, dotted, ->] (10,3.5) -- (6.5,1);
  \draw[color=white] (0,0.5) node[color=black,right] {~~$\N=2$ $d=4$};
  \draw[color=white] (11,0.5) node[color=black] {$g=\text{deform.}\, (2e^{+}e^{-} - S^2)$};
  \draw[color=white] (3.5,0.5) node[color=black,right] {$g= \left( e^{0}\right)^{2} -{\mathsf{GT}^{}}_{(3)}$};
\end{tikzpicture}
\caption{
\label{fig:SusyQMMap}
A graphical depiction of the relations between the fSUGRA theories and their fake-supersymmetric solutions, based on dimensional reduction over a spacelike circle. Vertical full lines indicate dimensional reduction over a $S^{1}$ in the basespace, which changes the characteristics of the basespace. Dotted lines represent dimensional reduction over a spacelike circle in the null-cone, a reduction which does not change the characteristics of the basespace. In the fakeSupergravities considered in this thesis the geometry of the base-spaces are all Einstein-Weyl manifolds.}
\end{figure}

Furthermore, as was shown in \mycite{Gibbons:2007zu} and illustrated in section \ref{sec:PotIsNul}, time-dependent solutions of the kind found in the timelike case can be obtained by dimensional reduction of spaces with $\Sim$ holonomy. Moreover, they can also be obtained from solutions in the five-dimensional timelike case. This strongly suggests that the ordinary hierarchy of supersymmetric solutions and the geometric structures appearing in them, between theories in six, five and four dimensions with eight supercharges \mycite{Meessen:2007te}, has a fake analogue. A graphical description of what this would be like is given in fig.~(\ref{fig:SusyQMMap}).

As a matter of fact, in chapter \ref{5dminimal} we will investigate the solutions to five-dimensional minimal fakeSupergravity, and we shall see that those precise relations hold between the five-dimensional null case and the structures obtained in this chapter. In fact, the maps give relations between Einstein-Weyl spaces, since a HKT space is nothing more than a four-dimensional Weyl geometry admitting covariantly constant spinors. In view of this, we then postulate that the relations in fig.~(\ref{fig:SusyQMMap}) also hold for a $N=(1,0)$ $d=6$ fakeSUGRA, which to the best of our knowledge has not yet been constructed, due to the inherent ungaugeability of the regular minimal SUGRA theory.\clearpage

\phantom{a}\vspace{\baselineskip}
\vspace{\stretch{1}}

\cleardoublepage

\renewcommand{\leftmark}{\MakeUppercase{Chapter \thechapter. $N=1$ $d=5$ minimal fakeSUGRA}}
\chapter{\texorpdfstring{$\N=1$ $d=5$ minimal fakeSUGRA}{N=1 d=5 minimal fakeSUGRA}}
\label{5dminimal}
This chapter studies the classification of solutions to $\N=1$ $d=5$ minimal fakeSUGRA, using the bilinear method. In particular, it gives details corresponding to the null case of such a theory, since the timelike case was studied in \mycite{Grover:2008jr}. For an equivalent analysis using spinorial geometry methods one can consult \mycite{Grover:2009ms}.

The outline of the chapter is the following: in section \ref{sec:FakeSetUp} we introduce a fake Killing spinor equation (fKSE) and use a subset of its integrability equations to see that they give rise to relations between the equations of motion. The introduction of bilinears constructed out of the Killing spinors allows us to introduce a five-dimensional frame, w.r.t.~which the implications of the integrability conditions are discussed. In section \ref{sec:DifConst} we analyse the differential constraints on the bilinears and obtain a necessary set of demands on the five-dimensional metric and the gauge field for the existence of a Killing spinor; in \ref{sec:suff} we show that said constraints are also sufficient, furthermore showing that we are dealing with solutions that are half fake-BPS. In section \ref{sec:EOMimp} we discuss the constraints imposed by the equations of motion on the configurations, to which we give some simple solutions in sec.~\ref{sec:sols}. Section \ref{sec:DR} considers the dimensional reduction of the theory to $\N=2$ $d=4$ fSUGRA, and studies the relation between the general null solution in five dimensions and the general timelike solution in four. At last, section \ref{sec:Concl5d} contains a summary of the chapter. The conventions used are given in appendix \ref{sec-d4conventions}, and \ref{sec:bilinears} contains the bilinears which will be used extensively in section \ref{sec:DifConst}. The interested reader will find technical information on the solution geometry in appendix (\ref{sec:Spin}), and a small introduction to Gauduchon-Tod spaces in appendix \ref{sec:EWspaces}.   

\section{Fake KSE and first implications}
\label{sec:FakeSetUp}
The bosonic field content of minimal $\N =1$ $d=5$ Supergravity comprises only of the metric $g_{\mu\nu}$ and one vector field $A_{\mu}$. It is this simplicity of the field content which allows for a clear derivation of the relevant geometrical structures, which rely on the form of the fake gravitino equation. In chapter \ref{5dgauged}, when we also consider the coupling to matter multiplets, these structures will be somewhat obfuscated.

Since fSUGRA gauges an $\mathbb{R}$-symmetry, we define the following connection on the spinors
\begin{equation}
  \label{eq:1m}
  \mathbb{D}_{a}\epsilon^{i} \; =\; \nabla_{a} \epsilon^{i} \; -\; \xi\,A_{a}\epsilon^{i} \ .
\end{equation}
Using the above definition for the gauge-covariant derivative, we can write the fake supersymmetry rule as
\begin{equation}
  \label{eq:2m}
  \mathbb{D}_{a}\epsilon^{i} \; =\; {\textstyle\frac{1}{4\sqrt{3}}}\ \left[_{a}\slashed{F} \; +\; 2\slashed{F}_{a}\right]\, \epsilon^{i}\; +\; \textstyle{1\over \sqrt{3}}\xi\, \gamma_{a}\epsilon^{i} \ ,
\end{equation}
where $_{a}\slashed{F}\equiv \frac{1}{2}\gamma_{abc}F^{bc}$ and $\slashed{F}_{a}\equiv -F_{ab}\gamma^b$. The integrability condition for this fKSE can be seen to be \mycite{Bellorin:2006yr}
\begin{equation}
  \label{eq:KSI}
  \tilde{\mathcal{E}}_{ab}\, \gamma^{b}\epsilon^{i} \; =\; -\frac{\sqrt{3}}{2}\, \mathcal{M}_{a}\epsilon^{i}\ ,
\end{equation}
where we have defined
\begin{eqnarray}
  \tilde{\mathcal{E}}_{ab} & =& \mathcal{E}_{ab} \; -\; \frac{1}{2}\, g_{ab}\, \mathcal{E}_{c}{}^{c} \ ,\\
  \label{eq:3m}
  \mathcal{E}_{ab} & =& R_{ab} \; -\; \textstyle{ 16\over 3}\xi^{2}\ \eta_{ab} \; -\; \textstyle{1\over 2}\left(F_{ac}F_{b}{}^{c}-\textstyle{1\over 6}\eta_{ab}F^{2}\right)\ , \\
  \label{eq:3am}
  \star\mathcal{M} & =& d\star F \; -\; \textstyle{1\over\sqrt{3}}\ F\wedge F \ .
\end{eqnarray}
As before, the above integrability condition leads to relations between components of the EOMs, and in the (null) case at hand they can be found by using the bilineals of \mycite{Bellorin:2006yr}. In general, there are three types of bilinears that can be constructed out of the spinors $\epsilon^{i}$: a scalar $f=i\overline{\epsilon}_{i}\epsilon^{i}$, a vector $V_{a}=i\overline{\epsilon}_{i}\gamma_{a}\epsilon^{i}$ and three 2-forms $\Phi^{x}_{ab}= (\sigma^{x})_{i}{}^{j}\ \overline{\epsilon}_{j}\gamma_{ab}\epsilon^{i}$ ($x=1,2,3$). As the timelike case corresponding to $f\neq 0$ was already treated in \mycite{Grover:2008jr}, we shall restrict our attention to the null case and put $f=0$. Furthermore, we shall again rename $V$ as $L$(ightlike).

The Fierz identities imply various relations between the bilinears, which in this case read \mycite{Bellorin:2006yr}
\begin{eqnarray}
  \label{eq:Fid1}
  \imath_{L}\hat{L} & =& 0 \ ,\\
  \label{eq:Fid2}
  \imath_{L}\Phi^{x} & =& 0\ ,\\
  \label{eq:Fid3}
  \hat{L}\wedge\Phi^{x} & =& 0\ ,\\
  \label{eq:Fid4}
  \delta^{xy}\ L_{a}L_{b} & =& \Phi^{x}_{a}{}^{c}\Phi^{y}_{cb}\ ,\\
  \label{eq:Fid5}
  \Phi^{x}\wedge\Phi^{y} & =& 0\ .
\end{eqnarray}
Eq.~(\ref{eq:Fid1}) of course implies that $L$ is a null vector, and eqs.~(\ref{eq:Fid2}) and (\ref{eq:Fid3}) imply that 
\begin{equation}
  \label{eq:Base1}
  \Phi^{x} \; =\; \hat{L}\wedge E^{x} \qquad\mbox{with}\quad \imath_{L}E^{x} \; =\; 0\ ,
\end{equation}
for some 1-forms $E^{x}$, which automatically satisfy (\ref{eq:Fid5}). A minor calculation shows that (\ref{eq:Fid4}) implies that the three 1-forms are actually orthonormal, in the sense that
\begin{equation}
  \label{eq:Base2}
  g^{-1}\left(\ E^{x}\, ,\, E^{y}\ \right) \; =\; -\delta^{xy} \ ,
\end{equation}
which implies that the $E^{x}$ can actually be used to build up a F\"unfbein. This we do by introducing the missing linearly independent 1-form $N$, normalised such that $\imath_{L}N =1$ and $\imath_{N^{\flat}}E^{x} =0$. The five-dimensional metric is then
\begin{equation}
  \label{eq:met1}
  ds^{2} \; =\; \hat{L}\otimes N \; +\; N\otimes \hat{L} \, -\, E^{x}\otimes E^{x} \ .
\end{equation}

Using the above base one can now express the implications of the integrability condition (\ref{eq:KSI}) as
\begin{eqnarray}
  \label{eq:hop1}
  0 & =& \mathcal{M}\wedge \hat{L} \ ,\\
  0 & =& \tilde{\mathcal{E}}_{ab}\ L^{b}\ ,\\
  0 & =& \tilde{\mathcal{E}}_{a}{}^{b}\ \Phi^{x}_{cb} \quad\longrightarrow\quad 0\ =\ \tilde{\mathcal{E}}_{a}{}^{b}E^{x}_{b} \ .
\end{eqnarray}
These imply that, given a solution to eq.~(\ref{eq:2m}), one only needs to ensure that 
\begin{equation}
\begin{array}{rcl}
\mathcal{E}_{++}&\equiv& N^{a}N^{b}\mathcal{E}_{ab}\ =0\ \ ,\\
\mathcal{M}_{+}&\equiv& N^{a}\mathcal{M}_{a}\ =\ 0\ ,
\end{array}
\end{equation}
to have a bona fide fake-supersymmetric solution.

In this section we have analysed the non-differential constraints on the bilinears. We now proceed to discuss the differential ones.

\section{Differential constraints}
\label{sec:DifConst}
Using the fermionic rule (\ref{eq:2m}), one can derive the following constraints
\begin{eqnarray}
\label{eq:DCscal}
\imath_{L}F & =& -2\xi\ \hat{L} \ ,\\
\label{eq:DCvect}
\mathbb{D}_{a}L_{b} & =& -\textstyle{1\over 2\sqrt{3}}\, \left(\star F\right)_{ab}{}^{c}L_{c}\ ,\\
\label{eq:DC2form}
\mathbb{D}_{a}\Phi^{x}_{bc} & =& -\textstyle{2\over \sqrt{3}}\xi\ \left(\star\Phi^{x}\right)_{abc}\; -\; \textstyle{1\over 2\sqrt{3}}\, F^{de}\left(\star\Phi^{x}\right)_{de[b}\eta_{c]a}\nonumber\\
& & +\textstyle{1\over \sqrt{3}}\ F_{a}{}^{d}\left(\star\Phi^{x}\right)_{bcd}\; +\; \textstyle{1\over \sqrt{3}}\left(\star\Phi^{x}\right)_{ad[b}F_{c]}{}^{d} \ .
\end{eqnarray}
Eq.~(\ref{eq:DCvect}) can actually be rewritten in a more suggestive form
\begin{equation}
\label{eq:Hol1}
\mathtt{D}_{a}L_{b}\, \equiv\, \nabla_{a}L_{b} \; -\; S_{ab}{}^{c}L_{c} \; =\; 2\xi\ A_{a}\ L_{b} \ ,\quad\mbox{with}\quad S_{abc} \; =\; \textstyle{-1\over 2\sqrt{3}}\, \left(\star F\right)_{abc} \ .
\end{equation}
As described in chapter \ref{4d}, from this one can see that if the $S$-contribution were absent, the space-time would have local holonomy contained in $\mathfrak{sim}(3)$ \mycite{Gibbons:2007zu}. Moreover, $S$ can be seen as a totally antisymmetric torsion which is metric-compatible\footnote{A torsionful connection is said to be {\em metric} if the torsion tensor $S(X,Y,Z)\equiv g(S_{X}Y,Z)$ is antisymmetric in the last two entries, {\em i.e.\/} $S(X,Y,Z)=-S(X,Z,Y)$, in fact implying that $S$ is a 3-form. Furthermore, the torsionful connection is said to be {\em strong} if $S$ is closed, {\em i.e.\/} $dS=0$. In the case at hand we have that $S\sim\star F$, whence $dS\sim d\star F \sim \alpha F\wedge F$ which generally non-vanishing, so that our torsion-structure is not strong.}, and thus $\text{Hol}(\mathtt{D})\subseteq \Sim(3)$. We will actually see later that if the field strength satisfies the radiation condition $F\wedge \hat{L}=0$, the space has again local holonomy contained in $\mathfrak{sim}(3)$. This is another way of stating that the {\em generalised holonomy} of the fKSE is contained in $\Sim(3)$, which is precisely what is needed for the existence of Killing spinors. 

Using the base given above we can express the fieldstrength $F$ in terms of it, and the constraint (\ref{eq:DCscal}) prescribes
\begin{equation}
  \label{eq:FBase1}
  F \; =\; 2\xi\, \hat{L}\wedge N \; +\; \varrho_{x}\, \hat{L}\wedge E^{x} \; +\; \textstyle{1\over 2}\, f_{xy}\, E^{x}\wedge E^{y} \ ,
\end{equation}
for some (still) undetermined entities $\varrho_{x}$ and $f_{xy}$. Using that $\imath_{L}\star F=\star\left( \hat{L}\wedge F\right)$ one can then write the antisymmetrised version of eq.~(\ref{eq:DCvect}) as
\begin{equation}
  \label{eq:DCv2}
  \mathbb{D}\hat{L} \; =\; -\textstyle{1\over \sqrt{3}}\, \star\left( \hat{L}\wedge F\right)\ .
\end{equation}
Since $L$ is a null vector we know that\footnote{Where we have chosen the orientation $\varepsilon^{+-xyz}=\varepsilon^{xyz}$, for $\varepsilon^{xyz}$ the three-dimensional Riemannian totally antisymmetric tensor.}
\begin{equation}\label{eq:DefAleph}
\textstyle{1\over 2}\varepsilon^{xyz}f_{yz}\,\equiv\ 2\sqrt{3}\xi\ \aleph_{x}\quad\longrightarrow\quad \star\left( \hat{L}\wedge F\right) \;\equiv\; -2\sqrt{3}\xi\, \hat{L}\wedge \aleph\ ,
\end{equation}
where we have decomposed $\aleph =\aleph_{x}\ E^{x}$ as follows from (\ref{eq:FBase1}). This implies that $\hat{L}$ is hypersurface-orthogonal, {\em i.e.\/} $\hat{L}\wedge d\hat{L}=0$, which in its turn implies the existence of two functions $Y$ and $u$ such that $\hat{L}=Ydu$; but since $\hat{L}$ has $\mathbb{R}$-weight 2, we can gauge-fix $Y=1$, and thus $\hat{L}=du$. Plugging all of this into eq.~(\ref{eq:DCv2}) then implies that 
\begin{equation}
\label{eq:DCv3}
A \; =\; \Upsilon\ \hat{L}\, +\, \aleph \ ,
\end{equation}
for some function $\Upsilon$. This form of the vector potential has $\imath_{L}A=0$, which after contracting eq.~(\ref{eq:DCvect}) with $L^{a}$ implies that $L$ is a geodesic vector, {\em i.e.\/} $\nabla_{L}L =0$.\vspace{\baselineskip}

Let us further introduce a coordinate $v$ adapted to the vector $L$ by $L=\partial_{v}$, and three more coordinates $y^{m}$ ($m=1,2,3$) such that the F\"unfbein can be taken to be
\begin{equation}
  \label{eq:Coord1}
  \begin{array}{lclclcl}
    E^{+} & =& \hat{L} \ =\ du\ , &\qquad& \theta_{+} & =& N^{\flat} \; =\; \partial_{u}\; -\; H\partial_{v} \ ,\\
    E^{-} & =& N \ =\ dv + Hdu +\omega_{m}dy^{m}\ , &\qquad & \theta_{-} & =& L \; =\; \partial_{v} \ , \\
    E^{x} & =& E^{x}_{m}\ dy^{m}\ , & \qquad&\theta_{x} & =& E_{x}{}^{m}\left(\partial_{m}\; -\; \omega_{m}\partial_{v}\right)\ ,
  \end{array}
\end{equation}
where we have defined $E^{A}\left(\theta_{B}\right) =\delta^{A}{}_{B}$ and also the Dreibein $E^{x}_{m}$, which leads to the three-dimensional metric $h_{mn}\equiv E_{m}^{x}E^{x}_{n}$. The resulting line element then takes on the Walker-form \mycite{Walker:1950}
\begin{equation}
\label{eq:Coord2}
ds^{2} \; =\; 2du\left( dv +Hdu+\omega\right) \; -\; h_{mn}dy^{m}dy^{n} \ .
\end{equation}
Given this base and the result (\ref{eq:DCv3}), we see that the symmetrised version of eq.~(\ref{eq:DCvect})
implies that 
\begin{eqnarray}
\label{eq:vDep1}
\partial_{v}H & =& 2\xi\ \Upsilon \ , \\
\label{eq:vDep2}
\partial_{v}\omega_{m} & =& 2\xi\ \aleph_{m} \ ,\\
\partial_{v}h_{mn} & =& 0\ .
\end{eqnarray}

In order to fix the $v$-dependence of the metric we need the $v$-dependence of the vector potential $A$. This can be obtained by calculating
\begin{eqnarray}
  \label{eq:Ares1}
  \pounds_{L}A & =& \imath_{L}F \, =\, -2\xi\ \hat{L} \nonumber \\
              & =& \partial_{v}\Upsilon\, \hat{L} \, +\, \pounds_{L}\aleph 
              \ =\ \partial_{v}\Upsilon\ \hat{L} \, +\, \left(\partial_{v}\aleph_{m}\right)\ dy^{m}\ ,
\end{eqnarray}
which implies that $\aleph$ is $v$-independent, and also that
\begin{eqnarray}
  \label{eq:vDep2a}
  \Upsilon & =& -2\xi\ v \, +\, \Upsilon_{1}(u,y) \ ,\\
  \label{eq:vDep2b}
  H & =& -2\xi^{2}\ v^{2} \, +\, 2\xi\ \Upsilon_{1}\ v\, +\, \Upsilon_{0}(u,y) \ ,\\
  \label{eq:vDep2c}
  \omega_{m} & =& 2\xi\ v\ \aleph_{m} \, +\, \varpi_{m}(u,y)\ . 
\end{eqnarray}
As discussed in \mycite{Gibbons:2007zu}, $\Upsilon_{1}$ can be made to vanish by means of a suitable coordinate transformation $v\rightarrow v+U(u,y)$, and thus from now on we shall be taking $\Upsilon_{1}=0$.

Having now the explicit coordinate dependence, we can proceed to find the expressions for $\varrho_{x}$ and $f_{xy}$ in the field strength. 
\begin{equation}
  \label{eq:lambdax}
  \varrho_{x} \; =\; E_{x}{}^{m}\left(\dot{\aleph}_{m}\; -\; 2\xi\, \omega_{m}\right) \ .
\end{equation}
It is clear from eq.~(\ref{eq:DCv3}) that $f_{mn}=2\partial_{[m}\aleph_{n]}$, or equivalently $f=\eth\aleph$, if we introduce the three-dimensional exterior derivative $\eth = dy^{m}\partial_{m}$. This implies that the definition of $\aleph$ in eq.~(\ref{eq:DefAleph}) is actually a {\em constraint on} $\aleph$. In fact this type of constraint is part of the definition of our previously considered Gauduchon-Tod spaces \mycite{Gauduchon:1998}, which appear naturally as the geometric structure of the three-dimensional base-space in $\N=2$ fakeSupergravities, as discussed in chapter \ref{4d}. To confirm it, we investigate the totally antisymmetric form of the constraint (\ref{eq:DC2form}). In form-notation this reads
\begin{equation}
\label{eq:GTrest}
\mathbb{D}\Phi^{x} \; =\; -2\sqrt{3}\xi\, \star\Phi^{x} \ .
\end{equation}
Using the definition (\ref{eq:Base1}) we can rewrite the LHS as
\begin{equation}
\mathbb{D}\Phi^{x}\; =\; -\hat{L}\wedge\left( dE^{x}\; -\; 2\xi\, \aleph\wedge E^{x}\right)\ ,
\end{equation}
so that eq.~(\ref{eq:GTrest}) can be recast as
\begin{equation}
\label{eq:GTrest2}
0 \, =\, \hat{L}\wedge\left( dE^{x} -2\xi\, \aleph\wedge E^{x} \; +\; \sqrt{3}\xi\ \varepsilon^{xyz}E^{y}\wedge E^{z}\right) \ .
\end{equation}
Ignoring the possible $u$-dependence, we can recast the above equation in terms of purely three-dimensional objects as
\begin{equation}
  \label{eq:GTrest4}
  \eth E^{x} \; =\; 2\xi\, \aleph\wedge E^{x} \; -\; \sqrt{3}\xi\, \varepsilon^{xyz}E^{y}\wedge E^{z} \ ,
\end{equation}
which offers a way to define Gauduchon-Tod spaces \mycite{Gauduchon:1998}. The conclusion then is that the geometry of the transverse space is a Gauduchon-Tod manifold, albeit with a possible $u$-dependence.\vspace{\baselineskip} 

Furthermore, eq.~(\ref{eq:DC2form}) provides us with an additional constraint. Let us calculate
\begin{equation}
\label{eq:Euh1}
\hat{L}\wedge\nabla_{N}E^{x} \; =\; \mathbb{D}_{N}\Phi^{x} \; -\; \mathbb{D}_{N}\hat{L}\wedge E^{x} \ ,
\end{equation}
and use\footnote{This can be obtained from eq.~(\ref{eq:DCvect}).} 
\begin{equation}
\label{eq:Euh2}
\mathbb{D}_{N}\hat{L} \; =\; \xi\, \aleph \ ,
\end{equation}
and eq.~(\ref{eq:DC2form}) to arrive at
\begin{equation}
\begin{array}{ccl}
\label{eq:Euh3}
0& =& \hat{L}\wedge\left[\nabla_{N}E^{x} \; -\; \xi\,\aleph_{x}\, N \; +\; \frac{\sqrt{3}}{2}\varepsilon^{xyz}\varrho_{y}E^{z}\right] \\
& =& E^{+}\wedge\left[\nabla_{+}E^{x} \; -\; \xi\,\aleph_{x}\, E^{-} \; +\; \frac{\sqrt{3}}{2}\varepsilon^{xyz}\varrho_{y}E^{z}\right] \ .
\end{array}
\end{equation}
As the $E$s form a frame, one can use the spin connection in $\nabla_{+}E^{x}=\Omega_{+,}{}^{x}{}_{a}E^{a}$ to rewrite the above as
\begin{equation}
  \label{eq:Euh4}
  \left(\ \Omega_{+,-x}-\xi\,\aleph_{x}\right)\, E^{+}\wedge E^{-} \; =\;\left[\Omega_{+,xy}\; +\; \frac{\sqrt{3}}{2}\varepsilon^{xyz}\varrho_{z}\right]\, E^{+}\wedge E^{x} \ .
\end{equation}
The fSUSY thus imposes the constraints
\begin{eqnarray}
  \label{eq:Euh4a}
  \Omega_{+,-x} & =& \xi\, \aleph_{x} \; =\; \textstyle{1\over 2}\theta_{-}\omega_{x} \ ,\\
  \label{eq:Euh4b}
  \Omega_{+,xy} & =& -\frac{\sqrt{3}}{2}\varepsilon^{xyz}\varrho_{z} \ ,
\end{eqnarray}
where eq.~(\ref{eq:vDep2}) has been used in the second equality on (\ref{eq:Euh4a}). In fact, comparing it with eq.~(\ref{eq:SPcon-x}), we see that they agree automatically, and thus this gives us no further information. The situation w.r.t.~(\ref{eq:Euh4b}) is different, however, as it imposes a constraint whose consequences we now investigate. By using the explicit forms in eqs.~(\ref{eq:lambdax}) and (\ref{eq:SPconxy}), we see that the $v$-dependent part drops out and one finds
\begin{equation}
\label{eq:Kutje1}
2\sqrt{3}\xi\ \varpi_{x} \; -\; \textstyle{1\over 2}\varepsilon^{xyz}B_{yz}\; =\; \sqrt{3}E_{x}{}^{m}\dot\aleph_{m} \; +\; \varepsilon^{xyz}E_{y}^{m}\dot{E}_{zm}\ ,
\end{equation}
where we have defined
\begin{equation}
\label{eq:DefB}
B \;\equiv\; \eth\varpi \; +\; 2\xi\, \aleph\wedge\varpi \ .
\end{equation}

As we now see, the constraints derived thus far are enough to have have a fake-supersymmetric configuration.

\subsection{Sufficiency of the derived constraints}
\label{sec:suff}
So far we have derived necessary constraints on a configuration to be fake-supersymmetric. In this subsection we shall show that they are actually sufficient; we do this by plugging them into the fKSE (\ref{eq:2m}), and confirming that no further constraints arise.

Imposing\footnote{Note that the Fierz identity \mycite{Bellorin:2006yr}{eq.~(A.27)} instructs us to do so.} $\gamma^{+}\epsilon^{i}=0$ implies that the $\epsilon^{i}$ are $v$-independent. Using this fact and the chosen orientation one can then see that
\begin{equation}
  \label{eq:GammaRel}
  \gamma^{xyz}\epsilon^{i} \; =\; \varepsilon^{xyz}\epsilon^{i}\ ,\qquad \gamma^{xy}\epsilon^{i} \; =\; -\varepsilon^{xyz}\gamma^{z}\epsilon^{i}\ ,\qquad \gamma^{x}\epsilon^{i}\; =\; -\textstyle{1\over 2}\varepsilon^{xyz}\gamma^{yz}\epsilon^{i}\ .
\end{equation}
We can use these relations to massage the fKSE in the $+$ direction into the form
\begin{equation}
  \label{eq:fKSE+}
  \theta_{+}\epsilon^{i}\; =\; -\textstyle{1\over 4}\,\left(\sqrt{3}\varrho_{x}\; +\; \varepsilon^{xyz}\Omega_{+,yz}\right)\, \gamma^{x}\epsilon^{i} \; =\; 0 \ ,
\end{equation}
which vanishes due to eq.~(\ref{eq:Euh4b}). Hence, the fake Killing spinor is both $u$- and $v$-independent.

Doing a similar thing to the fKSE in the $x$-directions gives
\begin{equation}
  \label{eq:fKSExRes}
  0 \; =\; \nabla^{(\lambda )}_{x}\epsilon^{i} \; +\; \xi\ \gamma^{xy}\aleph_{y}\ \epsilon^{i}\; +\; \textstyle{\sqrt{3}\over 2}\xi\ \gamma^{x}\epsilon^{i}\ ,
\end{equation}
where $\nabla^{(\lambda )}$ is the three-dimensional covariant derivative for the spin-connection $\lambda$ (cf. appendix \ref{sec:Spin}). In \mycite{Buchholz:2000} equations like these are called {\em Killing equations in Weyl geometry}\footnote{The identification is obvious from \mycite{Buchholz:2000}{Th.~1.2} once one sees that \mycite{Buchholz:2000}{eq.~(1)} implies that $\left\{ \gamma^{i},\gamma^{j}\right\} = -\delta^{ij}$, and furthermore that the 1-form used differs from the one here by $\theta_{\text{\mycite{Buchholz:2000}}}=-2\xi\ \aleph$.}, and one can see that eq.~(\ref{eq:fKSExRes}) corresponds to a Killing equation for a three-dimensional weightless spinor. The existence theorem for solutions to the above equation is found in \mycite{Buchholz:2000b}, where it is shown that the above equation has solutions if and only if the Weyl geometry is Gauduchon-Tod.

The amount of preserved fSUSY remains to be studied. For this we can use eq.~(\ref{eq:GT0conn}) to derive
\begin{equation}
  \nabla_{x}^{(\lambda )}\epsilon^{i} \; \equiv \; \partial_{x}\epsilon^{i} \; -\; \textstyle{1\over 4}\Omega_{x,yz}\gamma^{yz}\epsilon^{i} \; =\; \partial_{x}\epsilon^{i} \; -\; \xi\, \gamma^{xy}\aleph_{y}\, \epsilon^{i}\; -\; \textstyle{\sqrt{3}\over 2}\xi\, \gamma^{x}\epsilon^{i}\ ,
\end{equation}
where we made use of the relations in (\ref{eq:GammaRel}). Comparing this equation with (\ref{eq:fKSExRes}), one sees that $\partial_{x}\epsilon^{i}=0$, so that the Killing spinor is constant. As the only restriction imposed has been $\gamma^{+}\epsilon^{i}=0$, the configuration thus breaks half of the fakeSupersymmetry.
\section{Solving the EOMs}
\label{sec:EOMimp}
In order to study the Maxwell equation of motion we first calculate
\begin{equation}
  \label{eq:starF}
  \star F \; =\; 2\sqrt{3}\xi\, E^{+}\wedge E^{-}\wedge \aleph
          \; -\; \textstyle{1\over 2}\varepsilon^{xyz}\, \varrho_{x}\ E^{y}\wedge E^{z}\wedge E^{+}
          \; -\; \textstyle{2\xi\over 3!}\ \varepsilon^{xyz}\, E^{x}\wedge E^{y}\wedge E^{z}\ .
\end{equation}
With this expression, the Maxwell EOM can be seen to give
\begin{eqnarray}
  \label{eq:MEOM}
  \nabla_{x}^{(\lambda )}\varrho_{x} & =& 4\xi \varrho_{x}\aleph_{x} 
            \; +\; 24\xi^{3}\, v\, \aleph_{x}\aleph_{x}\; +\; \sqrt{3}\xi\, \aleph_{x}\varepsilon^{xyz}B_{yz}
            \; +\; \omega_{x}\partial_{v}\varrho_{x} \; +\; 2\xi\ E^{xm}\theta_{+}E_{m}^{x}\; .
\end{eqnarray}
We shall have a look at some particular subcases.

Consider first the $u$-independent case with $\varpi =0$. The above equation implies that $\nabla_{x}^{(\lambda )}\aleph_{x}=0$, whence the three-dimensional metric $h$ is Gauduchon, which was already implied by eq.~(\ref{eq:DefAleph}). If we relax the condition that $\varpi =0$, we find that 
\begin{equation}
  \label{eq:MEOMuind}
  \nabla_{x}^{(\lambda )}\varpi_{x} \; =\; -\sqrt{3}\xi\,\left( \star_{(3)}B_{x} \; -\; 2\sqrt{3}\xi\, \varpi_{x}\right)\,\aleph_{x} \ .
\end{equation}
Combining this with eq.~(\ref{eq:Kutje1}) for the $u$-independent case, we see that one arrives at
\begin{equation}
  \label{eq:MEOMuind2}
  \nabla_{x}^{(\lambda )}\varpi_{x} \; =\; 0\ .
\end{equation}

As indicated by fSUSY, the only component of the Einstein equations that needs to be checked explicitly is $\mathcal{E}_{++}$. In the $u$-independent case this is easily calculated to give
\begin{equation}
  \label{eq:EEOMuind1}
  0 \; =\; \nabla_{x}^{(\lambda )}\left(\partial_{x}\; +\; 2\xi\aleph_{x}\right)\, \Upsilon_{0} \ .
\end{equation}
This means that in the $u$-independent case the full solution is given by a Gauduchon-Tod space determining the pair $(E^{x},\aleph)$ through eq.~(\ref{eq:GTrest4}), an $\Upsilon_{0}$ which is determined by eq.~(\ref{eq:EEOMuind1}) and a $\varpi$ determined through eq.~(\ref{eq:Kutje1}), which in this case reads
\begin{equation}
2\sqrt{3}\xi\ \varpi_{x} \; =\; \textstyle{1\over 2}\varepsilon^{xyz}\left(\eth\varpi \; +\; 2\xi\ \aleph\wedge\varpi  \right)_{yz}\ .
\end{equation}
Knowing $(E^{x},\aleph ,\Upsilon_{0}, \varpi )$, the full solution can be written down using eqs.~(\ref{eq:vDep2a}) - (\ref{eq:vDep2c}) and (\ref{eq:DCv3}), (\ref{eq:Coord1}). An example is given next.

\subsection{Explicit solutions: five-dimensional Nariai and squashed Nariai cosmoses}
\label{sec:sols}
A first explicit solution is given by
\begin{equation}
\label{eq:NariaiG}
\begin{array}{ccl}
ds^{2} & =& 2du\left( dv \; -\; 2\xi^{2}v^{2}\, du\right) \; -\; \frac{1}{3\xi^{2}}\ dS^{3} \ ,\\
\\
A      & =& -2\xi\ v\ du\ ,
\end{array}
\end{equation}
where $dS^{3}$ is the standard metric on the 3-sphere. The above background is an electrically-charged five-dimensional Nariai cosmos \mycite{Cardoso:2004uz}. Observe that as this solution satisfies $F\wedge \hat{L}=0$, it has $\text{Hol}(\nabla )\subseteq \Sim(3)$; in fact it has $\text{Hol}(\nabla )= \mathfrak{so}(1,1)\oplus \mathfrak{so}(3)\subset\Sim(3)$.

We can of course also use the Berger sphere as a GT structure, \emph{i.e.}~employ eqs.~(\ref{eq:W6a}) to construct a different solution, which we shall call the \emph{squashed} Nariai cosmos. Explicitly, the solution reads
\begin{eqnarray}
  \label{eq:NariaiS1}
  ds^{2} & =& 2du\left( dv \; -\; 2\xi^{2}v^{2}\ du\; +\; \sin (\mu)\, v\, \left[ d\chi -\cos (\phi) d\varphi\right]
                 \right) \; -\; \frac{\cos^{2}(\mu)}{12\xi^{2}}\, dB_{[\chi ,\phi ,\varphi ]}^{3} \ ,\\
       A & =& -2\xi\, v\, du\ +\ \frac{\sin (\mu)}{2\xi}\, \left( d\chi \; -\; \cos (\phi) d\varphi\right)\ ,
\end{eqnarray}
where $dB_{[\chi ,\phi ,\varphi ]}^{3}$ is the metric of the Berger sphere.

Note that in the squashed case $\aleph\nsim 0$, so that we are dealing with a background where the relevant holonomy group is w.r.t.~the connection $\mathtt{D}$. 
Also observe that the Berger sphere is the only compact GT space that is not an Einstein space and has non-vanishing\footnote{This is because the Weyl scalar is constrained to be $\mathtt{W}=-18\xi^{2}$, which is non-vanishing.} Weyl-scalar \mycite{Gauduchon:1998}{prop.~6}. Thus the only Einstein space that can be used to construct a solution background is the 3-sphere, which leads to the Nariai solution in eq.~(\ref{eq:NariaiG}); this of course can also be obtained from the squashed Nariai presented here with the choice $\mu =0$. 

\section{\texorpdfstring{Dimensional reduction and link with the solution to $d=4$ fakeSUGRA}{Dimensional reduction and link with the solution to d=4 fakeSUGRA}}
\label{sec:DR}
As is well known, minimal $\N=1$ $d=5$ Supergravity can be dimensionally reduced to four dimensions, where it can be identified with $\N=2$ $d=4$ Supergravity coupled to one vector-multiplets. The field content is a metric $g$, two vector fields $A^{\Lambda}$ ($\Lambda =0,1$), and one complex scalar $Z$. The resulting Special Geometry is governed by the cubic prepotential
\begin{equation}
\label{eq:DR1}
\mathcal{F}\left(\mathcal{X}\right) \; =\; -\textstyle{1\over 8}\frac{\left(\mathcal{X}^{1}\right)^{3}}{\mathcal{X}^{0}} \ .
\end{equation}
The explicit KK-Ansatz necessary for obtaining this identification reads
\begin{eqnarray}
  \label{eq:DR2a}
  ds^{2}_{(5)} & =& k^{-1}\, ds^{2}_{(4)} \; -\; k^{2}\left( dy \; +\; A^{0}\right)^{2}\ ,\\
  \label{eq:DR2b}
  \hat{A}^{1} & =& -\sqrt{3}\left[ A^{1} \; -\; B^{1}\left( dy \; +\; A^{0}\right) \right]\ ,\\ 
  \label{eq:DR2c}
  Z & =& B^{1}\; +\; ik\ .
\end{eqnarray}

Along the same lines, the theory studied in this chapter can also be dimensionally reduced to four dimensions, and using the above KK-Ansatz the resulting four-dimensional theory falls into the class studied in chapter \ref{4d}, with a potential $\mathtt{V}$ that is given by
\begin{equation}
  \label{eq:DR3}
  \mathtt{V} \; =\; 16\xi^{2}\, k^{-1} \; =\; 16\xi^{2}\, \mathrm{Im}^{-1}\left(Z\right)\ .
\end{equation}
Comparing it to the general potential given in eq.~\ref{eq:25j} of chapter \ref{4d}, we see that
\begin{equation}
  \label{eq:DR4}
  \mathtt{C}_{0}\; =\; 0\ ,\qquad (\mathtt{C}_{1})^{2}\; =\; 24\xi^{2} \ , 
\end{equation}
where the coupling constant $g$ appearing in eq.~(\ref{eq:25j}) has been absorbed into $\mathtt{C}$ to avoid confusion.

Since this is an analysis of the null case, there are two possible directions over which one can reduce: over a space-like direction in the null-plane, or over a direction in the Gauduchon-Tod space. For a dimensional reduction in the first class, the prime candidate is reduction along the $u$-direction, which implies that one must take the solution to be $u$-independent. We can thus reduce the general solution to
\begin{eqnarray}
  \label{eq:DR5a}
  ds^{2}_{(4)} & =& k^{-1}\left( dv\; +\;\omega\right)^{2}\; -\; k\, h_{mn}dy^{m}dy^{n} \ ,\\
  \label{eq:DR5b}
  A^{0} & =& -k^{-2}\left( dv\; +\; \omega\right)\ ,\\
  \label{eq:DR5c}
  A^{1} & =& -3^{-1/2}\, \left(\aleph\;+\;2\xi\, v\, A^{0}\,\right)\ ,
\end{eqnarray}
where the scalars are given by
\begin{equation}
  \label{eq:DR6}
  k^{2} \; =\; 4\xi^{2}\, v^{2}\; -\; 2\Upsilon_{0}\ ,\qquad B^{1} \; =\; -\textstyle{2\xi\over\sqrt{3}}\, v\ .
\end{equation}
These solutions have a stationarity vector $\omega$ that is at most linear in $\tau$, and the geometry of the transverse space is a Gauduchon-Tod space. As expected, they thus resemble those found in the timelike class in four dimensions (cf.~section \ref{sec:VectBil}), where $v$ plays the r\^ole of the time coordinate $\tau$. 

\section{Summary of the chapter}
\label{sec:Concl5d}
This chapter has analysed the characterisation of solutions to five-dimensional fSUGRA with one sympletic-Majorana spinor, also commonly referred to as $\N=1$ $d=5$ De Sitter SUGRA. Our main result was that all solutions to the theory admitting fake-Killing spinors, from which a null vector field can be constructed, fall into the following family of backgrounds
\begin{eqnarray}
\label{solution}
ds^2 &=& 2 du \bigg( dv + \big(\Upsilon_{0}-2\xi^2 v^2\big) du + 2\xi v \aleph+ \varpi \bigg) - ds_{GT}^2  \ ,\\
\label{solutionbis}
\displaystyle F & =& {\chi \over 4} du \wedge dv+\frac{1}{2}d{{\cal{\aleph}}}\ ,
\end{eqnarray}
where $8\xi^2$ is the cosmological constant, $GT$ is a $u$-dependent Gauduchon-Tod space \mycite{Gauduchon:1998}, and $\Upsilon_0$, $\aleph$ and $\varpi$ are, respectively, a function and two 1-forms on $GT$ which may also depend on $u$ (but not on $v$). 

Gauduchon-Tod spaces were initially discussed in the context of hyper-hermitian spaces admitting a tri-holomorphic Killing vector field \cite{Gauduchon:1998}. They are special types of Einstein-Weyl 3-spaces, obeying the constraint (\ref{eq:GTrest4}), and we have seen that they play a r\^ole in the solutions to both four- and five-dimensional fSUGRA. In this sense, they were used \emph{e.g.}~in \mycite{Grover:2008jr} to construct examples of timelike solutions of $d=5$ minimal fSUGRA for which the base-space is not conformally hyper-K\"ahler. In $d=4$, it was shown in chapter \ref{4d} that the timelike solutions are defined by a base-space which is GT. But whereas the Ricci curvature of the Weyl connection is always non-flat in the solutions we have described in this chapter, the four-dimensional timelike solutions also allow flat GT spaces. 

As for the null supersymmetric solutions of minimal five-dimensional ungauged and gauged SUGRA theories, the family of backgrounds \eqref{solution} admits a geodesic, expansion-free, twist-free and shear-free null vector field $N$. To see this, consider the null vector field $N=\partial/\partial v$. The congruence of integral curves affinely parametrised by $v$ fulfills $\nabla_N N=0$ (geodesic). $N$ is hypersurface orthogonal, {\em i.e.}~${\bf{e}}^- \wedge d {\bf{e}}^-=0$,  which means that the congruence is twist-free. It is also non-expanding, $\nabla_\mu N^\mu=0$, and shear-free, $\nabla_{(\mu} N_{\nu)} \nabla^\mu N^\nu=0$. As we saw in section \ref{sec:Null}, such geometries are dubbed Kundt metrics in four-dimensional General Relativity \mycite{Kundt:1961}. It is a special case of the higher-dimensional metrics considered in \mycite{Coley:2005sq}{Brannlund:2008zf,Podolsky:2008ec}{Coley:2009ut}.

But $N$ has distinct properties in the De Sitter theory, as compared with the Minkowski or AdS theories. In these latter, the null vector is always Killing and for some special cases it becomes covariantly constant. Then the Kundt geometries become plane-fronted waves with parallel rays (\textit{pp-waves}). This is not the case for the De Sitter theory. For the special case with $\aleph=0$, however, the null vector acquires an interesting property; it becomes \textit{recurrent}, that is, it obeys 
\begin{equation}
\label{recurrence0}
\nabla_\mu N^\nu = C_\mu N^\nu \ , 
\end{equation}
for some non-trivial, recurrence one form $C_{\mu}$. This means that the geometries \eqref{solution} have special holonomy $\Sim(3)$, which is the maximal proper subgroup of the Lorentz group $SO(4,1)$. Some technical information on the $\Sim$ group, holonomy and recurrency can be found in appendix \ref{appsec:Sim}.  

The four parameter Similitude group, $\Sim(2)$, became a focus of interest due to the proposal of \textit{Very Special Relativity} (VSR) \mycite{Cohen:2006ky}. Cohen and Glashow asked the question if the exact symmetry group of Nature could be isomorphic to a proper subgroup of the Poincar\'e group, rather than the Poincar\'e group itself. The proper subgroup they considered was $\ISim(2)$, obtained by adjoining the maximal proper subgroup of the Lorentz group, $\Sim(2)$, with spacetime translations. The theory based on this symmetry group, VSR, actually implies Special Relativity if a discrete symmetry, namely CP, is also added. But since the latter is broken in nature, VSR is necessarily distinct from Special Relativity. 
Additionally, there has been other subsequent developments on theories employing the Sim and ISim groups, including the \textit{General Very Special Relativity} of \mycite{Gibbons:2007iu}. In this sense, studies of $d$-dimensional Lorentzian geometries with $\Sim(d-2)$ holonomy have been carried out recently \mycite{Gibbons:2007zu}. The resulting geometries have interesting properties, such as the possibility of vanishing quantum corrections \mycite{Coley:2008th}. Possible connections to Supersymmetry were also hinted in \mycite{Brannlund:2008zf}. In this chapter, as well as in chapter \ref{4d}, we have shown how indeed these geometries emerge in an explicit supersymmetric computation.

In the next chapter we shall extend the investigations to include the coupling to $\N =1$ $d=5$ Abelian matter.

\cleardoublepage

\renewcommand{\leftmark}{\MakeUppercase{Chapter \thechapter. $N=1$ $d=5$ fakeSUGRA coupled to Abelian vectors}}
\chapter{\texorpdfstring{$\N=1$ $d=5$ fakeSUGRA coupled to Abelian vectors}{N=1 d=5 fakeSUGRA with vectors}}
\label{5dgauged}
This chapter studies the classification of the null class solutions to $\N=1$ $d=5$ fSUGRA coupled to Abelian vector multiplets. Its content is essentially that of \mycite{Gutowski:2011gc}, on which it is modelled. It is a generalisation of the analysis of chapter \ref{5dminimal}, and one can indeed see that those solutions (where only the gravity supermultiplet was present) are a limiting case of the ones presented here. To obtain this theory we analytically continue the supersymmetry transformations of the gravitino, as well as those for the gauginos, of the regular SUGRA theory. The vanishing of these transformations produces fake Killing spinor equations, and we consider fake-supersymmetric solutions which admit (non-trivial) spinors satisfying the equations.

The outline of this chapter is the following: section \ref{sec:fSUGRA} gives a summary of the basic equations of the theory. In section \ref{subsec:KSE} we analyse them, focusing on the case where the 1-form spinor bilinear is null. The conditions obtained from the gravitino equation are derived from the analysis of the minimal fake-supersymmetric solutions in chapter \ref{5dminimal}. We also introduce local coordinates, and show that, not too surprisingly, the solutions are given in terms of a one-parameter family of three-dimensional Gauduchon-Tod (GT) spaces. Imposing fSUSY together with the Bianchi identity and the gauge field equations is sufficient to ensure that all the remaining equations, with the exception of one component of the Einstein equations, hold automatically. Section \ref{sec:examples} gives some simple examples of our solutions, including some near-horizon geometries and an explicit model with one gauge multiplet, which under vanishing of the potential provides non-BPS solutions to a SUGRA theory. In section \ref{subsec:Berger} we provide solutions where the GT space is the Berger sphere. Section \ref{sec:properties} considers the conditions for which the null Killing spinor 1-form is recurrent, and investigate the properties of various scalar curvature invariants. Finally, section \ref{secgaugednulld5:summary} has the summary of the chapter.

\section{\texorpdfstring{$\N=1$ $d=5$ fSUGRA and Killing spinors}{N=1 d=5 fSUGRA and Killing spinors}}
\label{sec:fSUGRA}
The theory we start from is $\N=1$, $d=5$ gauged Supergravity coupled to Abelian vector multiplets \mycite{Gunaydin:1984ak}. Apart from the fields already present in its minimal version, this theory also has $n_{v}$ vector fields and $n_{v}$ real scalar fields. As it is customary, the $\bar{n}=n_{v}+1$ vector fields are denoted jointly
by $A^{I}$ ($I=0,\ldots ,n$). Similarly to four-dimensional supergravity, the scalar self-interactions and their interaction with the vector fields can be derived from a geometrical structure, which in this case is known as {\em Real Special geometry}\footnote{The modifier \emph{Real} is used to stress the fact that the scalar fields are now real-valued. A small introduction to this structure is given in appendix \ref{sec:RealSG}. Even though the notation used there is different from the one used in this section, the essence of the structure remains the same.}. The bosonic action is given by
\begin{equation}
S={\frac{1}{16\pi G}}\int \left( -R+2g^{2}{\mathcal{V}}\right){\mathcal{\ast }}1-Q_{IJ}\left( -dX^{I}\wedge \star dX^{J}+F^{I}\wedge \ast F^{J}\right) -{\frac{C_{IJK}}{6}}F^{I}\wedge F^{J}\wedge A^{K}\ ,
\label{action}
\end{equation}
where $I,J,K$ take values $1,\ldots ,n$ and $F^{I}=dA^{I}$ are 2-forms representing gauge field strengths (one of the gauge fields corresponds to the graviphoton). The constants $C_{IJK}$ are symmetric in $\{I,J,K\}$, and we shall be assuming that $Q_{IJ}$ (the gauge coupling matrix) is positive-definite and invertible, with inverse $Q^{IJ}$. The $X^{I}$ are scalar fields subject to the constraint
\begin{equation}
\label{eqn:conda}
{\frac{1}{6}}C_{IJK}X^{I}X^{J}X^{K}=X_{I}X^{I}=1\ .
\end{equation}
The fields $X^{I}$ can thus be regarded as being functions of $(n-1)$ unconstrained scalars $\phi ^{r}$. We list some useful relations associated with $\N=2$, $d=5$ gauged Supergravity
\begin{equation}
\begin{array}{rcl}
Q_{IJ} &=&\displaystyle{\frac{9}{2}}X_{I}X_{J}-{\frac{1}{2}}C_{IJK}X^{K}\ ,
\\
\\
\text{}Q_{IJ}X^{J} &=&\displaystyle{\frac{3}{2}}X_{I}\ ,\qquad Q_{IJ}dX^{J}= -{\frac{3}{2}}dX_{I}\ ,\\ \\
{\mathcal{V}} &=&\displaystyle{9V_{I}V_{J}(X^{I}X^{J}-{\frac{1}{2}}Q^{IJ})} \ ,
\end{array}
\end{equation}
where $V_{I}$ are constants. The De Sitter supergravity theory is obtained by sending $g^{2}$ to $-g^{2}$ in eq.~(\ref{action}).

Fake-supersymmetric De Sitter solutions admit a Dirac spinor $\epsilon$ satisfying a gravitino and the gauginos fake Killing spinor equations. The gravitino fKSE reads
\begin{equation}
\label{grav}
\left[ \nabla _{M}+{\frac{1}{8}}\gamma _{M}H_{N_{1}N_{2}}\gamma^{N_{1}N_{2}}-{\frac{3}{4}}H_{M}{}^{N}\gamma _{N}-
g(\frac{1}{2}X\gamma _{M}-\frac{3}{2}A{}_{M})\right] \epsilon =0\ ,  
\end{equation}
where we have defined
\begin{equation}
V_{I}X^{I}=X\ ,\qquad V_{I}A^{I}{}_{M}=A{}_{M}\ ,\qquad X_{I}F^{I}{}_{MN}=H_{MN}\ .
\end{equation}
The gaugino fKSE is given by
\begin{equation}
\left( (-F_{MN}^{I}+X^{I}H_{MN})\gamma ^{MN}+2\nabla _{M}X^{I}\gamma^{M}-4gV_{J}(X^{I}X^{J}-{\frac{3}{2}}Q^{IJ})\right) \epsilon =0 \ .
\label{dkse}
\end{equation}
We adopt a mostly minus signature for the metric, which is written in a null frame as
\begin{eqnarray}
ds^2 = 2 {\bf{e}}^+ {\bf{e}}^- - \delta_{ij} {\bf{e}}^i {\bf{e}}^j\ ,
\end{eqnarray}
for $i,j,k=1,2,3$.

\section{Analysis of gravitino Killing spinor equation}
\label{subsec:KSE}
We proceed with the analysis of the fake Killing spinor equations, focusing on the case for which the 1-form spinor bilinear generated from the Killing spinor is null; our basis is chosen such that this bilinear 1-form is given by ${\bf{e}}^-$.

The analysis of the gravitino fKSE has already been completed in the case of minimal $d=5$ De Sitter Supergravity in chapter \ref{5dminimal}. But for the present purposes, in which we are employing spinorial geometry techniques (cf. section \ref{spinorialgeometry}), a more suitable analysis is given in \mycite{Grover:2009ms}. The conditions on the geometry and the fluxes obtained from eq.~(\ref{grav}) can thus be read off from the results in \mycite{Grover:2009ms}{sec.~3}, which are listed in equations (3.1)-(3.20). We shall not be incorporating any of the conditions obtained from the Bianchi identity in \mycite{Grover:2009ms} here, because the 2-form flux $H$ which appears in eq.~(\ref{grav}) is {\it not} the exterior derivative of $A$, in contrast to the minimal theory. Also, to establish the correspondence between the fKSE solved in \mycite{Grover:2009ms} and (\ref{grav}) one makes the following replacements
\begin{eqnarray}
\label{replacements}
F \rightarrow {\sqrt{3} \over 2} H\ , \qquad \chi \rightarrow - 2 \sqrt{3} g X\ , \qquad \chi A \rightarrow -3g A\ ,
\end{eqnarray}
where the quantities on the LHS of these expressions are the field strength, the cosmological constant and the gauge potential of the minimal theory, using the conventions of \mycite{Grover:2009ms}.

Following the reasoning given there, one can without loss of generality work in a gauge for which the conditions on the geometry are
\begin{eqnarray}
\label{geo1}
d {\bf{e}}^- &=&0\ ,\\
\label{geo2}
d {\bf{e}}^+ &=& -3g {\bf{e}}^+ \wedge A - \omega_{-,-i} {\bf{e}}^- \wedge {\bf{e}}^i - \omega_{[i,|-| j]} {\bf{e}}^i \wedge {\bf{e}}^j\ ,\\
\label{geo3}
d {\bf{e}}^i &=& 2 \omega_{[-,j]}{}^i {\bf{e}}^- \wedge {\bf{e}}^j + {\cal{B}} \wedge {\bf{e}}^i + 3gX \star_3 {\bf{e}}^i\ ,
\\
\label{geo4}
\cL_N {\bf{e}}^i &=&0\ ,
\end{eqnarray}
where $N$ is the vector field dual to ${\bf{e}}^-$, and $\omega$ is the five-dimensional spin connection.
${\cal{B}}$ is a 1-form given in terms of the spin connection by
\begin{eqnarray}
{\cal{B}}= -2 \omega_{i,+-} {\bf{e}}^i \ .
\end{eqnarray}
In addition, the 1-form $A$ satisfies
\begin{eqnarray}
\label{flux1}
-3g A = - \omega_{-,+-} {\bf{e}}^- + {\cal{B}}\ ,
\end{eqnarray}
and the 2-form flux $H$ satisfies
\begin{eqnarray}
\label{flux2}
H = \star_3 {\cal{B}} + gX {\bf{e}}^+ \wedge {\bf{e}}^- +{1 \over 3} {\bf{e}}^- \wedge \star_3 (\omega_{-,ij} {\bf{e}}^i \wedge {\bf{e}}^j) \ .
\end{eqnarray}
Here $\star_3$ denotes the Hodge dual taken w.r.t.~the 1-parameter family of 3-manifolds $E$ equipped with metric
\begin{eqnarray}
ds_E^2 = \delta_{ij} {\bf{e}}^i {\bf{e}}^j\ ,
\end{eqnarray}
whose volume form $\epsilon^{(3)}$ satisfies
\begin{eqnarray}
\gamma_{ijk}\, \epsilon = {\epsilon^{(3)}}_{ijk}\, \epsilon \ .
\end{eqnarray}
Furthermore, in this gauge, the spinor $\epsilon$ can be taken to be a constant, which satisfies
\begin{eqnarray}
\gamma_+ \epsilon =0\ ,
\end{eqnarray}
or equivalently
\begin{eqnarray}
\gamma_{+-} \epsilon = \epsilon \ .
\end{eqnarray}

\subsection{Analysis of the gaugino Killing spinor equation}
\label{subsec:gauginoKSE}
The gaugino fake Killing spinor equation (\ref{dkse}) can be rewritten as
\begin{eqnarray}
\label{aux1}
\bigg(2 F^I_{+-} -2 F^I_{+i} \gamma_- \gamma^i - F^I_{ij} \epsilon^{ij}{}_k \gamma^k -2 X^I H_{+-} + 2 X^I H_{+i} \gamma_- \gamma^i + X^I H_{ij} \epsilon^{ij}{}_k \gamma^k\ ,\nn
+2 \nabla_+ X^I \gamma_- + 2 \nabla_i X^I \gamma^i -4gXX^I +6g Q^{IJ} V_J \bigg) \epsilon =0\ ,
\end{eqnarray}
where we have made use of the identities
\begin{eqnarray}
\gamma^{ij} \epsilon = {\epsilon^{(3)\,ij}}_k \gamma^k \epsilon
\end{eqnarray}
and $\gamma^i = - \delta^{ij} \gamma_j$. Acting on eq.~(\ref{aux1}) with the projectors ${1 \over 2}(1 \pm \gamma_{+-})$ one obtains two equations of the form
\begin{eqnarray}
(\alpha + \beta_i \gamma^i)\epsilon =0\ ,
\end{eqnarray}
for real $\alpha$, $\beta_i$ coefficients. As $\epsilon$ is non-zero, the only solution of such an equation is $\alpha=0\,$, $\beta_i=0\,$, and on evaluating the resulting conditions on the fluxes and scalars obtained from these equations, one finds that
\begin{equation}
\label{sc1}
\cL_N X^I=0
\end{equation}
and
\begin{eqnarray}
\label{flux3}
F^I_{+-}&=&3g(X X^I-Q^{IJ} V_J)\ , \\
F^I_{ij} &=& X^I (\star_3 {\cal{B}})_{ij} + \epsilon_{ij}{}^k \nabla_k X^I\ , \\
F^I_{+i} &=&F^I_{-i}\ =\ 0\ ,
\end{eqnarray}
where we have adopted the convention that $(\epsilon^3)_{ijk}=(\epsilon^3)_{ij}{}^k$, \emph{i.e.}~indices on the volume form are raised with the metric of signature $(+,+,+)$.

\subsection{Introduction of local coordinates}
\label{subsec:adaptation}
As ${\bf{e}}^-$ is a closed form, one can introduce local coordinates $u, v, y^\alpha$ for $\alpha=1,2,3$ such that
\begin{eqnarray}
{\bf{e}}^- = du\ , \qquad N={\partial \over \partial v}\ , \qquad {\bf{e}}^i = e^i{}_\alpha dy^\alpha \ .
\end{eqnarray}
Observe that possible $du$ terms in ${\bf{e}}^i$ can be removed without loss of generality by using a gauge transformation which leaves the spinor $\epsilon$ invariant, as described in \mycite{Grover:2009ms}. The three-dimensional Dreibein $e^i{}_{\alpha}$ does not depend on $v$, but in general it depends on $y^\alpha$ and $u$, as do the scalars $X^I$. 

We start by finding the $v$-dependence of ${\bf{e}}^+$. Note that eq.~(\ref{geo2}) implies that
\begin{eqnarray}
\label{aux2}
\cL_N {\bf{e}}^+ = -3gA\ ,
\end{eqnarray}
and eqs.~(\ref{flux1}) and (\ref{flux3}) that
\begin{eqnarray}
\label{aux3}
\cL_N A = 3g (X^2-Q^{IJ} V_I V_J) {\bf{e}}^-\ .
\end{eqnarray}
Moreover, eq.~(\ref{aux3}) together with eq.~(\ref{flux1}) implies that one can take
\begin{eqnarray}
\label{Aflux}
A = 3g (X^2-Q^{IJ} V_I V_J) v du -{1 \over 3g} {\cal{B}} \ .
\end{eqnarray}
Note also that if ${\tilde{d}}$ denotes the exterior derivative restricted to hypersurfaces of constant $u, v$, eq.~(\ref{geo3}) then implies that
\begin{eqnarray}
\label{ews}
{\tilde{d}} {\bf{e}}^i = {\cal{B}} \wedge {\bf{e}}^i +3gX \star_3 {\bf{e}}^i \ .
\end{eqnarray}
This is again the Gauduchon-Tod structure found in the study of a special class of Einstein-Weyl spaces \mycite{Gauduchon:1998,Dunajski:1999qs} (cf. appendix \ref{sec:Weylgeometry}). 

Furthermore, eqs.~(\ref{geo4}) and (\ref{sc1}) imply that
\begin{eqnarray}
\cL_N {\cal{B}}=0\ ,
\end{eqnarray}
and likewise eq.~(\ref{ews}) implies that
\begin{eqnarray}
{\tilde{d}} {\cal{B}} +3g \star_3 (X {\cal{B}} +{\tilde{d}} X ) =0 \ .
\end{eqnarray}
It is then straightforward to integrate eq.~(\ref{aux2}) up to find
\begin{eqnarray}
{\bf{e}}^+ = dv -{9 \over 2} g^2 (X^2-Q^{IJ} V_I V_J) v^2 du + W du + v {\cal{B}} + \phi_i {\bf{e}}^i\ ,
\end{eqnarray}
where $W$ is a $v$-independent function, and $\phi = \phi_i {\bf{e}}^i$ is a $v$-independent 1-form.

Next we wish to determine the $v$-dependence of the field strengths $F^I$. Note first that
\begin{eqnarray}
\label{aux4}
\cL_N F^I = d (i_N F^I) = {\tilde{d}} (3g(X X^I - Q^{IJ} V_J)) \wedge du\ ,
\end{eqnarray}
on making use of the Bianchi identity $dF^I=0$. From eq.~(\ref{flux3}) one also finds that
\begin{eqnarray}
F^I = 3g (X X^I - Q^{IJ} V_J) {\bf{e}}^+ \wedge {\bf{e}}^- + \star_3 (X^I {\cal{B}}+ {\tilde{d}} X^I) + {\bf{e}}^- \wedge S^I\ ,
\end{eqnarray}
where
\begin{eqnarray}
S^I = S^I{}_j {\bf{e}}^j \ .
\end{eqnarray}
One thus takes the Lie derivative of this expression and compares with eq.~(\ref{aux4}), to find that
\begin{eqnarray}
\cL_N S^I = 3g (X X^I - Q^{IJ} V_J) {\cal{B}} -3g {\tilde{d}} (X X^I - Q^{IJ} V_J)\ ,
\end{eqnarray}
and whence
\begin{eqnarray}
\label{fluxfin}
F^I &=& 3g (X X^I -Q^{IJ}V_J) (v {\cal{B}} + dv + \phi) \wedge du + \star_3 (X^I {\cal{B}}+ {\tilde{d}} X^I)\ ,
\nn
&+& du \wedge \bigg( 3gv \big(  (X X^I - Q^{IJ} V_J) {\cal{B}} - {\tilde{d}} (X X^I - Q^{IJ} V_J)  \big) + T^I \bigg)\ ,
\end{eqnarray}
where
\begin{eqnarray}
T^I = T^I{}_j {\bf{e}}^j
\end{eqnarray}
are $v$-independent 1-forms on $E$.\vspace{\baselineskip}

Having this expression for the flux, we continue by imposing two consistency conditions. The first is $V_I F^I = dA$, where $F^I$ is given by eq.~(\ref{fluxfin}) and $A$ is obtained from eq.~(\ref{Aflux}). It gives the following condition on the 1-forms $T^I$,
\begin{eqnarray}
\label{cc1}
V_I T^I = -{1 \over 3g} {\dot{\cal{B}}} +3g (X^2-Q^{IJ} V_I V_J) \phi\ ,
\end{eqnarray}
where ${\dot{{\cal{B}}}}=\cL_{{\partial \over \partial u}} {\cal{B}}$. We also impose $X_I F^I = H$, where $H$ is given in
eq.~(\ref{flux2}), and we note that
\begin{eqnarray}
\omega_{-,ij} = -{1 \over 2} (v {\tilde{d}} {\cal{B}} + {\tilde{d}} \phi - \phi \wedge {\cal{B}})_{ij} +{1 \over 2} ({\dot{{\bf{e}}}}^i)_j - {1 \over 2} ({\dot{{\bf{e}}}}^j)_i \ .
\end{eqnarray}
This gives an additional condition on $T^I$
\begin{eqnarray}
\label{cc2}
X_I T^I = -{1 \over 3} \star_3 \bigg( {\tilde{d}} \phi - \phi \wedge {\cal{B}} + \delta_{ij} {\dot{{\bf{e}}}}^i \wedge {\bf{e}}^j \bigg) \ .
\end{eqnarray}

Finally, we impose the Bianchi identities $dF^I=0$,  which give the conditions
\begin{equation}
\label{cc3}
{\tilde{d}} T^I = \cL_{{\partial \over \partial u}} \star_3 (X^I {\cal{B}}+ {\tilde{d}} X^I) +3g {\tilde{d}}(X X^I - Q^{IJ} V_J) \wedge \phi +3g (X X^I-Q^{IJ} V_J) {\tilde{d}} \phi\ ,
\end{equation}
\begin{equation}
\label{cc4}
0=\tilde{d} \star_3 \big( X^I {\cal{B}} + {\tilde{d}} X^I \big) \ .
\end{equation}
This exhausts the content of the fake Killing spinor equations.

\subsection{Equations of motion}
\label{subsec:EOMs}
In addition to conditions imposed by fSUSY, we require that our configurations solve the field equations. We start by evaluating the gauge field equations
\begin{eqnarray}
\label{geq}
d \star (Q_{IJ} F^J) +{1 \over 4} C_{IJK} F^J \wedge F^K =0\ ,
\end{eqnarray}
where the five-dimensional volume form $\epsilon^5$ is related to the three-dimensional volume form $\epsilon^3$ by
\begin{eqnarray}
\epsilon^5 = {\bf{e}}^+ \wedge {\bf{e}}^- \wedge \epsilon^3 \ .
\end{eqnarray}
One obtains the following condition
\begin{eqnarray}
{\tilde{d}} \star_3 (Q_{IJ} T^J) &=& -3g \cL_{{\partial \over \partial u}} \big(({3 \over 2} X X_I - V_I) \epsilon^3 \big)
+{3 \over 2} \phi \wedge {\cal{B}} \wedge {\tilde{d}} X_I +{3 \over 2} {\tilde{d}} \phi \wedge ({\cal{B}} X_I - {\tilde{d}} X_I)\nn
&& +3g \phi \wedge \star_3 \big( ({3 \over 2} X X_I - V_I){\cal{B}} -{3 \over 2} X_I {\tilde{d}} X +{3 \over 2} X {\tilde{d}}X_I + Q_{IJ} {\tilde{d}} Q^{JN} V_N \big)\nn
&&+{1 \over 2} C_{IJK} T^J \wedge \star_3 (X^K {\cal{B}}+{\tilde{d}} X^K) \ .
\end{eqnarray}
If one considers the limit of this equation in the pure supergravity case (\emph{i.e.}~having only the gravity supermultiplet), it becomes satisfied automatically. This is the rather peculiar feature of the \emph{null} case of minimal fSUGRA theories, which was noticed before \mycite{Grover:2009ms,Gutowski:2009vb}.  
 
Next, consider the Einstein field equations. It is straightforward to show that the integrability conditions of the
fKSEs imply that all components of the Einstein equations hold automatically, with the exception of the ``$--$" component. The field equations are
\begin{eqnarray}
0 &=& R_{\alpha \beta} + Q_{IJ} F^I_{\alpha \mu} F^J_\beta{}^\mu - Q_{IJ} \nabla_\alpha X^I \nabla_\beta X^J \nn
&& +g _{\alpha \beta} \bigg(-{1 \over 6} Q_{IJ} F^I_{\beta_1 \beta_2} F^{J \beta_1 \beta_2}-6g^2 ({1 \over 2} Q^{IJ}-X^I X^J) V_I V_J \bigg)\ , 
\end{eqnarray}
and hence the ``$--$" component is
\begin{eqnarray}
R_{--} - Q_{IJ} F^I_{-i} F^J_{-j} \delta^{ij} - Q_{IJ} \nabla_- X^I \nabla_- X^J=0 \ .
\end{eqnarray}
This equation imposes the additional condition
\begin{eqnarray}
0 & = & {\tilde{\nabla}}^2 W + {\tilde{\nabla}}^i (W {\cal{B}}_i) - {\tilde{\nabla}}^i {\dot{\phi}}_i -3g \phi^i V_I (T^I)_i
- (\ddot{{\bf{e}}}^i)_i -3 (\dot{{\bf{e}}}^j)_i X_I (\star_3 T^I)^i{}_j \nn
& & +{1 \over 2} C_{IJK} X^K \big( (T^I)_i (T^J)^i + {\dot{X}}^I \dot{X}^J \big)\ .
\end{eqnarray}
We also require that the solution satisfies the scalar field equations. However, the integrability conditions of the fake Killing spinor equations, together with the gauge field equations, imply that the scalar field equations hold with no additional conditions.

\subsection{Summary of results obtained}
\label{subsec:summary}
In order to construct a supersymmetric solution in the null class, we introduce local coordinates $u, v, y^\alpha$, together with a family of Gauduchon-Tod 3-manifolds ${\rm GT}$, and write the metric as
\begin{eqnarray}
\label{Kundtmetric}
ds^2 = 2 {\bf{e}}^+ {\bf{e}}^- - ds_{{\rm GT}}^2\ , \qquad ds_{{\rm GT}}^2 = \delta_{ij} {\bf{e}}^i {\bf{e}}^j\ ,
\end{eqnarray}
with
\begin{equation}
\begin{array}{lll}
{\bf{e}}^+ &=& dv -{9 \over 2} g^2 v^2 (X^2-Q^{IJ} V_I V_J) du +  W du + v {\cal{B}} + \phi_i {\bf{e}}^i\ ,\\
{\bf{e}}^- &=& du\ , \\
{\bf{e}}^i &=& {e_\alpha}^i dy^\alpha\ ,
\end{array}
\end{equation}
where the basis elements ${\bf{e}}^i$ do not depend on $v$, but can depend on $u$, $W$ is a $v$-independent function and ${\cal{B}}= {\cal{B}}_i {\bf{e}}^i$, $\phi = \phi_i {\bf{e}}^i$ are $v$-independent 1-forms. The metric (\ref{Kundtmetric}) is a Kundt wave, whose recurrency properties we shall analyse in section \ref{sec:properties}. Because the base-space is Gauduchon-Tod, the basis elements ${\bf{e}}^i$ have to satisfy
\begin{eqnarray}
\label{GT}
{\tilde{d}} {\bf{e}}^i = {\cal{B}} \wedge {\bf{e}}^i +3gX \star_3 {\bf{e}}^i\ ,
\end{eqnarray}
where ${\tilde{d}}$ is the exterior derivative restricted to hypersurfaces of constant $v,u$ and $\star_3$ is the Hodge dual on the GT. The scalar fields satisfy
\begin{eqnarray}
\label{cond1}
\tilde{d} \star_3 \big( X^I {\cal{B}} + {\tilde{d}} X^I \big) &=&0\ ,
\end{eqnarray}
and
\begin{eqnarray}
\label{cond2}
{\tilde{d}} {\cal{B}} +3g \star_3 (X {\cal{B}} +{\tilde{d}} X ) &=&0 \ .
\end{eqnarray}
The field strengths are given by
\begin{equation}
\label{fs}
\begin{array}{ccl}
F^I &=& 3g (X X^I -Q^{IJ}V_J) ( dv + \phi) \wedge du +  \star_3 (X^I {\cal{B}}+ {\tilde{d}} X^I)\ , \\
\\
 && +du \wedge \left(- 3gv{\tilde{d}} (X X^I - Q^{IJ} V_J)  + T^I \right)\ ,
\end{array}
\end{equation}
where
\begin{eqnarray}
T^I = T^I{}_j {\bf{e}}^j
\end{eqnarray}
are $v$-independent 1-forms on the GT. The 1-forms $T^I$ must further satisfy
\begin{equation}
\label{gcond}
\begin{array}{lll}
\hspace{-7pt}{\tilde{d}} \star_3 (Q_{IJ} T^J) &=& -3g \cL_{{\partial \over \partial u}} \big(({3 \over 2} X X_I - V_I) {\rm dvol}_{{\rm GT}} \big)+{3 \over 2} \phi \wedge {\cal{B}} \wedge {\tilde{d}} X_I\\
\\
&&+{3 \over 2} {\tilde{d}} \phi \wedge ({\cal{B}} X_I - {\tilde{d}} X_I)+{1 \over 2} C_{IJK} T^J \wedge \star_3 (X^K {\cal{B}}+{\tilde{d}} X^K)\\
\\
&&+ 3g \phi \wedge \star_3 \big( ({3 \over 2} X X_I - V_I){\cal{B}} -{3 \over 2} X_I {\tilde{d}} X+{3 \over 2} X {\tilde{d}}X_I + Q_{IJ} {\tilde{d}} Q^{JN} V_N \big)\ ,
\end{array}
\end{equation}
as well as 
\begin{eqnarray}
\label{VT}
V_I T^I = -{1 \over 3g} {\dot{\cal{B}}} +3g (X^2-Q^{IJ} V_I V_J) \phi
\end{eqnarray}
and
\begin{eqnarray}
\label{XT}
X_I T^I = -{1 \over 3} \star_3 \bigg( {\tilde{d}} \phi - \phi \wedge {\cal{B}} + \delta_{ij} {\dot{{\bf{e}}}}^i \wedge {\bf{e}}^j \bigg)
\end{eqnarray}
and
\begin{eqnarray}
\label{dT}
{\tilde{d}} T^I = \cL_{{\partial \over \partial u}} \star_3 (X^I {\cal{B}}+ {\tilde{d}} X^I) +3g {\tilde{d}} \bigg( (X X^I - Q^{IJ} V_J) \phi \bigg) \ .
\end{eqnarray}
Finally, the function $W$ is found by solving
\begin{eqnarray}
\label{W}
{\tilde{\nabla}}^2 W + {\tilde{\nabla}}^i (W {\cal{B}}_i) - {\tilde{\nabla}}^i {\dot{\phi}}_i -3g \phi^i V_I (T^I)_i
- (\ddot{{\bf{e}}}^i)_i -3 (\dot{{\bf{e}}}^j)_i X_I (\star_3 T^I)^i{}_j &&\nn
+{1 \over 2} C_{IJK} X^K \big( (T^I)_i (T^J)^i + {\dot{X}}^I \dot{X}^J \big) &=&0 \ .
\end{eqnarray}
We remark that eqs.~(\ref{gcond}), (\ref{dT}) and (\ref{W}) always admit solutions, however it is not apparent a priori that eqs.~(\ref{VT}) and (\ref{XT}) can always be solved.

\section{Some simple examples}
\label{sec:examples}
We now present some simple solutions to the system prescribed above. 
\subsection{Near-horizon geometries}
\label{subsec:nearhorizon}
The near-horizon geometries found in \mycite{Gutowski:2010qv} are in fact all examples of fake-supersymmetric solutions in the null class. We take $\partial/{\partial u}$ as a symmetry of the full solution, and set the  $X^I$ to be constant, as well as $W=0$, $\phi =0$, $T^I=0$. The remaining conditions on the geometry thus simplify considerably, and one finds
\begin{eqnarray}
{\bf{e}}^- &=& du\ , \\
{\bf{e}}^+ &=& dv -{9 \over 2} g^2 (X^2-Q^{IJ} V_I V_J)\, v^2 du + v {\cal{B}}\ ,\\
{\bf{e}}^{\,i\,} &=& {e_\alpha}^i\, dy^\alpha\ , 
\end{eqnarray}
where
\begin{eqnarray}
{\tilde{d}} {\bf{e}}^i = {\cal{B}} \wedge {\bf{e}}^i +3gX \star_3 {\bf{e}}^i\ ,
\end{eqnarray}
and the gauge field strengths are
\begin{eqnarray}
F^I = 3g (X X^I -Q^{IJ} V_J)\,  dv \wedge du + X^I \star_3 {\cal{B}}\ ,
\end{eqnarray}
with the 1-form ${\cal{B}}$ satisfied
\begin{eqnarray}
{\tilde{d}} {\cal{B}} + 3gX \star_3 {\cal{B}}=0\ .
\end{eqnarray}
This solution can be interpreted as the (fake-supersymmetric) near-horizon geometry of a (possibly non-fake-supersymmetric) black hole. The case for which  $V_IX^I=0\,,\ {\cal{B}}\neq 0$  is of particular interest, as the spacetime geometry is $M_3 \times S^2$, where $M_3$ is a $U(1)$ vibration over $AdS_2$ related to the near-horizon extremal Kerr solution, and the spatial cross-sections of the event horizon are $S^1 \times S^2$.

\subsection{A small model of Real Special geometry}
\label{sec:simplemodel}
A particularly simple class of solutions can be constructed with only one vector multiplet. Since the five-dimensional gravity multiplet does not contain scalars, there is only one physical scalar field $\varphi$, and we choose the only non-vanishing value for the symmetric constant $C_{IJK}$ to be given by \mbox{$C_{122}=1$}. With this choice,

\begin{equation}
\label{eq:modelconstraints}
\begin{array}{c}
\displaystyle{
X^I=\left(\hspace{-2mm}\begin{tabular}{c}$\varphi^{-2}$\ ,\\
$\sqrt{2}\varphi$\end{tabular}\hspace{-2mm}\right)\ ,\quad X_I=\frac{1}{3}\left(\hspace{-2mm}\begin{tabular}{c}$\varphi^2$\ ,\\ 
$\sqrt{2}\varphi^{-1}$\end{tabular}\hspace{-2mm}\right)\ ,}\\ \\
Q_{IJ}=\frac{1}{2}\text{diag}\,(\varphi^4,\varphi^{-2})\ ,\quad Q^{IJ}=2\text{diag}\,(\varphi^{-4},\varphi^{2})\ .
\end{array}
\end{equation}
The equations resulting from the classification of the theory, summarised in subsection \ref{subsec:summary}, must also be satisfied, however we shall not present such analysis here.

This model is interesting as it provides a simple setting for non-supersymmetric solutions to a supersymmetric theory. The potential is given by
\begin{equation}
{\mathcal{V}}=9V_2(V_2 \varphi^2+2\sqrt{2}V_1 \varphi^{-1})\ ,
\end{equation}
and one can immediately see that if $V_2=0$ the theory is supersymmetric ({\em i.e.}~it was the definite-positiveness of $\mathcal{V}$ that granted us a De Sitter-like fSUGRA structure). This solution, however, is non-BPS, since the presence of $V_1$ means its KSE is not that of standard five-dimensional SUGRA. This kind of behaviour was already present in the classification of four-dimensional fSUGRA (see section \ref{sec:PotIsNul}), and thus one can also make use here of the oxidation/ dimensional reduction relations between supergravity theories (see {\em e.g.}~\mycite{LozanoTellechea:2002pn} or \mycite{Gutowski:2003rg}{sec.~(5.3)}) to obtain solutions to minimal $\N=(2,0)$ $d=6$ Supergravity.

To achieve this, we make use of the results developed in \mycite{LozanoTellechea:2002pn}, which provide the five-dimensional action (obtained through a KK compactification over an $S^1$) and compare it to the action for our model. The fields in these actions, however, do not promptly correspond, and they have to be appropiately identified. The dimensionally-reduced action is given by\footnote{We have adapted the conventions of \mycite{LozanoTellechea:2002pn} to those of here. In particular, the Riemmann tensor has the opposite defining sign.}
\begin{equation}
\label{reducedaction}
S=\int d^5 x\, \sqrt{|g|}\,k \left( -R-\frac{1}{4}k^2F^2(A)-\frac{1}{4}k^{-2}F^2(B)+\frac{\epsilon^{\mu\nu\rho\sigma\tau}}{8\sqrt{|g|}}k^{-1}F(A)_{\mu\nu} F(B)_{\rho\sigma} B_\tau\right)\ .
\end{equation}
The kinetic term for the graviton in this action does not have a canonical form, so we proceed to rescale the metric by a scalar $k$, and hence unveil the kinetic term for the scalars hidden in the Einstein-Hilbert term
\begin{equation}
\label{metricrescaling}
g_{\mu\nu} \rightarrow k^{\frac{-2}{3}} g_{\mu\nu}\ .
\end{equation}
The action hence becomes
\begin{equation} 
\label{eq:scaledaction}
S=\int d^5x\, \sqrt{|g|} \left( -R+\frac{4}{3}k^{-2} (\partial k)^2 -\frac{1}{4}k^{\frac{8}{3}} F^2(A)-\frac{1}{4}k^{-\frac{4}{3}} F^2(B)+\frac{\epsilon}{8\sqrt{|g|}} F(A)F(B)B \right)\ .
\end{equation}

We then compare this action with that of our model
\begin{eqnarray}
\label{actionmodel}
S&=&\int  d\text{vol}\left( -R+3\varphi^{-2}(\partial\varphi)^2-\frac{1}{4}\varphi^{4}(F^{1})^2- \frac{1}{4}\varphi^{-2} (F^{2})^2\right.\ ,\nn
&&\hspace{4em}\left. -\frac{\epsilon}{24\sqrt{|g|}} F^{2}F^{2}A^{1}-\frac{\epsilon}{12\sqrt{|g|}} F^{1}F^{2}A^{2}\right)\ ,
\end{eqnarray}
where the topological term is integrated by parts to identify the gauge fields. Upon inspection, $k=a\varphi^{\frac{3}{2}}$, where $a$ is just a real constant of integration, and
\begin{eqnarray}
A & = & -a^{-\frac{4}{3}} A^1\ , \\
B & = & \pm a^{\frac{2}{3}} A^2\ .
\end{eqnarray}
To identify the remaining elements we consider the supersymmetric variation of the gaugino, \emph{i.e.}~eq.~(\ref{dkse}), and that obtained from the dimensional reduction of the gravitino KSE of the six-dimensional theory (see \emph{e.g.}~\mycite{LozanoTellechea:2002pn}{eq.~(1.15)}). We obtain that $A_\mu=-A^1_\mu$, $B_\mu=-A^2_\mu$ and $k=-\varphi^{\frac{3}{2}}$.

This gives the identification of fields, and one can use the equations of the reduction over a circle to obtain the six-dimensional ones\footnote{In flat indices, the six-dimensional space is labelled by $a=\{0,1,2,3,4\}$ and $\sharp$, where $a$ spans the five-dimensional space.} 
\begin{eqnarray}
\label{eq:6dsolutions}
\begin{tabular}{c}
$g^{(6)}_{\mu\nu}=g^{(5)}_{\mu\nu}- \varphi^3 A^1_\mu A^1_\nu\ ,\hspace{1cm}g^{(6)}_{\mu\sharp}=-\varphi^3 A^1_\mu\ ,\hspace{1cm}g^{(6)}_{\sharp \sharp}=-\varphi^3\ ,$\\
\\
$H^-_{ab\sharp}=\varphi^{-\frac{3}{2}} F^2_{ab}\ ,\hspace{1cm} H^-_{abc}=-\varphi^{-\frac{3}{2}}\left(\star_\text{(6)}(e^\sharp \wedge F^2)\right)_{abc}\ .$\\
\end{tabular}
\end{eqnarray}
The geometries obtained in this manner, lifted from solutions to the model (\ref{eq:modelconstraints}) that fulfill the constraining equations of subsection \ref{subsec:summary}, are solutions to minimal $\N=(2,0)$ $d=6$ SUGRA with (the bosonic part of the) action given by
\begin{eqnarray}
\label{6daction}
\int d^6x \sqrt{|g|} \left( R+\frac{1}{24}(H^-)^2\right)\ ,
\end{eqnarray}
where as usual one considers the antiself-duality of $H^-$ as an additional constraint on the theory, rather than on the actual action.

\subsection{Embeddings of the Calderbank-Tod wave}
\label{subsec:CTwave}
Consider now the Gauduchon-Tod space presented in \mycite{Calderbank:1999ad} \footnote{See also the end of appendix \ref{sec:EWspaces}.}, embedded into an fSUGRA background as
\begin{equation}
ds^2=2du(dv -{9 \over 2} g^2 v^2 (X^2-Q^{IJ} V_I V_J) du + W du + v {\cal{B}} + \phi_i\,{\bf{e}}^i)-dx^2-|x+h|^2\:ds^2_{S^2}
\end{equation}
\begin{equation}
\label{CTwave} B=\frac{2x+h+\bar{h}}{|x+h|^2}\,dx\ ,\qquad X^I=-\frac{i(h-\bar{h})\,C^I}{3g|x+h|^2}\ ,
\end{equation}
where $C^I$ are constants such that $V_I\,C^I=1$, and $h=h(z)$ is a holomorphic and monotonic 
function. This wave fulfills eq.~(\ref{cond1}). 
However, the non-constancy of the $X^I$ spoils the identities in (\ref{eqn:conda}), and hence this GT space (without rescaling) is not a good cross-section for a solution to the fSUGRA theory. 

Alternatively, one might consider scaling the metric so that eq.~(\ref{ews}) is compatible with eq.~(\ref{eq:GTa}). This is
\begin{eqnarray}
ds^2&=&2du(dv -{9 \over 2} g^2 v^2 (X^2-Q^{IJ} V_I V_J) du + W du + v {\cal{B}} + \phi_i\,{\bf{e}}^i)\nn
&&\hspace{20mm}+\frac{(h-\bar{h})^2}{9|x+h|^4}\frac{1}{g^2 X^2}\left(dx^2+|x+h|^2\:ds^2_{S^2}\right)\ .
\end{eqnarray}
This inmediately says that the choice $h=\bar{h}$, which was prohibited in the minimal case \mycite{Grover:2009ms}, also cannot be used here. Furthermore $X\neq 0$, and it is constrained by eqs.~(\ref{cond1}) and (\ref{cond2}). 

\section{The Berger sphere provides a solution}
\label{subsec:Berger}
In this section we concentrate on the case for which the GT space is compact and without boundary for all $u$, and such that $X^I, T^I, \phi, W$ are smooth. The near-horizon geometries of the previous section are special examples of these solutions.

Consider eq.~(\ref{cond1}) and contract it with $X^I$. One finds that
\begin{equation}
{\tilde{\nabla}}^i {\cal{B}}_i +{2 \over 3} Q_{IJ} {\tilde{\nabla}}^i X^I {\tilde{\nabla}}_i X^J =0 \ .
\end{equation}
On integrating over ${\rm GT}$ one finds the constraint
\begin{equation}
\int_{{\rm GT}} Q_{IJ} {\tilde{\nabla}}^i X^I {\tilde{\nabla}}_i X^J =0\ ,
\end{equation}
which, assuming that $Q_{IJ}$ is positive-definite, implies that $X^I = X^I(u)$ and $X=X(u)$.

We shall consider eq.~(\ref{GT}) with $X \neq 0$ and take GT to be the Berger sphere{\footnote{If $X=0$ then the GT is either $S^1 \times S^2$ or $T^3$, according whether ${\cal{B}} \neq 0$ or ${\cal{B}}=0$, respectively. We do not consider those cases here.}} \mycite{Berger:1961,Gauduchon:1998}. One can then write
\begin{eqnarray}
\label{berg1}
ds^2_{{\rm GT}} & = & \displaystyle{\cos^2 (\mu) \over 9 g^2 X^2} \bigg( \cos^2 \mu (\sigma^3_L)^2  + (\sigma^1_L)^2+(\sigma^2_L)^2 \bigg)\ , \\
{\cal{B}} & = &  \sin(\mu)\cos(\mu)\,\sigma^3_L\ ,
\end{eqnarray}
where $\mu=\mu(u)$, and $\sigma^i_L$ are the left-invariant 1-forms on $SU(2)$ satisfying
\begin{equation}
\tilde{d}\sigma_L^i=-\frac{1}{2}\epsilon^{ijk}\,\sigma_L^j\wedge \sigma_L^k\ .
\end{equation}
Note that eqs.~(\ref{VT}) and (\ref{dT}) can be solved by making use of eq.~(\ref{cond2}), to give
\begin{equation}
\label{texp}
T^I = - {\cal{L}}_{\partial \over \partial u} \bigg( {X^I \over 3gX} {\cal{B}} \bigg) +3g (X X^I - Q^{IJ} V_J ) \phi + \Theta^I\ ,
\end{equation}
where $\Theta^I$ are 1-forms on ${\rm GT}$ satisfying
\begin{equation}
\label{tq1}
V_I \Theta^I =0
\end{equation}
and
\begin{equation}
\label{tq2}
{\tilde{d}} \Theta^I =0 \ .
\end{equation}

Next, simplify eq.~(\ref{gcond}) using eqs.~(\ref{XT}) and (\ref{dT}), using the identity
\begin{eqnarray}
{\tilde{d}} \bigg(  \delta_{ij} {\dot{{\bf{e}}}}^i \wedge {\bf{e}}^j  \bigg) =2 {\cal{B}} \wedge  \bigg(  \delta_{ij} {\dot{{\bf{e}}}}^i \wedge {\bf{e}}^j  \bigg) +9 g {\dot{X}} {\rm dvol_{GT}} +3 gX {\cal{L}}_{\partial \over \partial u} {\rm dvol_{GT}} \ .
\end{eqnarray} 
After some manipulation, one finds that (\ref{gcond}) can be rewritten as
\begin{eqnarray}
{\tilde{d}} \bigg( {3 \over 4} X_I   \delta_{ij} {\dot{{\bf{e}}}}^i \wedge {\bf{e}}^j -{1 \over 3gX} Q_{IJ} T^J \wedge {\cal{B}} -{1 \over X} V_I \phi \wedge {\cal{B}} \bigg)+{9 \over 2} g \big({1 \over 2}  {\dot{X}} X_I -X {\dot{X}}_I \big) {\rm dvol_{GT}}\phantom{aaaa}\nonumber \\
+ \big(3g V_I -{9 \over 4} g X X_I \big) {\cal{L}}_{\partial \over \partial u} {\rm dvol_{GT}}+{1 \over 2gX} \bigg( - {\dot{X}}_I {\star_3 {\cal{B}}} \wedge {\cal{B}} + X_I ({\cal{L}}_{\partial \over \partial u} \star_3 {\cal{B}}) \wedge {\cal{B}} \bigg) = 0 \ .
\end{eqnarray}
For the Berger sphere the second and third lines of this expression can be written in the form $Q^I(u) {\rm dvol_{GT}}\,$,
and hence on integrating over the GT one obtains two separate conditions
\begin{eqnarray}
\label{ccx1}
{\tilde{d}} \bigg( {3 \over 4} X_I   \delta_{ij} {\dot{{\bf{e}}}}^i \wedge {\bf{e}}^j -{1 \over 3gX} Q_{IJ} T^J \wedge {\cal{B}} -{1 \over X} V_I \phi \wedge {\cal{B}} \bigg) =0
\end{eqnarray}
and
\begin{eqnarray}
\label{ccx2}
{9 \over 2} g \big({1 \over 2}  {\dot{X}} X_I - X {\dot{X}}_I \big) {\rm dvol_{GT}}+ \big(3g V_I -{9 \over 4} g X X_I \big) {\cal{L}}_{\partial \over \partial u} {\rm dvol_{GT}}&&\nn
+{1 \over 2gX} \bigg( - {\dot{X}}_I {\star_3 {\cal{B}}} \wedge {\cal{B}} + X_I ({\cal{L}}_{\partial \over \partial u} 
\star_3 {\cal{B}}) \wedge {\cal{B}} \bigg) &=& 0 \ .
\end{eqnarray}
On using eq.~(\ref{texp}), eqs.~(\ref{ccx1}) and (\ref{XT}) can be rewritten as
\begin{eqnarray}
\label{ddx1}
{\tilde{d}} \bigg( \star_3 \Theta^I +{1 \over 3gX} \Theta^I \wedge {\cal{B}}
+3g (X X^I - Q^{IJ} V_J) \star_3 \phi +{1 \over 2} X^I   \delta_{ij} {\dot{{\bf{e}}}}^i \wedge {\bf{e}}^j  \bigg) =0
\end{eqnarray}
and
\begin{eqnarray}
\label{XT2}
-{\cal{L}}_{\partial \over \partial u} \bigg( {1 \over 3gX} {\cal{B}} \bigg) +gX \phi + X_I \Theta^I
= -{1 \over 3} \star_3 \bigg({\tilde{d}} \phi - \phi \wedge {\cal{B}} +   \delta_{ij} {\dot{{\bf{e}}}}^i \wedge {\bf{e}}^j  \bigg) \ .
\end{eqnarray}
On contracting eq.~(\ref{ddx1}) with $X_I$, and using eq.~(\ref{XT2}), one finds
\begin{eqnarray}
{\cal{L}}_{\partial \over \partial u} {\rm dvol_{GT}} = -3{{\dot{X}} \over X} {\rm dvol_{GT}}\ ,
\end{eqnarray}
which for the Berger sphere implies that the squashing of the $S^3$ is $u$-independent, \emph{i.e.}~$\mu$ is constant in 
eq.~(\ref{berg1}), and the $u$-dependence of the metric on the GT is inside the overall conformal factor of $X^{-2}$. 
It follows that
\begin{eqnarray}
{\cal{L}}_{\partial \over \partial u} \star_3 {\cal{B}} = -{{\dot{X}} \over X} \star_3 {\cal{B}}\ ,
\end{eqnarray}
and hence eq.~(\ref{ccx2}) can be simplified to give
\begin{eqnarray}
\label{ccx3}
{\dot{X}} X_I -{1 \over 2} X {\dot{X}}_I - {{\dot{X}} \over X} V_I +{1 \over 18g^2X^2} \big(- X {\dot{X}}_I - {\dot{X}} X_I \big) {\cal{B}}^2 =0 \ .
\end{eqnarray} 
On contracting this expression with $X^I$ and $Q^{IJ} V_J$ one finds
\begin{eqnarray}
{\dot{X}} {\cal{B}}^2 =0\ , \qquad (X^2-Q^{IJ} V_I V_J) {\dot{X}} =0 \ .
\end{eqnarray}
\phantom{a}

Suppose first that $X^2- Q^{IJ} V_I V_J \neq 0$. Then ${\dot{X}}=0$, and eq.~(\ref{ccx3}) further implies that 
\begin{eqnarray}
{\dot{X}}_I =0
\end{eqnarray}
Furthermore, contracting eq.~(\ref{ddx1}) with $V_I$ gives
\begin{eqnarray}
{\tilde{d}} \star_3 \phi =0\ ,
\end{eqnarray}
so eq.~(\ref{ddx1}) now reads
\begin{eqnarray}
{\tilde{d}} \star_3 \Theta^I + \Theta^I \wedge \star_3 {\cal{B}}=0 \ .
\end{eqnarray}
As ${\tilde{d}} \Theta^I=0$, and the Berger sphere is simply connected, this equation implies that the $\Theta^I$ are exact
\begin{eqnarray}
\Theta^I = {\tilde{d}} H^I\ ,
\end{eqnarray}
so that
\begin{eqnarray}
{\tilde{\nabla}}^2 H^I + {\cal{B}}^i {\tilde{\nabla}}_i H^I =0 \ .
\end{eqnarray}
It follows that ${\tilde{d}} H^I=0$, and so is $\Theta^I =0$. Also, eq.~(\ref{W}) implies that $W=W(u)$. It remains to consider the condition \ref{XT2}
\begin{eqnarray}
\label{auxcc}
gX \phi = -{1 \over 3} \star_3 \big( {\tilde{d}} \phi - \phi \wedge {\cal{B}} \big) \ .
\end{eqnarray}
%
After some manipulation, it can be shown that it implies
\begin{eqnarray}
{\tilde{\nabla}}^2 \phi^2 + {\cal{B}}^i {\tilde{\nabla}}_i \phi^2 = 2 {\tilde{\nabla}}^{(i} \phi^{j)} {\tilde{\nabla}}_{(i} \phi_{j)} + 3 \bigg( {\cal{B}}^2 \phi^2 - ({\cal{B}} \phi)^2 \bigg) \ .
\end{eqnarray}
As the RHS of this expression is a sum of two non-negative terms, it follows from the maximum principle that $\phi^2$ is constant, and
\begin{eqnarray}
{\tilde{\nabla}}_{(i} \phi_{j)} =0\ , \qquad  {\cal{B}}^2 \phi^2 - ({\cal{B}} \phi)^2 =0 \ .
\end{eqnarray}
These conditions imply that one can take, without loss of generality,
\begin{eqnarray}
\phi = k\, \sigma^3_L
\end{eqnarray}
for a constant $k$, irrespectively of whether ${\cal{B}}$ vanishes or not. Also note that by making a coordinate transformation of the form
\begin{eqnarray}
{\hat{u}}= f(u)\ , \qquad v = h(u) {\hat{v}} + g(u)\ , \qquad \psi = {\hat{\psi}} + \ell(u)\ ,
\end{eqnarray}
where we have taken the vector field dual to $\sigma^3_L$ to be $\partial/ \partial \psi$, one can choose the functions $f, h, g, \ell$ such that the form of the metric and gauge field strengths is preserved, and in the new coordinates $W=0$.

To summarise, if $X^2-Q^{IJ} V_I V_J \neq 0$, then the Berger sphere squashing-parameter $\mu$ and the $X^I$ are constant, and the background is given by 
\begin{eqnarray}
ds^2 &=& 2 du \left(dv -{9 \over 2} g^2 v^2 (X^2-Q^{IJ} V_I V_J) du + (v \sin \mu \cos \mu + k) \sigma^3_L\right)\nonumber\\
&& -{\cos^2 \mu \over 9 g^2 X^2} \bigg( \cos^2 \mu\, (\sigma^3_L)^2  + (\sigma^1_L)^2+(\sigma^2_L)^2 \bigg)\ , \\
F^I &=& 3g (X X^I -Q^{IJ}V_J)\, dv \wedge du + {X^I \over 3gX} \sin \mu \cos \mu \  \sigma^1_L \wedge \sigma^2_L \ ,
\end{eqnarray}
where $k$ is a constant.

In the special case $X^2-Q^{IJ} V_I V_J=0$ there are two possibilities. If ${\cal{B}} \neq 0$ then eq.~(\ref{ccx3}) implies that the $X^I$ are again constant, whereas if ${\cal{B}}=0$ then it implies that
\begin{eqnarray}
X_I = {2 \over 3} X^{-1} V_I + X^2 Z_I\ ,
\end{eqnarray}
for constant $Z_I$. Also, eqs.~(\ref{ddx1}) and (\ref{XT2}) imply
\begin{eqnarray}
H^I -{3 \over X} (X X^I-Q^{IJ}V_J) X_N H^N = L^I
\end{eqnarray}
and
\begin{eqnarray}
\phi = -{1 \over gX} \bigg( {\tilde{d}} (X_I H^I) + X_I H^I {\cal{B}} \bigg) + k(u) \sigma^3_L\ ,
\end{eqnarray}
where $\Theta^I ={\tilde{d}} H^I$, $H^I$ are functions, and $L^I=L^I(u)$ satisfy $X_I L^I =0$. Of course a simplified eq.~(\ref{W}) still needs to be solved, determining the function $W$.

\section{Solutions with a recurrent vector field}
\label{sec:properties}
The geometry we have found, eq.~(\ref{Kundtmetric}), is again a five-dimensional Kundt wave \mycite{Kundt:1961} (see also appendix \ref{appsec:Kundt}). Thus it admits a null vector generating a geodesic null congruence that is hypersurface orthogonal, non-expanding and shear-free. 
As in the case of minimal De Sitter $d=5$ Supergravity, the null vector field $N$ is not Killing. We shall consider the necessary and sufficient conditions for $N$ to be recurrent, which places additional restrictions on the holonomy of the Levi-Civit\`a connection. Recurrency with respect to this connection is defined as 
\begin{eqnarray}
\nabla_\mu N^\nu=C_\mu N^\nu \ ,
\end{eqnarray}
where $C$ is the recurrent one-form \mycite{Gibbons:2007zu}. $d$-dimensional geometries that allow recurrent vector fields have holonomy group contained in the Similitude group $\Sim(d-2)$. This is a $(d^2-3d+4)/2$-dimensional subgroup of the Lorentz group $SO(d-1,1)$, and it is isomorphic to the Euclidean group $E(d-2)$ augmented by homotheties. It is also the maximal proper subgroup of the Lorentz group, and hence connections admitting $Sim(d-2)$ have a minimal (non-trivial) holonomy reduction. See appendix \ref{appsec:Sim} for some technical information about this group.

As commented in chapter \ref{5dminimal}, theories with the Similitude group have received some attention in the past few years, as they have been shown to hold interesting physical features. They are linked to theories with vanishing quantum corrections \mycite{Coley:2008th} and to the recently proposed theories of \emph{Very Special Relativity} \mycite{Cohen:2006ky} and \emph{General Very Special Relativity} \mycite{Gibbons:2007iu}. For our solutions, note that
\begin{eqnarray}
\nabla_- N_j = {1 \over 2} {\cal{B}}_j\ ,
\end{eqnarray}
and so a necessary condition for the $N$ to be recurrent is ${\cal{B}}=0$. In fact, this is also sufficient, and one finds that if ${\cal{B}}=0$ then
\begin{eqnarray}
\nabla_\mu N^\nu = -9g^2 (X^2-Q^{IJ} V_I V_J) v N_\mu N^\nu \ .
\end{eqnarray}
For the remainder of this section we take ${\cal{B}}=0$, and investigate the resulting conditions imposed on the geometry.

Consider first eq.~(\ref{GT}). If $X \neq 0$ then the GT space is $S^3$, whereas if $X=0$ it is flat.
Next consider eq.~(\ref{cond1}), on contracting with $X_I$ this condition is equivalent to
\begin{eqnarray}
Q_{IJ} {\tilde{\nabla}}^i X^I {\tilde{\nabla}}_i X^J =0\ ,
\end{eqnarray}
and since $Q_{IJ}$ is positive-definite this implies that $X^I=X^I(u)$. Eqs.~(\ref{VT}) and (\ref{dT}) then further imply that
\begin{eqnarray}
T^I = 3g (X X^I - Q^{IJ} V_J) \phi + K^I\ ,
\end{eqnarray}
where $K^I$ are 1-forms on the GT satisfying
\begin{eqnarray}
{\tilde{d}} K^I = 0\ , \qquad V_I K^I =0\ ,
\end{eqnarray}
and eq.~(\ref{XT}) simplifies to
\begin{eqnarray}
gX \phi + X_I K^I = -{1 \over 3} \star_3 {\tilde{d}} \phi \ .
\end{eqnarray}
The conditions obtained from eq.~(\ref{gcond}) are 
\begin{eqnarray}
{\dot{X}}=0\ ,
\end{eqnarray}
and
\begin{eqnarray}
{\tilde{d}} \star_3 K^I = -3g (X X^I-Q^{IJ} V_J) {\tilde{d}} \star_3 \phi +3g X {\dot{X}}^I {\rm dvol_{GT}} \ .
\end{eqnarray}
In particular, note that if $X^2 - Q^{IJ} V_I V_J \neq 0$, then on contracting this condition with $V_I$ one finds that
\begin{eqnarray}
{\tilde{d}} \star_3 \phi =0\ , \qquad {\tilde{d}} \star_3 K^I = 3g X {\dot{X}}^I {\rm dvol_{GT}} \ .
\end{eqnarray}
The function $W$ must satisfy 
\begin{eqnarray}
\label{wrecur}
\hspace{-14pt}{\tilde{\nabla}}^2 W  & = & {\tilde{\nabla}}^i {\dot{\phi}}_i +9g^2 (X^2 - Q^{IJ}V_I V_J)\, \phi\cdot \phi -{1 \over 2} C_{IJK} X^K \big( (T^I)_i (T^J)^i + {\dot{X}}^I \dot{X}^J \big)\ ,
\end{eqnarray}
as a consequence of eq.~(\ref{W}).

Thus given $K^I, \phi, X^I, W$ satisfying these conditions, the metric and field strengths are
\begin{eqnarray}
\label{Bnullsol}
ds^2 & = & 2 du\left(dv -{9 \over 2} g^2 v^2 (X^2-Q^{IJ} V_I V_J) du +  W du +  \phi_i {\bf{e}}^i\right)- ds_{{\rm GT}}^2\ ,\\ 
F^I  & = & 3g (X X^I -Q^{IJ}V_J)\, dv \wedge du - K^I  \wedge du \ .
\end{eqnarray}
As a simple example, take $X \neq 0$, $X^I$ constant and $K^I=0$. Then the Gauduchon-Tod space is $S^3$. All of the conditions are satisfied if one takes $\phi=\xi_i\,\sigma_L^i$, where $\xi_i=\xi_i(u)$. With this choice of $\phi$, eq.~(\ref{wrecur}) implies that $W$ is a ($u$-dependent) harmonic function on $S^3$. This solution describes gravitational waves propagating through a generalized squashed Nariai universe. Generically, these waves will be plane-fronted waves, as $N$ is not a Killing vector. However, if $X^2-Q^{IJ}V_I V_J=0$ they are pp-waves.

Alternatively, taking again $X^I$ constant with $X \neq 0$, one can instead set $\phi=0$ and 
\begin{eqnarray}
K^I = K^I_i(u) \sigma_L^i \ .
\end{eqnarray}
Then the conditions which must be satisfied are
\begin{eqnarray}
V_I K^I= X_I K^I =0\ , \qquad {\tilde{\nabla}}^2 W = Q_{IJ} (K^I)_i (K^J)^i \ .
\end{eqnarray}
If $K^I \neq 0$, one must thus have non-vanishing $W$.\vspace{\baselineskip}

In \mycite{Grover:2009ms} it was shown that, for recurrent solutions in the minimal theory, all scalar curvature invariants constructed purely algebraically from the Riemann tensor are constant. By computing the Ricci scalar for the recurrent solutions constructed here, one can see that a necessary and sufficient condition for the Ricci scalar to be constant is that $Q^{IJ} V_I V_J$ is constant. This condition is also sufficient to ensure that all the other algebraic scalar curvature invariants are also constant. To see this, define
\begin{eqnarray}
\psi_{\mu \nu \lambda} &=& {2 \over 3} \nabla_\mu \nabla_{[\nu} \phi_{\lambda ]}+ {1 \over 3} \nabla_\nu \nabla_{[\mu} \phi_{\lambda ]} - {1 \over 3} \nabla_\lambda \nabla_{[\mu} \phi_{\nu ]}\ , \\
\theta_{\mu \nu} &=& \nabla_\mu \nabla_\nu W +9 g^2(X^2-Q^{IJ} V_I V_J) v \nabla_{(\mu}\phi_{\nu )} +{1 \over 4} (d \phi)_{\mu \lambda} (d \phi)_{\nu}{}^\lambda \ ,
\end{eqnarray}
and note that
\begin{eqnarray}
N^\mu \psi_{\mu \nu \lambda}=0\ , \qquad N^\mu \theta_{\mu \nu}=0 \ .
\end{eqnarray}
The Riemann tensor satisfies
\begin{eqnarray}
R_{\mu \nu \lambda \tau} = (R^0)_{\mu \nu  \lambda \tau} +4 N_{[\mu} \theta_{\nu ] [ \lambda} N_{ \tau]} +N_\mu \psi_{\nu \lambda \tau} - N_\nu \psi_{\mu \lambda \tau} + N_\lambda \psi_{\tau \mu \nu} -N_\tau \psi_{\lambda \mu \nu}\ ,
\end{eqnarray}
where $R^0$ is the Riemann tensor for the metric $g^0$ (obtained from the metric in eq.~(\ref{Bnullsol}) by setting $W=0$, $\phi=0$, where $(R^0)_{\mu \nu \lambda \tau} \equiv g^0_{\mu \kappa} (R^0)^\kappa{}_{\nu \lambda \tau}$). $g^0$ is the metric on $AdS_2 \times \text{GT}$, $\bR^{1,1} \times \text{GT}$ or $dS_2 \times {\rm GT}$ according to whether $X^2-Q^{IJ} V_I V_J$ is negative, zero or positive. It then follows, from exactly the same reasoning as set out for the minimal case, that all algebraic scalar curvature invariants constructed from the metric $g$ and the Riemann tensor $R$ are identical to the same invariants constructed from $g^0$ and $R^0$, and hence they are constant. The status of scalar curvature invariants constructed from covariant derivatives of the Riemann tensor remains to be determined.\vspace{\baselineskip}

For ${\cal{B}}\ne 0$, one can also provide the construction with a $\Sim$-holonomy structure. This is relevant \emph{e.g.}~for the embedding of the Berger sphere considered in section \ref{subsec:Berger}. By considering the gravitino fake Killing spinor equation, one obtains 
\begin{equation}
\nabla_\mu N_\nu=-3gA_\mu N_\nu + \frac{1}{2}X_I\, [\star (N\wedge F^I)]_{\mu\nu} \ .
\end{equation}
So define a new covariant derivative ${\cal{D}}$ by
\begin{equation}
\mathcal{D}_\mu N_\nu\equiv \nabla_\mu N_\nu- {S_{\mu\nu}}^\rho N_\rho= -3gA_\mu N_\nu\ ,\qquad \text{where}\ \  S=\frac{1}{2}\star (X_I F^I)
\end{equation}
can be interpreted as a totally antisymmetric torsion 3-form, and thus the new connection is metric-compatible. It is clear that that $N$ is then recurrent w.r.t.~the connection ${\cal{D}}$, and consequently its holonomy is a subgroup of $\Sim(3)$ \mycite{Gibbons:2007zu}.

\section{Summary of the chapter}
\label{secgaugednulld5:summary}
In this chapter we have analysed the geometric structure of the null case solutions of $\N=1$ $d=5$ fSUGRA coupled to Abelian vector multiplets. The general fake-supersymmetric solution is given by the Kundt wave
\begin{equation}
\label{Kundtmetricsummary}
ds^2=2du(dv -{9 \over 2} g^2 v^2 (X^2-Q^{IJ} V_I V_J) du +  W du + v {\cal{B}} + \phi_i {\bf{e}}^i)-h_{mn} dy^m\,dy^n\ ,
\end{equation}
where $h_{mn}=e_m^i\,e_n^i$ is the metric on the Gauduchon-Tod 3-space, $W$ is a $v$-independent function and ${\cal{B}}= {\cal{B}}_i {\bf{e}}^i$, $\phi = \phi_i {\bf{e}}^i$ are $v$-independent 1-forms. The field strengths are given by 
\begin{eqnarray}
\label{fssummary}
F^I &=& 3g (X X^I -Q^{IJ}V_J) ( dv + \phi) \wedge du +  \star_3 (X^I {\cal{B}}+ {\tilde{d}} X^I) \nn
 &&+ du \wedge \left(- 3gv{\tilde{d}} (X X^I - Q^{IJ} V_J)  + T^I \right)\ ,
\end{eqnarray}
where $T^I = T^I{}_j {\bf{e}}^j$ are $v$-independent 1-forms on the GT.

Furthermore, we have studied the conditions for which the 1-form bilinear is recurrent, so the holonomy of the Levi-Civit\`a connection is inside $\Sim(3)$, and investigated the properties of various scalar curvature invariants. We have found that recurrency is obtained by setting ${\cal{B}}=0$, and that depending on whether the norm of $X$ is vanishing or not, the GT space is either $\mathbb{R}^3$ or $S^3$, respectively. In addition, we find that our general recurrent solutions include plane-fronted waves propagating through a generalised squashed Nariai cosmos. For $X^2-Q^{IJ}V_I V_J=0$ these actually become pp-waves. Moreover, having $X^2-Q^{IJ}V_I V_J$ equal to a constant guarantees that all scalar curvature invariants constructed algebraically from the metric and the Riemann tensor are constant, and thus the ideas of \mycite{Coley:2008th} also apply in this set-up.

\cleardoublepage

\renewcommand{\leftmark}{\MakeUppercase{Chapter \thechapter. `Supersymmetric' Einstein-Weyl spaces}}
\chapter{\texorpdfstring{Classification of `Supersymmetric' Einstein-Weyl spaces}{'Supersymmetric' Einstein-Weyl spaces}}
\label{EWspaces}
The work in this chapter is similar in construction to that of previous ones, where we also employ techniques inherited from the classification of supersymmetric solutions to SUGRA theories to attack a problem of a different nature. We consider a `novel' KSE (in the sense that such KSE is not related \textit{a priori} to any supersymmetric setting), whose relevance becomes apparent once we analyse its integrability condition. This is the same as in previous chapters, but our motivation now is different from that of characterisation of solutions to fSUGRA theories. We are interested in classifying Lorentzian Einstein-Weyl spaces of arbitrary dimension, and the KSE is chosen in such a way that the integrability condition matches the geometric constraint for a manifold to be of Einstein-Weyl type. The chapter follows \mycite{Meessen:2011wb}, which contains the original work.

As said, the tools use here are the same ones employed in the programme of classification of solutions to supergravity theories, and we thus split the problem at hand according to whether they employ a timelike or null vector field. The characterisation we give is that of those EW spaces that arise from the existence of a \textit{Killing spinor}, \emph{i.e.}~a spinor that fulfills the KSE we propose, and it is in this sense that we refer to them as supersymmetric geometries.

The outline of the chapter is the following: section \ref{sec:maths} introduces the spinorial rule, its integrability condition (which resembles the geometric constraint for Einstein-Weyl spaces) and a short manipulation on a vector bilinear valid for all dimensions and cases. Section~\ref{sec:timelike} analyses all possible timelike cases, showing their triviality. Section~\ref{sec:N1D4} describes the null solutions for the $\N=1$, $d=4$ case, while section~\ref{sec:D6chiral} treats the $d=6$ null case, and section~\ref{sec:general} includes the remaining ones. Section~\ref{sec:conclusions} has a summary of the chapter. For the interested reader, appendix~\ref{sec:Weylgeometry} gives some information on Weyl geometry and Einstein-Weyl spaces and appendix~\ref{appsec:spinors} presents the spinorial notation we employ. A little description on the geometry of Kundt waves, which show up as solutions, is given in appendix~\ref{appsec:Kundt}.

\section{Covariant rule and the Einstein-Weyl condition}
\label{sec:maths} 
Consider the following rule for the covariant derivative of some spinor $\epsilon$, which we shall take to be Dirac\footnote{In order to construct bilinears, it is useful to impose a bit more structure on the spinor. This naturally leads to question the compatibility of eq.~(\ref{eq:1}) with the conditions for the existence of different kinds of spinors (Weyl, Majorana, Majorana-Weyl, etc.). Indeed, such additional constrains are in fact compatible with the given rule of parallelity of $\epsilon$.}

\begin{equation}
\label{eq:1}
\nabla_{a}\epsilon \; =\; \textstyle{\frac{4-d}{4}}\ A_{a}\epsilon \; +\; \textstyle{1\over 2} \gamma_{ab}A^{b}\epsilon \ ,
\end{equation}
where $d$ is the number of spacetime dimensions and $A$ is just some real 1-form, which at this point is completely unconstrained. This equation is related to the fKSE (\ref{eq:fKSExRes}) in chapter \ref{5dminimal}, in that the latter can be obtained from the former (for $d=4$) by applying the Jones-Tod reduccion mechanism \mycite{Tod:1985}. We shall call the solutions $\epsilon$ to eq.~(\ref{eq:1}) Killing spinors, and the corresponding metric and 1-form a supersymmetric field configuration. Observe that with our choice of Dirac conjugate, the above rule implies
\begin{equation}
  \label{eq:38}
  \nabla_{a}\overline{\epsilon} \; =\; \textstyle{\frac{4-d}{4}}\, A_{a}\overline{\epsilon} \; -\; \textstyle{1\over 2}A^{b}\; \overline{\epsilon}\gamma_{ab} \ .
\end{equation}

A straightforward calculation of the integrability condition leads to
\begin{equation}
\label{eq:2}
\textstyle{1\over 2} \gamma_{a}\slashed{F}\,\epsilon \, =\, \textstyle{1\over 2}\, \mathrm{W}_{(ab)}\gamma^{b}\epsilon\ ,
\end{equation}
where $F\equiv dA$ is called the Faraday tensor and 
\begin{equation}
  \label{eq:14}
  \mathrm{W}_{(ab)} \; =\; \mathtt{R}(g)_{ab}\,-\, (d-2) \nabla_{(a}A_{b)} \, -\, (d-2)\, A_{a}A_{b}\, -\, g_{ab}\, \left(\nabla_{c}A^{c} \, -\, (d-2)\,A_{c}A^{c} \right)\ ,
\end{equation}
which is readily identified with (the symmetric part of) the Ricci tensor in Weyl geometry (see appendix~\ref{sec:EWspaces} for a small introduction). Contracting the above integrability condition with $\gamma^{a}$ one finds that
\begin{equation}
  \label{eq:3}
  d\slashed{F}\epsilon \; =\; \mathrm{W}\ \epsilon \ ,
\end{equation}
which when combined with eq.~(\ref{eq:2}) leads to 
\begin{equation}
  \label{eq:4}
  \textstyle{1\over 2}\left(\mathrm{W}_{(ab)} \; -\; \textstyle{1\over d}\,\eta_{ab}\mathrm{W}\right)\, \gamma^{b}\epsilon \; =\; 0\ .
\end{equation}
In the Riemannian setting the above is enough to conclude that if we find a spinor $\epsilon$ satisfying eq.~(\ref{eq:1}), then the underlying geometry is Einstein-Weyl. In the non-Riemannian setting this conclusion is however not true; as in the classification of supersymmetric solutions to supergravity theories, there are two quite different cases to be considered, namely the timelike or the null case. The minimal set of equations of motion that need to be imposed in order to guarantee that all EOMs are satisfied is different in each case: in the timelike case a supersymmetric field configuration automatically satisfies the EW condition, whereas in the null case the minimal set consists of only one component of the EW condition, namely the one lying in the double direction of the null vector bilinear.

Seeing the similarity of the integrability condition of the spinorial rule with the geometric constraint for EW spaces, it should not come as a surprise that eq.~(\ref{eq:1}) is invariant under the following Weyl transformations
\begin{equation}
  \label{eq:16}
  \begin{array}{lclclcl}
    g & =& e^{2w}\tilde{g}\ , & \qquad& e^{a} & =& e^{w}\tilde{e}^{a} \ ,\\
    A & =& \tilde{A}+dw\ , & & \theta_{a} & =& e^{-w}\tilde{\theta}_{a} \ , \\
    \epsilon & =& e^{\alpha w}\tilde{\epsilon}\ , & & \alpha & =& \textstyle{\frac{4-d}{4}} \ .
  \end{array}
\end{equation}
This symmetry can in fact be used to obtain the RHS of eq.~(\ref{eq:1}), which would otherwise have to be wild-guessed: since by definition the structure of all EW spaces will contain it, the appropiate connection to use is the Weyl one, eq.~(\ref{eq:W2}), which in the spinorial representation is given by $\mathtt{D}_a=\nabla_a-\frac{1}{2}\gamma_{ab}A^b\,$. Furthermore, an additional $\alpha A_a$-term serves to preserve the spinor transformation rule, and the constant $\alpha$ is chosen by demanding that the integrability condition of the resulting equation includes the criterion for EW spaces. This thus allows us to formulate the parallelity equation as $\mathtt{D}_a\epsilon=\frac{4-d}{4}\,A_a\,\epsilon$. In other words, we have the structure of a weighted Killing spinor in Weyl geometry. 

The next step is to define the bilinear $\hat{L} = L_{\mu}dx^{\mu} = \overline{\epsilon}\gamma_{\mu}\epsilon\ dx^{\mu}$, which (as shown in appendix~\ref{appsec:spinors}) is a real 1-form and for a Lorentzian spacetime is either timelike, \emph{i.e.}~$g(L,L)>0$ in our conventions, or null $g(L,L)=0$.  In any case, one can always derive from the spinorial equation (\ref{eq:1}) the following differential rule for the bilinear
\begin{equation}
  \label{eq:33}
  \nabla_{a}L_{b} \; =\; \frac{4-d}{2}\, A_{a}L_{b} \; -\; L_{a}A_{b} \; +\; \imath_{L}A\ g_{ab}\ ,
\end{equation}
whose totally antisymmetric part reads
\begin{equation}
  \label{eq:6}
  d\hat{L} \; =\; \frac{6-d}{2}\ A\wedge \hat{L} \ ,
\end{equation}
singling out the $d=6$ case as special, as $\hat{L}$ is then closed.

We start the analysis by considering the timelike case.
\section{Timelike solutions}
\label{sec:timelike}
Suppose that $L$ is timelike and define $f\equiv g(L,L)$. We can straightforwardly use eq.~(\ref{eq:33}) to find
\begin{equation}
  \label{eq:34}
   df \; =\; (4-d)A\, f\ .
\end{equation}
This implies that, as long as $d\neq 4$, the Weyl structure is exact and any supersymmetric EW space is equivalent to a metrical space allowing for a parallel spinor (w.r.t.~the Levi-Civit\`a connection). Bryant classified all the pseudo-Riemannian spaces admitting covariantly constant spinors for a different number of dimensions \mycite{Bryant:2000}. Thus, this prescribes the \emph{timelike} Einstein-Weyl metrics with Lorentzian signature in dimensions three (flat), five and six ($g=\mathbb{R}^{1,d-5} \times \tilde{g}$, where $\tilde{g}$ is a four-dimensional Ricci-flat K\"ahler manifold). A general study for the remaining dimensions is still an open problem, as far as we know. However, Galaev and Leistner provide a partial answer by giving a blueprint for the geometry of simply-connected, complete Lorentzian spin manifolds that admit a Killing spinor \mycite{Galaev:2008}{Th.~1.3}.

For the $d=4$ case, we use the same building blocks as in chapter \ref{4d} to set up the whole calculus of spinor bilinears. We deal with the spinor structure of $\N=2$ $d=4$ supersymmetry, which allows us to decompose a Dirac spinor as a sum of two Majorana spinors, which we can then project onto its anti-chiral part, denoted $\epsilon_{I}$ ($I=1,2$), and its chiral part $\epsilon^{I}$. Note that here the position of the $I$-index indicates exclusively the chirality, and these are interchanged by complex conjugation, {\em i.e.\/} $(\epsilon_{I})^{*}=\epsilon^{I}$, so the theory has two independent spinorial fields. Doing this decomposition, the rule (\ref{eq:1}) can then be written as
\begin{equation}
  \label{eq:35}
  \nabla_{a}\epsilon_{I} \; =\; \textstyle{1\over 2}\gamma_{ab}A^{b}\epsilon_{I} \qquad\mbox{and}\qquad \nabla_{a}\epsilon^{I} \; =\; \textstyle{1\over 2}\gamma_{ab}A^{b}\epsilon^{I} \ .
\end{equation}
Using the spinors one can then construct (cf. chapter \ref{4d}) a complex scalar $X\equiv\frac{1}{2}\varepsilon^{IJ}\bar{\epsilon}_I\epsilon_J$, three complex 2-forms $\Phi^{x}$ ($x=1,2,3$) that will not play any r\^{o}le in what follows, and four real 1-forms $V^{a}=i\bar{\epsilon}^I \gamma^a \epsilon_I$. These latter ones form a linearly independent base and can be used to write the metric $g$ as
\begin{equation}
  \label{eq:36}
  4|X|^{2}\, g \; =\; \eta_{ab}\, V^{a}\otimes V^{b} \ ,
\end{equation}
whence $V^{0}\sim L$. Given the definitions of the bilinears we can calculate
\begin{eqnarray}
\label{eq:37}
dX  &=& 0\ ,\\
\label{eq:37a}
dV^{a} &=& A\wedge V^{a} \ ,
\end{eqnarray}
meaning that $X$ is just a complex constant. The integrability condition of eq.~(\ref{eq:37a}) is $F\wedge V^{a} = 0$ which, due to the linear-independency of the $V^{a}$, implies that $F=0$. Locally, then, we can transform $A$ to zero and introduce coordinates $x^{a}$ such that $V^{a}= 4|X|^{2}\,dx^{a}$, resulting in a Minkowski metric. Thus, a \emph{timelike} supersymmetric four-dimensional Lorentzian EW space is locally conformal to Minkowski space.

The conclusion then w.r.t.~the \emph{timelike} solutions to the rule (\ref{eq:1}) is that they are trivial in the sense that they are always related by a Weyl transformation to a Lorentzian space admitting parallel spinors, {\em i.e.\/} spinors satisfying the rule $\nabla_{a}\epsilon =0$.

The analysis of the null cases is more involved, mainly due to a lack of systematics in the bilinears\footnote{The exception is the vector bilinear $L$, as one can see from eq.~(\ref{eq:33}).}, but also because the bilinear approach to the classification of supersymmetric solutions becomes unwieldy for $d>6$. Instead of attempting to do a complete analysis in all the cases where the bilinear approach can be applied, we shall analyse the cases $d=4$ and $d=6$ explicitly, and then give some generic comments about the rest in section (\ref{sec:general}).

\section{\texorpdfstring{Null $\N=1$ $d=4$ solutions}{Null N=1 d=4 solutions}}
\label{sec:N1D4}
In view of the explicit case treated in the foregoing section, the natural starting point for this analysis would be the null case in $\N=2$ $d=4$. Prior experience with this case in Supergravity, however, shows that this is related to the simpler case of $\N=1$ $d=4$ Supergravity \mycite{Ortin:2008wj}, a theory for which the vector bilinear $L$ is automatically a null vector, and the spinor is of Weyl type. As commented above, the KSE (\ref{eq:1}) is in fact compatible with the truncation of $\epsilon$ to a chiral spinor, and thus for the rest of this section we shall take $\epsilon$ to be a Weyl spinor.

The first rule one can derive for the bilinear is
\begin{equation}
  \label{eq:5}
  \nabla_{a}L_{b} \; =\; - L_{a}A_{b} \; +\; \imath_{L}A\, g_{ab} \ ,
\end{equation}
which is enough to see that $L^{\flat}$ is a geodesic null vector. Its antisymmetric and symmetric parts read
\begin{eqnarray}
  \label{eq:7}
  d\hat{L} & =& A\wedge \hat{L} \ , \\
  \label{eq:8}
  \nabla_{(a} L_{b)} & =& - A_{(a}L_{b)}  \; +\; \textstyle{1\over 3}\, \nabla\cdot L\ g_{ab}\ .
\end{eqnarray}
Another bilinear that can be constructed is a 2-form defined as $\Phi_{ab}=\overline{\epsilon}\gamma_{ab}\epsilon$ \mycite{Ortin:2008wj}. By using the propagation rule one can deduce
\begin{equation}
  \label{eq:17}
  \nabla_{a}\Phi_{bc} \; =\; 2\Phi_{a[b}A_{c]} \; -\; 2 g_{a[b}\Phi_{c]d}A^{d} \ ,
\end{equation}
which through antisymmetrisation gives rise to
\begin{equation}
  \label{eq:18}
  d\Phi \; =\; 2A\wedge\Phi \ .
\end{equation}

Eq.~(\ref{eq:7}) implies that $\hat{L}\wedge d\hat{L} =0$, whence $\hat{L}$ is hypersurface orthogonal, and one can use the Frobenius theorem to introduce two real functions $u$ and $P$ such that $\hat{L}=e^{P}du$. Since by eq.~(\ref{eq:7}) $\hat{L}$ has gauge charge 1 under $A$, one can perform a Weyl-gauge transformation to take $P=0$ and thus obtain $\hat{L}=du$. This further implies that $A=\Upsilon\,\hat{L}$, where $\Upsilon$ is a real function whose coordinate dependence needs to be deduced, and also that $\imath_{L}A=0$. Furthermore, we see that $d^{\dagger} \hat{L}=0$ and $\nabla_{L}L=0$, {\em i.e.\/} $L$ is the tangent vector to an affinely parametrised null geodesic.

Observe that one can apply the same reasoning for eq.~(\ref{eq:6}) in dimensions different from six: as long as $d\neq 6$ we can always use a Weyl transformation to fix $\hat{L}=du$ and write $A=\Upsilon\,\hat{L}$. The fact that in $d=6$ the 1-form $\hat{L}$ is automatically closed has profound implications, as will be shown in section~(\ref{sec:D6chiral}).\vspace{\baselineskip}

Having fixed the Weyl symmetry, we can now introduce a normalised null tetrad \mycite{Penrose:1985jw} and a corresponding coordinate representation by
\begin{equation}
\label{eq:19}
\begin{array}{lclllcl}
\hat{L} & = &  du\ , & & L & =& \partial_{v} \ , \\
\hat{N} & = & dv + Hdu +\varpi dz + \bar{\varpi}d\bar{z}\ ,&\qquad &N & =& \partial_{u}\ -\ H\partial_{v} \ , \\
\hat{M} & = & Udz\ , & & M & =& -\bar{U}^{-1}\left( \partial_{\bar{z}} \; -\; \bar{\varpi}\partial_{v}\right) \ ,\\
\hat{\overline{M}} & = & \bar{U}d\bar{z}\ ,& &\overline{M} & =& -U^{-1}\left( \partial_{z} \; -\; \varpi\partial_{v}\right) \ ,
\end{array}
\end{equation}
for which the metric reads
\begin{eqnarray}
\label{eq:20}
g & = & \hat{L}\otimes \hat{N} \ +\ \hat{N}\otimes \hat{L} \ -\hat{M}\otimes\hat{\overline{M}} \ -\ \hat{\overline{M}}\otimes \hat{M} \nonumber \\
& = & 2du\left( dv + Hdu + \varpi dz + \bar{\varpi}d\bar{z}\right) \ -\ 2|U|^{2} dzd\bar{z} \ .
\end{eqnarray}
A straightforward calculation shows that the constraint (\ref{eq:8}) implies that 
\begin{equation}
\label{eq:21}
\Upsilon = -\partial_{v}H\ , \qquad\partial_{v}\varpi \; =\; 0\ ,\qquad\partial_{v}\bar{\varpi} \; =\; 0\ ,\qquad \partial_{v}|U|^{2} \; =\; 0 \ ,
\end{equation}
so that the only $v$-dependence resides in the function $H$, and we have thus determined the gauge field $A$ in terms of it. Moreover, in $\N=1$ $d=4$ theory one can see that $\Phi = \hat{L}\wedge\hat{\overline{M}}$ (see {\em e.g.\/} eq.~(\ref{eq:2j}) in chapter \ref{4d}). Combining this with eq.~(\ref{eq:18}) one has
\begin{equation}
  \label{eq:22}
  0 \; =\; \hat{L}\wedge d\hat{\overline{M}} \; =\; d\bar{U}\wedge d\bar{z}\wedge du\ ,\qquad\mbox{whence}\quad  \bar{U} \; =\; \bar{U}(u,\bar{z}) \ . 
\end{equation}
This result means that we can take $U=1$ by a suitable coordinate transformation $Z=Z(u,z)$ with $\partial_{z}Z = U$, which leaves the chosen form of the metric invariant.

To finish the analysis we shall investigate eq.~(\ref{eq:17}). As $A\sim \hat{L}$ we have that
\begin{equation}
\imath_{A}\Phi \sim \imath_{L}\Phi = 0\ ,
\end{equation}
and we find that $\nabla_{a}\Phi_{bc} = 2\Upsilon\; \Phi_{a[b}L_{c]}$. Combining this with $\Phi_{ab} = 2L_{[a}\overline{M}_{b]}$ we have
\begin{equation}
  \label{eq:23}
  0 \; =\; L_{[b|}\nabla_{a}\overline{M}_{|c]} \ ,
\end{equation}
which can be evaluated on the chosen coordinate basis to give
\begin{equation}
  \label{eq:24}
  0\; =\; \partial_{\bar{z}}\varpi \; -\; \partial_{z}\bar{\varpi}\ ,
\end{equation}
which implies
\begin{equation}
\varpi \; =\; \partial_{z}B \ ,\qquad \bar{\varpi}\; =\; \partial_{\bar{z}}B \ ,
\end{equation}
for $B$ a real function. As is well known, one can then get rid of $\varpi$ altogether by a suitable shift of the coordinate $v \rightarrow v-B$.

The end result is that, given a Weyl spinor $\epsilon$, any solution\footnote{By solution we refer to a geometry that arises from the existence of a spinor that fulfills eq.~(\ref{eq:1}).} to the equation~(\ref{eq:1}) is related by a Weyl transformation to 
\begin{eqnarray}
\label{eq:10}
&ds^{2}_{(4)} = 2du\left( dv \; +\; Hdu\right) \; -\; 2dzd\bar{z} \ , \\
& A = -\partial_{v}H\, du \ .
\end{eqnarray}
As a matter of fact, this metric is a special case of a more general metric referred to as a Kundt metric\footnote{Moreover, it is in the Walker form \mycite{Walker:1950}.} in the Physics literature (see appendix \ref{appsec:Kundt} for more information). This kind of metric appears naturally in the null case of not only Supergravity solutions \mycite{Brannlund:2008zf}, but also fakeSupergravity ones, as was seen for example in chapters \ref{4d} and \ref{5dminimal}, or in \mycite{Gutowski:2009vb}.\vspace{\baselineskip}

We now return to the topic of pseudo-Riemannian signatures and certain EOMs (the EW conditions in this case) being automatically satisfied. Since we are trying to give a prescription for EW spaces, we are obviously bound to satisfy eq.~(\ref{eq:W5}). An explicit calculation shows that the integrability conditions (\ref{eq:4}) are automatically satisfied, with the only non-trivial component being $\mathrm{W}(N,N) L_{c}\gamma^{c}\epsilon$. Adapting the Fierz identities to the null case scenario one obtains the constraint $L_{c}\gamma^{c}\epsilon =0$ (see {\em e.g.}~\mycite{Bellorin:2005zc}{eq.~(5.12)}), satisfying this way all of (\ref{eq:4}) \footnote{Aditionally, one reaches the same conclusion by analysing the sufficiency of the derived constraints arising from the KSE.}.

However, one still needs to ensure that the local geometry (\ref{eq:10}) indeed solves all EW conditions (\ref{eq:W5}), and we must therefore impose by hand that $\mathrm{W}(N,N)=0$. A small calculation shows that this implies that $H$
must satisfy the following differential equation
\begin{equation}
  \label{eq:11}
  \partial_{u}\partial_{v}H \; -\; H\partial_{v}^{2}H  \; =\; \partial\bar{\partial}H \ . 
\end{equation}
One can easily see that the above equation is invariant under the substitutions $u\rightarrow u$, $v\rightarrow \lambda^{2}v$, $z\rightarrow \lambda z$, $\bar{z}\rightarrow \lambda\bar{z}$ and $H\rightarrow \lambda^{2}H$, which implies that for cases for which $H$ is a weighted-homogeneous function, {\em i.e.\/}
\begin{equation}
\label{eq:42}
H\left(u,\lambda^{2}v,\lambda z,\lambda\bar{z}\right) \; =\; \lambda^{2}\, H(u,v,z,\bar{z}) \ ,
\end{equation}
there exists a homothety $K$
\begin{equation}
\label{eq:13}
K \; =\; 2v\partial_{v} \; +\; z\partial \; +\; \bar{z}\bar{\partial} \quad \longrightarrow\quad\pounds_{K}g \; =\; 2\, g \ .
\end{equation}
It should be noted that it is not the case that every solution to eq.~(\ref{eq:11}) is homogeneous, as exemplified by the following three solutions.
\begin{equation}
\label{eq:12a}
\begin{array}{c}
\mathtt{i)}\; H \; =\; uv + |z|^{2} \ , \hspace{2cm} \mathtt{ii)}\; H \; =\; -\displaystyle\frac{v^{2}}{2|z|^{2}} \ ,\\
\\
\mathtt{iii)}\; H\;=\; v\partial F +v\bar{\partial} \bar{F}+\bar{z}\partial_u F +z \partial_u \bar{F}\ , \qquad \text{where } F=F(u,z)\ .
\end{array}
\end{equation}
The first two can be seen to be homogeneous, while the third one would only be so for the special case that $F=zf(u)$, for any function $f(u)$.

Moreover, one can see that the field $A$ in example $\mathtt{i)}$ is pure gauge, namely $A = d( -u^{2}/2)$, so one can do a conformal rescaling $g\rightarrow e^{2\omega}\tilde{g}$ and $A=\tilde{A}+d\omega$ such that $\tilde{A}=0$, {\em i.e.\/}~we take $\omega = -u^{2}/2$.  The coordinate transformations $v=e^{\omega}t$, $z=e^{\omega}y$ followed by $t=s-ue^{\omega}|y|^{2}$, then takes the metric $\tilde{g}$ to a Minkowski metric. Case $\mathtt{i)}$ is thus conformally Minkowski. This is however only a particular example, and one cannot generically say that all solutions that exhibit an exact Weyl structure ({\em i.e.}~trivial solutions) are conformally related to flat space. For illustrative purposes note that for $A=0$ the wave profile is of the type $H= f(u,z)+\bar{f}(u,\bar{z})$; if one then calculates the Riemannian curvature one obtains $R_{+\circ +\circ}=\partial^{2}f$ and $R_{+\bullet +\bullet}=\bar{\partial}^{2}\bar{f}$, hence it is not generically a flat space. In the general case $A=dw$, the wave profile is prescribed to be $H=-vf(u)+K(u,z,\bar{z})$. The requirement for being an EW space dictates that $\partial_u f(u)=-\partial\bar{\partial}K$, and the non-vanishing components of the Weyl tensor are
\begin{eqnarray}
C_{+\circ + \circ}=\partial^2 K\ ,\qquad C_{+\bullet + \bullet}=\bar{\partial}^2 K\ ,
\end{eqnarray}
whence only for cases where $\partial^2 K=0=\bar{\partial}^2 K$ is our space conformal to Minkowski space. Case $\mathtt{i)}$ falls under these conditions.

Furthermore, example $\mathtt{iii)}$ above is a Weyl-scalar-flat background. It is a four-dimensional generalisation of the three-dimensional Weyl-flat EW geometries obtained in \mycite{Calderbank:2001}. It gives rise to a non-trivial EW space as long as $\partial^2 F\neq 0$. More about this correspondence will be said in section (\ref{sec:general}).

\section{\texorpdfstring{Null $\N=(1,0)$ $d=6$ solutions}{Null N=(1,0) d=6 solutions}}
\label{sec:D6chiral}
As in the foregoing section, we shall be considering a chiral spinor $\epsilon$. The Fierz identities imply that the vector bilinear is null. Moreover, one can also make use of \mycite{Gutowski:2003rg}, where the supersymmetric solutions of ungauged chiral supergravity in six dimensions, {\em i.e.\/} minimal $\N=(1,0)$ $d=6$ Supergravity \mycite{Cariglia:2004kk}, are prescribed. This theory is in itself quite curious, and so are the spinor bilinears: there is only a null vector $L$ and a triplet of self-dual 3-forms $\Phi^{r}_{(3)}$ ($r=1,2,3$). The bilinears are defined by
\begin{equation}
\label{eq:40}
\begin{array}{lclclcl}
L_{a} & \equiv& -\varepsilon^{IJ}\ \epsilon_{I}^{c}\gamma_{a}\epsilon_{J}\ , &\qquad& \epsilon_{I}^{c}\gamma_{a}\epsilon_{J} & =& -\textstyle{1\over 2}\ \varepsilon_{IJ}\ L_{a} \ ,\\
&& && && \\
\Phi^{r}_{abc} &\equiv& i\left[\sigma^{r}\right]^{IJ}\,\epsilon_{I}^{c}\gamma_{abc}\epsilon_{J}\ , &&\epsilon_{I}^{c}\gamma_{abc}\epsilon_{J} & =& \textstyle{i\over 2}\; \left[\sigma^{r}\right]_{IJ}\, \Phi^{r}_{abc} \ , 
\end{array}
\end{equation}
where $\epsilon^{c}=\epsilon^{T}\mathcal{C}$ denotes the Majorana conjugate. These bilinears satisfy the following Fierz relations
\begin{eqnarray}
  \label{eq:39}
  L_{a}L^{a} &=& 0 \ , \\
  \label{eq:39a}
  \imath_{L}\Phi^{r}_{(3)} &=& 0 \qquad\longrightarrow\qquad \hat{L}\wedge\Phi^{r}_{(3)} \; =\; 0 \ , \\
  \label{eq:39b}
  \Phi^{r\ fab}\Phi^{s}_{fcd} &=& 4\delta^{rs}\ L^{[a}L_{[c}\ \eta^{b]}_{d]} \; -\; \varepsilon^{rst}L^{[a|}\Phi^{t\ |b]}{}_{cd}\; +\; \varepsilon^{rst}L_{[c}\Phi^{t\ ab}{}_{d]} \ .
\end{eqnarray}
With means of eqs.~(\ref{eq:39a}) and (\ref{eq:39b}) one finds that $\Phi^{r}_{(3)} =\hat{L}\wedge\mathsf{K}^{r}_{(2)}$, with $\imath_{L}\mathsf{K}_{(2)}^{r}=0$.

We use the rule (\ref{eq:1}) to calculate the effect of parallel-transporting the bilinears. The result is that for an arbitrary vector field $X$ we have
\begin{eqnarray}
\label{eq:43}
\nabla_{X}\hat{L} & = & -\imath_{X}A\, \hat{L} \; -\; \imath_{X}\hat{L}\, A \; +\; \imath_{L}A\ \hat{X} \ ,\\
\label{eq:43a}
\nabla_{X}\Phi^{r} & = & -\imath_{X}A\, \Phi^{r} \; +\; \hat{X}\wedge\imath_{A^{\flat}}\Phi^{r} \; -\; A\wedge\imath_{X}\Phi^{r} \ .
\end{eqnarray}
From eq.~(\ref{eq:43}) it is clear that $L$ is a null geodesic, {\em i.e.\/} $\nabla_{L}L=0$, and, as we already knew from eq.~(\ref{eq:6}), that $d\hat{L}=0$.

Moreover, we introduce a Vielbein, adapted to the null nature of $L$, in terms of the natural coordinates $u$, $v$ and $y^{m}$ ($m=1,\ldots ,4$)
\begin{equation}
\label{eq:KundtCurv1a}
\begin{array}{lclclcl}
E^{+} & = & du\ , &\qquad& \theta_{+} & =& \partial_{u} \; -\;H \partial_{v} \ , 
\\
E^{-} & = & dv + Hdu + S_{m}dy^{m}\ , & & \theta_{-} & =& \partial_{v} \ ,
\\
E^{i} & = & {e_{m}}^{i}\ dy^{m}\ , & & \theta_{i} & =& {e_{i}}^{m}\left[ \partial_{m} \; -\; S_{m}\partial_{v}\right] \ ,
\end{array}
\end{equation}
where $\hat{L}\equiv E^{+}$ and $L\equiv \theta_{-}$. As usual we can then define the metric on the base-space by $h_{mn}\equiv{e_{m}}^{i}{e_{n}}^{i}$, and write the full six-dimensional Kundt metric as
\begin{equation}
  \label{eq:44}
  ds_{(6)}^{2} \; =\; 2du\left( dv\; +\; Hdu\; +\; \hat{S}\right) \; -\; h_{mn}\,dy^{m}dy^{n} \ .
\end{equation}
We expand the 2-forms as $2\; \mathsf{K}^{r}\equiv \mathsf{K}^{r}_{ij} E^{i}\wedge E^{j}\,$ w.r.t.~the above Vielbein, and by choosing the light-cone directions such that $\varepsilon^{+-1234}=1=\varepsilon^{1234}$ we see that $\star_{(4)}\mathsf{K}^{r} = -\mathsf{K}^{r}$. Defining the $(1,1)$-tensors $\mathsf{J}^{r}$ through $h(\mathsf{J}^{r}X,Y)\equiv
\mathsf{K}^{r}(X,Y)$, we can see that eq.~(\ref{eq:39b}) implies
\begin{equation}
  \label{eq:41}
  \mathsf{J}^{r}\mathsf{J}^{s} \; =\; -\delta^{rs} \; +\; \varepsilon^{rst}\, \mathsf{J}^{t} \ ,
\end{equation}
so that the four-dimensional base-space is an {\em almost quaternionic manifold}.

At this point we shall fix part of the Weyl gauge symmetry by imposing the gauge-fixing condition $\imath_{L}A=0$; consequently, we can expand the gauge field as
\begin{equation}
  \label{eq:45}
  A \; =\; \Upsilon\ \hat{L} \; +\; \mathsf{A}_{m}dy^{m} \ .
\end{equation}
Using this expansion and the explicit form of the Vielbein in terms of the coordinates, we can analyse eq.~(\ref{eq:43}), resulting in
\begin{eqnarray}
  \label{eq:46}
  \Upsilon & =& -\textstyle{1\over 2}\, \partial_{v}H \ , \\
  \label{eq:46a}
  \partial_{v}\hat{S} & =& -2\, \mathsf{A}\ ,\\
  \label{eq:46b}
  0 & =& \partial_{v}h_{mn} \ .
\end{eqnarray}
Contrary to what is usually the case in (fake)supergravities, we do not know the full $v$-dependence of $H$, and therefore we cannot completely fix the $v$-dependence of the unknowns. The above results comprise all the information contained in eq.~(\ref{eq:43}).\vspace{\baselineskip}

In order to analyse the content of eq.~(\ref{eq:43a}) we first take $X=L$, to find that
\begin{equation}
\nabla_{L}\Phi^{r}=0\ .
\end{equation}
When this is evaluated in the chosen coordinate system it implies that $\partial_{v}\mathsf{K}^{r}_{mn}=0$. This apparently innocuous result fixes however the $v$-dependence of $\mathsf{A}$: from the totally antisymmetric part of eq.~(\ref{eq:43a}) one obtains
\begin{equation}
  \label{eq:47}
  d\Phi^{r}\; =\; 2A\wedge\Phi^{r} \quad\longrightarrow\quad 0\; =\; \hat{L}\wedge\left( \mathsf{d}\mathsf{K}^{r}\; -\; 
2\mathsf{A}\wedge\mathsf{K}^{r}\, \right) \ ,
\end{equation}
where we have introduced the exterior derivative on the base-space $\mathsf{d}\equiv dy^{m}\partial_{m}$.  As the $\mathsf{K}^r$ are $v$-independent and $\hat{L}=du$, we see that the consistency of the above equation also requires $\mathsf{A}$ to be $v$-independent. Thus, we also obtain from eq.~(\ref{eq:46a}) that
\begin{equation}
\label{eq:48}
\hat{S} \; =\; -2v\, \mathsf{A} \; +\; \varpi \ ,
\end{equation}
with $\varpi$ a 1-form living on the base-space such that $\partial_{v}\varpi_{m}=0$. It should be clear from eq.~(\ref{eq:47}) that the $y$-dependence of the $\mathsf{K}^r$ is given by the equation
\begin{equation}
\label{eq:49}
\mathsf{d}\mathsf{K}^{r}\; =\; 2\mathsf{A}\wedge\mathsf{K}^{r}\ ,
\end{equation}
whose integrability condition reads
\begin{equation}
\mathsf{F}\wedge\mathsf{K}^{r} \; =\; 0\ ,
\end{equation}
where we have defined $\mathsf{F}=\mathsf{d}\mathsf{A}$. Actually, this last equation implies that $\mathsf{F}$ is self-dual, {\em i.e.\/} $\star_{(4)}\mathsf{F}=\mathsf{F}$, whence $\mathsf{A}$ is a self-dual connection or, in physics-speak, an $\mathbb{R}$-instanton.

The analysis of eq.~(\ref{eq:43a}) in the direction $X=\theta_{+}$ is straightforward, and leads to the following constraints on the spin connection
\begin{eqnarray}
\label{eq:51}
\omega_{+-k}\ \mathsf{K}^{r}_{kj} & = & -\mathsf{A}_{k}\, \mathsf{K}^{r}_{kj} \ ,
\\
\label{eq:51a}
0 & = & \omega_{+i}{}^{k}\; \mathrm{K}^{r}_{kj} \; +\; \omega_{+j}{}^{k}\; \mathrm{K}^{r}_{ik}\ .
\end{eqnarray}
By using the results in appendix~\ref{appsec:Kundt}, we see that eq.~(\ref{eq:51}) is automatically satisfied. A small investigation of eq.~(\ref{eq:51a}) shows that it implies the base-space 2-form $\omega_{+ij}\ E^{i}\wedge E^{j}$ is self-dual. Coupling this observation with eq.~(\ref{eq:KundtCurv2d}), and taking into account $\mathsf{F}$'s self-duality,
we see that the base-space 2-form $2\Omega = \Omega_{ij}\ E^{i}\wedge E^{j}$, whose components are defined by
\begin{equation}
\label{eq:52}
\Omega_{ij}\; \equiv\; 2\mathsf{D}_{[i}\varpi_{j]} \; +\; 2e_{[i}{}^{m}\partial_{u}e_{j]m}\ , \qquad \mbox{where}\quad \mathsf{D}\varpi \equiv \mathsf{d}\varpi\; -\; 2\mathsf{A}\wedge\varpi\ ,
\end{equation}
has to be self-dual, {\em i.e.\/} $\star_{(4)}\Omega = \Omega$.

In order to completely drain eq.~(\ref{eq:43a}) we need to consider $X$ lying on the base-space. Let $\mathsf{X}$ be such a vector. Then, we find that
\begin{equation}
\label{eq:50}
\nabla^{(\lambda )}_{\mathsf{X}}\, \mathrm{K}^{r} \; =\; \mathsf{X}^{\sharp}\wedge\, \star_{(4)}\left[\mathsf{A}\wedge\mathsf{K}^{r}\right]\; -\; \mathsf{A}\wedge \imath_{\mathsf{X}}\mathsf{K}^{r} \ ,
\end{equation}
where $\nabla^{(\lambda )}$ is the ordinary spin connection on the base-space using the $\lambda$ in eq.~(\ref{eq:KundtCurv2d}). Following \mycite{Grover:2008jr} we can then introduce a torsionful connection $\overline{\nabla}_{\mathsf{X}}\mathsf{Y}\equiv \nabla^{(\lambda)}_{\mathsf{X}}\mathsf{Y}\; -\; \mathsf{S} _{\mathsf{X}}\mathsf{Y}$, with the
torsion being totally antisymmetric and proportional to the Hodge dual of the $\mathbb{R}$-gauge field, {\em i.e.\/}
\begin{equation}
\label{eq:53}
h\left( \mathsf{S}_{\mathsf{X}}\mathsf{Y},\mathsf{Z}\right) \; \equiv\; -\left[\star_{(4)}\mathsf{A}\right]\ \left( \mathsf{X},\mathsf{Y},\mathsf{Z}\right) \ ,
\end{equation}
such that eq.~(\ref{eq:50}) can be written compactly as $\overline{\nabla}\mathsf{K}^{r}=0$. Almost quaternionic manifolds admitting a torsionful connection parallelising the almost quaternionic structure are called {\em hyper-K\"ahler Torsion} manifolds (HKT), a name that first appeared in \mycite{Howe:1996kj} to describe the geometry of supersymmetric sigma model manifolds with torsion \mycite{Gates:1984nk}.

As pointed out in \mycite{Grover:2008jr}, one can make use of the residual Weyl symmetry in eq.~(\ref{eq:16}) with $w=w(y)$, {\em i.e.\/} a Weyl transformation depending only on the coordinates of the base-space, to gauge-fix the condition $\mathsf{d}^{\dagger}\mathsf{A}=0$.  This immediately implies that the torsion $\mathsf{S}$ is closed, and the resulting four-dimensional structure is called a {\em closed HKT} manifold. Let us mention, even though it will not be needed, that the coordinate transformation $v\rightarrow v+\Lambda (y)$, induces the transformation $\varpi\rightarrow \varpi + \mathsf{D}\Lambda$.

Thus far, the analysis has shown that the pair $(g,A)$ admits a solution to eq.~(\ref{eq:1}) iff $g$ is the metric of a Kundt wave whose base-space is a $u$-dependent family of HKT spaces. Given such a family of spaces we can find the 1-form $\varpi$ by imposing self-duality of the 2-form $\Omega$ in eq.~(\ref{eq:52}), and then the only indeterminate element of the metric is the wave profile $H$. This study has given us the necessary conditions for the existence of a non-trivial spinor satisfying eq.~(\ref{eq:1}). It remains to be checked that they are also sufficient, which we do by direct substitution into (\ref{eq:1}).\vspace{\baselineskip}

A quick calculation of the $(-)$ component leads to $\theta_{-}\epsilon =0$, whence the spinor is $v$-independent. The $(+)$-component, after using the constraint $\gamma^{+}\epsilon =0$, leads to
\begin{equation}
\label{eq:25}
\partial_{u}\epsilon \; =\; -\textstyle{1\over 4}\, T_{ij}\, \gamma^{ij}\epsilon \; =\; 0 \ ,
\end{equation}
where the last step follows from the self-duality of $T$ (cf. eq.~(\ref{eq:9bis})) 
\begin{equation}
T_{ij} \; =\; v\, \mathsf{F}_{ij} \; -\; \textstyle{1\over 2}\, \left[ \mathsf{D}\varpi\right]_{ij}\ ,
\end{equation}
and the chirality of the spinor $\epsilon$. We thus conclude that the spinor is also $u$-independent. Giving the $i$ components of eq.~(\ref{eq:1}) a similar treatment we end up with
\begin{equation}
\label{eq:55}
\nabla_{i}^{(\lambda )}\epsilon \; =\; \textstyle{1\over 2}\, \tilde{\gamma}_{ij}\, \mathsf{A}_{j}\epsilon \ ,
\end{equation}
where we have defined $\tilde{\gamma}^{i}\equiv i\gamma^{i}$, so $\{\tilde{\gamma}^{i},\tilde{\gamma}^{j}\} = 2\delta^{ij}$, in order to obtain a purely Riemannian spinorial equation.

As one can readily see from eq.~(\ref{eq:1}), the above equation is nothing more than its Riemannian version for four-dimensional spaces. This kind of spinorial equations was studied by Moroianu in \mycite{Moroianu:1996}, who investigated Riemannian Weyl geometries admitting spinorial fields parallel w.r.t.~the  Weyl connection. For $d\neq 4$ it was found that any such Weyl structure was closed, whereas in $d=4$ the HKT structure outlined above was prescribed. Furthermore, it was shown that if the four-dimensional space is compact, then the HKT structure is conformally related to either a flat torus, a K3 manifold or the Hopf surface $S^{1}\times S^{3}$ with the standard locally-flat metric (see {\em e.g.\/} \mycite{Gibbons:1997iy}).

The integrability condition for eq.~(\ref{eq:55}) implies that the Ricci tensor of the metric $h$ has to satisfy
\begin{equation}
  \label{eq:58}
  \mathsf{R}(h)_{ij} \; =\; 2\nabla^{(\lambda )}_{(i}\mathsf{A}_{j)} \; +\; 2\mathsf{A}_{i}\mathsf{A}_{j}
              \; +\; h_{ij}\left(\nabla^{(\lambda )}_{i}\mathsf{A}_{i}\; -\; 2\mathsf{A}^{2}\right) \ ,
\end{equation}
which by comparison with eq.~(\ref{eq:14}) it is equivalent to saying that the pair $(h,\mathsf{A})$ forms a Ricci-flat Weyl geometry {\em i.e.\/} $\mathrm{W}_{(ij)}=0$.\vspace{\baselineskip}

As we did in section~(\ref{sec:N1D4}), we impose the Einstein-Weyl equations in those directions in which they are not trivially satisfied, {\em i.e.}~in the ($++$)-direction. This fixes the function $H$, which was otherwise
unknown. Furthermore, we would like to impose the simplifying restriction that the HKT structure on the base-space does not depend on $u$. We take this route because of the difficulty of finding analytic solutions to the differential equation resulting from a $u$-dependent base-space. A calculation of the $(++)$-components of the EW equations then shows that
\begin{equation}
  \label{eq:59}
  2\theta_{+}\theta_{-}H + \left(\theta_{-}H\right)^{2} \; =\; \left(\nabla^{(\lambda )}_{i} - S_{i}\theta_{-} - 4\mathsf{A}_{i}\right)\left(\partial_{i} - S_{i}\theta_{-} -2\mathsf{A}_{i}\right)\, H \ ,
\end{equation}
where we have allowed for a $u$-dependence of $H$.

To summarise, any solution to the $\N=(1,0)$ $d=6$ null problem is once again prescribed by a Kundt wave of the form eq.~(\ref{eq:44}) constrained by eqs.~(\ref{eq:48}), (\ref{eq:52}) and (\ref{eq:59}), whose four-dimensional base-space is given by a $v$-independent closed HKT manifold subject to eqs.~(\ref{eq:58}), and the gauge connection being an $\mathbb{R}$-instanton.

\section{Remaining null cases}
\label{sec:general}
Having treated the null cases in $d=4$ and $d=6$, we now make some general comments for the null class in other dimensions. Since performing the complete bilinear analysis is a daunting task, and in view of the wave-like nature of the null case solutions above, we shall write down a similar Ansatz for the metric. Furthermore, as was pointed out in section~(\ref{sec:N1D4}), as long as $d\neq 6$ we can use a Weyl transformation to introduce a coordinate $u$ such that $\hat{L}=du$ and $A=\Upsilon\,\hat{L}$. Choosing then the coordinate $v$ to be aligned with the flow of $L\; (=\partial_{v})$, one can introduce base-space coordinates $y^{m}$ ($m=1,\ldots , d-2$) and a Vielbein similar to the one in eq.~(\ref{eq:KundtCurv1}), so that the metric is of the form
\begin{equation}
\label{eq:15}
ds^{2}_{(d)} \; =\; 2du\left( dv \ +\ Hdu\ +\ S_{m}dy^{m}\right) \; -\; h_{mn}\,dy^{m}dy^{n}\ ,
\end{equation}
where $h_{mn}\equiv e^i_m e^i_n$. This is by definition a Kundt metric, and evaluating the symmetric part of eq.~(\ref{eq:33}) in this coordinate system, we get the following restrictions
\begin{equation}
  \label{eq:26}
  \Upsilon \; =\; -\frac{2}{d-2}\ \partial_{v}H\ ,  \qquad \partial_{v}S_{m} \; =\; 0\ , \qquad \partial_{v}h_{mn} \; =\; 0 \ ,
\end{equation}
so that the whole $v$-dependence resides exclusively in $H$ and $\Upsilon$, which implies that eq.~(\ref{eq:15}) is actually in the Walker form \mycite{Walker:1950}. Following the notation of section \ref{sec:D6chiral}, we shall call the $v$-independent part of $\hat{S}$ by $\varpi$, so that in the $d\neq 6$ case we have $\hat{S}=\varpi$.

With this information and the constraint of $u$-independence imposed, we can proceed to analyse the spinorial rule. The KSE in the $v$-direction is automatically satisfied, {\em i.e.}~$\partial_v \epsilon=0$, and the remaining directions read
\begin{eqnarray}
\label{eq:56}
0 & =& \nabla_{i}^{(\lambda )}\, \epsilon \ , \\
\label{eq:56a}
\partial_{u}\epsilon & =& \textstyle{1\over 8}\, \left[ \mathsf{d}\varpi\right]_{ij}\gamma^{ij}\epsilon \ .
\end{eqnarray}
Eq.~(\ref{eq:56}) clearly states that the base-space must be a Riemannian manifold of special holonomy.  The integrability condition of the above two equations is
\begin{equation}
  \label{eq:54}
  0\; =\; \left[ \nabla^{(\lambda )}_{i} (\mathsf{d}\varpi )_{kl}\right]\ \gamma^{kl}\epsilon\ ,
\end{equation}
which implies
\begin{equation}
\left(\mathsf{d}^{\dagger}\mathsf{d}\varpi\right)_{i}\gamma^{i}\epsilon \; =\; 0 \ ,
\end{equation}
so that\footnote{The same constraint can be obtained through an explicit evaluation of the Einstein-Weyl equations.} $\mathsf{d}^{\dagger}\mathsf{d}\varpi =0$. Using the coordinate transformation $v\rightarrow v +\Lambda (y)$, one can always take $\mathsf{d}^{\dagger}\varpi =0$, whence $\varpi\in \mathrm{Harm}^{1}(\mathcal{B})$, {\em i.e.}~$\varpi$ is a harmonic 1-form on the base-space\footnote{Bochner's theorem states that any harmonic 1-form on a compact oriented Ricci-flat manifold is parallel, which by eq.~(\ref{eq:56a}) implies that in such a case the Killing spinor is $u$-independent. In the non-compact case however there is no such a theorem, as can be seen by taking the base-space to be $\mathbb{R}^{d-2}$ and $2\varpi \equiv f_{mn}x^{m}dx^{n}$, where the $f_{mm}$ are constants.}.

Given this input, the condition for such a pair $(g,A)$ to be an Einstein-Weyl manifold is
\begin{equation}
\label{eq:57}
2\partial_{u}\partial_{v}H \; -\; 2H\partial_{v}^{2}H \; +\; 2\,{d-4\over d-2}\, \left(\partial_{v}H\right)^{2} \; =\; -\left( \nabla^{(\lambda )} - \varpi\right)^{i}\, \theta_{i}H  \ .
\end{equation}
The factor on the RHS of the above equations becomes, in the $\varpi =0$ limit, the d'Alembertian on the base-space, and it makes contact with eq.~(\ref{eq:11}). This shows that the $d=4$ case is a subcase of the general one studied in this section, where one was allowed to use the 2-form $\Phi$ to get rid of $\hat{S}$. $d=6$, however, is an independent case, where the characteristic behaviour of the theory in that dimension (see {\em e.g.}~eq.~(\ref{eq:6})) nurtures the closed HKT structure.

\section{Summary of the chapter}
\label{sec:conclusions}
In this chapter we have presented a characterisation of supersymmetric Einstein-Weyl spaces with Lorentzian signature in arbitrary dimensions. We have done this by making use of the techniques developed for the classification of SUGRA solutions. In particular, we assumed the existence of a spinor $\epsilon$ satisfying eq.~(\ref{eq:1}). It is in this sense that our solutions have a supersymmetric character. We then proceeded to build and analyse the bilinears that can be constructed from $\epsilon$, which shape the resulting geometry.

We have found that for most dimensions those spaces arising from a vector bilinear which is timelike are trivial, in the sense that they are conformally related to a space admitting a parallel spinor. The odd duck in the pond is the four-dimensional case, for which the only timelike solution actually turns out to be Minkowski space, which coincides with which was already known \mycite{Bryant:2000}. The null class solutions are given by a Kundt metric and a prescribed Weyl gauge field. For the $d=6$ case, we also find that the base-space is a closed Hyper-K\"ahler Torsion manifold. Furthermore, it is inspiring that our $d=3$ characterisation contains a solution to a three-dimensional Weyl-scalar-flat Lorentzian EW space that was previously presented in \mycite{Calderbank:2001}.

In said article, Calderbank and Dunajski derived the possible forms of three-dimensional scalar-flat Lorentzian EW spaces, and found that, up to Weyl and coordinate transformations, there are only two possible metric classes. In our construction, due to eq.~(\ref{eq:3}) and the results obtained so far, one can see that the scalar curvature is given by
\begin{equation}
\label{eq:60}
\mathrm{W} \; =\; \frac{2d}{d-2}\; \partial_{v}^{2}H \ ,  
\end{equation}
so that the scalar-flat supersymmetric Lorentzian EW spaces are given by a function $H$ which is at most linear in $v$. It then follows that the $d=3$, $W=0$ approximation of eqs.~(\ref{eq:15}) and (\ref{eq:26}) is one of the cases presented by the authors, namely the $h_{1}$ in \mycite{Calderbank:2001}{Prop.~2}. As shown in \mycite{Dunajski:2010}{sec.~10.3.1.3}, this is the unique class of (1,2)-dimensional EW spaces admitting a weighted covariantly constant null vector. This supersymmetric class can be obtained by the Jones-Tod construction \mycite{Tod:1985} on a conformal space of neutral signature admitting an antiself-dual Null-K\"ahler structure \mycite{Dunajski:2010}, a geometric structure which admits a Killing spinor. Furthermore, once one takes into account that one can perform coordinate transformations such that $h=1$ and $\varpi =0$, eq.~(\ref{eq:57}) for $d=3$ then corresponds to the dispersionless Kadomtsev-Petviashvili equation \mycite{Kadomtsev:1970xx,Kodama:1988}. 

The fact that we only obtain one of the solution classes from said article is indicative that the world of Lorentzian EW spaces is richer than the ones obtainable as solutions to eq.~(\ref{eq:1}) ({\em i.e.}~that there are also `non-supersymmetric' EW spaces), and hence our classification does not aim at solving the mathematical problem of characterising the whole spectrum. A natural question is whether there are equations like (\ref{eq:1}) whose integrability condition leads to more general EW spaces. In any case, the construction shows that the supersymmetric formalism allows us to generalise and extend solutions that were obtained through other geometric methods.

\cleardoublepage


\addtocontents{toc}{\newpage}
\addtocontents{toc}{\vspace{15pt}}
\addtocontents{toc}{\contentsline{chapter}{\numberline {\small{ANNEXE}}}{}{}}
\renewcommand{\chaptername}{Anexo}
\renewcommand{\thechapter}{I}
\renewcommand{\leftmark}{\MakeUppercase{\chaptername\ \thechapter. Introducci\'on}}
\chapter{Introducci\'on}
\label{resumen}
El trabajo recogido en esta tesis doctoral ha estado centrado en la obtenci\'on de soluciones a varias teor\'ias de inter\'es f\'isico-te\'orico y matem\'atico, a trav\'es de t\'ecnicas e ideas caracter\'isticas del campo de la Supergravedad (SUGRA). Estas t\'ecnicas surgen originariamente del programa de clasificaci\'on de soluciones supersim\'etricas a las mismas, que se explican brevemente en el anexo \ref{conclusiones}. Pasamos ahora a ofrecer una breve introducci\'on sobre la SUGRA, en el contexto de la f\'isica contempor\'anea de Gravitaci\'on y Part\'iculas.

De forma muy b\'asica, la Supergravedad se puede definir como una construcci\'on te\'orica donde se junta la Relatividad General con la Supersimetr\'ia (SUSY). La Relatividad General es la teor\'ia de gravitaci\'on que Einstein propuso en 1915, donde se enfatiza el papel que la Geometr\'ia (Matem\'aticas) juega en la F\'isica, y en que las transformaciones de coordenadas en cada punto (difeomorfismos locales) son esencia misma de la teor\'ia. La SUSY fue introducida de forma independiente por los investigadores Golfand \& Likhtman, Volkov \& Akulov y Wess \& Zumino entre 1971 y 1974, y es una simetr\'ia que relaciona part\'iculas bos\'onicas (aquellas que tienen un n\'umero de esp\'in entero) con part\'iculas fermi\'onicas (con n\'umero de esp\'in=$n/2$, donde $n$ es un n\'umero impar, e.g.~1/2, 3/2, ...).

Las primeras construcciones de SUGRA se pueden trazar en torno a 1973-1975, cuando los cient\'ificos Volkov, Akulov y Soroka, a lo largo de varios art\'iculos, propusieron unos modelos para gaugear las transformaciones supersim\'etricas. Esto significa que estas transformaciones, que hasta ese momento eran globables o r\'igidas, pasan a ser locales, de forma que en cada punto del espacio-tiempo se puede hacer una transformaci\'on distinta. Esto, unido a un mecanismo de Higgs que desarrollaron para SUGRA (en que los `fermiones de Goldstone', de esp\'in 1/2, son comidos) da lugar a una part\'icula masiva de esp\'in 3/2, llamada el campo de Rarita-Schwinger, y supone una realizaci\'on no-lineal de las transformaciones. El primer modelo de SUGRA cuatri-dimensional lineal lo construyeron en 1976
Ferrara, Freedman y Van Nieuwenhuizen, y tambi\'en Deser y Zumino. En sus respectivos art\'iculos,estudiaron la teor\'ia de campos (con interacciones) de lo que especularon ser\'ian la part\'icula que mediar\'ia en la interacci\'on gravitacional, el gravit\'on, de esp\'in 2. \'Este ya hab\'ia aparecido con anterioridad en la teor\'ia de representaciones del \'algebra global de Super-Poincar\'e (que conten\'ia adem\'as su compa\~nero supersim\'etrico de esp\'in 3/2, tambi\'en conocido como gravitino), pero lo novedoso en este caso es que adem\'as demostraban que su acci\'on eran invariante bajo transformaciones locales de supersimetr\'ia.

La SUGRA gan\'o mucha fama cuando se pens\'o que su versi\'on en 11 dimensiones, propuesta por Cremmer, Julia y Scherk, pod\'ia servir como candidata a una \emph{Teor\'ia del Todo}. En \'ultima instancia se demostr\'o que esto no pod\'ia ser as\'i, y que en realidad la SUGRA ten\'ia que ser un l\'imite de bajas energ\'ias de la \emph{Teor\'ia de Cuerdas}. En concreto, las SUGRAs en 10 y 11 dimensiones son teor\'ias de campos efectivas para part\'iculas sin masa, y a nivel \'arbol, y por tanto sus soluciones describen el comportamiento de materia cuerdosa en distancias largas (\emph{i.e.}~baja energ\'ia). Es por esto que el estudio de las construcciones de SUGRA han sido, y siguen siendo, muy importantes en Teor\'ia de Cuerdas, ya que sus soluciones son un pilar fundamental para el desarollo de la misma. Pero incluso de forma independiente de las cuerdas, las teor\'ias de SUGRA parecen ser de gran utilidad. Investigaciones recientes parecen sugerir la viabilidad de determinadas teor\'ias extendidas de SUGRA como teor\'ias de gravedad cu\'antica \mycite{Bern:2007hh}. Esto ser\'ia un gran avance, ya que el problema de la gravedad cu\'antica lleva intentando resolverse desde tiempos de Einstein. Es por tanto que la SUGRA, y sus soluciones, siguen ofreciendo una puntera l\'inea de investigaci\'on a trav\'es de la cual poder dilucidar algunos de los misterios de la Naturaleza.

Por ejemplo, las teor\'ias de Supergravedad se han analizado para obtener soluciones cl\'asicas y supersim\'etricas, que representan agujeros negros en 4 y 5 dimensiones. Estas teor\'ias (de las que el agujero negro es soluci\'on) se pueden relacionar con teor\'ias de SUGRA de dimensiones m\'as altas por medio de una t\'ecnica conocida como oxidaci\'on dimensional. Seg\'un hemos explicado, estas SUGRAs de dimensiones altas (d=10 o d=11, generalmente) representan un l\'imite de bajas energ\'ias de teor\'ias de supercuerdas de la misma dimensionalidad, lo que permite relacionar la soluci\'on del agujero negro original con un vac\'io de cuerdas. Operar en esta teor\'ia de supercuerdas (que es una teor\'ia cu\'antica de campos) no es una tarea sencilla, ya que las correciones cu\'anticas que surgen en dicha teor\'ia lo complica enormemente. Tanto que a d\'ia de hoy todav\'ia no se sabe bien c\'omo contar el n\'umero de estados microsc\'opicos de sus vac\'ios. Sin embargo, las soluciones supersim\'etricas salen al rescate, ya que precisamente \'estas son estables bajo dichas correciones cu\'anticas, lo que permite poder contar el n\'umero de estados microsc\'opicos de tal vac\'io. Esto es un muy notable c\'alculo de la entrop\'ia microsc\'opica de agujeros negros supersim\'etricos, que supuso un verdadero hito cuando se realiz\'o para una clase de agujeros negros extremos en 5 dimensiones, ya que coincid\'ia con el valor de la entrop\'ia macrosc\'opica predicha por Bekenstein y Hawking \mycite{Strominger:1996sh}. En vista de esto, y otras impactantes e influyentes ideas, como la existencia de dualidades no perturbativas entre teor\'ias de supercuerdas, o la conjetura $AdS/CFT$ \footnote{\'Esta establece una correspondencia entre una teor\'ia de cuerdas sobre una soluci\'on maximalmente sim\'etrica y maximalmente supersim\'etrica (el producto geom\'etrico entre el espacio $AdS_5$ y una $5$-esfera), y una teor\'ia maximalmente supersim\'etrica de tipo Yang-Mills sobre el producto de una l\'inea y una $3$-esfera, que uno identifica como el contorno del espacio $AdS_5$. Se la llama tambi\'en la correspondencia \emph{Gravedad/ Teor\'ias gauge}, o \emph{Correspondencia de Maldacena} \mycite{Maldacena:1997re}.} y sus generalizaciones, es l\'ogico que a lo largo de los a\~nos se hayan realizado numerosos esfuerzos en el estudio de teor\'ias de SUGRA.

Para m\'as m\'erito, si la Supersimetr\'ia es una simetr\'ia real de la Naturaleza\footnote{Aunque \'esta est\'e rota a nuestra escala de energ\'ias.}, algo de lo que probablemente haya indicios en el LHC del CERN en los pr\'oximos a\~nos, la Supergravedad gozar\'ia de un papel fundamental. En dicho caso las soluciones supersim\'etricas sean seguramente buenas aproximaciones al vac\'io o los vac\'ios de la Teor\'ia de Cuerdas que describan la realidad a escalas de energ\'ias altas, y esto har\'ia adem\'as de la SUGRA una pieza clave en la construcci\'on de modelos fenomenol\'ogicos de F\'isica de Part\'iculas, y en la unificaci\'on de las interacciones.
\cleardoublepage

\renewcommand{\thechapter}{II}
\renewcommand{\leftmark}{\MakeUppercase{\chaptername\ \thechapter. Resumen}}
\chapter{Resumen}
\label{conclusiones}
En este anexo comentamos brevemente las t\'ecnicas e ideas que dan lugar al trabajo expuesto en esta tesis doctoral. Comenzamos con un resumen sobre el `formalismo bilineal' y el `formalismo de la geometr\'ia espinorial', que motivan los problemas estudiados durante mi doctorado, para terminar describiendo los resultados de los cap\'itulos \ref{4d}, \ref{5dminimal}, \ref{5dgauged} y \ref{EWspaces}.

En el anexo \ref{resumen} hemos discurrido sobre el inter\'es de obtener soluciones supersimetr\'icas. En el art\'iculo \mycite{Gauntlett:2002nw} se propuso un m\'etodo para obtener este tipo de soluciones en la teor\'ia de SUGRA cinco-dimensional m\'inima, bas\'andose en la existencia de un espinor de Killing (tal espinor no debe entenderse como un campo fermi\'onico, si no que se trata de un espinor cl\'asico, de car\'acter geom\'etrico). \'Este sirve de par\'ametro respecto al cual la variaci\'on supersim\'etrica del gravitino es cero, dando lugar a la conocida como `ecuaci\'on del espinor de Killing' (KSE). En t\'erminos puramente geom\'etricos, \'esta no es m\'as que una condici\'on de parelizaci\'on del espinor, que surge a ra\'iz del \'algebra de SUSY. El siguiente paso es construir campos bilineales en tal espinor, y usar las identidades de Fierz para obtener relaciones algebraicas entre esos campos. Esto permite obtener la forma m\'as general de la geometr\'ia y los flujos de las soluciones. \'Estas son supersim\'etricas por construcci\'on (ya que asumen la existencia de al menos un espinor de Killing), y est\'an completamente determinadas en funci\'on de dichas condiciones. Se puede encontrar un resumen de este m\'etodo en la secci\'on \ref{bilinearformalism} de estas p\'aginas.

As\'i mismo, en \mycite{Gillard:2004xq} se propuso un m\'etodo cuyo objetivo era el mismo, pero en el que se elimina la complejidad que supone usar matrices Gamma en escenarios multi-dimensionales. Este formalismo interpreta los campos espinoriales de la teor\'ia en funci\'on de formas diferenciales (m\'as una acci\'on de las matrices Gamma sobre ellas, tambi\'en expresada en funci\'on de formas). En lugar de construir bilineales de los que extraer las condiciones de geometrizaci\'on de las soluciones, el m\'etodo implica resolver las KSEs directamente, y leer de ellas la geometr\'ia y los flujos resultantes. La secci\'on \ref{spinorialgeometry} ofrece un resumen de este m\'etodo, adem\'as de contener una resoluci\'on del mismo problema descrito arriba, esta vez usando esta alternativa. Se puede ver que los resultados obtenidos son los mismos. 

Ambos m\'etodos han sido extensamente usados para el estudio y clasificaci\'on de soluciones supersim\'etricas de m\'ultiples teor\'ias, con y sin mater\'ia, dando lugar a un buen n\'umero de resultados, \emph{e.g.}~\mycite{Gauntlett:2002fz,Gauntlett:2003fk,Gutowski:2003rg,Gauntlett:2003wb,Gran:2005wn}{Gran:2005wu,Gran:2005ct}{Gran:2005wf,Meessen:2006tu,Huebscher:2006mr,Gran:2006dq}{Gran:2007fu,Cacciatori:2007vn}{Gran:2008vx,Ortin:2008wj,Elvin:2009xx,Akyol:2012cq}. Sin embargo, el uso de estos m\'etodos durante mis estudios doctorales ha estado encaminado a objetivos distintos. En lugar de obtener Ans\"atze de soluciones supersim\'etricas a partir de la KSE, proponemos una condici\'on de paralelizaci\'on alternativa para los espinores. \'Esta es elegida de modo que la acci\'on de la teor\'ia tenga una constante cosmol\'ogica de tipo De Sitter (signo positivo), en lugar de las Minkowski (valor cero) y AdS (signo negativo) que, salvo casos contados, son las \'unicas posibilidades que hay en las teor\'ias de SUGRA genuinas \mycite{Pilch:1984aw,Lukierski:1984it}. Esto implica que las teor\'ias detr\'as de estas construcci\'ones no son supersim\'etricas, y han sido previamente bautizadas como Teor\'ias de Supergravedad falsas (fakeSUGRA o fSUGRA). Sus soluciones, que no son supersim\'etricas, son en cualquier caso de gran inter\'es f\'isico, ya que resuelven las ecuaciones de Einstein-Maxwell-De Sitter. Esto es interesante puesto que resolver estas ecuaciones no es una tarea sencilla; tanto que a pesar de que el problema se conoce desde hace m\'as de 60 a\~nos, apenas hay ejemplos que lo hagan. Adem\'as, seg\'un modelos cosmol\'ogicos recientes, el Universo estar\'ia descrito por un Lagrangiano con una constante cosmol\'ogica de este tipo, con un valor absoluto muy peque\~no.

En este sentido, hemos usado ambos formalismos para caracterizar soluciones de teor\'ias en cuatro y cinco dimensiones. Al igual que ocurre con la clasificaci\'on de soluciones supersim\'etricas en SUGRA, tambi\'en aqu\'i separamos nuestro estudio en funci\'on de la norma del vector que se construye como un bilineal de los espinores que satisfacen la KSE. As\'i, distinguimos entre el caso \emph{timelike}, cuando la norma del vector es positiva, o caso \emph{nulo}, cuando la misma es cero. En el caso de fSUGRA cuatri-dimensional acoplada a vectores no-Abelianos, estudiada en el cap\'itulo \ref{4d}, encontramos que las soluciones para el caso timelike vienen dadas por una m\'etrica de tipo conforma-estacionaria
\begin{equation}
ds^{2} \; =\; 2|X|^{2}\left( d\tau \ +\ \omega\right)^{2}-\textstyle{\frac{1}{2|X|^2}}\ h_{mn} dy^{m}dy^{n}\ ,
\end{equation}
con la particularidad de que \'esta conlleva una dependencia en el tiempo. Dicha dependencia coincide con la que propusieron Behrndt y Cveti\v{c} en \mycite{Behrndt:2003cx} para generar soluciones a partir de las supersim\'etricas conocidas en $\N=2$ $d=4$ SUGRA. Adem\'as, encontramos que el espacio-base con m\'etrica $h_{mn}$ es una subclase de espacios Einstein-Weyl tri-dimensionales, llamados espacios hyper-CR o Gauduchon-Tod \mycite{Behrndt:2003cx}. Para el caso nulo encontramos que las soluciones tienen holonom\'ia dentro del grupo $\Sim(2)$, cuyo inter\'es est\'a fundamentado en que las soluciones puramente gravitacionales con esta caracter\'istica no reciben correciones cu\'anticas \mycite{Coley:2008th}.
  
As\'i mismo, en el cap\'itulo \ref{5dminimal} se estudia la clasificaci\'on del caso nulo a la teor\'ia de fSUGRA m\'inima en cinco dimensiones. La soluci\'on viene dada por las ecuaciones (\ref{solution}), (\ref{solutionbis}), y, al igual que en el caso nulo en cuatro dimensiones, \'esta incluye el caso especial donde el vector nulo $N$ es recurrente, \emph{i.e.}
\begin{equation}
\nabla_\mu N^\nu=C_\mu N^\mu\ ,
\end{equation} 
que implica que la geometr\'ia resultante (la conexi\'on de Levi-Civit\`a) tiene holonom\'ia especial $\Sim(3)$. Por su parte, el cap\'itulo \ref{5dgauged} contiene el an\'alisis de la clasificaci\'on de la teor\'ia de fSUGRA cinco-dimensional acoplada a vectores Abelianos. \'Esta es una generalizaci\'on de la estudiada en el cap\'itulo anterior, y en este sentido la contiene. Obtenemos que la estructura geom\'etrica viene determinada por las ecuaciones (\ref{Kundtmetricsummary}), (\ref{fssummary}), y que la condicion de recurrencia para $\nabla^{\text{L.C.}}$ est\'a dada por ${\cal{B}}=0$. En dicho caso, el valor de la norma de $X$ determina si el espacio Gauduchon-Tod tri-dimensional es plano o por el contrario la 3-esfera. Adem\'as, encontramos que para un valor de $X^2-Q^{IJ}V_I V_J$ constante volvemos a tener invariantes escalares algebraicos, en el sentido de \mycite{Coley:2008th}. 

Para finalizar, discurrimos sobre c\'omo utilizar el mismo concepto en otros contextos, cambiando la condici\'on de paralelizaci\'on del espinor para obtener distintos resultados. En este sentido, el cap\'itulo \ref{EWspaces} contiene una clasificaci\'on de espacios Lorentzianos de tipo Einstein-Weyl (EW), que son de inter\'es entre la comunidad matem\'atica. Nuestras soluciones se basan en la existencia de un espinor paralelo bajo una determinada conexi\'on, de tal modo que su condici\'on de integrabilidad da lugar a la ecuaci\'on para espacios EW. Aplicamos entonces las mismas t\'ecnicas (caso timelike y caso nulo), que nos permiten as\'i encontrar que para el caso timelike en cualquier dimensi\'on menos en $d=4$, estos son conformes a espacios Lorentzianos que admiten espinores paralelos. Estos fueron parcialmente clasificados por Bryant en \mycite{Bryant:2000}. El caso cuatri-dimensional viene dado por $\mathbb{R}^{1,3}$, como ya se explicaba en dicha referencia. Las soluciones en la clase nula vienen dadas por m\'etricas Kundt y un vector gauge de tipo Weyl, distintos en funci\'on de la dimensionalidad de la teor\'ia. 

Un punto interesante es que nuestro an\'alisis en tres dimensiones contiene una generalizaci\'on de una de las dos m\'etricas con escalar de Weyl plano estudiadas por Calderbank y Dunajski en \mycite{Calderbank:2001}. No contiene sin embargo la otra, ya que nuestro m\'etodo se basa en la existencia de un espinor paralelo, y esto limita el espectro de soluciones que se pueden obtener. En cualquier caso, se tratan de resultados nunca antes publicados, a los que se ha podido acceder por medio de t\'ecnicas caracter\'isticas del estudio de soluciones supersim\'etricas a teor\'ias de SUGRA.

El ap\'endice \ref{conventions} contiene las convenciones usadas en los cap\'itulos centrales, tanto la relativa al c\'alculo tensorial como sobre las estructuras espinoriales y los bilineales espinoriales e identidades de Fierz. El ap\'endice \ref{scalargeometries} ofrece informaci\'on sobre las distintas variedades de escalares relevantes en las teor\'ias estudiadas. El \ref{someusefulgeometry} contiene detalles t\'ecnicos sobre la geometr\'ia de los casos nulos, as\'i como sobre las m\'etricas Kundt, que aparecen de forma reiterada como soluciones en estas clases. El ap\'endice \ref{sec:Weylgeometry} ofrece una peque\~na introducci\'on a la geometr\'ia de Weyl y a los espacios Einstein-Weyl y Gauduchon-Tod, mientras que el \ref{appsec:Sim} contiene detalles t\'ecnicos sobre el grupo $\Sim$, as\'i como sobre su relaci\'on con teor\'ias con vectores recurrentes, cuya conexi\'on tiene un grupo de holonom\'ia que es un subgrupo del mismo. Por \'ultimo, el ap\'endice \ref{appsec:lorentz} es una peque\~na introducci\'on a los grupos de Lorentz y esp\'in, y su relaci\'on 2 a 1. 

\cleardoublepage

\cleardoublepage

\addtocontents{toc}{\vspace{15pt}}
\addtocontents{toc}{\contentsline{chapter}{\numberline {\small{APPENDICES}}}{}{}}
\renewcommand{\chaptername}{Appendix}
\appendix
\renewcommand{\leftmark}{\MakeUppercase{\chaptername\ \thechapter. Conventions}}
\chapter{Conventions}
\label{conventions}
The conventions and definitions used in this thesis are taken from the respective research articles which conform the central chapters of this work. They are contained in this appendix for reference and completeness. Appendix \ref{tensorconventions} gives those related to the tensorial calculus. Appendix \ref{spinorialstructures} presents the spinorial structures of the theories considered. In particular, it contains the conventions used for the gamma matrices and the spinors. The spinor bilinears and Fierz identities, which are fundamental to the first of the formalisms, are presented in appendix \ref{app:bilinearsandfierzs}.

\section{Tensor conventions}
\label{tensorconventions}
Different notations and conventions have been used for different chapters. This is due to having used different methods to attack the classification of fSUGRAs. While in principle there is no relation between the method employed and the notations/ conventions used, the analyses considered previous published works in order to reduce the load of work. Hence the conventions taken are those of the previous articles, which unfortunately do not agree. We present both sets for completeness.

\subsection{\texorpdfstring{Tensor conventions for the bilinear formalism}{For the bilinear formalism}}
\label{sec-d4conventions}
Chapters \ref{4d}, \ref{5dminimal} and \ref{EWspaces} all employ the notation of \mycite{Andrianopoli:1996cm}. The conventions, however, are slightly different. Those of chapter \ref{4d} are taken from \mycite{Bellorin:2005zc}, which in turn are based on \mycite{Ortin:2004ms}, to which we have adapted the formulae of \mycite{Andrianopoli:1996cm}. The difference between the two is the sign of the spin connection, of the completely antisymmetric tensor $\epsilon^{abcd}$ and of $\gamma_{5}$. Thus, chiralities are reversed and self-dual tensors are replaced by antiself-dual tensors, and viceversa. The curvatures are identical. Also, all fermions and supersymmetry parameters from \mycite{Andrianopoli:1996cm} have been rescaled by a factor of $1/2$, which introduces additional factors of $1/4$ in all the supersymmetry transformations of the bosonic fields. Furthermore, the normalization of forms differs by a factor of $n!$ (see eq.~(\ref{eq:ourchoiceofnormalisationofforms}) below for our choice), which induces an additional factor of 2 in the supersymmetry transformations of the vectors.

The five-dimensional conventions of chapter \ref{5dminimal} are those used in \mycite{Bellorin:2006yr}. Those, in turn, come from \mycite{Bergshoeff:2004kh}, having changed the sign of the metric to have mostly minus signature, multiplying all the gamma matrices $\gamma^a$ by $+i$ and all the $\gamma_a$ by $-i$ and setting $\kappa=1/\sqrt{2}$. They are essentially the same as in four dimensions above, including the additional space coordinate. We proceed to describe them.\vspace{\baselineskip}

We use Greek letters $\mu,\nu,\rho,\ldots$ to denote \textit{curved} tensor indices in a coordinate basis, and Latin letters $a,b,c\ldots$ as \textit{flat} tensor indices in a Vielbein basis.  We symmetrize $(~)$ and antisymmetrize $[~]$ with weight one, \emph{i.e.}~dividing by $n!$, where $n$ is the number of indices inside the bracket, \emph{e.g.}
\begin{equation}
\begin{array}{lll}
A_{[a_1\,a_2}\,B_{b_1]}&\equiv\frac{1}{3!}\Big(&\hspace{-10pt}A_{a_1\,a_2}\,B_{b_1}+A_{a_2\,b_1}\,B_{a_1}+A_{b_1\,a_1}\,B_{a_2}\\
&&-A_{a_2\,a_1}\,B_{b_1}-A_{a_1\,b_1}\,B_{a_2}-A_{b_1\,a_2}\,B_{a_1}\Big)\ . 
\end{array}
\end{equation}
We use a mostly minus signature $\text{diag}(+,[-]^{d-1})$. $\eta$ is the Minkowski metric and we denote a general metric by $g$. Flat and curved indices are related by tetrads $e_{a}{}^{\mu}$ and their inverses $e^{a}{}_{\mu}$, thus satisfying
\begin{equation}
e_{a}{}^{\mu}e_{b}{}^{\nu}g_{\mu\nu}=\eta_{ab}\ ,\quad e^{a}{}_{\mu}e^{b}{}_{\nu}\eta_{ab}=g_{\mu\nu}\ .  
\end{equation}
We define the (Hodge) dual of a completely antisymmetric tensor of rank $k$, $F_{(k)}$, by
\begin{equation}
\star F_{(k)}{}^{\mu_{1}\cdots \mu_{(d-k)}} = {\textstyle\frac{1}{k! \sqrt{|g|}}} \epsilon^{\mu_{1}\cdots\mu_{(d-k)}\mu_{(d-k+1)}\cdots\mu_{d}} F_{(k)\mu_{(d-k+1)}\cdots\mu_{d}}\ .
\end{equation}
Differential forms of rank $k$ are normalized as follows:
\begin{equation}
\label{eq:ourchoiceofnormalisationofforms}
F_{(k)}\equiv {\textstyle\frac{1}{k!}} F_{(k)\, \mu_{1}\cdots \mu_{k}} dx^{\mu_{1}}\wedge\cdots \wedge dx^{\mu_{k}}\ .
\end{equation}

$\nabla$ is the total (general and Lorentz) covariant derivative, whose action on tensors and spinors $\psi$ we choose to be given by
\begin{equation}
\begin{array}{rcl}
\nabla_{\mu}\phi & =  & \partial_{\mu}\phi\ , \\
& & \\
\nabla_{\mu}\xi^{\nu} & =  & \partial_{\mu}\xi^{\nu}+\Gamma_{\mu\rho,}{}^{\nu}\,\xi^{\rho}\ , \\
& & \\
\nabla_{\mu}w_{\nu} & =  & \partial_{\mu}w_{\nu}-w_{\rho}\,\Gamma_{\mu\nu,}{}^{\rho}\ ,\\
& & \\
\end{array}
\end{equation}
\begin{equation}
\begin{array}{rcl}
\nabla_{\mu}\xi^{a}  & =  &  \partial_{\mu}\xi^{a} +\omega_{\mu, b}{}^{a}\, \xi^{b}\ ,\\
& & \\
\nabla_{\mu}w_{a}  & =  &  \partial_{\mu}w_{a} -w_b\,\omega_{\mu, a}{}^{b}\ ,\\
& & \\
\nabla_{\mu}\psi & = & \partial_{\mu} \psi -{\textstyle\frac{1}{4}} \omega_{\mu,}{}^{ab}\,\gamma_{ab}\psi\ ,\\
\end{array}
\end{equation}
where $\gamma_{ab}$ is the antisymmetric product of two gamma matrices, \emph{i.e.}
\begin{equation}
\gamma_{ab}\equiv \gamma_{[a}\,\gamma_{b]}=\frac{1}{2}(\gamma_a\,\gamma_b-\gamma_b\,\gamma_a)\ ,
\end{equation}
$\omega_{\mu, b}{}^{a}$ is the spin connection (antisymmetric in the last two indices) and $\Gamma_{\mu\rho,}{}^{\nu}$ is the affine connection (symmetric in the first two indices). From this point on we shall not include the comma, in order to avoid cluttering of indices.

The respective curvatures, for torsionless connections, are defined through the Ricci identities
\begin{equation}
\label{eq:Ricciidentities}
  \begin{array}{rcl}
\left[ \nabla_{\mu} , \nabla_{\nu} \right]\ \xi^{\rho} & = & R_{\mu\nu\sigma}{}^{\rho}(\Gamma)\, \xi^{\sigma} \; ,\\
& & \\
\left[ \nabla_{\mu} , \nabla_{\nu} \right]\, \xi^{a} & = & R_{\mu\nu b}{}^{a} (\omega)\xi^{b}\ ,\\
& & \\
\left[ \nabla_{\mu} , \nabla_{\nu} \right]\, \psi & = & -\frac{1}{4}R_{\mu\nu}{}^{ab}(\omega) \gamma_{ab}\psi\ ,\\
\end{array}
\end{equation}
and given in terms of the connections by
\begin{equation}
\label{eq:curvatures}
\begin{array}{rcl}
R_{\mu\nu\rho}{}^{\sigma}(\Gamma) & = & 
2\partial_{[\mu}\Gamma_{\nu]\rho}{}^{\sigma} + 2\Gamma_{[\mu|\lambda}{}^{\sigma} \Gamma_{\nu]\rho}{}^{\lambda}\, ,\\
& & \\
R_{\mu\nu a}{}^{b} (\omega) & = & 2\partial_{[\mu}\, \omega_{\nu] a}{}^{b} -2\omega_{[\mu| a}{}^{c}\,\omega_{|\nu]c}{}^{b}\, .\\
\end{array}  
\end{equation}
These two connections are related by 
\begin{equation}
\omega_{\mu a}{}^{b} = \Gamma_{\mu a}{}^{b} +e_{\nu}{}^{b}\partial_{\mu}e_{a}{}^{\nu}\ ,
\end{equation}
where the Vielbein postulate 
\begin{equation}
\nabla_{\mu}e_{a}{}^{\nu}=0
\end{equation}
has been used. This, in turn, implies that the curvatures are related by
\begin{equation}
R_{\mu\nu\rho}{}^{\sigma}(\Gamma) = e_{\rho}{}^{a} e^{\sigma}{}_{b} R_{\mu\nu a}{}^{b}(\omega)\ .
\end{equation}
Finally, metric compatibility and torsionlessness fully determine the connections to be of the form
\begin{equation}
  \begin{array}{rcl}
\Gamma_{\mu\nu}{}^{\rho} & = &
{\textstyle\frac{1}{2}}g^{\rho\sigma\vspace{\baselineskip}} \left\{\partial_{\mu}g_{\nu\sigma} +\partial_{\nu}g_{\mu\sigma}
-\partial_{\sigma}g_{\mu\nu} \right\}\ ,\\
& & \\
\omega_{abc} & = & -\Omega_{abc}+\Omega_{bca} -\Omega_{cab}\ ,\quad \text{where}\qquad \Omega_{ab}{}^{c} = e_{a}{}^{\mu}e_{b}{}^{\nu} \partial_{[\mu}e^{c}{}_{\nu]}\ .\\
\end{array}
\end{equation}

The indices used in these chapters are explained in the following table.
\begin{table}[h]
\centering
\begin{tabular}{|l|l|}
\hline
 Type & Associated structure \\
\hline\hline
 $\mu$, $\nu$, $\ldots$  &  Curved space \\
\hline
 $a$, $b$, $\ldots$  &  Tangent space \\
\hline
 $m,n,\ldots$ & Cartesian $\mathbb{R}^{3}$ indices\\
\hline
 $i,j,\ldots$; $i^{*},j^{*},\ldots$ & Complex scalar fields and their conjugates.\\
\hline
 $\Lambda ,\Sigma ,\ldots$ & $\mathfrak{sp}(n+1)$ indices\\
\hline
 $I,J ,\ldots$ & $\N$ spinor indices \\
\hline
\end{tabular}
\caption{Meaning of the indices employed in the bilinear formalism.}
\label{tab-indices}
\end{table}

\subsubsection{\texorpdfstring{Further conventions for $d=4$}{Further conventions for d=4}}
\label{app-conventions}
The particularities of the four-dimensional theory induce the existence of (anti) self-dual forms. This implies that there are certain formulae which only hold in this case, and which have been used in chapter \ref{4d}. The four-dimensional fully antisymmetric tensor is defined in flat indices by
\begin{equation}
\epsilon^{0123}\; \equiv\; +1\quad \Rightarrow \quad\epsilon_{0123} = -1\ ,
\end{equation}
and in curved indices by
\begin{equation}
\epsilon^{\mu_{1}\cdots \mu_{4}} = \sqrt{|g|}\, e^{\mu_{1}}{}_{a_{1}} \cdots e^{\mu_{4}}{}_{a_{4}}\, \epsilon^{a_{1}\cdots a_{4}}\ ;
\end{equation}
thus, with upper indices, it is independent of the metric and has the same value as with flat indices.

For any 2-form, we define the imaginary (anti-)selfdual 2-forms
\begin{equation}
F^{\pm}\equiv {\textstyle\frac{1}{2}}(F\pm i\,\star \!F)\ ,\qquad\pm i \star  \!F^{\pm}=F^{\pm}\ .
\end{equation}
For any two 2-forms $F$, $G$ we have 
\begin{equation}
F^{\pm}{}_{\mu\nu} G^{\mp\, \mu\nu}=0\ ,\qquad F^{\pm}{}_{[\mu}{}^{\rho}G^{\mp}{}_{\nu]\rho}=0\ .
\end{equation}
A small identity that comes in useful when considering the stress-energy tensor for a number of field strengths is
\begin{equation}
\label{eq:StressTensDecomp}
T^{(\Lambda \Sigma )}_{\mu\nu}\; \equiv\;
F^{(\Lambda}_{\mu\rho}\,F^{\Sigma )}_{\nu}{}^{\rho}-\textstyle{1\over 4}\,g_{\mu\nu}\,F^{\Lambda}_{\sigma\rho}\,F^{\Sigma{\sigma\rho}} = 2F^{\Lambda +}_{(\mu |\rho}\ F^{\Sigma -}_{|\nu )}{}^{\rho}\; .
\end{equation}
Given any non-null 2-form $F$, {\em i.e.\/} $F_{\mu\nu}F^{\mu\nu}\neq 0$,  and a non-null 1-form $\hat{V}=V_{\mu}dx^{\mu}$, we can express $F$ in the form
\begin{equation}
\label{eq:decomposition1}
F=-V^{-2}[E\wedge \hat{V} -\star (B\wedge \hat{V})]\ ,
\end{equation}
where
$E\equiv \imath_{V}F\ (\equiv V^{\mu}F_{\mu\nu}dx^{\nu})$ and $B\equiv \imath_{V}\star F$. For the complex combinations $F^{\pm}$ we have
\begin{equation}
\label{eq:decomposition2}
F^{\pm}=-V^{-2}[C^{\pm}\wedge \hat{V}\pm i\,\star (C^{\pm}\wedge \hat{V})]\ ,
\end{equation}
with $C^{\pm}\equiv \imath_{V}F^{\pm}$.

If we have a (real) null vector $l^{\mu}$, we can always add three more null vectors $n^{\mu},m^{\mu},m^{*\, \mu}$ to construct a complex null tetrad such that the local metric in this basis takes the form
\begin{equation}
\label{eq:nulltetradmetric}
 \left(\begin{array}{rrrr}
     0 & 1 & 0 & 0 \\
     1 & 0 & 0 & 0 \\
     0 & 0 & 0 & -1 \\
     0 & 0 & -1 & 0 \\
 \end{array}\right)
\end{equation}
with the ordering $\left(l,n,m,m^{*}\right)$. For the local volume element we obtain $\epsilon^{lnmm^{*}}=i$.  With the dual basis of 1-forms $\left(\hat{l},\hat{n},\hat{m},\hat{m}^{*}\right)$ we can construct three independent complex self-dual 2-forms that we choose to normalize as follows:
\begin{equation}
\label{eq:2-formbasis}  
\begin{array}{rcl}
\hat{\Phi}^{(1)}& = & \hat{l}\wedge \hat{m}^{*}\ ,\\
& & \\  
\hat{\Phi}^{(2)} & = &{\textstyle\frac{1}{2}} [\hat{l}\wedge \hat{n} +\hat{m}\wedge \hat{m}^{*}]\ ,\\
& &  \\  
\hat{\Phi}^{(3)} & = & -\hat{n}\wedge \hat{m}\ .
\end{array}
\end{equation}
Any self-dual 2-form $F^{+}$ can be written as a linear combination of these, with complex coefficients
\begin{equation}
F^{+} = c_{i}\hat{\Phi}^{(i)}\ .  
\end{equation}
The coefficients $c_{i}$ can be found by contracting $F^{+}$ with $l^{\mu},n^{\mu},m^{\mu},m^{*\mu}$, \emph{i.e.}
\begin{equation}
\begin{array}{rcl}
l^{\nu}F^{+}{}_{\nu\mu} & = & -\frac{1}{2}c_{2}l_{\mu}-c_{3}m_{\mu}\ ,\\
& & \\
n^{\nu}F^{+}{}_{\nu\mu} & = & c_{1}m^{*}_{\mu}+\frac{1}{2}c_{2}n_{\mu}\ ,\\
& & \\
m^{\nu}F^{+}{}_{\nu\mu} & = & c_{1}l_{\mu}+\frac{1}{2}c_{2}m_{\mu}\ ,\\
& & \\
m^{*\nu}F^{+}{}_{\nu\mu} & = & -\frac{1}{2}c_{2}m^{*}_{\mu}-c_{3}n_{\mu}\ .
\end{array}
\end{equation}

\subsection{\texorpdfstring{Tensor conventions for the spinorial geometry formalism}{For the spinorial geometry formalism}}
\label{sec-d5conventions}
Chapter \ref{5dgauged} employs the spinorial geometry method.
The notation is different from that of the spinor-bilinear case. In particular, the following equivalences hold
\begin{table}[!ht]
  \renewcommand{\arraystretch}{1.1}
  \centering
  \begin{tabular}{|c|c|}
   	\hline
    \multicolumn{1}{|c|}{\textbf{\quad Spinorial geometry notation\quad\phantom{i}}}&
    \multicolumn{1}{c|}{\textbf{\quad Spinor-bilinear notation \quad\phantom{i}}}\\
	\hline \hline
    index $M$& index $a$\\
	\hline   
    index $I$& index $\Lambda$\\
	\hline
    $C_{IJK}$&$(-16/\sqrt{3})\,C_{IJK}$\\
	\hline
    $X_I$&$(-2/\sqrt{3})\,h_I$\\
	\hline
    $X^I$&$(-\sqrt{3}/2)\,h^I$\\
	\hline
    $Q_{IJ}$&$2\,a_{IJ}$\\
	\hline
    $Q^{IJ}$&$(1/2)a^{IJ}$\\
	\hline
    $V^I$&$\nexists$\\
	\hline
    $g\,V_I$&$-(4g/3)\xi_I$\\
	\hline
    $F_{ab}$&$(1/2)F_{ab}$\\
	\hline
   \end{tabular}
\caption{Equivalence between notation employed in the two formalisms.}
\label{tab-indicesspinorial}
\end{table}

The conventions used are those of \mycite{Hawking:1973uf}. In particular, the form of the covariant derivative takes on the form
\begin{equation}
\nabla_\mu \xi^\nu=\partial_\mu \xi^\nu + {\Gamma^\nu}_{\rho\mu} \xi^\rho\ ,
\end{equation}
where 
\begin{equation}
{\Gamma^\nu}_{\rho\mu}= {1 \over 2} g^{\nu\sigma} (\partial_\mu g_{\rho\sigma} + \partial_\rho g_{\mu\sigma} - \partial_\sigma g_{\rho\mu})\ ,
\end{equation}
and the symmetry is clearly on the last two indices. The Riemann curvature tensor is defined as 
\begin{equation}
{R^\rho}_{\sigma\mu\nu} = \partial_\mu {\Gamma^\rho}_{\nu\sigma} - \partial_\nu {\Gamma^\rho}_{\mu\sigma} + {\Gamma^\rho}_{\mu\lambda} {\Gamma^\lambda}_{\nu\sigma} - {\Gamma^\rho}_{\nu\lambda} {\Gamma^\lambda}_{\mu\sigma}\ . 
\end{equation}
This implies that the curvature scalar $R$ has the opposite sign to that of chapters \ref{4d}, \ref{5dminimal} and \ref{EWspaces}.

\section{Spinorial structures}
\label{spinorialstructures}
We give the conventions used for the spinorial structure of a general Lorentzian manifold of dimension $(1,d-1)$. This is general for all chapters, and especially important for those that make use of the bilinear formalism.
In this sense, there is also a general proof for the positive-semidefiniteness (for Lorentzian spacetimes) of the vector bilinear, which in turn explain why we consider timelike and null cases in our classification.
Notice that chapter \ref{EWspaces} has a Dirac conjugation matrix equal to $\mathcal{D}=\gamma_0$, while in chapters \ref{4d} and \ref{5dminimal} this is $\mathcal{D}=i\gamma^0$.
The particularities that apply to these latter cases are given in appendix \ref{appsec:spinors4d} for $d=4$, and appendix \ref{app-spinors} for $d=5$.\vspace{\baselineskip}

\label{appsec:spinors}
On $\mathbb{R}^{1,n-1}$ we shall put the mostly-negative metric $\eta=\mathrm{diag}(+,[-]^{n-1})$ and take the Gamma matrices to satisfy
\begin{equation}
  \label{eq:27}
  \left\{\gamma_{a},\gamma_{b}\right\} \; =\; 2\eta_{ab}\;\mathbbm{1}\ .
\end{equation}
We use a unitary representation for these matrices, which implies that $\gamma_{0}^{\dagger} =\gamma_{0}$ and $\gamma_{i}^{\dagger} =-\gamma_{i}$.\vspace{\baselineskip}

In chapter \ref{EWspaces} we have chosen the Dirac conjugation matrix $\mathcal{D}=\gamma_{0}$, and we define the Dirac conjugate of a spinor $\psi$ by $\overline{\psi}\equiv \psi^{\dagger}\mathcal{D}$, to obtain 
\begin{equation}
  \label{eq:28}
  \mathcal{D}\gamma_{a}\mathcal{D}^{-1} \; =\; \gamma_{a}^{\dagger}\ ,\qquad   \mathcal{D}\gamma_{ab}\mathcal{D}^{-1} \; =\; -\gamma_{ab}^{\dagger}\ .
\end{equation}
Defining the 1-form $L=L_{a}\,e^{a}$ by means of $L_{a}\equiv \overline{\psi}\gamma_{a}\psi$, it is then automatically real, \emph{i.e.}
\begin{equation}
  \label{eq:29}
  L_{a}^{*} \ =\  \psi^{T} \left(\mathcal{D}\gamma_{a}\right)^{*}\ \psi^{*}
             \ =\  \overline{\ \overline{\psi}\gamma_{a}\psi\ }\,\mathcal{D}
              \ =\ \psi^{\dagger} \left(\mathcal{D}\gamma_{a}\right)^{\dagger}\ \psi
              \ =\ \overline{\psi}\mathcal{D}^{-1}\gamma_{a}^{\dagger}\mathcal{D}^{\dagger}\psi
              \ =\ \overline{\psi}\gamma_{a}\psi \ =\ L_{a}\ ,
\end{equation}
where a perhaps expected $-1$ sign in the third step is absent as we are dealing with classical (\emph{i.e.}~commuting) spinors. In terms of the components we have that $L_{a}=\epsilon^{\dagger}\mathcal{D}\gamma_{a}\epsilon$ and it is clear that $L_{0}=\epsilon^{\dagger}\epsilon$. Furthermore, we can always rotate the spatial components of $L$ in such a way that only the first component is non-vanishing. This then implies that
\begin{equation}
  \label{eq:31}
  g(L,L) \ =\ L_{0}^{2}\ -\ L_{1}^{2}\ .
\end{equation}
$L_{1}=\epsilon^{\dagger}\gamma_{01}\epsilon$ and if we combine this with $\gamma_{01}^{\dagger}=\gamma_{01}$, $\gamma_{01}^{2}=1$ and $\mathrm{Tr}(\gamma_{01} )=0$ we can use a $\mathrm{SO}(\lfloor n/2\rfloor )$ rotation to write $\gamma_{01}= \mathrm{diag}([+]^{ \lfloor n/2\rfloor } , [-]^{\lfloor n/2\rfloor })$. Decomposing the spinor w.r.t.~the structure of
$\gamma_{01}$ as $\epsilon^{t}=(v,w)$, where $v$ and $w$ are vectors in $\mathbb{C}^{\lfloor n/2\rfloor }$, we see that
\begin{equation}
  \label{eq:32}
  L_{0} \ =\ |v|^{2}+|w|^{2} \ ,\quad L_{1} \ =\ |v|^{2}-|w|^{2} \;\;\longrightarrow\;\; g(L,L) \ =\ 4|v|^{2}|w|^{2} \ ,
\end{equation}
which implies the positive-definiteness of $|L|^2$.

\subsection{\texorpdfstring{Gamma matrices and spinors in $\mathrm{SO}(1,3)$}{Gamma matrices and spinors in SO(1,3)}}
\label{appsec:spinors4d}
We work with a unitary, purely imaginary representation of the Gamma matrices $\gamma^a$ 
\begin{equation}
{(\gamma^a)}^*= -\gamma^{a}\ ,  
\end{equation}
and the same convention for their anticommutator as above. Thus, 
\begin{equation}
\gamma^{0}\gamma^{a}\gamma^{0}= \gamma^{a\, \dagger}= (\gamma^{a})^{-1}\equiv\gamma_{a}\ .
\end{equation}
Because of the even dimensionality of the theory, there is a chirality matrix, which is defined by
\begin{equation}
\gamma_{5}\equiv -i\gamma^{0}\gamma^{1}\gamma^{2}\gamma^{3}={\textstyle\frac{i}{4!}} \epsilon_{abcd} \gamma^{a}\gamma^{b}\gamma^{c}\gamma^{d}\ ,
\end{equation}
and satisfies
\begin{equation}
\gamma_{5}{}^{\dagger}=-\gamma_{5}{}^{*}=\gamma_{5}\ ,\quad (\gamma_{5})^{2}=1\ .
\end{equation}
With such a chirality matrix, we have the identity
\begin{equation}
\label{eq:dualgammaidentityind4}
\gamma^{a_{1}\cdots a_{d}} =\frac{(-1)^{\lfloor d/2\rfloor}i}{(4-d)!} \epsilon^{a_{1}\cdots a_{d}\,b_{1}\cdots b_{4-d}} \gamma_{b_{1}\cdots b_{4-d}} \gamma_{5}\ .
\end{equation}

Our convention for Dirac conjugation is
\begin{equation}
\psi^{\text{D}} \equiv \bar{\psi}=\psi^\dagger \mathcal{D}\, ,
\end{equation}
where in this case we have taken $\mathcal{D}=i\gamma_0$. We use 4-component chiral spinors. In $N=2$ theory, the chirality is related to the position of the $SU(2)$-indices and $Sp(n_h/2)$-indices as follows (see table \ref{tab-indices}):
\begin{equation}
\gamma_{5}\psi_{I\, \mu} = - \psi_{I\, \mu}\ ,
\hspace{1cm}
\gamma_{5}\lambda^{Ii} = +\lambda^{Ii}\ ,
\hspace{1cm}
\gamma_{5}\epsilon_{I} = - \epsilon_{I}\ .
\end{equation}
Both (chirality and position of the index) are reversed under complex conjugation:
\begin{equation}
\gamma_{5}\psi^{I}{}_{\mu} = \psi^{I}{}_{\mu}\ ,
\hspace{1cm}
\gamma_{5}\lambda_{I}{}^{i^{*}} = -\lambda_{I}{}^{i^{*}}\ ,
\hspace{1cm}
\gamma_{5}\zeta^{\alpha} = +\zeta^{\alpha}\ ,
\hspace{1cm}
\gamma_{5}\epsilon^{I} = \epsilon^{I}\ .
\end{equation}
We take this fact into account when Dirac-conjugating chiral spinors:
\begin{equation}
\bar{\epsilon}^{I}\equiv i(\epsilon_{I})^{\dagger}\gamma_{0}\ ,\quad \bar{\epsilon}^{I}\gamma_{5}=+\bar{\epsilon}^{I}\ ,\,\,\,\, {\rm etc.}
\end{equation}

\subsection{\texorpdfstring{Gamma matrices and spinors in $\mathrm{SO}(1,4)$}{Gamma matrices and spinors in SO(1,4)}}
\label{app-spinors}
In five dimensions, the first four of the Gamma matrices are taken to be identical to the four-dimensional purely imaginary Gamma matrices $\gamma^{0},...,\gamma^{3}$ satisfying
\begin{equation}
\{\gamma^{a},\gamma^{b}\}=2\eta^{ab}\;\mathbbm{1}\ ,  
\end{equation}

\noindent and the fifth is $\gamma^{4}=-\gamma^{0123}$, so it is purely real. The Clifford algebra anticommutator (\ref{eq:27}) is still valid for $a=0,\cdots,4$. Furthermore, $\gamma^{a_{1}\cdots a_{5}}=\epsilon^{a_{1}\cdots a_{5}}$ and in general
\begin{equation}
\label{eq:identidaddegammas}
\gamma^{a_{1}\cdots a_{d}} =\frac{(-1)^{\lfloor d/2 \rfloor}}{\left(5-d\right)!} \epsilon^{a_{1}\cdots a_{d}b_{1}\cdots b_{d-5}} \gamma_{b_{1}\cdots b_{5-d}}\ .   
\end{equation}
$\gamma^{0}$ is Hermitean and the other Gammas are anti-Hermitean, thus having a unitary representation.

To explain our convention for symplectic-Majorana spinors, let us start by defining the Dirac, complex and charge conjugation matrices $\mathcal{D}_{\pm},\mathcal{B}_{\pm},\mathcal{C}_{\pm}$. By definition they satisfy \mycite{Ortin:2004ms}
\begin{equation}
\mathcal{D}_{\pm}\, \gamma^{a}\, \mathcal{D}_{\pm}^{-1} = \pm\gamma^{a\, \dagger}\ ,\quad \mathcal{B}_{\pm}\, \gamma^{a}\, \mathcal{B}_{\pm}^{-1} = \pm\gamma^{a\, *}\ ,\quad \mathcal{C}_{\pm}\, \gamma^{a}\, \mathcal{C}_{\pm}^{-1} = \pm\gamma^{a\, T}\ .
\end{equation}
The natural choice for a real Dirac conjugation matrix is
\begin{equation}
\mathcal{D}=i\gamma^{0}\ ,  
\end{equation}
which corresponds to $\mathcal{D}=\mathcal{D}_{+}$. The other conjugation matrices are related to it by 
\begin{equation}
\label{eq:eqrelatingmatrices}
\mathcal{C}_{\pm}=\mathcal{B}^{T}_{\pm}\mathcal{D}\ ,  
\end{equation}
but it can be shown that in this case only $\mathcal{C}=\mathcal{C}_{+}$ and $\mathcal{B}=\mathcal{B}_{+}$ exist and are both antisymmetric. We take them to be
\begin{equation}
\mathcal{C}=i\gamma^{04}\ ,\quad \mathcal{B}=\gamma^{4}\quad \Rightarrow \mathcal{B}^{*}\mathcal{B}=-1\ .
\end{equation}
The Dirac conjugate is defined by
\begin{equation}
\psi^\text{D}\equiv \bar{\psi}=\psi^{\dagger}\mathcal{D}=i\psi^{\dagger}\gamma^{0}\ ,
\end{equation}
and the Majorana conjugate by
\begin{equation}
\psi^\text{M}\equiv\psi^{T}\mathcal{C}=i\psi^{T}\gamma^{04}\ .
\end{equation}
The Majorana condition (Dirac conjugate = Majorana conjugate) cannot be consistently imposed because it requires $\mathcal{B}^{*}\mathcal{B}=+1$. Therefore, we introduce the symplectic-Majorana conjugate in pairs of spinors by using the corresponding symplectic matrix, \emph{e.g.}~$\varepsilon_{ij}$ in
\begin{equation}
\psi^{i\,\text{SM}}\equiv \varepsilon_{ij}\psi^{j\, T}\mathcal{C}\ ;
\end{equation}
then the symplectic-Majorana condition is
\begin{equation}
\psi^{i\, *}=\varepsilon_{ij}\gamma^{4}\psi^{j}\ .
\end{equation}
To impose the symplectic-Majorana condition on hyperinos $\zeta^{A}$, the only thing we have to do is to replace the matrix $\varepsilon_{ij}$ by $\mathbb{C}_{AB}$, which is the invariant metric of $Sp(n_{h})$.

Our conventions on $SU(2)$ indices are intended to keep manifest the $SU(2)$-covariance.
In $SU(2)$, besides the preserved metric, there is the preserved tensor $\varepsilon_{ij}$.
We also introduce $\varepsilon^{ij}$, normalised as $\varepsilon_{12}=\varepsilon^{12} =+1$.
Therefore, we may construct new covariant objects by using $\varepsilon_{ij}$ and $\varepsilon^{ij}$, \emph{e.g.}~$\psi_{i}\equiv \varepsilon_{ij}\psi^{j}$, $\psi^{j} = \psi_{i}\varepsilon^{ij}$.
Using this notation the symplectic-Majorana condition can be simply stated as
\begin{equation}
\psi^{i\, *}=\gamma^{4}\psi_{i}\ .
\end{equation}
We use the bar on spinors to denote the (single) Majorana conjugate:
\begin{equation}
\bar\psi^{i} \equiv {\psi^{i}}^T\mathcal{C}\ ,
\end{equation}
which transforms under $SU(2)$ in the same representation as $\psi^{i}$ does. We can also lower its $SU(2)$ index as $\bar\psi_{i} \equiv \varepsilon_{ij}\bar\psi^{j}$. In terms of single Majorana conjugates the symplectic Majorana condition reads
\begin{equation}
\left(\bar\psi^{i}\right)^* = \bar\psi_{i}\gamma^4\ .
\end{equation}
Finally, observe that after imposing the symplectic Majorana condition the following simple relation between the single Dirac and Majorana conjugates holds:
\begin{equation}
{\psi^{i}}^\dagger\mathcal{D} = \bar\psi_{i}\ ,
\end{equation}

\noindent which is very useful if one prefers to use the Dirac conjugate instead of the Majorana one.

\section{Spinor bilinears and Fierz identities}
\label{app:bilinearsandfierzs}
In this appendix we provide information on the spinor bilinears constructed out of the $\epsilon$'s. This is primary in the characterisation of the solutions based on the spinor-bilinear method used in chapters \ref{4d}, \ref{5dminimal} and \ref{EWspaces}. We give as well the Fierz identities, with a proof of their derivation.

\subsection{\texorpdfstring{Spinor bilinears in $d=4$}{Spinor bilinears in d=4}}
\label{sec:Bil}
These bilinears are based on (but not equal to) those of \mycite{Meessen:2006tu}. The scalar ones are defined by
\begin{equation}
  \label{eq:Bil1}
  \begin{array}{lclclcl}
    X & =& \textstyle{1\over 2}\varepsilon^{IJ}\ \bar{\epsilon}_{I}\epsilon_{J} 
      &,\hspace{.5cm}\hspace{.5cm}&
    \bar{\epsilon}_{I}\epsilon_{J} & =& \varepsilon_{IJ}\ X\ ,\\
     & & & & & &\\
    \overline{X} & =& \textstyle{1\over 2}\varepsilon_{IJ}\ \bar{\epsilon}^{I}\epsilon^{J} 
      &,\hspace{.5cm}\hspace{.5cm}&
    \bar{\epsilon}^{I}\epsilon^{J} & =& \varepsilon^{IJ}\ \overline{X}\ .
  \end{array}
\end{equation}
The vector bilinears are defined by
\begin{equation}
  \label{eq:Bil2}
  V_{a}^{I}{}_{J} \;\equiv\; i\bar{\epsilon}^{I}\gamma_{a}\epsilon_{J}
                 \; =\; \textstyle{1\over 2}\ V_{a}\ \delta^{I}{}_{J} 
                 \ +\ \textstyle{1\over 2}\ V^{x}_{a}\ \left(\sigma^{x}\right)^{I}{}_{J}\ ,
\end{equation}
which can be inverted to
\begin{equation}
  \label{eq:Bil2a}
  V_{a} \; =\; V_{a}^{I}{}_{I} \hspace{1cm}\mbox{and}\hspace{1cm}
  V^{x}_{a} \; =\; \left(\sigma^{x}\right)_{I}{}^{J}\ V_{a}^{I}{}_{J} \ .
\end{equation}
{}Finally we have 3 imaginary-selfdual 2-forms defined by
\begin{equation}
  \label{eq:Bil3}
  \Phi_{IJ\ ab} \;\equiv\; \bar{\epsilon}_{I}\gamma_{ab}\epsilon_{J}
               \; =\; \Phi^{x}_{ab}\ \textstyle{i\over 2}\left(\sigma^{x}\right)_{IJ}
  \;\; \longrightarrow\;\;
  \Phi^{x} \; =\; i\left(\sigma^{x}\right)^{IJ}\ \Phi_{IJ} \ . 
\end{equation}
The anti-imaginary-self-dual 2-forms are defined by complex conjugation.
\par
{}From the Fierz identities eq.~(\ref{eq:N=2Fierzidentities}) we can then derive that 
\begin{equation}
  \label{eq:Bil4}
  \eta_{ab} \; =\; \frac{1}{4|X|^{2}}\left[
                      V_{a}V_{b} \; -\; V^{x}_{a}V^{x}_{b}
                  \right] \ ,
\end{equation}
and consistently with the above that 
\begin{equation}
  \label{eq:Bil5}
  \imath_{V}V^{x}\ =\ 0 \;\; ,\;\;
  g\left( V,V\right) \ =\ 4|X|^{2} \;\; ,\;\;
  g\left( V^{x},V^{y}\right) \ =\ -4|X|^{2}\ \delta^{xy} \ .
\end{equation}
A result that is harder to be found is 
\begin{equation}
  \label{eq:Bil6}
  \overline{X}\ \Phi^{x}_{ab}\; =\; -i\ 
      \left[
         V_{[a}V_{b]}^{x} \ +\ \textstyle{i\over 2}\varepsilon_{ab}{}^{cd}V_{c}V_{d}^{x}\
      \right] \ ,
\end{equation}
which translates to
\begin{equation}
  \label{eq:12j}
  \overline{X}\ \Phi^{x} \; =\; \frac{1}{2i}\; \left[
                      V\wedge V^{x} \ +\ i\ \star\left( V\wedge V^{x}\right)
                  \right]
\end{equation}
in form notation.
\par
In the null case, {\em i.e.\/} for $X=0$, the $V^{x}$ are proportional to $V$ and the $\Phi$s become linear dependent, severely limiting the utility of the bilinears. 
In section (\ref{sec:Null}), we will, following \mycite{Tod:1983pm}, introduce an auxiliar spinor which leads to Fierz identities similar to the ones above.

\subsection{\texorpdfstring{Spinor bilinears in $d=5$}{Spinor bilinears in d=5}}
\label{sec:bilinears}
With one commuting symplectic-Majorana spinor $\epsilon^{i}$ we can construct the following independent, $SU(2)$-covariant bilinears:
\begin{enumerate}
\item[$\bar{\epsilon}_{i}\ \epsilon^{j}\;\;\;\;$:] It is easy to see that 
\begin{equation}
\begin{array}{rcl}
\bar{\epsilon}_{i}\epsilon^{j} & = & -\varepsilon^{jk}  (\bar{\epsilon}_{k}\epsilon^{l}) \varepsilon_{li}\, , \\
 & & \\
(\bar{\epsilon}_{i}\epsilon^{j})^{*} & = & -\bar{\epsilon}_{j}\epsilon^{i}\, ,\\
\end{array}
\end{equation}
The first equation implies that this matrix is proportional to $\delta_{i}{}^{j}$ and the second equation implies that the constant is purely imaginary. Thus, we define the $SU(2)$-invariant scalar
\begin{equation}
\label{eq:defoff}
\f \ \equiv\ i \bar{\epsilon}_{i}\epsilon^{i} \ =\ i \bar{\epsilon}\sigma^{0}\epsilon
\hspace{.5cm},\hspace{.5cm} \bar{\epsilon}_{i}\epsilon^{j} \; =\; 
-\textstyle{i\over 2}\ \f\ \delta_{i}{}^{j} \; . 
\end{equation}
All the other scalar bilinears $i \bar{\epsilon}\sigma^{r}\epsilon$ ($r=1,2,3$) vanish identically.

\item[$\bar{\epsilon}_{i}\gamma^{a}\epsilon^{j}\;$:] This matrix satisfies the same properties as $\bar{\epsilon}_{i}\epsilon^{j}$, and so we define the vector bilinear
\begin{equation}
V^{a} \ \equiv\ i \bar{\epsilon}_{i}\gamma^{a}\epsilon^{i} \ =\ 
i \bar{\epsilon}\gamma^{a}\sigma^{0}\epsilon \hspace{.5cm},\hspace{.5cm} \bar{\epsilon}_{i}\gamma^{a}\epsilon^{j} 
\; =\; -\textstyle{i\over 2}\  \delta_{i}{}^{j}\ V^{a} \ .
\end{equation}
which is also $SU(2)$-invariant, the other vector bilinears being automatically zero.

\item[$\bar{\epsilon}_{i}\gamma^{ab}\epsilon^{j}$:] In this case
\begin{equation}
\begin{array}{rcl}
\bar{\epsilon}_{i}\gamma^{ab}\epsilon^{j} & = & +\varepsilon^{jk}  (\bar{\epsilon}_{k}\gamma^{ab}\epsilon^{l}) \varepsilon_{li}\ , \\
& & \\
(\bar{\epsilon}_{i}\gamma^{ab}\epsilon^{j})^{*} & = & \bar{\epsilon}_{j}\gamma^{ab}\epsilon^{i}\ ,\\
\end{array}
\end{equation}
which means that these 2-form matrices are traceless and Hermitean and we have three non-vanishing real 2-forms
\begin{equation}
\Phi^{r\, ab} \ \equiv\  \sigma^{r}{}_{i}{}^{j}\ \bar{\epsilon}_{j}\gamma^{ab}\epsilon^{i} \hspace{.5cm},\hspace{.5cm}  \bar{\epsilon}_{i}\gamma^{ab}\epsilon^{j} \; =\; \textstyle{1\over 2}{\sigma^{r}}_{i}{}^{j}\ \Phi^{r\, ab} \ .
\end{equation}
for $r=1,2,3$, which transform as a vector in the adjoint representation of $SU(2)$, and the fourth is $\bar{\epsilon}\gamma^{ab}\sigma^{0}\epsilon=0$.
\end{enumerate}

Using the Fierz identities eq.~(\ref{eq:N=2Fierzidentities}) for commuting spinors we get, among other identities, 
 \begin{eqnarray}
 \label{fsquared}
 V^{a}V_{a} & = & \f^{2}\ ,\\
 & & \nonumber \\
 V_{a}V_{b} & = & \eta_{ab}\f^{2}
              +{\textstyle\frac{1}{3}}\sum_{r}\Phi^{r}{}_{a}{}^{c}\Phi^{r}{}_{cb}\ ,\\
 & & \nonumber \\
 \label{VPhi}
 V^{a}\Phi^{r}{}_{ab} & = & 0\ ,\\
 & & \nonumber \\ 
 \label{VStarPhi}
 V^{a}({}^{\star}\Phi^{r}){}_{abc} & = & -\f\Phi^{r}{}_{bc}\ ,\\
 & & \nonumber \\
 \label{quaternions}
 \Phi^{r}{}_{a}{}^{c}\Phi^s{}_{cb} & = &  -\delta^{rs}(\eta_{ab}\f^{2}
 -V_{a}V_{b})-\varepsilon^{rst}\f\Phi^t{}_{ab}\ ,\\
 & & \nonumber \\ 
 \Phi^{r}{}_{[ab}\Phi^s{}_{cd]} & = & -{\textstyle\frac{1}{4}}\f\delta^{rs}
 \varepsilon_{abcde}V^{e}\ , \\
 & & \nonumber   \\
 \label{vepsilon}
         V_a\gamma^a\epsilon^{i} &=& \f\epsilon^{i}\ ,                \\
 & & \nonumber   \\
 \label{phiepsilon}
 \Phi^{r}_{ab}\gamma^{ab}\epsilon^{i} &=& 4i\f\epsilon^{j}\sigma^{r}{}_{j}{}^{i}\ .
 \end{eqnarray}
\subsection{Fierz identities}
\label{app-idFierz}
The bilinears that can be constructed from Killing spinors in our four- and five-dimensional studies are $2\times 2$ matrices that can be written as linear combinations of the Pauli matrices $\sigma^{\hat r}$ ($\hat{r}=0,\ldots ,3$) where $\sigma^{0}=\mathbbm{1}_{2}$. It is natural, then, that we want to use the equation first derived by Fierz to help us obtain important identities on the bilinears.

The general Fierz identity arises from the completeness of the antisymmetric product of Dirac gamma matrices $\gamma^a$ as a basis for $2^{\lfloor d/2\rfloor}\times 2^{\lfloor d/2\rfloor}$ matrices. The $2^{2\lfloor d/2\rfloor}$-dimensional canonical basis for the vector space of these matrices is
\begin{equation}
\{\mathcal{O}^I\}=\{\mathcal{O}^0,\mathcal{O}^1,...,\mathcal{O}^d\}=r^I\{\mathbbm{1},\gamma^a,\gamma^{a_1\,a_2},...\,,\gamma^{a_1\,a_2\,...\,a_d}\}\ ,
\end{equation}
where $r^I=1$ or $r^I=i$ according to whether $\lfloor I/2\rfloor$ is (respectively) even or odd. We can construct an orthogonal dual basis
\begin{equation}
\{\mathcal{O}_I\}=r_I\{\mathbbm{1},\gamma_a,\gamma_{a_1\,a_2},...\,,\gamma_{a_1\,a_2\,...\,a_d}\}\ ,
\end{equation}
where $r_I=r^I$; one can see that $\mathcal{O}^I\mathcal{O}_J\equiv\text{tr}(\mathcal{O}_I\mathcal{O}_J)=2^{\lfloor d/2\rfloor} \delta_{IJ}$, such that any $2^{\lfloor d/2\rfloor}\times 2^{\lfloor d/2\rfloor}$ matrix $A$ can be written as a linear combination of the elements in $\mathcal{O}^I$. This implies that $A=a_I \mathcal{O}^I\,,$ and hence $\text{tr}(\mathcal{O}_I A)=2^{\lfloor d/2\rfloor} a_I$\,; furthermore
\begin{equation}
\label{eq:AFierzmatrix}
{A^\alpha}_\beta=2^{-\lfloor d/2\rfloor} \sum_I {A^\sigma}_\rho {(\mathcal{O}_I)^\rho}_\sigma {{(\mathcal{O}^I)}^\alpha}_\beta\ .
\end{equation}
By considering a general bilinear of spinors
\begin{equation}
\left(\bar{\lambda}M\varphi\right)\left(\bar{\psi}N\chi\right)\equiv\bar{\lambda}_\alpha {M^\alpha}_\beta \varphi^\beta \bar{\psi}_\sigma {N^\sigma}_\rho \chi^\rho
\end{equation}
and eq.~(\ref{eq:AFierzmatrix}) on ${M^\alpha}_\beta {N^\sigma}_\rho$ as a ${A^\sigma}_\beta$-matrix, we obtain the Fierz identity
\begin{equation}
\left(\bar{\lambda}M\varphi\right)\left(\bar{\psi}N\chi\right)=p\,2^{-\lfloor d/2\rfloor}\sum_I (\bar{\lambda}M\mathcal{O}_I N \chi)(\bar{\psi}\mathcal{O}^I \varphi)\ ,
\end{equation}
where $p=+1$ for \textit{commuting} spinors, and $p=-1$ for \textit{anti-commuting} spinors.\vspace{\baselineskip}

Moreover, to obtain the general form in $d=4$, we consider eq.~(\ref{eq:dualgammaidentityind4}), that relates products of $n$ matrices with products of $(d-n)$ matrices. This gives,
\begin{equation}
\label{eq:Fierzidentities}
  \begin{array}{rcl}
p(\bar{\lambda} M\chi) (\bar{\psi} N \varphi ) & = &
{\textstyle\frac{1}{4}} (\bar{\lambda} M N \varphi) (\bar{\psi} \chi )
+{\textstyle\frac{1}{4}} (\bar{\lambda} M \gamma^{a}N \varphi) 
(\bar{\psi} \gamma_{a}\chi ) 
-{\textstyle\frac{1}{8}} (\bar{\lambda} M \gamma^{ab}N \varphi) 
(\bar{\psi} \gamma_{ab}\chi )
\\
& & \\
& & 
-{\textstyle\frac{1}{4}} (\bar{\lambda} M \gamma^{a}\gamma_{5}N \varphi) 
(\bar{\psi} \gamma_{a}\gamma_{5}\chi )
+{\textstyle\frac{1}{4}} (\bar{\lambda} M \gamma_{5}N \varphi) 
(\bar{\psi}\gamma_{5}\chi )\ .\\
\end{array}
\end{equation}
For $d=5$ we use eq.~(\ref{eq:identidaddegammas}), as well as an extended basis of Pauli matrices ($\hat{r}=\{0,1,2,3\}$) to express the $Sp(1)$-structure succinctly. This gives
 \begin{equation}
 \label{eq:N=2Fierzidentities}
 \begin{array}{rcl}
 \left(\bar{\lambda}M\varphi\right)\left(\bar{\psi}N\chi\right)
 & = & 
 \displaystyle\frac{p}{8} \sum_{\hat{r}=0}^3\bigg( 
  \left(\bar{\lambda}M\sigma^{\hat{r}} N \chi\right)
 \left(\bar{\psi}\sigma^{\hat{r}}\varphi\right)
 +
 \left(\bar{\lambda}M\gamma^{a}\sigma^{\hat{r}} N \chi\right)
 \left(\bar{\psi}\gamma_{a}\sigma^{\hat{r}}\varphi\right)
 \bigg. \\
 & &  \\
 & &
 \bigg.
 \phantom{\displaystyle\frac{p}{8} \sum_{\hat{r}}\bigg( }-{\textstyle\frac{1}{2}}
 \left(\bar{\lambda}M\gamma^{ab}\sigma^{\hat{r}} N \chi\right)
 \left(\bar{\psi}\gamma_{ab}\sigma^{\hat{r}}\varphi\right)
 \bigg)\ .
 \end{array}
 \end{equation}

\cleardoublepage

\renewcommand{\leftmark}{\MakeUppercase{\chaptername\ \thechapter. Scalar manifolds}}
\chapter{Scalar manifolds}
\label{scalargeometries}
In this appendix we present the technical information about the scalar manifolds which play a part in the theories studied. These geometries are different according to the dimensionality of the theory in question, and its field content. K\"ahler geometry, for example, is relevant when considering a four-dimensional sigma model with $\N=1$ global supersymmetry \mycite{Zumino:1979et}. To introduce local SUSY, however, ones needs to consider the subclass of K\"ahler-Hodge manifolds (see {\em e.g.}~\mycite{Bagger:1984ge}{ch.~3}). An introduction to these, including the conventions used, is given in appendix \ref{app-kahlergeometry}.

For a lagrangian in $d=4$ with $\N=2$ global SUSY, containing only massless scalars, the couplings among these fields are given in terms of a hyper-K\"ahler manifold \mycite{AlvarezGaume:1981hm}. Adding gravity implies that the scalar fields will parametrise instead a quaternionic-K\"ahler manifold \mycite{Bagger:1983tt}{deWit:1983rz}{deWit:1984px}. This geometry is treated at some length in appendix \ref{app:QuatSpace}. When introducing vector multiplets as well, the self-couplings between the complex scalars, and those relating complex scalars and vectors are dictated by Special K\"ahler geometry \mycite{deWit:1984pk,Cremmer:1984hj}, which is discussed in appendix \ref{appsec:SpecGeom}. The complete target space of the non-linear sigma model for the scalar fields is thus given by the product of a Special K\"ahler manifold and a quaternionic-K\"ahler manifold.

Real Special geometry, on the other hand, arises when considering the coupling of pure $d=5$ SUGRA to vector multiplets \mycite{deWit:1992cr}. This is given in appendix \ref{sec:RealSG}. A more involved theory not considered here is that formed by also considering hyperscalar multiplets. The complete target space is then given by the product of a Real Special and a quaternionic-K\"ahler manifold.

Before starting with K\"ahler geometry we give a short scholium on complex and hyper-complex (quaternionic) structures.

\subsection*{Complex and quaternionic structures}
\label{appsec:complexstructures}
We say that a manifold $M$ is almost-complex when it is endued with an almost-complex structure $\mathcal{J}$. This is a linear map defined at each patch of the manifold such that 
\begin{equation}
\mathcal{J}:T_p M \rightarrow T_p M\ ,\qquad \text{with}\qquad \mathcal{J}^2=-1\ .
\end{equation}
We call it almost-complex, as opposed to regularly complex, since we lack an integrability condition connecting the different patches of $M$. Usually such integrability condition can be linked to the vanishing of the Nijenhuis tensor. In the context of supergravity theories, it is common to find the condition of parallelity of $\mathcal{J}$ w.r.t.~the Levi-Civit\`a connection
\begin{equation}
\label{eq:JparallelwrtLC}
\nabla_a \mathcal{J}_{bc}=0\ .
\end{equation} 
Eq.~(\ref{eq:JparallelwrtLC}) implies the vanishing of the Nijenhuis tensor, whence $M$ is a complex manifold.

This almost-complex structure $\mathcal{J}$ need not necessarily be compatible with the metric $\mathsf{H}$, such that
\begin{equation}
\mathcal{J}\circ \mathsf{H}(\partial_a,\partial_b)=\mathsf{H}(\mathcal{J}\circ \partial_a,\partial_b)+\mathsf{H}(\partial_a,\mathcal{J}\circ \partial_b)=\mathcal{J}_{ab}+\mathcal{J}_{ba}=0\ .
\end{equation}
If it is, we call the metric Hermitean and we define a 2-form $\mathsf{K}(X,Y) \equiv \mathsf{H}(\mathcal{J}\circ X,Y)$, which is identified with the K\"ahler 2-form. If the form $\mathsf{K}$ is closed, \emph{i.e.}~$d\mathsf{K}=0$, then $M$ is said to be a K\"ahler manifold. Clearly this is the case if $\mathcal{J}$ is parallel \`a la (\ref{eq:JparallelwrtLC}).\vspace{\baselineskip}

If instead of one $\mathcal{J}$ we have a triplet of them $\mathcal{J}^r\ (r=1,2,3)$, that satisfy the algebra of imaginary unit quaternions, \emph{i.e.}
\begin{equation}
{(\mathcal{J}^r\cdot \mathcal{J}^s)_a}^b=-\delta^{rs} {\delta_a}^b+\varepsilon_{rst} {(\mathcal{J}^t)_a}^b\ ,
\end{equation}
we have what is called an almost-quaternionic structure (it is also often called an almost-hyper-complex structure).
Again, if the three maps are compatible with the metric, in which case the metric is said to be hyper-Hermitean,
we can define the three 2-forms $\mathsf{K}^r(X,Y) \equiv \mathsf{H}(\mathcal{J}^rX,Y)$. 
If the condition $d\mathsf{K}^r=0$ is satisfied, then the two-forms are closed and $M$ is said to be a hyper-K\"ahler manifold.

\section{K\"ahler geometry}
\label{app-kahlergeometry}
A K\"ahler manifold $\mathcal{M}$ is a complex manifold on which there exist complex coordinates $Z^{i}$ and $Z^{*\, i^{*}} =(Z^{i})^{*}$ and a real function $\mathcal{K}(Z,Z^{*})$, called the {\em K\"ahler potential}, such that the line element is
\begin{equation}
\label{eq:Kmetric}
ds^{2} = 2 \mathcal{G}_{ii^{*}}\ dZ^{i}\,dZ^{*\, i^{*}}\ ,\qquad \text{with}\ \mathcal{G}_{ii^{*}} = \partial_{i}\partial_{i^{*}}\mathcal{K}\ .
\end{equation}
The \textit{K\"ahler (connection) 1-form}  $\mathcal{Q}$ is defined by
\begin{equation}
\label{eq:K1form}
\mathcal{Q} \equiv {\textstyle\frac{1}{2i}}(dZ^{i}\partial_{i}\mathcal{K} - dZ^{*\, i^{*}}\partial_{i^{*}}\mathcal{K})\ ,
\end{equation}
and the \textit{K\"ahler 2-form} $\mathsf{K}$ is its exterior derivative
\begin{equation}
\label{eq:K2form}
\mathsf{K} \equiv d\mathcal{Q} = i\mathcal{G}_{ii^{*}} dZ^{i}\wedge dZ^{*\, i^{*}}\ .
\end{equation}
The choice of complex coordinates is such that the complex structure is thus trivial, \emph{i.e.}~${\mathcal{J}_i}^i=-{\mathcal{J}_{i^*}}^{i^*}=i$. The Levi-Civit\`a connection on a K\"ahler manifold can be shown to be
\begin{equation}
\label{eq:KCCChrisSymb}
\Gamma_{jk}{}^{i} = \mathcal{G}^{ii^{*}}\partial_{j}\mathcal{G}_{i^{*}k}\ ,\quad \Gamma_{j^{*}k^{*}}{}^{i^{*}}  = \mathcal{G}^{i^{*}i}\partial_{j^{*}}\mathcal{G}_{k^{*}i}\ .
\end{equation}
The only non-vanishing components of the Riemann curvature tensor are given by $R_{ij^{*}kl^{*}}$, but we shall not be needing its explicit expression. The Ricci tensor is given by
\begin{equation}
\label{eq:KCCRicci}
R_{ij^{*}}  = \partial_{i}\partial_{j^{*}} \left(\textstyle{1\over 2}\log\det\mathcal{G}\right) \ .
\end{equation}

As can be easily seen from inspection of eq.~(\ref{eq:Kmetric}), the K\"ahler potential is not unique; it is defined up to \textit{K\"ahler transformations} of the form
\begin{equation}
\label{eq:Kpotentialtransformation}
\mathcal{K}^{\prime}(Z,Z^{*})=\mathcal{K}(Z,Z^{*})+f(Z)+f^{*}(Z^{*})\ , 
\end{equation}
where $f(Z)$ is a holomorphic function of the complex coordinates $Z^{i}$. Under these transformations, the K\"ahler metric and K\"ahler 2-form are invariant, while the components of the K\"ahler connection 1-form transform according to
\begin{equation}
\label{eq:K1formtransformation}
\mathcal{Q}^{\prime}_{i} =\mathcal{Q}_{i} -{\textstyle\frac{i}{2}}\partial_{i}f\ .
\end{equation}
By definition, objects with K\"ahler weight $(q,\bar{q})$ transform under the above K\"ahler transformations with a factor
$e^{-(qf+\bar{q}f^{*})/2}$ and the K\"ahler-covariant derivative $\mathcal{D}$ acting on them is given by
\begin{equation}
\label{eq:Kcovariantderivative}
\mathcal{D}_{i} \equiv \nabla_{i} +iq \mathcal{Q}_{i}\ ,\quad \mathcal{D}_{i^{*}} \equiv \nabla_{i^{*}} -i\bar{q} \mathcal{Q}_{i^{*}}\ ,
\end{equation}
where $\nabla$ is the covariant derivative associated to the Levi-Civit\`a connection on $\mathcal{M}$. The Ricci identity for this covariant derivative, for a weight $(q,\bar{q})$ scalar object, reads
\begin{equation}
[\mathcal{D}_{i},\mathcal{D}_{j^{*}}]\, \phi^{(q,\bar{q})}\; =\; -{\textstyle\frac{1}{2}}\ (q-\bar{q})\ \mathcal{G}_{ij^{*}} \,\phi^{(q,\bar{q})}\ .
\end{equation}

\subsection{K\"ahler-Hodge manifolds}
When $(q,\bar{q})=(1,-1)$, this defines a complex line bundle $L^{1}\rightarrow \mathcal{M}$ over the K\"ahler manifold $\mathcal{M}$, whose first (and only) Chern class equals the K\"ahler 2-form $\mathsf{K}$. A complex line bundle with this property is known as a \textit{K\"ahler-Hodge} (KH) manifold. These are the manifolds parametrized by the complex scalars
of the chiral multiplets of $\N=1,d=4$ Supergravity. Furthermore, objects such as the superpotential and the spinors of the theory have a well-defined K\"ahler weight. On the other hand, manifolds parametrized by the complex scalars of the vector multiplets of $\N=2,d=4$ Supergravity are also KH manifolds, but must satisfy further constraints that define what is known as \textit{Special K\"ahler Geometry} (or Special Geometry for short), described in appendix~\ref{appsec:SpecGeom}.

If one is interested on the spacetime pullback of the K\"ahler-covariant derivative on tensor fields with K\"ahler weight $(q,-q)$ (weight $q$, for short), this takes the simple form
\begin{equation}
\label{eq:Kcovariantderivative2}
\mathfrak{D}_{\mu}= \nabla_{\mu} +iq\mathcal{Q}_{\mu}\ ,
\end{equation}
where $\nabla_{\mu}$ is the standard spacetime covariant derivative plus possibly the pullback of the Levi-Civit\`a connection on $\mathcal{M}$, and $\mathcal{Q}_{\mu}$ is the pullback of the K\"ahler 1-form, {\em i.e.}
\begin{equation}
\label{eq:KahlerConPB}
\mathcal{Q}_{\mu} =  {\textstyle\frac{1}{2i}}(\partial_{\mu}Z^{i}\partial_{i}\mathcal{K} - \partial_{\mu}Z^{*\, i^{*}}\partial_{i^{*}}\mathcal{K})\ .
\end{equation}
\subsection{Gauging holomorphic isometries}
\label{app-gaugingKH}
We now proceed to review some of the basics of the gauging of holomorphic isometries of K\"ahler-Hodge manifolds that occur in $\N=1$ and $\N=2,d=4$ supergravities. We will first study the general problem in complex manifolds. This is enough for purely bosonic theories, in which only the complex structure is relevant. In the presence of fermions, however, the K\"ahler-Hodge structure becomes necessary, and only those transformations that preserve it will be symmetries (of the full theory) that can be gauged. We study this problem next. 

The special-K\"ahler structure is necessary in $\N=2,d=4$ Supergravity and, again, only those transformations that preserve it are symmetries that can be gauged. This problem will be studied in appendix~\ref{sec:SGisom}, after we define special-K\"ahler manifolds.

\subsubsection{Gauging in complex manifolds}
\label{app-complex}
We start by assuming that the Hermitean metric $\mathcal{G}_{ij^{*}}$ (we will use the K\"ahler-Hodge structure later) admits a set of Killing vectors\footnote{The index $\Lambda$ always takes values from $1$ to $n_{V}$ ($\bar{n}=n_{V}+1$) in $\N=1$ ($\N=2$) Supergravity , but some (or all) the Killing vectors may actually be zero.}  $\{K_{\Lambda}= k_{\Lambda}{}^{i}\partial_{i}+k^{*}_{\Lambda}{}^{i^{*}}\partial_{i^{*}}\}$ satisfying the Lie algebra
\begin{equation}
\label{eq:Liealgebra}
[K_{\Lambda},K_{\Sigma}]= -f_{\Lambda\Sigma}{}^{\Omega} K_{\Omega}\ ,
\end{equation}
of the group $G_{V}$ that we want to gauge. Hermiticity implies that the components $k_{\Lambda}{}^{i}$ and $k^{*}_{\Lambda}{}^{i^{*}}$ of the Killing vectors are, respectively, holomorphic and antiholomorphic, and satisfy (separately) the above Lie algebra. Once (anti-)holomorphicity is taken into account, the only non-trivial components of the Killing equation are
\begin{equation}
\label{eq:Killingeq}
{\textstyle\frac{1}{2}}\pounds_{\Lambda}\mathcal{G}_{ij^{*}}= \nabla_{i^{*}}k^{*}_{\Lambda\, j}+ \nabla_{j}k_{\Lambda\, i^{*}} = 0\ ,
\end{equation}
where $\pounds_{\Lambda}$ stands for the Lie derivative w.r.t.~$K_{\Lambda}$. 

The standard $\sigma$-model kinetic term $\mathcal{G}_{ij^{*}}\partial_{\mu}Z^{i} \partial^{\mu}Z^{*j^{*}}$ is then automatically invariant under infinitesimal reparametrizations of the form
\begin{equation}
\label{eq:deltazio}
\delta_{\alpha} Z^{i} = \alpha^{\Lambda} k_{\Lambda}{}^{i}(Z)\ ,  
\end{equation}
where the infinitesimal variation parameters $\alpha^{\Lambda}$ are constants.
If instead they become arbitrary functions of the spacetime coordinates, \emph{i.e.}~$\alpha^\Lambda=\alpha^{\Lambda}(x)$, we need to introduce a covariant derivative using as connection the vector fields present in the theory.
We write the covariant derivative as
\begin{equation}
\label{eq:nablazio}
\mathfrak{D}_{\mu} Z^{i} = \partial_{\mu} Z^{i}+gA^{\Lambda}{}_{\mu}k_{\Lambda}{}^{i}\ , 
\end{equation}
which transforms as
\begin{equation}
\label{eq:igaugetrans}
\delta_{\alpha}\mathfrak{D}_{\mu} Z^{i}= \alpha^{\Lambda}(x) \partial_{j} k_{\Lambda}{}^{i}\mathfrak{D}_{\mu} Z^{j}
=-\alpha^{\Lambda}(x)(\pounds_{\Lambda}-K_{\Lambda})\mathfrak{D}_{\mu} Z^{j} \ ,
\end{equation}
provided that the gauge potentials transform as
\begin{equation}
\label{eq:gaugepotentialtransformations}
\delta_{\alpha} A^{\Lambda}{}_{\mu} = -g^{-1}\mathfrak{D}_{\mu}\alpha^{\Lambda} \equiv -g^{-1}(\partial_{\mu}\alpha^{\Lambda} +gf_{\Sigma\Omega}{}^{\Lambda} A^{\Sigma}{}_{\mu} \alpha^{\Omega})\ .
\end{equation}
The gauge field strength is given by 
\begin{equation}
\label{eq:Fdef}
F^{\Lambda}{}_{\mu\nu} = 2\partial_{[\mu}A^{\Lambda}{}_{\nu]}+gf_{\Sigma\Omega}{}^{\Lambda}A^{\Sigma}{}_{[\mu}A^{\Omega}{}_{\nu]}\ , 
\end{equation}
and changes under gauge transformations as
\begin{equation}
\delta_{\alpha} F^{\Lambda}{}_{\mu\nu} = -\alpha^{\Sigma}(x)f_{\Sigma\Omega}{}^{\Lambda} F^{\Omega}{}_{\mu\nu}\ .
\end{equation}
Now, to make the $\sigma$-model kinetic term gauge invariant, it is enough to replace the partial derivatives by covariant derivatives
\begin{equation}
\mathcal{G}_{ij^{*}}\partial_{\mu}Z^{i} \partial^{\mu}Z^{*j^{*}}\quad \longrightarrow \mathcal{G}_{ij^{*}}\mathfrak{D}_{\mu}Z^{i} \mathfrak{D}^{\mu}Z^{*j^{*}}\ .
\end{equation}

For any tensor field $\Phi$ \footnote{Where spacetime ($\mu,\nu,\ldots$), gauge ($\Lambda,\Sigma,\ldots$) and target space tensor ($i,i^{*},\ldots$) indices are not explicitly shown.} transforming covariantly under gauge transformations, \emph{i.e.}~transforming as
\begin{equation}
\delta_{\alpha}\Phi = -\alpha^{\Lambda}(x)(\mathbb{L}_{\Lambda} -K_{\Lambda}) \Phi\ ,  
\end{equation}
where we have defined the \textit{Lie covariant derivative} $\mathbb{L}_\Lambda$ as\footnote{We will extend this definition to fields with non-zero K\"ahler weight after we study the symmetries of the K\"ahler structure. For the moment we only consider tensors of the Hermitean space with metric $\mathcal{G}_{ij^{*}}$, possibly with gauge and spacetime indices.}
\begin{equation}
\mathbb{L}_{\Lambda} \equiv \pounds_{\Lambda} -\mathcal{S}_{\Lambda}\ ,   
\end{equation}
and $\mathcal{S}_{\Lambda}$ represents a symplectic rotation, the gauge covariant derivative is given by 
\begin{equation}
\mathfrak{D}_{\mu}\Phi = \{\nabla_{\mu} +\mathfrak{D}_{\mu}Z^{i}\Gamma_{i}+\mathfrak{D}_{\mu}Z^{*i^{*}}\Gamma_{i^{*}}  
-gA^{\Lambda}{}_{\mu}(\mathbb{L}_{\Lambda}-K_{\Lambda})\}\Phi\ .
\end{equation}
In particular, on $\mathfrak{D}_{\mu}Z^{i}$
\begin{equation}
\mathfrak{D}_{\mu}  \mathfrak{D}_{\nu} Z^{i}  =  \nabla_{\mu}\mathfrak{D}_{\nu} Z^{i}+\Gamma_{jk}{}^{i}\mathfrak{D}_{\mu} Z^{j}\mathfrak{D}_{\nu} Z^{k} +gA^{\Lambda}{}_{\mu}\partial_{j}k_{\Lambda}{}^{i}\mathfrak{D}_{\nu} Z^{j}\ ,
\end{equation}
which implies
\begin{equation}
\label{eq:bianchidz}
\left[ \mathfrak{D}_{\mu},\mathfrak{D}_{\nu} \right]Z^{i} =  gF^{\Lambda}{}_{\mu\nu} k_{\Lambda}{}^{i}\ .
\end{equation}
\vspace{\baselineskip}

An important case is that of fields $\Phi$ which only depend on the spacetime coordinates through the complex scalars $Z^{i}$ and their complex conjugates, so that
\begin{equation}
\nabla_{\mu} \Phi = \partial_{\mu} \Phi = \partial_{\mu} Z^{i}\partial_{i} \Phi +\partial_{\mu} Z^{*i^{*}} \partial_{i^{*}} \Phi\ .
\end{equation}
$\Phi$ is then an \textit{invariant field} if\footnote{Alternatively, we could say that it is a field invariant under reparametrizations up to rotations.}
\begin{equation}
\mathbb{L}_{\Lambda}\Phi  \equiv (\pounds_{\Lambda}-\mathcal{S}_{\Lambda}) \Phi =0\ .
\end{equation}
Only if all the fields present in the theory are invariant fields, can the theory be gauged. Only in such a case $\nabla_{\mu} \Phi = \partial_{\mu} \Phi = \partial_{\mu} Z^{i}\partial_{i} \Phi +\partial_{\mu} Z^{*i^{*}} \partial_{i^{*}} \Phi$ can be true irrespectively of gauge transformations. These fields transform under gauge transformations according to
\begin{equation}
\delta_{\alpha}\Phi =-\alpha^{\Lambda}(\mathbb{L}_{\Lambda}-K_{\Lambda})\Phi = \alpha^{\Lambda}K_{\Lambda}\Phi\ ,
\end{equation}
and their covariant derivative is given by
\begin{equation}
\mathfrak{D}_{\mu}\Phi = \{\partial_{\mu} +\mathfrak{D}_{\mu}Z^{i}\Gamma_{i} +\mathfrak{D}_{\mu}Z^{*i^{*}}\Gamma_{i^{*}}  
+gA^{\Lambda}{}_{\mu}K_{\Lambda}\} \Phi\ ,
\end{equation}
which is always the covariant pullback of the target covariant derivative:
\begin{equation}
\mathfrak{D}_{\mu}\Phi =\mathfrak{D}_{\mu}Z^{i}\nabla_{i} \Phi +\mathfrak{D}_{\mu}Z^{*i^{*}}\nabla_{i^{*}} \Phi\ .  
\end{equation}

As an example, consider the holomorphic \textit{kinetic matrix} $f_{\Lambda\Sigma}(Z)$ in $\N=1$, $d=4$ Supergravity or the \textit{period matrix} $\mathcal{N}_{\Lambda\Sigma}(Z,Z^{*})$ in $\N=2$, $d=4$ Supergravity, both of which are symmetric matrices that codify the couplings between the complex scalars and the vector fields. These matrices transform under global rotations of the vector fields
\begin{equation}
\delta_{\alpha}A^{\Lambda}{}_{\mu} =-\alpha^{\Sigma}f_{\Sigma\Omega}{}^{\Lambda}A^{\Omega}{}_{\mu}  
\end{equation}
according to
\begin{equation}
\delta_{\alpha}f_{\Lambda\Sigma} \equiv -\alpha^{\Omega}\mathcal{S}_{\Omega}f_{\Lambda\Sigma}= 2\alpha^{\Omega}f_{\Omega(\Lambda}{}^{\Pi}f_{\Sigma)\Pi}\ ,  
\end{equation}
(analogously for $\mathcal{N}_{\Lambda\Sigma}$) and under the reparametrizations of the complex scalars, eq.~(\ref{eq:deltazio}), as
\begin{equation}
\delta_{\alpha}f_{\Lambda\Sigma} = -\alpha^{\Omega}\pounds_{\Omega}f_{\Lambda\Sigma}-\alpha^{\Omega}k_{\Omega}{}^{i}\partial_{i}f_{\Lambda\Sigma}\ .  
\end{equation}
These transformations will only be a symmetry of the theory if their values coincide, \emph{i.e.}~if
\begin{equation}
(\pounds_{\Omega}-\mathcal{S}_{\Omega}) f_{\Lambda\Sigma} =\mathbb{L}_{\Omega}\,f_{\Lambda\Sigma}=0\ ,
\end{equation}
this is, only if $f_{\Lambda\Sigma}(Z)$ is an invariant field according to the above definition. Its covariant derivative is given by 
\begin{equation}
\label{eq:covariantderivativefls} 
\mathfrak{D}_{\mu}f_{\Lambda\Sigma} = \mathfrak{D}_{\mu}Z^{i}\partial_{i}f_{\Lambda\Sigma}\ , 
\end{equation}
on account of its holomorphicity.
\subsubsection{Gauging in K\"ahler-Hodge manifolds}
\label{app-KH}
Let us now assume that the scalar manifold is not just Hermitean, but rather K\"ahler-Hodge, and proceed to study how the K\"ahler structure is preserved. The transformations generated by the Killing vectors will preserve the K\"ahler structure if they leave the K\"ahler potential invariant up to K\"ahler transformations, \emph{i.e.}~for each Killing vector $K_{\Lambda}$
\begin{equation}
\label{eq:Kconservation}
\pounds_{\Lambda}\mathcal{K}\equiv k_{\Lambda}{}^{i}\partial_{i}\mathcal{K}+ k^{*}_{\Lambda}{}^{i^{*}}\partial_{i^{*}}\mathcal{K} =\lambda_{\Lambda}(Z)+ \lambda^{*}_{\Lambda}(Z^{*})\ .
\end{equation}
From this condition it follows that 
\begin{equation}
\label{eq:lambdaalgebra}
\pounds_{\Lambda}\lambda_{\Sigma} -\pounds_{\Sigma}\lambda_{\Lambda} = -f_{\Lambda\Sigma}{}^{\Omega}\lambda_{\Omega}\ .
\end{equation}

On the other hand, the preservation of the K\"ahler structure implies the conservation of the K\"ahler 2-form $\mathsf{K}$ 
\begin{equation}
\label{eq:Jconservation}
\pounds_{\Lambda}\mathsf{K}=0\ .
\end{equation}
The closedness of $\mathsf{K}$ implies that $\pounds_{\Lambda}\mathsf{K}= d(i_{k_{\Lambda}}\mathsf{K})$ and therefore the preservation of the K\"ahler structure implies the existence of a set of real 0-forms $\mathcal{P}_{\Lambda}$ known as \textit{momentum maps}, such that
\begin{equation}
\label{eq:KMomMap}
i_{k_{\Lambda}}\mathsf{K}= d\mathcal{P}_{\Lambda}\ .
\end{equation}
A local solution for this equation is provided by
\begin{equation}
i\mathcal{P}_{\Lambda}=k_{\Lambda}{}^{i}\partial_{i}\mathcal{K} -\lambda_{\Lambda}\ ,   
\end{equation}
which, on account of eq.~(\ref{eq:Kconservation}), it is equivalent to
\begin{equation}
i\mathcal{P}_{\Lambda}=-(k^{*}_{\Lambda}{}^{i^{*}}\partial_{i^{*}}\mathcal{K} -\lambda^{*}_{\Lambda})\ ,
\end{equation}
or
\begin{equation}
\mathcal{P}_{\Lambda}= i_{k_{\Lambda}}\mathcal{Q}-{\textstyle\frac{1}{2i}}(\lambda_{\Lambda}-\lambda^{*}_{\Lambda})\, .
\end{equation}
The momentum map can be used as a prepotential from which the Killing vectors can be derived
\begin{equation}
\label{eq:prepo}
k_{\Lambda\, i^{*}} =i\partial_{i^{*}}\mathcal{P}_{\Lambda}\ .  
\end{equation}
This is why they are sometimes called \textit{Killing prepotentials}.\vspace{\baselineskip}

In principle, the momentum maps are defined up to an additive real constant. In $\N=1$, $d=4$ theories (but not in $\N=2$, $d=4$) it is possible to have non-vanishing, constant momentum maps with $i\mathcal{P}_{\Lambda}=-\lambda_{\Lambda}$ for vanishing Killing vectors.  In this case, no isometry is gauged; instead it is the $U(1)$ symmetry associated to K\"ahler transformations\footnote{Cf. eq.~(\ref{eq:K1formtransformation}).} (in K\"ahler-Hodge manifolds) that is gauged. These constant momentum maps are called \textit{D-} or \textit{Fayet-Iliopoulos} terms, and appear in the supersymmetry transformation rules of gaugini, in the potential and in the covariant derivatives of sections that we now discuss.

Using Eqs.~(\ref{eq:Liealgebra}),(\ref{eq:Kconservation}) and (\ref{eq:lambdaalgebra}) one finds
\begin{equation}
\label{eq:momentummaptransformationrule}
\pounds_{\Lambda}\mathcal{P}_{\Sigma} = 2i k_{[\Lambda}{}^{i}k^{*}_{\Sigma]}{}^{j^{*}}\mathcal{G}_{ij^{*}} = -f_{\Lambda\Sigma}{}^{\Omega} \mathcal{P}_{\Omega}\ .  
\end{equation}
This equation fixes the additive constant of the momentum map in directions in which a non-Abelian group is going to be gauged. The gauge transformation rule for a section $\Phi$ of K\"ahler weight $(p,q)$ is\footnote{Again, spacetime and target space tensor indices are not explicitly shown. Symplectic indices are not shown, either.}
\begin{equation}
\delta_{\alpha}\Phi = -\alpha^{\Lambda}(x)(\mathbb{L}_{\Lambda}-K_{\Lambda}) \Phi\ ,  
\end{equation}
where $\mathbb{L}_{\Lambda}$ stands for the symplectic and K\"ahler-covariant Lie derivative w.r.t.~$K_{\Lambda}$, and it is given by
\begin{equation}
\mathbb{L}_{\Lambda} \Phi \equiv \{\pounds_{\Lambda} -[\mathcal{S}_{\Lambda}-{\textstyle\frac{1}{2}}(p\lambda_{\Lambda}+q\lambda^{*}_{\Lambda})]\} \Phi\ ,  
\end{equation}
where the $\mathcal{S}_{\Lambda}$ are $\mathfrak{sp}(\bar{n})$ matrices that provide a representation of the Lie algebra of the gauge group $G_{V}$ acting on the section $\Phi$:
\begin{equation}
\label{eq:SLiealgebra}
[\mathcal{S}_{\Lambda},\mathcal{S}_{\Sigma}]= +f_{\Lambda\Sigma}{}^{\Omega} \mathcal{S}_{\Omega}\ .
\end{equation}
The gauge covariant derivative acting on these sections is given by 
\begin{equation}
  \begin{array}{rcl}
\mathfrak{D}_{\mu}\Phi & = & \{\nabla_{\mu} +\mathfrak{D}_{\mu}Z^{i}\Gamma_{i}  +\mathfrak{D}_{\mu}Z^{*i^{*}}\Gamma_{i^{*}} +{\textstyle\frac{1}{2}}(pk_{\Lambda}{}^{i}\partial_{i}\mathcal{K}+ qk^{*}_{\Lambda}{}^{i^{*}}\partial_{i^{*}}\mathcal{K})  \\
& & \\
& & +gA^{\Lambda}{}_{\mu}[\mathcal{S}_{\Lambda} +{\textstyle\frac{i}{2}}(p-q)\mathcal{P}_{\Lambda} -(\pounds_{\Lambda} -K_{\Lambda})]\}\Phi\ .\\
\end{array}
\end{equation}

Invariant sections are then those for which 
\begin{equation}
\mathbb{L}_{\Lambda} \Phi =0\quad \Rightarrow\quad \pounds_{\Lambda}\Phi =[\mathcal{S}_{\Lambda} -{\textstyle\frac{1}{2}}(p\lambda_{\Lambda}+q\lambda^{*}_{\Lambda})] \Phi\ ,
\end{equation}
and their gauge covariant derivatives are, again, the covariant pullbacks of the K\"ahler-covariant derivatives
\begin{equation}
\mathfrak{D}_{\mu}\Phi \; =\; \mathfrak{D}_{\mu}Z^{i}\, \mathcal{D}_{i}\Phi \; +\; \mathfrak{D}_{\mu}Z^{*i^{*}}\, \mathcal{D}_{i^{*}}\Phi\ .  
\end{equation}
The prime example of an invariant field is the covariantly holomorphic section $\mathcal{L}(Z,Z^{*})$ of $\N=1$, $d=4$ theories. This is a K\"ahler weight $(1,-1)$ section, related to the holomorphic superpotential $W(Z)$ by 
\begin{equation}\label{eq:N1defL}
\mathcal{L}(Z,Z^{*})\equiv W(Z) e^{\mathcal{K}/2}\ ,
\end{equation}
and its covariant holomorphicity follows from the holomorphicity of $W$
\begin{equation}
\mathcal{D}_{i^{*}}\mathcal{L} = (\partial_{i^{*}}+i\mathcal{Q}_{i^{*}})\mathcal{L}=e^{\mathcal{K}/2}\partial_{i^{*}}( e^{-\mathcal{K}/2}\mathcal{L})=e^{\mathcal{K}/2}\partial_{i^{*}}W=0\ .
\end{equation}

In order for the global transformation eq.~(\ref{eq:deltazio}) to be a symmetry of the full theory that we can gauge, $\mathcal{L}$ must be an invariant section, \emph{i.e.}
\begin{equation}
\mathbb{L}_{\Lambda}\mathcal{L}= \{\pounds_{\Lambda}+{\textstyle\frac{1}{2}}(\lambda_{\Lambda}-\lambda^{*}_{\Lambda})\}\mathcal{L}=0\,\quad \Rightarrow\quad K_{\Lambda}\mathcal{L} =  -{\textstyle\frac{1}{2}}(\lambda_{\Lambda}-\lambda^{*}_{\Lambda})\mathcal{L}\ .
\end{equation}
Then under gauge transformations it will transform according to 
\begin{equation}
\delta_{\alpha}\mathcal{L} =-{\textstyle\frac{1}{2}}\alpha^{\Lambda}(x)(\lambda_{\Lambda}-\lambda^{*}_{\Lambda})\mathcal{L}\ ,
\end{equation}
and its covariant derivative will be given by 
\begin{equation}
\label{eq:covariantderivativeL}
\mathfrak{D}_{\mu}\mathcal{L} = (\partial_{\mu}+i\hat{\mathcal{Q}}_{\mu})\mathcal{L}=\mathfrak{D}_{\mu}Z^{i}\mathcal{D}_{i}\mathcal{L}\ ,
\end{equation}
where we have defined
\begin{equation}\label{eq:DefQhat}
\hat{\mathcal{Q}}_{\mu}\equiv \mathcal{Q}_{\mu}+gA^{\Lambda}{}_{\mu}\mathcal{P}_{\Lambda}\ .  
\end{equation}
Observe that this 1-form is, in general, different from the ``covariant pullback'' of the K\"ahler 1-form, which is
\begin{equation}
{\textstyle\frac{1}{2i}} \mathfrak{D}_{\mu}Z^{i}\partial_{i}\mathcal{K}+\mathrm{c.c.}\ .  
\end{equation}
The difference between this and the correct one is 
\begin{equation}
{\textstyle\frac{1}{2i}} \mathfrak{D}_{\mu}Z^{i}\partial_{i}\mathcal{K} +\mathrm{c.c.} -\hat{\mathcal{Q}}_{\mu} =
g A^{\Lambda}{}_{\mu}\Im {\rm m}\lambda_{\Lambda}\ , 
\end{equation}
and only vanishes when the isometries that have been gauged leave the K\"ahler potential exactly invariant, \emph{i.e.}~for $\lambda_{\Lambda}=0$.\vspace{\baselineskip}

It should be evident that $\mathcal{D}_{i}\mathcal{L}$ is also an invariant field, and therefore the part of the $\N=1$, $d=4$ Supergravity potential that depends on the superpotential 
\begin{equation}
-24 |\mathcal{L}|^{2} +8\mathcal{G}^{ij^{*}}\mathcal{D}_{i}\mathcal{L}\mathcal{D}_{j^{*}}\mathcal{L}^{*} 
\end{equation}
is automatically exactly invariant. On the other hand eq.~(\ref{eq:momentummaptransformationrule}) proves that the momentum map itself is an invariant field. Then,
\begin{equation}
\label{eq:covariantderivativepl}
\begin{array}{rcl}
\delta_{\alpha}\mathcal{P}_{\Lambda} & = & -\alpha^{\Sigma}(x)f_{\Sigma\Lambda}{}^{\Omega}\mathcal{P}_{\Omega} \ ,\\
& & \\
\mathfrak{D}_{\mu}\mathcal{P}_{\Lambda} & = & \partial_{\mu}\mathcal{P}_{\Lambda} +gf_{\Lambda\Sigma}{}^{\Omega}A^{\Sigma}{}_{\mu}\mathcal{P}_{\Omega}\ ,\\
& & \\
\mathfrak{D}_{\mu}\mathcal{P}_{\Lambda} & = & \mathfrak{D}_{\mu}Z^{i}\partial_{i}\mathcal{P}_{\Lambda} +
\mathfrak{D}_{\mu}Z^{*i^{*}}\partial_{i^{*}}\mathcal{P}_{\Lambda}\ , 
\end{array}
\end{equation}
and the part of the $\N=1$ $,d=4$ Supergravity potential that depends on it, \emph{i.e.}
\begin{equation}
+{\textstyle\frac{1}{2}}g^{2}(\Im {\rm m}\,f)^{-1|\Lambda\Sigma}\mathcal{P}_{\Lambda}\mathcal{P}_{\Sigma}\ , 
\end{equation}
is also automatically invariant.

Finally, let us consider the spinors of the theory.
They have a non-vanishing K\"ahler weight which is $(-1/2,1/2)$ times their chirality.
For instance, the gravitino of $\N=1$, $d=4$ theories transform as
\begin{equation}
\begin{array}{rcl}
\delta_{\alpha} \psi_{\mu} & = &  -{\textstyle\frac{1}{4}}\alpha^{\Lambda}(x)(\lambda_{\Lambda}-\lambda_{\Lambda}^{*})\psi_{\mu}\ , 
\\
& & \\
\mathfrak{D}_{\mu}\psi_{\nu} & = & \{\nabla_{\mu}+{\textstyle\frac{i}{2}}\hat{\mathcal{Q}}\}\psi_{\nu}\ .   
\end{array}
\end{equation}

\section{Special geometry}
\label{appsec:SpecGeom}
In this short appendix\footnote{For a complete review refer to \mycite{Andrianopoli:1996cm}.} we shall discuss the geometric structure underlying the couplings of vector supermultiplets in $\N=2$ $d=4$ Supergravity, which has received the name of \emph{Special K\"ahler geometry} (usually just called Special geometry). The first articles introducing this structure were \mycite{deWit:1984pk,Cremmer:1984hj} and the formalisation was given in \mycite{Strominger:1990pd}.
The essential references are \mycite{Castellani:1990tp}{Craps:1997gp,Ceresole:1995ca,Fre:1995bc,Ceresole:1995jg}{Fre:1996xz}. After discussing the coordinate independent formulation of special geometry, we make contact with the original formulation in terms of the prepotential in \ref{LauwersdeWitSpgeom}. Appendix \ref{sec:SGisom} discusses the gauging of isometries, and how this is used in order to construct gauged supergravities. 

The formal starting point for the definition of a Special K\"ahler manifold lies in the definition of a K\"ahler-Hodge manifold. As explained in the section above, a KH-manifold is a complex line bundle over a K\"ahler manifold $\mathcal{M}$, such that the first Chern class of the line bundle equals the K\"ahler form. This then implies that the exponential of the K\"ahler potential can be used as a metric on the line bundle. Furthermore, the connection on the line bundle is $\mathcal{Q}= (2i)^{-1}( dz^{i}\partial_{i}\mathcal{K} - d\bar{z}^{\bar{\imath}}\partial_{\bar{i}}\mathcal{K})$. Let us denote the line bundle by $L^{1}\rightarrow \mathcal{M}$, where the superscript is there to indicate that the covariant derivative is $\mathfrak{D} = \nabla + i\mathcal{Q}$

Consider then a flat $2\bar{n}$ vector bundle $E\rightarrow\mathcal{M}$ with structure group $Sp(\bar{n};\mathbb{R})$, and take a section $\mathcal{V}$ of the product bundle $E\otimes L^{1}\rightarrow\mathcal{M}$ and its complex conjugate $\overline{\mathcal{V}}$, which is a section of the bundle $E\otimes L^{-1}\rightarrow \mathcal{M}$. A special K\"ahler manifold is then a bundle $E\otimes L^{1}\rightarrow\mathcal{M}$, for which there exists a section $\mathcal{V}$ such that
\begin{equation}
  \label{eq:SGDefFund}
  \mathcal{V} \ =\ \left(
                    \begin{array}{c}
                       \mathcal{L}^{\Lambda}\\
                       \mathcal{M}_{\Lambda}
                     \end{array}
                   \right) \quad \rightarrow \quad
  \left\{
     \begin{array}{lcl}
       \langle \mathcal{V}\mid\overline{\mathcal{V}}\rangle & \equiv&
                  \overline{\mathcal{L}}^{\Lambda}\mathcal{M}_{\Lambda} \ -\ \mathcal{L}^{\Lambda}\overline{\mathcal{M}}_{\Lambda}
                  \; =\; -i\ , \\
       & & \\
       \mathfrak{D}_{\bar{\imath}}\mathcal{V} & =& 0 \ ,\\
       & & \\
       \langle\mathfrak{D}_{i}\mathcal{V}\mid\mathcal{V}\rangle & =& 0 \ .

     \end{array}
  \right.
\end{equation}
By defining the objects
\begin{equation}
  \label{eq:SGDefU}
  \mathcal{U}_{i} \; \equiv\; \mathfrak{D}_{i}\mathcal{V} \; =\;
       \left(
         \begin{array}{c}
             f_{i}^{\Lambda}\\
             h_{\scriptscriptstyle{\Lambda}\ i}
         \end{array}
       \right) \ ,\quad \overline{\mathcal{U}}_{\bar{\imath}} \; =\; \overline{\mathcal{U}_{i}} \ ,
\end{equation}
it follows from the basic definitions that
\begin{equation}
  \label{eq:SGProp1}
  \begin{array}{lclclcl}
     \mathfrak{D}_{\bar{\imath}}\ \mathcal{U}_{i} & =& \mathcal{G}_{i\bar{\imath}}\ \mathcal{V}, &&
     \langle\mathcal{U}_{i}\mid\overline{\mathcal{U}}_{\bar{\imath}}\rangle & =& i\mathcal{G}_{i\bar{\imath}} \ , \\
     & & & & & & \\
     \langle\mathcal{U}_{i}\mid\overline{\mathcal{V}}\rangle & =& 0\ , & &
     \langle\mathcal{U}_{i}\mid\mathcal{V}\rangle & =& 0 \ .
  \end{array}
\end{equation}
Let us now focus on $\langle\mathfrak{D}_{i}\mathcal{U}_{j}\mid\mathcal{V}\rangle = -\langle\ \mathcal{U}_{j}\mid \mathcal{U}_{i}\rangle$, where we have made use of the third constraint. As one can see the r.h.s. is antisymmetric in $i$ and $j$, whereas the l.h.s. is symmetric. This then means that $\langle\mathfrak{D}_{i}\mathcal{U}_{j}\mid\mathcal{V}\rangle = \langle \mathcal{U}_{j}\mid \mathcal{U}_{i}\rangle = 0$. The importance of this last equation is that if we group together $\mathcal{E}_{\Lambda} = (\mathcal{V},\mathcal{U}_{i})$, one can see that $\langle \mathcal{E}_{\Sigma}\mid\overline{\mathcal{E}}_{\Lambda}\rangle$ is a non-degenerate matrix, which allows one to construct an identity operator for the symplectic indices, such that for a given section $\mathcal{A}\in\Gamma\left( E,\mathcal{M}\right)$ we have
\begin{equation}
  \label{eq:SGSymplProj}
  \mathcal{A} \ =\ i\langle\mathcal{A}\mid\overline{\mathcal{V}}\rangle\ \mathcal{V}
              \ -\ i\langle\mathcal{A}\mid\mathcal{V}\rangle\ \overline{\mathcal{V}}
              \ +\ i\langle\mathcal{A}\mid\mathcal{U}_{i}\rangle\mathcal{G}^{i\bar{\imath}}\ \overline{\mathcal{U}}_{\bar{\imath}}
              \ -\ i\langle\mathcal{A}\mid\overline{\mathcal{U}}_{\bar{\imath}}\rangle\mathcal{G}^{i\bar{\imath}}\mathcal{U}_{i} \ .
\end{equation}
Furthermore, the inner product with $\overline{\mathcal{V}}$ and $\overline{\mathcal{U}}_{\bar{\imath}}$ vanishes due to the basic properties.
Let us define the weight $(2,-2)$ object
\begin{equation}
  \label{eq:SGDefC}
  \mathcal{C}_{ijk} \; \equiv\; \langle \mathfrak{D}_{i}\ \mathcal{U}_{j}\mid \mathcal{U}_{k}\rangle \quad \rightarrow\quad
  \mathfrak{D}_{i}\ \mathcal{U}_{j} \; =\; i\mathcal{C}_{ijk}\mathcal{G}^{k\bar{l}}\overline{\mathcal{U}}_{\bar{l}} \ ,
\end{equation}
where the last equation is a consequence of eq.~(\ref{eq:SGSymplProj}). Since the $\mathcal{U}$ are orthogonal, one can see that $\mathcal{C}$ is completely symmetric in its three indices, and one can show that
\begin{equation}
  \label{eq:SGCProp}
  \mathfrak{D}_{\bar{\imath}}\ \mathcal{C}_{jkl} \; =\; 0 \ ,\quad \mathfrak{D}_{[i}\ \mathcal{C}_{j]kl} \; =\; 0\ .
\end{equation}
Let us then introduce the concept of a monodromy matrix $\mathcal{N}$, which can be defined through the relations
\begin{equation}
  \label{eq:SGDefN}
  \mathcal{M}_{\Lambda} \; =\; \mathcal{N}_{\Lambda\Sigma}\ \mathcal{L}^{\Sigma} \ ,\quad h_{\Lambda i} \; =\; \overline{\mathcal{N}}_{\Lambda\Sigma}\ f_{i}^{\Sigma} \ .
\end{equation}
The relations of $\langle\mathcal{U}_{i}\mid\overline{\mathcal{V}}\rangle =0$ then imply that $\mathcal{N}$ is a symmetric matrix, which hence automatically trivialises $\langle\mathcal{U}_{i}\mid\mathcal{U}_{j}\rangle =0$.

Observe that as $\mathrm{Im}\left(\mathcal{N}_{\Lambda\Sigma}\right)\equiv\mathrm{Im}\left(\mathcal{N}\right)_{\Lambda\Sigma}$ appears in the kinetic term of the ($\bar{n}=n_V+1$) vector fields it has to be negative definite, whence also invertible, in order for the kinetic term to be well-defined. One can see that this is implied by the properties of special 
geometry \mycite{Cremmer:1984hj}. As it is invertible, we can use it as a `metric' for raising and lowering $\Lambda$-indices, {\em e.g.\/} $\mathcal{L}^{\Lambda}\equiv\mathrm{Im}\left(\mathcal{N}\right)^{-1|\Lambda\Sigma}\mathcal{L}_{\Sigma}$. Likewise, we shall use $\mathcal{G}_{i\bar{\jmath}}$ to raise and lower K\"ahler-indices. Moreover, from the other basic properties in (\ref{eq:SGProp1}) we find
\begin{equation}
  \label{eq:SGProp1N}
  \mathcal{L}_{\Lambda}\overline{\mathcal{L}}^{\Lambda}\ =\ -\textstyle{1\over 2} \ ,\quad \mathcal{L}_{\Lambda}\ f^{\Lambda}_{i} \ =\ 0 \ ,\quad f_{\Lambda i}\ \bar{f}^{\Lambda}_{\bar{\jmath}} \ =\ -\textstyle{1\over 2}\mathcal{G}_{i\bar{\jmath}} \ .
\end{equation}
An important identity that one can derive, is given by
\begin{equation}
  \label{eq:SGImpId}
  U^{\Lambda\Sigma} \; \equiv\;  f_{i}^{\Lambda}\mathcal{G}^{i\bar{\imath}}\bar{f}^{\Sigma}_{\bar{\imath}}
                    \; =\; -\textstyle{1\over 2}\mathrm{Im}(\mathcal{N})^{-1|\Lambda\Sigma}
                    \  -\  \overline{\mathcal{L}}^{\Lambda}\mathcal{L}^{\Sigma} \; ,
\end{equation}
so that $\overline{U^{\Lambda\Sigma}}=U^{\Sigma\Lambda}$.\vspace{\baselineskip}

Let us construct the $(n_V+1)\times (n_V+1)$-matrices $M=(\mathcal{M}_{\Lambda},\bar{h}_{\Lambda\ \bar{\imath}})$ and $L=(\mathcal{L}^{\Lambda}, \bar{f}^{\Lambda}_{\bar{\imath}})$. With them we can write the defining relations for the monodromy matrix as $M_{\Lambda\Sigma} = \mathcal{N}_{\Lambda\Omega}L^{\Omega}{}_{\Sigma}$, a system which can be easily solved by putting $\mathcal{N}=ML^{-1}$, where $L^{-1}$ is the inverse of $L$. Formally, one finds
\begin{equation}
  \label{eq:SGLinverse}
  L^{-1} \; =\; -2 \left(
                 \begin{array}{c}
                   \mathcal{L}_{\Lambda}\\
                   f^{\bar{\imath}}_{\Lambda}
                 \end{array}
                \right) \ ,
\end{equation}
which is a recursive argument, but useful to derive
\begin{equation}
  \label{eq:SGdifiN}
  \partial_{\bar{\imath}}\overline{\mathcal{N}}_{\Lambda\Sigma} \; =\;
     -4i\ \left(
           \bar{f}_{\Lambda\bar{\imath}}\mathcal{L}_{\Sigma} \ +\ \mathcal{L}_{\Lambda}\bar{f}_{\Sigma\bar{\imath}}
       \right) 
\end{equation}
and
\begin{equation}
  \label{eq:SGdifbiN}
  \partial_{\bar{\imath}}\mathcal{N}_{\Lambda\Sigma} \; =\;
      4\ \overline{\mathcal{C}}_{\bar{\imath}\bar{\jmath}\bar{k}}\ f^{\bar{\jmath}}_{\Lambda}\ f^{\bar{k}}_{\Sigma} \ .
\end{equation}

\subsection{Prepotential: Existence and more formulae}
\label{LauwersdeWitSpgeom}
In explicit constructions of the models it is worthwhile to introduce the explicitly holomorphic section $\Omega = e^{-\mathcal{K}/2}\mathcal{V}$, which allows us to rewrite the system (\ref{eq:SGDefFund}) as
\begin{equation}
  \label{eq:SGDefFund2}
  \Omega \ =\ \left(
                     \begin{array}{c}
                       \mathcal{X}^{\Lambda}\\
                       \mathcal{F}_{\Sigma}
                     \end{array}
                   \right) \quad \rightarrow \quad
  \left\{
     \begin{array}{lcl}
       \langle \Omega \mid\overline{\Omega}\rangle & \equiv&
                  \overline{\mathcal{X}}^{\Lambda}\mathcal{F}_{\Lambda} \ -\ \mathcal{X}^{\Lambda}\overline{\mathcal{F}}_{\Lambda}
                  \; =\; -i\,e^{-\mathcal{K}}\ , \\
       & & \\
       \partial_{\bar{\imath}}\Omega & =& 0 \ ,\\
       & & \\
       \langle\partial_{i}\Omega\mid\Omega\rangle & =& 0 \ .
     \end{array}
  \right.
\end{equation}
If we now assume that $\mathcal{F}_{\Lambda}$ depends on $Z^{i}$ through the $\mathcal{X}$'s, then the last equation above implies that
\begin{equation}
  \partial_{i}\mathcal{X}^{\Lambda}\left[
       2\mathcal{F}_{\Lambda} \ -\ \partial_{\Lambda}\left( \mathcal{X}^{\Sigma}\mathcal{F}_{\Sigma}\right)
  \right] \; =\; 0 \ .
\end{equation}
If $\partial_{i}\mathcal{X}^{\Lambda}$ is invertible as a $n_V\times (n_V+1)$ matrix, then we must conclude that
\begin{equation}
  \label{eq:Prepot}
  \mathcal{F}_{\Lambda} \; =\; \partial_{\Lambda}\mathcal{F}(\mathcal{X}) \ ,
\end{equation}
where $\mathcal{F}$ is a homogeneous function of degree 2, baptised in the literature as the {\em prepotential}.
Should $\partial_{i}\mathcal{X}^{\Lambda}$ not be invertible, then, as shown in \mycite{Craps:1997gp}, one can do a symplectic transformation such that a prepotential exists.

Making use of the prepotential and the definitions (\ref{eq:SGDefN}), we then calculate
\begin{equation}
  \label{eq:PrepotN}
  \mathcal{N}_{\Lambda\Sigma} \; =\; \overline{\mathcal{F}}_{\Lambda\Sigma}
                  \ +\ 2i\,\displaystyle{
                    \mathrm{Im}(\mathcal{F})_{\Lambda\Lambda^{\prime}}\mathcal{X}^{\Lambda^{\prime}}
                     \mathrm{Im}(\mathcal{F})_{\Sigma\Sigma^{\prime}}\mathcal{X}^{\Sigma^{\prime}}
                               \over
                      \mathcal{X}^{\Omega}\mathrm{Im}(\mathcal{F})_{\Omega\Omega^{\prime}}\mathcal{X}^{\Omega^{\prime}}
                                         }\ ,
\end{equation}
which is manifestly symmetric. From the above expression we can obtain the sometimes useful result
\begin{equation}
  \label{eq:PrePotImNinv}
  \mathrm{Im}\left(\mathcal{N}\right)^{-1|\Lambda\Sigma} \; =\; 
        -\mathrm{F}^{-1|\Lambda\Sigma}
        \ -\ 2\mathcal{L}^{\Lambda}\overline{\mathcal{L}}^{\Sigma}
        \ -\ 2\overline{\mathcal{L}}^{\Lambda}\mathcal{L}^{\Sigma} \ ,
\end{equation}
where $\mathrm{F}^{-1}$ is the inverse of $\mathrm{F}_{\Lambda\Sigma}\equiv\mathrm{Im}\left(\mathcal{F}_{\Lambda\Sigma}\right)$. Also, having the explicit form of $\mathcal{N}$ we can derive an explicit representation for $\mathcal{C}$
\begin{equation}
  \label{eq:PrepC}
  \mathcal{C}_{ijk} \; =\; e^{\mathcal{K}}\
                           \partial_{i}\mathcal{X}^{\Lambda}\ \partial_{j}\mathcal{X}^{\Sigma}\
                           \partial_{k}\mathcal{X}^{\Omega}\
                           \mathcal{F}_{\Lambda\Sigma\Omega} \; ,
\end{equation}
so that the prepotential determines all the structures present in Special geometry.

\subsection{Gauging holomorphic Killing vectors in Special geometry}
\label{sec:SGisom}
We are now interested in holomorphic Killing vectors associated to the K\"ahler manifold with metric $\mathcal{G}$. This is relevant \emph{e.g.}~in chapter \ref{4d}, where we have considered gauged vector fields in four-dimensional fSUGRA. Consider the real Killing vector
\begin{equation}
  \label{eq:SGK1}
  \mathtt{K} \; =\; \mathtt{K}^{i}(Z)\ \partial_{i} \; +\; \bar{\mathtt{K}}^{\bar{\imath}}(\overline{Z})\ \partial_{\bar{\imath}}
      \;\; \longrightarrow\;\;  \pounds_{\mathtt{K}}\mathcal{G} \; =\; 0 \ .
\end{equation}
For ease of treatment, it is customary to put all the vectors fields and the graviphoton (which is inside the gravity multiplet) on the same footing. This means that we shall be labelling the Killing vectors by an index like $\Lambda$, which runs from $1$ to $\bar{n}\;(=n_{V}+1)$. One then imposes an additional constrain bringing the number of vectors back to $n_V$. In five-dimensional theory, for example, this is done by eq.~(\ref{eq:RSGconstraint}). Here we shall be using one of the fields to gauge the $\mathbb{R}$-symmetry, and thus we can at most use at most $n_V$ vectors to gauge isometries. In general, these Killing vectors define a non-Abelian algebra, which we take to be
\begin{equation}
  \label{eq:SGK2}
  \left[\mathtt{K}_{\Lambda},\mathtt{K}_{\Sigma}\right] \; =\;  -\mathtt{f}_{\Lambda\Sigma}{}^{\Gamma}\ \mathtt{K}_{\Gamma} \ .
\end{equation}

These isometries need not leave invariant the K\"ahler potential, but only up to a K\"ahler transformation, {\em i.e.\/}
\begin{equation}
  \label{eq:SGK3}
  \pounds_{\Lambda}\mathcal{K} \; \equiv\; \mathtt{K}_{\Lambda}\ \mathcal{K} \; =\;\lambda_{\Lambda}(Z) \; +\; \overline{\lambda_{\Lambda}(Z)} \ , 
\end{equation}
where we employ the notation $\pounds_{\Lambda}=\pounds_{\mathtt{K}_{\Lambda}}$. It is clear that the K\"ahler transformation parameters $\lambda$ have to form a representation under the group that we are gauging, and in fact one sees that
\begin{equation}
  \label{eq:SGK4}
  \pounds_{\Lambda}\lambda_{\Sigma} \, -\, \pounds_{\Sigma}\lambda_{\Lambda}\; =\; -\mathtt{f}_{\Lambda\Sigma}{}^{\Omega}\ \lambda_{\Omega} \ .
\end{equation}
If we also assume that the Killing vectors are compatible with the complex structure $\mathcal{J}$ defined
on the K\"ahler manifold, and therefore also with the K\"ahler form $\mathsf{K}(X,Y)\sim \mathcal{G}(\mathcal{J}X,Y)$, 
we can derive, analogously to eq.~(\ref{eq:KMomMap}) above
\begin{equation}\label{eq:SGK4a}
  \pounds_{\Lambda}\ \mathsf{K} \; =\; d\left( \imath_{\Lambda}K\right) \;\; \longrightarrow\;\; 2\pi\ \imath_{\Lambda}K \; =\; d\mathtt{P}_{\Lambda} \ ,  
\end{equation}
where the object $\mathtt{P}_{\Lambda}$ is called the {\em momentum map associated to }$\mathtt{K}_{\Lambda}$. A closed form for the momentum map can be easily seen to be 
\begin{equation}
  \label{eq:SGK5}
  i\mathtt{P}_{\Lambda} \; =\; \textstyle{1\over 2}\left(
             \mathtt{K}_{\Lambda}^{i}\ \partial_{i}\mathcal{K} \, -\,
             \mathtt{K}_{\Lambda}^{\bar{\imath}}\ \partial_{\bar{\imath}}\mathcal{K}
                              \ -\ \lambda_{\Lambda} \ +\ \overline{\lambda}_{\Lambda}
           \right)
    \; =\; \mathtt{K}_{\Lambda}^{i}\ \partial_{i}\mathcal{K}
                              \ -\ \lambda_{\Lambda} \ , 
\end{equation}
where we made use of eq.~(\ref{eq:SGK3}) and fixed a possible constant to be zero. Using this form and eq.~(\ref{eq:SGK4}), it is straightforward to show that
\begin{equation}
  \label{eq:SGK6}
  \pounds_{\Lambda}\mathtt{P}_{\Sigma} \; =\; -\mathtt{f}_{\Lambda\Sigma}{}^{\Omega}\ \mathtt{P}_{\Omega}\ ,
\end{equation}

The action of the Killing vector on the symplectic section is most easily described on the $(1,0)$-weight section $\Omega$. In fact, by consistency, it must transform as
\begin{equation}
  \label{eq:SGK7}
  \pounds_{\Lambda}\,\Omega \; =\; S_{\Lambda}\ \Omega \, -\, \lambda_{\Lambda}\ \Omega \ ,
\end{equation}
where $S\in\mathfrak{sp}(\bar{n};\mathbb{R})$ and forms a representation of the algebra we are gauging, {\em i.e.\/} $[S_{\Lambda},S_{\Sigma}]=\mathtt{f}_{\Lambda\Sigma}{}^{\Gamma}S_{\Gamma}$. The natural spacetime (not the K\"ahler) connection that acts on this symplectic section is
\begin{equation}
  \label{eq:SGK8}
  \mathtt{D}\Omega \; =\; \left( \nabla +\,\partial Z^{i}\,\partial_{i}\mathcal{K}+\,i\mathtt{g}\,A^{\Lambda}\,\mathtt{P}_{\Lambda}+\mathtt{g}\,A^{\Lambda}\,S_{\Lambda}\right)\,\Omega \ ,
\end{equation}
which is constructed in such a way that $\delta_{\alpha}\mathtt{D}\Omega = \alpha^{\Lambda}\left( S_{\Lambda} -\lambda_{\Lambda}\right)\mathtt{D}\Omega$. From the above equation, it is a small calculation to derive the covariant derivative on objects such as $\mathcal{V}$ or $\overline{\mathcal{V}}$. In fact, one can see that, if dealing with a symplectic $(p,q)$-weight object, one has
\begin{eqnarray}
\label{eq:SGK9}
\delta_{\alpha}\Phi^{(p,q)} &=& \alpha^{\Lambda}\left(S_{\Lambda} \ -\ p\ \lambda_{\Lambda}\ -\ q\ \bar{\lambda}_{\Lambda}\right)\ \Phi^{(p,q)}\ ,\\
&&\nonumber\\
\mathtt{D}\Phi^{(p,q)} & =& \left[\nabla\ +p\ \partial Z^{i}\ \partial_{i}\mathcal{K}\ + q\ \partial\overline{Z}^{\bar{\imath}}\ \partial_{\bar{\imath}}\mathcal{K} \ +i(p-q)\mathtt{g}\ A^{\Lambda}\ \mathtt{P}_{\Lambda} \ +\mathtt{g}\ A^{\Lambda}\ S_{\Lambda} \right]\ \Phi^{(p,q)} \ ,\quad\qquad\\
& & \nonumber \\
\delta_{\alpha}\mathtt{D}\Phi^{(p,q)} & =& \alpha^{\Lambda}\left(\ S_{\Lambda} \ -\ p\ \lambda_{\Lambda}\ -\ q\ \bar{\lambda}_{\Lambda}\right)\ \mathtt{D}\Phi^{(p,q)} \ .
\end{eqnarray}
Having defined the various covariant derivatives, one can go on to derive
\begin{equation}
  \label{eq:SGK10}
  \mathtt{K}^{i}\ \mathcal{U}_{i} \; =\; \left( S_{\mathtt{K}} \ +\ i\mathtt{P}_{\mathtt{K}}\right)\ \mathcal{V}
  \;\longrightarrow\;\mathtt{D}\mathcal{V} \; =\; \mathtt{D}Z^{i}\ \mathcal{U}_{i} \ ,
\end{equation}
which in turn can be used to obtain
\begin{equation}
  \label{eq:SGK11}
  \mathtt{D}\mathcal{U}_{i} \; =\; \mathtt{D}Z^{j}\ \mathfrak{D}_{j}\mathcal{U}_{i} \; +\;\mathtt{D}\overline{Z}^{\bar{\jmath}}\ \mathfrak{D}_{\bar{\jmath}}\mathcal{U}_{i} \quad\mbox{and}\quad \mathtt{D}\mathcal{N}\; =\; \mathtt{D}Z^{i}\ \partial_{i}\mathcal{N}\; +\;\mathtt{D}\overline{Z}^{\bar{\imath}}\ \partial_{\bar{\imath}}\mathcal{N}\ .
\end{equation}
Equation (\ref{eq:SGK10}) allows us to write down the following identities
\begin{equation}
  \label{eq:SGK12}
  \begin{array}{lclclcl}
    0 & =& \langle \mathcal{V}\mid S_{\Lambda}\mathcal{V}\rangle\ , &&\mathtt{P}_{\Lambda} & =& \langle \overline{\mathcal{V}}\mid S_{\Lambda}\mathcal{V}\rangle \ ,\\
      & & & & & & \\
    \mathtt{K}_{\Lambda\bar{\imath}} & =& i\langle \overline{\mathcal{U}}_{\bar{\imath}}\mid S_{\Lambda}\mathcal{V}\rangle\ ,&& 0 & =&   \langle \mathcal{U}_{i}\mid\ S_{\Lambda}\mathcal{V}\rangle \ . 
  \end{array}
\end{equation}

As done in \mycite{Huebscher:2008yz}, we consider only a subset the possible gaugings; we restrict to groups whose embedding into $\mathfrak{sp}(\bar{n};\mathbb{R})$ is given by
\begin{equation}
  \label{eq:SGK15}
  S_{\Lambda} \; =\; \left(
	  \begin{array}{lcl}
      \left[ S_{\Lambda}\right]^{\Sigma}{}_{\Omega} & 0 \\
      0 & -\left[ S_{\Lambda}\right]_{\Sigma}{}^{\Omega}
    \end{array}
   \right) \; =\;  
   \left(
    \begin{array}{lcl}
      \mathtt{f}_{\Lambda\Omega}{}^{\Sigma} & 0 \\
      0 & -\mathtt{f}_{\Lambda\Sigma}{}^{\Omega}
    \end{array}
   \right)\ .
\end{equation}
With this restriction on the gaugeable symmetries, we can then derive the following important identity
\begin{equation}
  \label{eq:SGK16}
  0 \; =\; \mathcal{L}^{\Lambda}\ \mathtt{K}_{\Lambda}^{i} \ .
\end{equation}
Further identities that follow are
\begin{equation}
  \label{eq:SGK17}
   \mathcal{L}^{\Lambda}\ \mathtt{P}_{\Lambda} \ =\ 0 \ ,\quad    \mathcal{L}^{\Lambda}\ \lambda_{\Lambda} \ =\ 0 \ ,\quad \bar{f}^{\Lambda\ i}\ \mathtt{P}_{\Lambda} \ =\ i\ \overline{\mathcal{L}}^{\Lambda}\ \mathtt{K}_{\Lambda}^{i} \ .
\end{equation}

\section{Hyper-K\"ahler and quaternionic-K\"ahler geometry}
\label{app:QuatSpace}
Quaternionic spaces arise naturally in Supergravity in the context of $\N=2$, $d=4$ theories with hyperscalar multiplets \mycite{Bagger:1983tt}. Although these do not make an explicit appearance in this work, they lie at the heart of the Fayet-Iliopoulos terms (which are used to obtain a positive cosmological constant in fSUGRA), and hence will be briefly discussed. Appendix \ref{appsec:qKIsom} discusses the gauging of isometries in quaternionic-K\"ahler spaces, and section \ref{sec:FIinfSUGRA} gives a short description of the r\^ole of the FI terms in fSUGRA.
 
A \textit{quaternionic-K\"ahler} manifold is a real $4m$-dimensional Riemannian manifold $\mathcal{HM}$ endowed with a triplet of complex structures $\mathsf{J}^{x}: T(\mathcal{HM})\rightarrow T(\mathcal{HM})\ (x=1,2,3)$ that satisfy the quaternionic algebra
\begin{equation}\label{eq:QK1}
\mathsf{J}^{x}\,\mathsf{J}^{y} \ =\ -\delta^{xy} \; +\; \varepsilon^{xyz}\ \mathsf{J}^{z}\ ,    
\end{equation}
and with respect to which the metric, denoted by $\mathsf{H}$, is Hermitean
\begin{equation}\label{eq:QK2}
\mathsf{H}(\mathsf{J}^{x} X,\mathsf{J}^{x}Y)\, =\, \mathsf{H}(X,Y)\ ,\quad \forall\, X,Y \in  T(\mathcal{HM})\ .
\end{equation}
This implies the existence of a triplet of 2-forms $\mathsf{K}^{x}(X,Y)\equiv \mathsf{H}(X,\mathsf{J}^{x}Y)$ globally known as the $\mathfrak{su}(2)$-valued \textit{hyper-K\"ahler 2-forms}. Observe that the foregoing definition on a real coordinate base means $\mathsf{K}^{x}_{uv}\ =\ \mathsf{H}_{uw}\,(\mathsf{J}^x)^{w}{}_{v}\ \equiv\ (\mathsf{J}^{x})_{uv}$. Most of the time we shall use an $\so{3}$-valued notation in order not to have the $x$-indices floating around; this implies writing {\em e.g.\/} $\mathsf{K}\ =\ \mathsf{K}^{x}\ T_{x}$ where the generators $T_{x}$ satisfy $\left[ T_{x},T_{y}\right]\ =\ \varepsilon_{xyz}T_{z}$.
\par
The structure of a quaternionic-K\"ahler manifold requires an $\SU{2}$ bundle to be constructed over $\mathcal{HM}$ with connection 1-form $\mathsf{A}^{x}$, with respect to which the hyper-K\"ahler 2-form is covariantly closed, {\em i.e.\/}
\begin{equation}\label{eq:QK3}
0\; =\; \mathsf{D}_{X}\mathsf{K}\; \equiv\; \nabla_{X}\mathsf{K}+\ \left[ \mathsf{A}_{X},\mathsf{K}\right]\; =\; 0\ .  
\end{equation}
Then, depending on whether the curvature of this bundle 
\begin{equation}\label{eq:QK4}
\mathsf{F}\ \equiv\ d\mathsf{A}\ +\ \mathsf{A}\wedge\mathsf{A} \;\;\;\longrightarrow\;\;\;
\mathsf{F}^{x}\; \equiv\; d\mathsf{A}^{x} \ +\ {\textstyle\frac{1}{2}}\varepsilon^{xyz}\ \mathsf{A}^{y} \wedge \mathsf{A}^{z}\ , 
\end{equation}
is zero or proportional to the hyper-K\"ahler 2-form\footnote{
  Note that in constructions of SUGRA it is the additional constraint given by having SUSY, not the actual geometry, which fixes the value of $\varkappa$
  to $\varkappa =-2$.
}, \emph{i.e.}
\begin{equation}\label{eq:QK5}
\mathsf{F}\; = \varkappa\  \mathsf{K}\ ,\quad \varkappa\in \mathbb{R}{/\{0\}}\ ,
\end{equation}
the manifold is a \textit{hyper-K\"ahler} manifold or a \textit{quaternionic-K\"ahler} manifold, respectively.

The $\SU{2}$-connection acts on objects with vectorial $\SU{2}$indices, such as chiral spinors, as follows\footnote{The convention for raising and lowering of $\SU{2}$ indices is given by 
\begin{equation}
\chi^I=\chi_J \epsilon^{JI}\ , \quad \psi_I=\epsilon_{IJ}\psi^J\ .
\end{equation}
}
\begin{equation}
\label{eq:QK6}
  \begin{array}{rclcrcl}
    \mathsf{D} \xi_{I} & \equiv  & d\xi_{I} +\mathsf{A}_{I}{}^{J}\xi_{J}\ ,
    &&
    \mathsf{F}_{I}{}^{J} & =& d\mathsf{A}_{I}{}^{J}\ +\ \mathsf{A}_{I}{}^{K}\wedge\mathsf{A}_{K}{}^{J} 
    \ ,\\
    & & & & & & \\
    \mathsf{D} \chi^{I} & \equiv  & d\chi^{I} +\mathsf{B}^{I}{}_{J}\chi^{J}\ ,
    &&
    \mathsf{G}^{I}{}_{J} & =& d\mathsf{B}^{I}{}_{J}\ +\ \mathsf{B}^{I}{}_{K}\wedge\mathsf{B}^{K}{}_{J} 
    \ .\\
\end{array}
\end{equation}
Consistency with the raising and lowering of vector $\SU{2}$ then indices implies that 
\begin{equation}
  \label{eq:QK7}
  \mathsf{B}^{I}{}_{J} \; =\; -\mathsf{A}^{I}{}_{J}\; \equiv\;-\varepsilon^{IK}\ \mathsf{A}_{K}{}^{L}\ \varepsilon_{LJ}\ ,
\end{equation}
whereas compatibility with the raising of indices due to complex conjugation implies
\begin{equation}
 \label{eq:QK8}
    \mathsf{B}^{I}{}_{J} = (\mathsf{A}_{I}{}^{J})^{*}\, . 
\end{equation}
These two things together thus means that $\mathsf{A}_{I}{}^{J}$ is an anti-Hermitean matrix, whence we expand
\begin{equation}
 \label{eq:QK9}
    \mathsf{A}_{I}{}^{J}\; =\; {\textstyle\frac{i}{2}}\ \mathsf{A}^{x}\ (\sigma^{x})_{I}{}^{J} \quad\mbox{and}\quad
    \mathsf{B}^{I}{}_{J} \; =\; -\textstyle{i\over 2}\ \mathsf{A}^{x}\ (\sigma^{x})^{I}{}_{J} \ ,   
\end{equation}
where the indices of the $\sigma$-matrices are raised/ lowered with $\epsilon$.\vspace{\baselineskip}

At this point, there remains a question about the normalisation of the Pauli matrices, which is readily fixed by imposing that 
\begin{equation}\label{eq:QK10}
  \mathsf{F}_{I}{}^{J} \; =\; \textstyle{i\over 2}\ \mathsf{F}^{x}\ (\sigma^{x})_{I}{}^{J} \ ,
\end{equation}
which implies that
\begin{equation}
\label{eq:QK11}
(\sigma^{x}\sigma^{y})_{I}{}^{J} \; =\; \delta^{xy}\ \delta_{I}{}^{J} \, -\, i\varepsilon^{xyz}\ (\sigma^{z})_{I}{}^{J}\ .
\end{equation}
Let us then write down the explicit form of the $\sigma$-matrices that fullfill the above defining relations
\begin{equation}
  \label{eq:QK12}
  \begin{array}{rcl}
    (\sigma^{x})_{I}{}^{J} & :& 
           \left(\begin{array}{cc} 0 & 1\\ 1 & 0\end{array}\right) \; ,\;
           \left(\begin{array}{cc} 0 & i\\ -i & 0\end{array}\right) \; ,\;
           \left(\begin{array}{cc} 1 & 0\\ 0 & -1\end{array}\right) \ ,
           \\
     & & \\
    (\sigma^{x})^{I}{}_{J}   & :& 
           \left(\begin{array}{cc} 0 & 1\\ 1 & 0\end{array}\right) \; ,\;
           \left(\begin{array}{cc} 0 & -i\\ i & 0\end{array}\right) \; ,\;
           \left(\begin{array}{cc} 1 & 0\\ 0 & -1\end{array}\right) \ ,
            \\
    & & \\
    (\sigma^{x})_{IJ}     & :&  
           \left(\begin{array}{cc} 1 & 0\\ 0 & -1\end{array}\right) \; ,\;
           \left(\begin{array}{cc} i & 0\\ 0 & i\end{array}\right) \; ,\;
           \left(\begin{array}{cc} 0 & -1\\ -1 & 0\end{array}\right) \ ,
           \\
    & & \\
    (\sigma^x)^{IJ}        & :&
           \left(\begin{array}{cc} -1 & 0\\ 0 & 1\end{array}\right) \; ,\;
           \left(\begin{array}{cc} i & 0\\ 0 & i\end{array}\right) \; ,\;
           \left(\begin{array}{cc} 0 & 1\\ 1 & 0\end{array}\right) \ ,
  \end{array}
\end{equation}
where the $x=1,2,3$ matrices are ordered from left to right. Observe that these matrices imply the following orthogonality and completeness rules
\begin{equation}
  \label{eq:QK13}
  \delta^{xy} \; =\; -\textstyle{1\over 2}\ \sigma^{x}_{IJ}\sigma^{yIJ}\ ,\quad
  \delta^{I}{}_{(K}\delta^{J}{}_{L)} \; =\; -\textstyle{1\over 2}\ \sigma^{xIJ}\ \sigma^{x}_{KL} \ .
\end{equation}

It is convenient to use a Vielbein on $\mathcal{HM}$ having as flat indices a pair $(\alpha, I)$ consisting of an $\SU{2}$ index $I=1,2$ and an $\Sp{m}$ index $\alpha = 1,\ldots,2m$
\begin{equation}\label{eq:QK14}
\mathsf{U}^{\alpha I} \; =\; \mathsf{U}^{\alpha I}{}_{u}\ dq^{u}\ ,
\end{equation}
where $q^{u}$ ($u\ =\ 1,\ldots ,4m$) are real coordinates on $\mathcal{HM}$. We shall refer to this Vielbein $\mathsf{U}^{\alpha I}$ as the Quadbein. The Quadbein is related to the metric $\mathsf{H}_{uv}$ by
\begin{equation}
\label{eq:QK15}
\mathsf{H}_{uv} \; =\; \mathsf{U}^{\alpha I}{}_{u}\ \mathsf{U}^{\beta J}{}_{v}\,\varepsilon_{IJ}\mathbb{C}_{\alpha\beta}\ ,
\end{equation}
where $\mathbb{C}$ is a real antisymmetric $2m\times 2m$ matrix and it is in fact the metric for the $\Sp{m}$ group. Furthermore one requires that, in concordance with our rules of raising and lowering indices,
\begin{equation}
 \label{eq:QK16}
 \mathsf{U}_{\alpha I} \; \equiv\; \varepsilon_{IJ}\mathbb{C}_{\alpha\beta}\ \mathsf{U}^{\beta J}
                      \; =\; \left(\mathsf{U}^{\alpha I}\right)^{\star} \;.
\end{equation}
The Quadbein thus satisfies all the usual relations that a Vielbein satisfies.

In thise sense, the Quadbein satisfies a \textit{Vielbein postulate}, {\em i.e.\/} they are covariantly constant with respect to the standard Levi-Civit\`a connection $\Gamma_{uv}{}^{w}$, the $\SU{2}$-connection and the $\Sp{m}$-connection $\Delta_{u}{}^{\alpha}{}_{\beta}$:
\begin{equation}
\label{eq:QK17}
0 \, =\, \mathsf{D}_{u}\ \mathsf{U}^{\alpha I}{}_{v}\; =\; \nabla_{u}\mathsf{U}^{\alpha I}{}_{v}\ +\ \mathsf{B}_{u}{}^{I}{}_{J}\ \mathsf{U}^{\alpha J}{}_{v}\ +\ \Delta_{u}{}^{\alpha}{}_{\beta}\ \mathsf{U}^{\beta I}{}_{v}\mathbb{C}_{\beta\gamma}=0\ .
\end{equation}
This postulate relates the three connections and the respective curvatures, leading to the statement that the holonomy of a quaternionic-K\"ahler manifold is contained in $\Sp{1}\cdot \Sp{m}$, {\em i.e.\/}
\begin{equation}
\label{eq:QK18}
R_{ts}{}^{uv}\ \mathsf{U}^{\alpha I}{}_{u}\ \mathsf{U}^{\beta J}{}_{v} \; =\;
        -\mathsf{G}_{ts}^{IJ}\ \mathbb{C}^{\alpha\beta} 
        \ -\ \overline{\mathsf{R}}_{ts}^{\ \alpha\beta}\ \varepsilon^{IJ} 
    \; =\;  \mathsf{F}_{ts}^{IJ}\ \mathbb{C}^{\alpha\beta} 
        \ -\ \overline{\mathsf{R}}_{ts}^{\ \alpha\beta}\ \varepsilon^{IJ} \ ,
\end{equation}
where we have defined the $\Sp{m}$-curvature $\overline{\mathsf{R}}$ as
\begin{equation}\label{eq:QK19}
  \overline{\mathsf{R}}^{\ \alpha}{}_{\beta} \; \equiv\; 
    d\Delta^{\alpha}{}_{\beta} \ +\ \Delta^{\alpha}{}_{\gamma}\wedge\Delta^{\gamma}{}_{\beta}\ .
\end{equation}
It is clear that on $\mathbb{R}^{4m}$ we can define a quaternionic structure which is covariantly constant. In this case, the fact that the Quadbein is covariantly constant means that on a quaternionic-K\"ahler space we can induce a covariantly constant quaternionic structure by inducing the one from the tangent space. In fact, this means that we have
\begin{equation}
  \label{eq:QK20}
  \mathsf{K}^{x}_{uv} \; =\; -i\ \sigma^{x}_{IJ}\ \mathbb{C}_{\alpha\beta}\ 
                              \mathsf{U}_{u}^{\alpha I}\ \mathsf{U}^{\beta J}_{v}\ .
\end{equation}
Using the above expression for the quaternionic-K\"ahler forms, we can then obtain the identity
\begin{equation}
  \label{eq:QK21}
  \mathsf{U}^{\alpha I}_{u}\ \mathbb{C}_{\alpha\beta}\ \mathsf{U}^{\beta J}_{v}
   \; =\; \textstyle{1\over 2}\ \mathsf{H}_{uv}\ \epsilon^{IJ} 
   \; -\; \textstyle{i\over 2}\ \mathsf{K}^{x}_{uv}\ \sigma^{x\ IJ} \ .
\end{equation}
As a result of this and eq.~(\ref{eq:QK5}), we can derive the identity
\begin{equation}
  \label{eq:QK22}
  \mathsf{F}_{uv\ I}{}^{J} \; =\; \varkappa\ \mathsf{U}_{\alpha I\ [u}\ \mathsf{U}^{\alpha J}_{v]} \; ,
\end{equation}
where once again we would like to point out that Supergravity fixes $\varkappa =-2$.
\subsection{Gauging isometries in quaternionic-K\"ahler spaces}
\label{appsec:qKIsom}
As in previous sections, we now discuss the possible isometries of quaternionic-K\"ahler spaces. W.r.t.~the real coordinates $q^{u}$ on $\mathcal{HM}$, the Killing vectors are given by
\begin{equation}
  \label{eq:QKI1}
  \mathtt{K}_{\Lambda} \; =\; \mathtt{K}_{\Lambda}^{u}\partial_{u} \ ,\quad \text{such that}\quad \pounds_{\Lambda}\ \mathsf{H} \; =\; 0 \ .
\end{equation}
We shall consider fields on $\mathcal{HM}$ that transform in the adjoint representation of $\SO{3}$. An example of such
objects is $\mathsf{K}$. Calling such a generic field $\Psi=\Psi^{x}\ T_{x}$, it transforms under $\so{3}$ as 
\begin{equation}
  \label{eq:QKI2}
  \delta_{\lambda}\Psi \; =\; -\left[\lambda,\Psi \right] \ ,
\end{equation}
where $\lambda$ is an $\so{3}$-valued transformation parameter. Of course, a covariant derivative is easily introduced by putting
\begin{equation}
  \label{eq:QKI3}
  \mathsf{D}_{X}\Psi \; =\; \nabla_{X}\Psi \; +\; \left[\mathsf{A}_{X},\Psi\right] \quad \mbox{so long as}\quad\delta_{\lambda}\mathsf{A} \; =\; \mathsf{D}\lambda \ .
\end{equation}
We define an $\SO{3}$-covariant Lie derivative, by  postulating
\begin{equation}
  \label{eq:QKI4}
  \mathbb{L}_{\Lambda}\Psi \; =\; \pounds_{\Lambda}\Psi \; +\; \left[\mathsf{W}_{\Lambda},\Psi\right]\ ,\quad\mbox{which must satisfy}\quad \left\{
    \begin{array}{rcl}
      \delta_{\lambda}\mathbb{L}_{\Lambda}\Psi & =& -\left[ \lambda\ ,\ \mathbb{L}_{\Lambda}\Psi \right] \ ,\\
      \left[ \mathbb{L}_{\Lambda},\mathbb{L}_{\Sigma}\right] & =& -\mathtt{f}_{\Lambda\Sigma}{}^{\Omega}\mathbb{L}_{\Omega} \ .
    \end{array}
  \right. 
\end{equation}
The last rule is nothing but the usual commutation relations for Lie derivatives, but where we have used eq.~(\ref{eq:SGK2}) to define the commutation relations for the Killing vectors. The first constraint implies $\delta_{\lambda}\mathsf{W}_{\Lambda} = \mathbb{L}_{\Lambda}\lambda$, whereas the second implies 
\begin{equation}
  \label{eq:QKI5}
  \pounds_{\Lambda}\mathsf{W}_{\Sigma} \ -\ \pounds_{\Sigma}\mathsf{W}_{\Lambda} \ +\ \left[ \mathsf{W}_{\Lambda},\mathsf{W}_{\Sigma}\right] \; =\; -\mathtt{f}_{\Lambda\Sigma}{}^{\Omega}\ \mathsf{W}_{\Omega} \ .
\end{equation}
We go on to introduce the notion of momentum map $\mathsf{P}_{\Lambda}$ by defining 
\begin{equation}
  \label{eq:QKI6}
  \mathsf{W}_{\Lambda} \; =\; \imath_{\Lambda} \mathsf{A} \; -\; \mathsf{P}_{\Lambda} \ .
\end{equation}
Substituting the above definition into eq.~(\ref{eq:QKI5}), one can see that the momentum map has to satisfy
\begin{equation}
  \label{eq:QKI7}
  \mathsf{D}_{\Lambda}\mathsf{P}_{\Sigma} \ -\ \mathsf{D}_{\Sigma}\mathsf{P}_{\Lambda} -\left[ \mathsf{P}_{\Lambda},\mathsf{P}_{\Sigma}\right] +\varkappa\ \imath_{\Lambda}\imath_{\Sigma}\mathsf{K} \; =\; -\mathtt{f}_{\Lambda\Sigma}{}^{\Omega}\ \mathsf{P}_{\Omega}\ ,
\end{equation}
where we have defined $\mathsf{D}_{\Lambda}=\mathtt{K}_{\Lambda}^{u}\mathsf{D}_{u}$.

So far we have discussed the Killing vectors and their transformations, and we shall now consider their compatibility with the complex structures. This means imposing
\begin{equation}
  \label{eq:QKI8}
  \mathbb{L}_{\Lambda}\mathsf{K} \ =\ 0 \quad\longrightarrow\quad \mathsf{D}\left(\imath_{\Lambda}\mathsf{K}\right) \; =\; \left[ \mathsf{P}_{\Lambda},\mathsf{K}\right] \ .
\end{equation}
The integrability condition for the above equation can be massaged to give
\begin{equation}
  \label{eq:QKI9}
  \mathsf{DP}_{\Lambda} \; =\; -\varkappa\ \imath_{\Lambda}\mathsf{K} \ ,
\end{equation}
which, in view of the similarity with the result in eq.~(\ref{eq:SGK4a}), justifies the use of the name of tri-holomorphic map for $\mathsf{P}_{\Lambda}$. The above definition implies 
\begin{equation}
  \label{eq:QKI10}
  \left[\mathsf{P}_{\Lambda},\mathsf{P}_{\Sigma}\right]\; +\; \varkappa\ \imath_{\Lambda}\imath_{\Sigma}\mathsf{K} 
  \; =\;  \mathtt{f}_{\Lambda\Sigma}{}^{\Omega}\ \mathsf{P}_{\Omega} \ ,
\end{equation}
where eq.~(\ref{eq:QKI7}) has been used. Another implication is that the tri-holomorphic map is an invariant field, \emph{i.e.}~that its covariant Lie derivative is zero
\begin{equation}
  \label{eq:QKI11}
  0 \; =\; \mathbb{L}_{\Lambda}\mathsf{P}_{\Sigma}\; =\; \pounds_{\Lambda}\mathsf{P}_{\Sigma}+\left[ \mathsf{W}_{\Lambda},\mathsf{P}_{\Sigma}\right]+\mathtt{f}_{\Lambda\Sigma}{}^{\Omega}\mathsf{P}_{\Omega} \ ,
\end{equation}
where the derivative includes $\SO{3}$ and $\mathrm{G}$ terms.
\subsection{A small discussion of the FI terms}
\label{sec:FIinfSUGRA}
This subsection discusses the relevance of hyperscalar multiplets in fakeSupergravity, even for theories that do not  explicitly contain them. This is because the Wick rotation of the FI term lies behind the positivity of the cosmological constant (minus sign in the action), giving rise to fSUGRA.

Consider the case in which there are no hyperscalars. Eq.~(\ref{eq:QKI10}) can be written in components as
\begin{equation}
  \label{eq:FIcond}
  \varepsilon_{xyz}\ \mathsf{P}_{\Lambda}^{x}\ \mathsf{P}_{\Sigma}^{y}\; =\; {f_{\Lambda\Sigma}}^{\Omega}\ \mathsf{P}_{\Omega}^{z} \; .
\end{equation}
This equation allows for two different solutions, namely $U(1)$ and $SU(2)$.
\begin{itemize}
\item If we take the gauge group to be Abelian, {\em i.e.\/} $f_{\Lambda\Sigma}{}^{\Omega}=0$, the $\mathsf{P}_{\Lambda}^{x}$ will be given by $\mathfrak{su}(2)$-valued tri-holomorphic momentum maps that commute. Thus without loss of generality we can take them to be $\mathsf{P}_{\Lambda}=\mathsf{P}_{\Lambda}^x\,T_x=\xi_{\Lambda}\,T_{3}$; this expression for the tri-holomorphic momentum maps is called the $U(1)$ FI term.

\item If the gauge group is chosen to be $SU(2)$, then $f_{\Lambda\Sigma}{}^{\Omega}=\varepsilon_{\Lambda\Sigma\Omega}$.
In this case a solution to eq.~(\ref{eq:FIcond}) is given by $\mathsf{P}_{\Lambda}^{x}=\delta_{\Lambda}^{x}$ and it is called the $SU(2)$ FI term. We shall ignore the $SU(2)$ FI term in the main text, as it induces non-Abelian terms that are difficult to work with.
\end{itemize}
The $U(1)$ FI term is paramount in constructing fSUGRA. Consider \emph{e.g.}~minimal $N=2$ $d=4$ SUGRA; this theory has $\bar{n}=1$, $n_v=0$ and it is defined by a prepotential that reads $\mathcal{F}=-\textstyle{i\over 4}\mathcal{X}^{2}$. Including the $U(1)$ FI term into the mix, we find a supersymmetric action that is\footnote{The coupling constant $g$ that usually appears in supersymmetric actions, see {\em e.g.\/} eqs.~(\ref{eq:VectAct}), (\ref{eq:Potential}) in section \ref{sec:DSVector}, has been absorved into the FI term and also into the structure constants, even if this is not visible here. This is desirable as this absorption allows for different coupling constants (hence multiple gauge groups), a fact which is not obvious when having only $g$.}
\begin{equation}
  \label{eq:2FI}
  S\; =\; \int_{4}\sqrt{g}\ \left[ R\ -\ F^{2}\ +\ \textstyle{3\over 2}\ \xi_{0}^{2}\ \right]\ ,
\end{equation}
which is the action governing Einstein-Maxwell-anti De Sitter. In Supergravity lingo, this theory is known as {\em minimal gauged} $N=2$ $d=4$ SUGRA.

In order for the above action to describe an Einstein-Maxwell-De Sitter theory, the cosmological constant has to change sign, which can be done by Wick rotating the FI term $\xi_{0}\rightarrow iC_{0}$; this leads to
\begin{equation}
  \label{eq:1FI}
  S\; =\; \int_{4}\sqrt{g}\ \left[ R\ -\ F^{2}\ -\ \textstyle{3\over 2}\ C_{0}^{2}\ \right] \ .
\end{equation}
If we reinterpret this Wick rotation from a group theory perspective, we thus have
\begin{equation}
\mathsf{P}_{0}=\xi_{0}\,T_{3}\rightarrow iC_{0}\,T_{3} = C_{0}\,\tilde{T}_{3}\ ,
\end{equation}
where in the last step Weyl's unitarity trick has been used to introduce a non-compact generator $\tilde{T}_{3}$. One can then see that after a Wick rotation we are no longer gauging a $U(1)$ group, but rather $\mathbb{R}$.

\section[Real special geometry]{Real Special K\"ahler geometry}
\label{sec:RealSG}
This appendix contains some useful information on \emph{Real Special K\"ahler geometry} (usually just referred to as Real Special geometry, or even Very Special geometry, following the original article \mycite{deWit:1992cr}). This becomes relevant when considering five-dimensional SUGRA with gauge vector fields, and thus applicable to chapter \ref{5dgauged}.

We consider a theory containing $n$ vector multiplets. As commented briefly above, the geometry of the $n$ physical scalars $\phi^{x}$ ($x=1,\ldots ,n$) in these multiplets is fully determined by a constant real symmetric tensor $C_{IJK}$ ($I,J,K=0,1,\ldots,\bar{n}\equiv n+1$). The scalars thus appear through $\bar{n}$ functions $h^{I}(\phi)$ constrained to satisfy
\begin{equation}
\label{eq:RSGconstraint}
C_{IJK}h^{I}h^{J}h^{K}\; =\; 1\ .  
\end{equation}
One defines 
\begin{equation}
h_{I}\, \equiv\, C_{IJK} h^{J}h^{K}\quad \rightarrow \quad h_{I}h^{I}\, =\, 1\ ,
\end{equation}
and a metric $a_{IJ}$ that can be use to raise and lower the $SO(\bar{n})$ index
\begin{equation}
h_{I}\, \equiv\,  a_{IJ} h^{J}\ , \quad h^{I}\, \equiv\, a^{IJ} h_{J}\ .  
\end{equation}
The definition of $h_{I}$ allows one to find 
\begin{equation}
a_{IJ}\; =\; -2C_{IJK}h^{K}\ +\ 3h_{I}h_{J}\ .  
\end{equation}

Next, one defines
\begin{equation}
h^{I}_{x}\, \equiv\, -\sqrt{3}\ h^{I}{}_{,x}\, \equiv\,  -\sqrt{3} \ \frac{\partial h^{I}}{\partial\phi^{x}}\ ,  
\end{equation}
along with 
\begin{equation}
h_{Ix}\, \equiv\, a_{IJ} h^{J}_{x}\, =\, +\sqrt{3}h_{I, x}\ ,
\end{equation}
which satisfy
\begin{equation}
h_{I}h^{I}_{x} = 0\ , \quad h^{I}h_{Ix} = 0\ ,   
\end{equation}
because of eq.~(\ref{eq:RSGconstraint}).The $h^{I}$ enjoy the following properties of closure and orthogonality
\begin{equation}
\left(
\begin{array}{c} 
h^{I} \\ 
h^{I}_{x} 
\end{array}
\right)
\left(
\begin{array}{cc} 
h_{I} &   h_{I}^{y} \\  
\end{array}
\right)
= 
\left(
\begin{array}{cc}
1 & 0              \\
0 & \delta_{x}^{y} \\
\end{array}
\right)
\ ,
\qquad
\left(
\begin{array}{cc} 
h_{I} &  h_{I}^{x} \\
\end{array}
\right)
\left(
\begin{array}{c}
h^{J}     \\
h^{J}_{x} \\
\end{array}
\right)
=
\delta_{I}^{J}\ .           
\end{equation}
Therefore any object with $SO(\bar n)$ index can be decomposed as
\begin{equation}
\label{eq:RSGdecomp1}
A^{I} = \left(h_{J}A^{J}\right)h^{I} + \left(h^{x}_{J}A^{J}\right)h^{I}_{x}\ .
\end{equation}

The metric on the scalar manifold, $g_{xy}(\phi)$, is the pullback of $a_{IJ}$
\begin{equation}
g_{xy}=a_{IJ}h^{I}_{x}h^{J}_{y}=-2 C_{IJK}h^{I}_{x}h^{J}_{y}h^{K}\ ,
\end{equation}
and can be used to raise/ lower $\{x,y\}$ indices. Other useful expressions are
\begin{equation}
\begin{array}{rcl}
a_{IJ} & = & h_{I}h_{J}+h^{x}_{I}h_{Jx}\ ,\\
& & \\
C_{IJK}h^{K} & = &  h_{I}h_{J}-{\textstyle\frac{1}{2}}h^{x}_{I}h_{Jx}\ ,
\end{array}
\qquad
\begin{array}{rcl}
h_{I}h_{J} & = & {\textstyle\frac{1}{3}a_{IJ}} +{\textstyle\frac{2}{3}}C_{IJK}h^{K}\ ,\\
& &  \\
h^{x}_{I}h_{Jx} & = & {\textstyle\frac{2}{3}}a_{IJ} -{\textstyle\frac{2}{3}}C_{IJK}h^{K}\, .
\end{array}
\end{equation}
We also introduce the Levi-Civit\`a covariant derivative associated to such a metric $g_{xy}$
\begin{equation}
h_{Ix;y}\equiv h_{Ix,y} -\Gamma_{xy}{}^{z}h_{Iz}\ .  
\end{equation}
One can show that 
\begin{eqnarray}
h_{Ix;y} & = & {\textstyle\frac{1}{\sqrt{3}}}(h_{I}g_{xy} +T_{xyz}h^{z}_{I})\ ,\\  
& & \nonumber \\
h^{I}_{x;y} & = & -{\textstyle\frac{1}{\sqrt{3}}}(h^{I}g_{xy} +T_{xyz}h^{Iz})\ ,\\  
& & \nonumber \\
\Gamma_{xy}{}^{z} & = & h^{Iz}h_{Ix,y} -{\textstyle\frac{1}{\sqrt{3}}}T_{xy}{}^{w} =  8h_{I}^{z}h^{I}_{x,y} +{\textstyle\frac{1}{\sqrt{3}}}T_{xy}{}^{w}\ ,
\end{eqnarray}
for 
\begin{equation}
T_{xyz}  =  \sqrt{3} h_{Ix;y}h^{I}_{z} = -\sqrt{3} h_{Ix}h^{I}_{y;z}\ .
\end{equation}

\cleardoublepage

\renewcommand{\leftmark}{\MakeUppercase{\chaptername\ \thechapter. Geometrical data for null case solutions}}
\chapter{Geometrical data for \emph{null case} solutions}
\label{someusefulgeometry}
Here we give some explicit geometric information which is of relevance in the classification of \emph{null case} solutions. Appendix \ref{sec:NullCurv} gives the spin connection and curvatures for the four-dimensional case, and hence applied in section \ref{sec:Null}. Appendix \ref{sec:Spin} contains the five-dimensional information, which is used in chapter \ref{5dminimal}. A short scholium on the Kundt wave metric is given in appendix \ref{appsec:Kundt}, since this background appears repeteadly as solution to the null class.

\section{\texorpdfstring{Spin connection and curvatures in $d=4$ fSUGRA}{Spin connection and curvatures in d=4 fSUGRA}}
\label{sec:NullCurv}
Let us set-up a null-Vierbein by
\begin{equation}
  ds^{2}_{null} \;=\; e^{+}\otimes e^{-} \; +\; e^{-}\otimes e^{+}\; -\; e^{\bullet}\otimes e^{\bar{\bullet}} \; -\; e^{\bar{\bullet}}\otimes e^{\bullet} \ ,
\end{equation}
and choose\footnote{We define the directional derivatives $\theta_{a}$ to be the duals of the frame 1-forms $E^{a}$, {\em i.e.\/} normalised such that $E^{a}(\theta_{b})= \delta^{a}{}_{b}$.  We reserve the notation $\partial_{x}$ for the directional derivative on the base-space, namely $\partial_{x}\equiv e_{x}{}^{m}\partial_{m}$.}
\begin{equation}
  \label{eq:NCtetrad}
  \begin{array}{lclclclclcl}
    e^{+} & =& L & =& du\ , &\qquad&\theta_{+} & =& N^{\flat} & =& \partial_{u} \ -\ H\partial_{v}\ ,\\
    e^{-} & =& N & =& dv +Hdu+ \varpi dz +\overline{\varpi}d\bar{z}\ ,&\qquad&\theta_{-} & =& L^{\flat}&=&\partial_{v}\ ,\\
   e^{\bullet} & =& M & =& e^{U}dz\ , & \qquad&\theta_{\bullet} & =& -\overline{M}^{\flat}& =& e^{-U}\left[ \partial_{z}-\varpi\partial_{v}\right]\ ,\\
    e^{\bar{\bullet}} & =& \overline{M} & =&  e^{U}d\bar{z} \ , &\qquad&\theta_{\bar{\bullet}} & =& -M^{\flat} & =& e^{-U}\left[ \partial_{\bar{z}}-\overline{\varpi}\partial_{v}\right] \ , 
  \end{array}
\end{equation}
where, conforming to the results of eq.~(\ref{eq:9j}), only $H=H(u,v,z,\bar{z})$ and $U$ and the $\varpi$s depend on $u$, $z$ and $\bar{z}$.

The non-vanishing components of the spin connection can be seen to be 
\begin{eqnarray}
  \label{eq:NCspincon}
  \omega_{+-} & =& -\theta_{-}H\ e^{+} \; , \\
  \omega_{+\bullet} & =& \left( e^{-U}\theta_{+}\varpi\; -\; \theta_{\bullet}H\right)\ e^{+}
                   \; -\; \left[\theta_{+}U 
                           \; +\; \textstyle{1\over 2} e^{-2U}\left(
                                \partial_{z}\overline{\varpi}-\partial_{\bar{z}}\varpi
                        \right)\right]\ e^{\bar{\bullet}} \ , \\
   \omega_{+\bar{\bullet}} & =& \left( e^{-U}\theta_{+}\overline{\varpi}\; -\; \theta_{\bar{\bullet}}H\right)\ e^{+}
                   \; -\; \left[\theta_{+}U \; -\; \textstyle{1\over 2} e^{-2U}\left(
                                \partial_{z}\overline{\varpi} -\partial_{\bar{z}}\varpi
                        \right)\right]\ e^{\bullet} \ , \\
   \omega_{\bullet\bar{\bullet}} & =& \textstyle{1\over 2}e^{-2U}
                        \left( \partial_{z}\overline{\varpi} -\partial_{\bar{z}}\varpi
                        \right)\ e^{+}\; -\; e^{\bullet}\theta_{\bullet}U \; +\; e^{\bar{\bullet}}\ \theta_{\bar{\bullet}}U \ .
\end{eqnarray}
A further calculation leads to the Ricci tensor, whose non-vanishing coefficients are
\begin{eqnarray}
  \label{eq:NCR+-}
  R_{+-} & =& -\theta_{-}^{2}H\ ,\\
  \label{eq:NCRzbz}
  R_{\bullet\bar{\bullet}} & =& 2e^{-2U}\ \partial_{z}\partial_{\bar{z}}U\ ,\\
  \label{eq:NCR+z}
  R_{+\bullet}  & =& e^{-U}\theta_{+}\partial_{z}U \; -\; \theta_{\bullet}\theta_{-}H\; +\; \textstyle{1\over 2}\theta_{\bullet}\left( e^{-2U}\left[\partial_{z}\overline{\varpi} -\partial_{\bar{z}}\varpi
                        \right]\right) \ , \\
  \label{eq:NCR+bz}
  R_{+\bar{\bullet}} & =& \overline{R_{+\bullet}} \ ,\\
  \label{eq:NCR++}
  R_{++} & =& 2e^{-U}\theta_{+}^{2}e^{U}\; +\; 2\theta_{-}H\, \theta_{+}U \; +\; \textstyle{1\over 2}e^{-4U}
                \left( \partial_{z}\overline{\varpi} -\partial_{\bar{z}}\varpi\right)^{2}\nonumber\\
        &  & -e^{-U}\theta_{\bullet}\left[ e^{U}\theta_{\bar{\bullet}}H\right]-e^{-U}\theta_{\bar{\bullet}}\left[ e^{U}\theta_{\bullet}H\right] + e^{-2U}\partial_{u}\left(\partial_{z}\overline{\varpi} +\partial_{\bar{z}}\varpi
              \right)\ .
\end{eqnarray}
Observe that the last term in eq.~(\ref{eq:NCR++}) can always be put to zero by the coordinate transformation $v\longrightarrow v+\rho (u,z,\bar{z})$.

\section{\texorpdfstring{Spin connection and curvatures in minimal $d=5$ fSUGRA}{Spin connection and curvatures in minimal d=5 fSUGRA}}
\label{sec:Spin}
Defining the spin connection $\Omega^{a}{}_{b}$ by means of $dE^{a}=\Omega^{a}{}_{b}\wedge E^{b}$ and imposing it to be metric compatible $\Omega_{(ab)}=0$ leads to
\begin{eqnarray}
  \label{eq:SPconpm}
  \Omega_{+-} & =& -\theta_{-}H\ E^{+} \; -\; \textstyle{1\over 2}\, \theta_{-}\omega_{x}\ E^{x} \ ,\\
  \label{eq:SPcon+x}
  \Omega_{+x} & =& -\left( \theta_{x}H \; -\; E_{x}^{m}\theta_{+}\omega_{m}\right)\, E^{+}\; +\; \textstyle{1\over 2}\theta_{-}\omega_{x}\, E^{-} \nonumber \\
         & & -\left[E_{[y}{}^{m}\theta_{x]}\omega_{m}\; +\; E_{(y}^{m}\theta_{+}E_{x)m}\right]\, E^{y}\ ,\\
  \label{eq:SPcon-x}
  \Omega_{-x} & =& \textstyle{1\over 2}\theta_{-}\omega_{x}\, E^{+} \ ,\\
  \label{eq:SPconxy}
  \Omega_{xy} & =& -\lambda_{zxy}\ E^{z} \; -\;  \left[E_{[x}^{m}\theta_{y]}\omega_{m}\; -\; E_{[x}^{m}\theta_{+}E_{y]m}
             \right]\, E^{+} \ , 
\end{eqnarray}
where we have defined $\eth E^{x} = \lambda^{x}{}_{y}\wedge E^{y}$ and $\lambda_{zy}=\delta_{zx}\lambda^{x}{}_{y}$, whereas $\Omega_{xy}=\eta_{xz}\Omega^{z}{}_{y}$, so that the sign difference is paramount. Observe that a similar condition holds for defining $E_{mx}=E_{m}{}^{x}$. Of course, $\lambda$ is fixed by eq.~(\ref{eq:GTrest4}) to be
\begin{equation}\label{eq:GT0conn}
  \lambda_{xy} \; =\; 2\xi\, \aleph_{x}E^{y} \; -\; 2\xi\, \aleph_{y}E^{x}\; +\; \sqrt{3}\xi\, \varepsilon^{xyz}E^{z}\ .
\end{equation}

As stated above, if $(g,A)$ solves the fKSE one only needs to demand $\mathcal{M}_{+}=0$ and $\mathcal{E}_{++}= R_{++}+\textstyle{1\over 2}\varrho_{x}\varrho_{x}=0$ in order to ensure that $(g,A)$ solves all the equations of motion.
We now treat a simplified case which shows how the GT-geometry appears in the EOMs.

\subsection{\texorpdfstring{The $u$-independent case with $\varpi=0$}{The u-independent case with w=0}}
\label{sec:RiccUind}
The non-vanishing components of the Ricci tensor are given by
\begin{eqnarray}
  \label{eq:SPRicpm}
  R_{+-} & =& -\theta_{-}^{2}H \; -\; \textstyle{1\over 2}\theta_{-}\omega_{x}\theta_{-}\omega_{x}\; -\; \textstyle{1\over 2}\nabla_{x}\theta_{-}\omega_{x} \ ,\\
  \label{eqSPRicpp}
  R_{++} & =& -\nabla_{z}\left[ \nabla_{z}\Upsilon_{0} \; +\; 2\xi\, \aleph_{z}\Upsilon_{0}\right] \; -\; 8\xi^{4}\, v^{2}\ \aleph_{x}\aleph_{x}\ ,\\
  \label{eq:SPRicpx}
  R_{+x} & =& 4\xi^{3}\, v\, \aleph_{x}\ ,\\
  \label{eq:SPRicxy}
  R_{xy} & =& R^{(\lambda )}_{xy} \; +\; 2\xi^{2}\, \aleph_{x}\aleph_{y} \; -\; 2\xi\, \nabla_{(x}\aleph_{y)} \ ,
\end{eqnarray}
where $R^{(\lambda )}$ is the Ricci tensor for the three-dimensional spin connection $\lambda$.

It is also an easy task to calculate the non-vanishing components of the Einstein field equation in (\ref{eq:3m}), which read
\begin{eqnarray}
  \label{eq:EOM1a}
  \mathcal{E}_{++} & =& -\nabla_{z}\left[ \nabla_{z}\Upsilon_{0} \; +\; 2\xi\, \aleph_{z}\Upsilon_{0}\right] \ ,\\
  \label{eq:EOM1b}
  \mathcal{E}_{xy} & =& R^{(\lambda )}_{xy} \; -\; 2\xi\ \nabla_{(x}\aleph_{y)}\; -\; 4\xi^{2}\, \aleph_{x}\aleph_{y}
                   \; +\; 4\xi^{2}\, \aleph_{z}\aleph_{z}\, \delta_{xy}\; +\; 6\xi^{2}\, \delta_{xy} \ . 
\end{eqnarray}
Comparing the last equation with the symmetric part of the Ricci tensor for the Weyl connection in eq.~(\ref{eq:W3b}) for $d=3$, and taking into account that we are dealing with a Gauduchon metric, one can see that upon identifying $\theta =2\xi\, \aleph$ we can rewrite eq.~(\ref{eq:EOM1b}) as
\begin{equation}
  \label{eq:EOM1c}
  \mathcal{E}_{xy} \; =\; \mathtt{W}_{(xy)} \; +\; 6\xi^{2}\, \delta_{xy} \ .
\end{equation}
Comparing this and eq.~(\ref{eq:DefAleph}) to the results in appendix \ref{sec:EWspaces}, we see that the three-dimensional manifold is a Gauduchon-Tod space with $\kappa =2\sqrt{3}\xi$.

\section{Kundt metrics}
\label{appsec:Kundt}
A Kundt wave \mycite{Kundt:1961} is a metric that allows for a non-expanding, shear-free and twist-free geodesic null vector $N$. That is, respectively,
\begin{eqnarray}
\nabla_\mu N^\mu&=&0\ ,\\
\nabla_{(\mu} N_{\nu)} \nabla^\mu N^\nu&=&0\ ,\\
\hat{N}\wedge d\hat{N}&=&0\ ,\\
\nabla_N N&=&0\ ,
\end{eqnarray}
where $\hat{N}$ is the 1-form dual to the vector field. They were first studied in the arbitrary $d$-dimensional case in \mycite{Coley:2005sq}{Brannlund:2008zf,Podolsky:2008ec}{Coley:2009ut}. The line element can always be taken to read
\begin{equation}
  \label{eq:KundtMetric}
ds^2=\hat{E}^+ \otimes \hat{E}^-+\hat{E}^- \otimes \hat{E}^+ - \hat{E}^x \otimes \hat{E}^x\ ,
\end{equation}
where we have generically introduced the light-cone frame by
\begin{equation}
  \label{eq:KundtCurv1}
  \left\{
  \begin{array}{lclclcl}
    E^{+} & =& du\ , &\qquad& \theta_{+} & =& \partial_{u} \; -\; H \partial_{v} \ , \\
    E^{-} & =& dv + Hdu + S_{m}dy^{m}\ , &\qquad& \theta_{-} & =& \partial_{v} \ ,\\
    E^{x} & =& {e_{m}}^{x}\, dy^{m}\ , & \qquad& \theta_{x} & =& {e_{x}}^{m}\left( \partial_{m} \; -\; S_{m}\partial_{v}\right) \ ,
  \end{array}\right.
\end{equation}
where the Vielbein on the base-space $e_{i}^{x}$ is independent of $v$. The only $v$-dependence resides in $H$ and $\hat{S}\equiv S_{m}dy^{m}$. This is the kind of metric that appears in the characterisation of the null cases studied above, eqs.~(\ref{eq:Walkerfourdmetric}), (\ref{solution}), (\ref{Kundtmetric}), (\ref{eq:44}) and (\ref{eq:15}).

Moreover, whenever $\hat{S}$ does not depend on $v$, it can be written in the Walker form 
\begin{equation}
\label{eq:walkerkundt}
ds^2=2du(dv+H(u,v,y^p)\,du+S_m(u,y^p)\,dy^m)+g_{mn}(u,x^p)\,dy^m\,dy^n\ ,
\end{equation}
where $g_{mn}\equiv {e_{m}}^{x}\,{e_{n}}^{x}$. Eq.~(\ref{eq:walkerkundt}) is the general $d$-dimensional metric of a space with holonomy contained in $\mathrm{Sim}(d-2)$ \mycite{Walker:1950}. 

Defining the spin connection $\omega^{a}{}_{b}\equiv E^{c}\,\omega_{c,}{}^{a}{}_{b}$ by means of $dE^{a}=\omega^{a}{}_{b}\wedge E^{b}$ and imposing it to be metric compatible, \emph{i.e.}~$\omega_{(ab)}=0$, leads to
\begin{eqnarray}
  \label{eq:KundtCurv2a}
  \omega_{+-} & =& -\theta_{-}H\ E^{+} 
             \, -\ \textstyle{1\over 2}\ \theta_{-}S_{x}\ E^{x} \; ,\\
  \label{eq:KundtCurv2b}
  \omega_{+x} & =& -\left( \theta_{x}H \ -\ e_{x}^{m}\theta_{+}S_{m}\right)\ E^{+}
             \, +\ \textstyle{1\over 2}\theta_{-}S_{x}\ E^{-} 
             \ - \left[\ T_{yx} 
                     + e_{(y}^{m}\theta_{+}e_{x)m}
              \right]\, E^{y}\; ,\\
  \label{eq:KundtCurv2c}
  \omega_{-x} & =& \textstyle{1\over 2}\theta_{-}S_{x}\ E^{+} \; ,\\
  \label{eq:KundtCurv2d}
  \omega_{xy} & =& -\lambda_{zxy}\ E^{z} 
       \ -\  \left[\ T_{xy}
               \ -\ e_{[x}^{m}\theta_{+}e_{y]m}
             \right]\ E^{+} \; , 
\end{eqnarray}
where we have defined $\mathsf{d} E^{x} = \lambda^{x}{}_{y}\wedge E^{y}$ and also $\lambda_{zy}=\delta_{zx}\lambda^{x}{}_{y}$, whereas $\omega_{xy}=\eta_{xz}\omega^{z}{}_{y}$, so again the sign difference is paramount\footnote{Observe that a similar condition holds for defining $e_{mx}=e_{m}{}^{x}$.}. Furthermore, we define
\begin{equation}
\label{eq:9}
T_{xy} \; \equiv\; e_{[x}{}^{m}\theta_{y]}S_{m}\ ,  
\end{equation}
which for $d=6$ reads
\begin{equation}
\label{eq:9bis}
T_{ij} \; =\; v\, \mathsf{F}_{ij} \; -\; \textstyle{1\over 2}\, \left[ \mathsf{D}\varpi\right]_{ij} \ .
\end{equation}

If we impose that the only $u$-dependency resides in $H$, the non-vanishing components of the Ricci tensor become
\begin{eqnarray}
  \label{eq:KundtCurv3a}
  R_{++} & =& -\nabla^{(\lambda )}_{x}\partial_{x}H 
           \ +\   \theta_{-}H\ \nabla^{(\lambda )}_{x}S_{x}
           \ -\ H\ \nabla^{(\lambda )}_{x}\theta_{-}S_{x} \nonumber \\
           & & +\ 2S_{x}\ \partial_{x}\theta_{-}H 
            \ -\ \theta_{-}S_{x}\ \partial_{x}H 
            \ -\ S_{x}S_{x}\ \theta_{-}^{2}H \; ,\\
  \label{eq:KundtCurv3b}
  R_{+-} & =& -\theta_{-}^{2}H\ -\ \textstyle{1\over 2}\theta_{-}S_{x}\theta_{-}S_{x} \ +\ \textstyle{1\over 2} \nabla^{(\lambda )}_{x}\ \theta_{-}S_{x} \; ,\\
  \label{eq:KundtCurv3c} 
  R_{+x} & =& -\theta_{x}\theta_{-}H
                     \ -\ \nabla^{(\lambda )}_{y}T_{xy} 
                     \ +\ S_{y}\theta_{-}T_{xy}
                     \ +\ T_{xy}\theta_{-}S_{y} \; , \\
  \label{eq:KundtCurv3d} 
  R_{xy} & =& \mathsf{R}(\lambda )_{xy} \ -\ \nabla^{(\lambda )}_{(x|}\ \theta_{-}S_{|y)} \ +\ \textstyle{1\over 2}\ \theta_{-}S_{x}\theta_{-}S_{y} \; ,
\end{eqnarray}
The Ricci scalar is then given by
\begin{equation}
  \label{eq:KundtCurv4}
  R \; =\; -2\theta_{-}^{2}H
        \ -\ \textstyle{3\over 2}\ \theta_{-}S_{x}\theta_{-}S_{x}
        \ +\ 2\nabla^{(\lambda )}_{x}\theta_{-}S_{x}
        \ -\ \mathsf{R}(\lambda ) \ ,
\end{equation}
where $R(\lambda)$ is the Ricci scalar curvature of $\lambda$.

\cleardoublepage

\renewcommand{\leftmark}{\MakeUppercase{\chaptername\ \thechapter. Weyl geometry}}
\chapter{Weyl geometry}
\label{sec:Weylgeometry}
In this appendix we present a brief introduction to Weyl geometry, which arises in the context of theories with a scaling symmetry. Furthermore, we also give a sketch of Einstein-Weyl manifolds, and the special class of Gauduchon-Tod spaces, which appear repeatedly in the characterisation of solutions to fakeSupergravity theories.

Weyl geometry appeared naturally in an attempt to couple gravity and electromagnetism \mycite{Weyl:1918ib}. A Weyl manifold is a manifold $\mathcal{M}$ of dimension $d$ together with a conformal class $[g]$ of metrics on $\mathcal{M}$ and a torsionless connection $\mathtt{D}$, which preserves the conformal class, {\em  i.e.\/}
\begin{equation}
  \label{eq:W1}
     \mathtt{D}\; g \; =\; 2\theta \otimes g \ , 
\end{equation}
for a chosen representative $g\in [g]$. Using the above definition, we can express the connection $\mathtt{D}_{X}Y$
as 
\begin{equation}
  \label{eq:W2}
      \mathtt{D}_{\mu}Y_{\nu} \; =\; \nabla^{g}_{\mu}Y_{\nu} \, +\,\gamma_{\mu\nu}{}^{\rho}\, Y_{\rho}\ ,\quad\mbox{with}\quad\gamma_{\mu\nu}{}^{\rho} \; =\; g_{\mu}{}^{\rho}\theta_{\nu} \, +\, g_{\nu}{}^{\rho}\theta_{\mu} \, -\, g_{\mu\nu}\theta^{\rho}\; ,
\end{equation}
where $\nabla^{g}$ is the Levi-Civit\`a connection for the chosen $g\in [g]$. We define the curvature of this connection through $\left[ \mathtt{D}_{\mu},\mathtt{D}_{\nu}\right] Y_{\rho} = -\mathtt{W}_{\mu\nu\rho}{}^{\sigma}Y_{\sigma}$, using which we define the associated Ricci curvature as $\mathtt{W}_{\mu\nu} \equiv \mathtt{W}_{\mu\rho\nu}{}^{\rho}$.
The Ricci tensor is not symmetric and we have
\begin{eqnarray}
\label{eq:W3a}
\mathtt{W}_{[\mu\nu]} & =& -\textstyle{d\over 2}\,F_{\mu\nu}\ ,\quad\mbox{where}\, F\equiv d\theta \; , \\
& & \nonumber \\
\label{eq:W3b}
\mathtt{W}_{(\mu\nu)} & =& \mathtt{R}(g)_{\mu\nu}\, -\, (d-2) \nabla_{(\mu}\theta_{\nu)}\, -\, (d-2)\, \theta_{\mu}\theta_{\nu}\, -\, g_{\mu\nu}\left(\nabla_{\rho}\theta^{\rho}\, -\, (d-2)\, \theta_{\rho}\theta^{\rho}\right)\ .
\end{eqnarray}
The Ricci-scalar is defined as $\mathtt{W}\equiv\mathtt{W}_{\rho}{}^{\rho}$, which explicitly reads
\begin{equation}
  \label{eq:W4}
      \mathtt{W} \; =\; \mathtt{R}(g) \, -\, 2(d-1)\, \nabla_{\rho}\theta^{\rho}\, +\, (d-1)(d-2)\,\theta_{\rho}\theta^{\rho} \; .
\end{equation}
The 1-form $\theta$ acts as gauge field gauging an $\mathbb{R}$-symmetry, which is why we have been talking about a conformal class of metrics on $\mathcal{M}$. In fact under a transformation $g\rightarrow e^{2w}\ g$ we have that
$\theta \rightarrow \theta +dw$ and $\mathtt{W}\rightarrow e^{-2w}\mathtt{W}$, whereas $\mathtt{W}_{\mu\nu\rho}{}^{\sigma}$ and $\mathtt{W}_{\mu\nu}$ are conformally-invariant. We shall call a Weyl structure trivial/ closed if its curvature tensor is trivially zero, \emph{i.e.}~$\theta=d\Lambda\,$, for $\Lambda$ a function. 

A metric $g$ in the conformal class $[g]$ is said to be {\em standard} (also referred to as {\em Gauduchon}) if it is such that 
\begin{equation}
  \label{eq:W4a}
  d\star\theta \ =\ 0\ ,\quad\mbox{or equivalently}\quad\nabla_{\rho}\theta^{\rho}\ =\ 0\; ,
\end{equation}
where the $\star$ is taken w.r.t.~the chosen metric $g$.

\section{Einstein-Weyl and Gauduchon-Tod spaces}
\label{sec:EWspaces}
Einstein-Weyl (EW) manifolds are a special generalisation of Einstein manifolds, \emph{i.e.}~manifolds $(M,g)$ that satisfy
\begin{equation}
R_{\mu\nu}\,=\,k\,g_{\mu\nu}\ ,
\end{equation}
in the context of Weyl geometry. We say a manifold of dimension $d$ is Einstein-Weyl if the curvature satisfies
\begin{equation}
  \label{eq:W5}
  \mathtt{W}_{(\mu\nu)} \, =\, \frac{1}{d}\ g_{\mu\nu}\ \mathtt{W} \ .
\end{equation} 
Gauduchon proved the existence of a standard metric on a compact EW manifold \mycite{Gauduchon:1984}, and Tod proved that (on compact EW manifolds) this implies that $\theta^{\flat}$ is a Killing vector of the standard metric $g$ \mycite{Tod:1992}.

Einstein-Weyl geometries appear repeateadly in the study of fakeSupergravity theories, in the form of Gauduchon-Tod spaces.
These are a subclass of EW manifolds where an additional geometric constraint is demanded. In \mycite{Gauduchon:1998}, Gauduchon and Tod studied the structure of four-dimensional hyper-Hermitian Riemannian spaces admitting a tri-holomorphic Killing vector, {\em i.e.}~a Killing vector that is compatible with the three almost complex structures of the hyper-Hermitian space. They found that the three-dimensional base-space is determined by a Dreibein, or orthonormal frame, $E^{x}$, a 1-form $\theta$ and a real function $\kappa$ satisfying
\begin{equation}
  \label{eq:GTa}
  dE^{x} \; =\; \theta\wedge E^{x} \ -\ \kappa\ \star_{\text{(3)}} E^{x} \; ,
\end{equation}
where $\star_{\text{(3)}}$ is to be taken w.r.t.~the Riemannian metric constructed out of the Dreibein. The underlying geometry imposed by the above equation is that of a specific type of three-dimensional EW space, called hyper-CR or Gauduchon-Tod (GT) space\footnote{Observe that the Jones-Tod construction implies that the three-dimensional GT space, orthogonal to a generic Killing vector on a four-dimensional hyper-Hermitian space, is always Einstein-Weyl \mycite{Tod:1985}.}. The restriction (\ref{eq:GTa}) can equivalently be given by\footnote{The sign difference between eq.~(\ref{eq:GTb}) and eq.~(S) in \mycite{Gauduchon:1998}{prop.~5} is due to a differing definition of the Riemann tensor.}
\begin{eqnarray}
  \label{eq:GTb}
  \mathtt{W} & =& -\frac{3}{2}\, \kappa^{2}\ ,\\
  \label{eq:GTc} 
  \star d\theta & =& d\kappa\, +\, \kappa\ \theta \ .
\end{eqnarray}
\vspace{-12pt}

The standard example of a GT space is the {\em Berger sphere} \mycite{Gauduchon:1998} 
\begin{equation}
\label{eq:W6a}
\begin{array}{ccl}
ds^{2} & =& d\phi^{2}\; +\; \sin^{2}(\phi )d\varphi^{2}\; +\; \cos^{2}(\mu)\left( d\chi\; +\; \cos(\phi)\,d\varphi \right)^{2} \; ,\\
\theta & =& \sin (\mu)\cos(\mu)\; \left( d\chi\; +\; \cos(\phi)\,d\varphi \right)\ ,
\end{array}
\end{equation}
which is the unique compact Riemannian GT manifold, and can be seen as a squashed $S^{3}$ or an ${SU}(2)$ group manifold with a ${U}(1)$-invariant metric. One can easily see that the metric is Gauduchon-Tod with $\kappa = \cos(\mu)$. Thus, in order to use it in the five-dimensional solutions of chapters \ref{5dminimal} and \ref{5dgauged}, it needs to be rescaled by a constant.

Another class of GT spaces, albeit not in the Gauduchon-gauge, was given in \mycite{Calderbank:1999ad}{Th. 1.3} and reads
\begin{eqnarray}
  \label{eq:Cald1}
  ds^{2}  &=& dx^{2} \, +\, 4\left| x+h\right|^{2}\, \frac{dzd\bar{z}}{(1+|z|^{2})^{2}} \ ,\\
  \label{eq:Cald2}
  \theta &=& 2\mathrm{Re}\left(\textstyle{1\over x+h}\right)\, dx\ ,\\
  \label{eq:Cald3}
  \kappa &=& 2\mathrm{Im}\left(\textstyle{1\over x+h}\right) \ ,
\end{eqnarray}
where $h$ is an arbitrary holomorphic function $h=h(z)$. This was used in section \ref{subsec:CTwave} on chapter \ref{5dgauged} to propose a solution to the theory. Note that the choice $h=-\bar{h}$ results in the 3-sphere, and $h=\bar{h}$ leads to the flat metric on $\mathbb{R}^{3}$ with $\kappa =0$.


\cleardoublepage

\renewcommand{\leftmark}{\MakeUppercase{\chaptername\ \thechapter. Similitude group and holonomy}}
\chapter{Similitude group and holonomy}
\label{appsec:Sim}
In this appendix we present some information on the \emph{Similitude} group, which appears in the main chapters of this thesis. We also provide with details on its relationship to the holonomy group, and \emph{recurrent} vector fields. For more information, see \mycite{Gibbons:2009zz} and references therein.

\section{Sim and ISim as subgroups of the Caroll and Poincar\'e groups}
The Galilean (or Galilei) group describes, in Classical Mechanics, how to transform coordinates which are measured in two references frames moving relative to each other, at constant speed. Of course we now know that these transformations are only valid in a regime of low velocities, being superseded by Lorentz transformations in most cases of interest for Particle Physics. This means that the Galilean group provides a non-relativistic limit to the Poincar\'e group. The Carroll group is constructured similarly, where one chooses a different coordinate to label time (see below for details).
 
Both the Galilean and the Carroll group are symmetry groups of a theory living on a Minkowski spacetime of $(1,1)$ dimensions more \mycite{Bacry:1986pm,Gibbons:2007zu}. Hence they are best understood as subgroups of the Poincar\'e group. The $(d^2+3d+2)/2$ generators of the Poincar\'e algebra $\mathfrak{iso}(1,d)$ are translations, rotations and boosts, subject to
\begin{equation}
\label{eq:Poincarealgebraconm}
\begin{array}{lll}
\lbrack M_{ij},M_{kl}]&=&i(\eta_{ik}M_{jl}-\eta_{il}M_{jk}-\eta_{jk}M_{il}+\eta_{jl}M_{ik})\ ,\\
\lbrack M_{ij},P_k]&=&i(\eta_{ik}P_j-\eta_{jk}P_i)\ ,\\
\lbrack P_{i},P_j]&=&0\ .
\end{array}
\end{equation}
However, an alternative description of these is more appropiate to describe the Galilean and Carroll subgroups. By going to the light-cone metric of Minkowski space
\begin{equation}
ds^2=-2dudv+dx^i dx^i\quad \text{for}\ i=2,...,d
\end{equation}
we find that the elements of the Poincar\'e algebra can be cast as space translations and rotations $P_i=\partial_i,\ M_{ij}=-i(x_i \partial_j - x_j \partial_i)$, two null translations $H=\partial_u,\ M=\partial_v$ and $2d-1$ boosts $N=-i(u\partial_u -v\partial_v),\ K_i=-i(u\partial_i+x_i\partial_v),\ V_i=-i(v\partial_i+x_i\partial_u)$.\vspace{\baselineskip}

The Bargmann algebra (see \mycite{Duval:1984cj}) is obtained by considering the subalgebra of $\mathfrak{iso}(1,d)$ that commutes with $M$; this is the reason why it is often referred to as the `central extension of the Galilean algebra', and equivalent to excluding generators $N$ and $V_i$. It has the following rules
\begin{equation}
\begin{array}{ll}
\lbrack M_{ij},M_{kl}]=i(\eta_{ik}M_{jl}-\eta_{il}M_{jk}-\eta_{jk}M_{il}+\eta_{jl}M_{ik})\ ,\qquad & 
                                                \lbrack M_{ij},P_k]=i(\eta_{ik}P_j-\eta_{jk}P_i)\ ,\\
\lbrack M_{ij},K_k]=i(\eta_{ik}K_j-\eta_{jk}K_i)\ , & \lbrack M_{ij},H]=0\ ,\\
\lbrack P_i,P_j]=0\ , & \lbrack P_i,H]=0 \ , \\
\lbrack K_i,K_j]=0\ , & \lbrack K_i,H]=iP_i\ , \\
\lbrack K_i,P_j]= i\eta_{ij}M\ . & 
\end{array}
\end{equation}
The Galilean group, in turn, is the Bargmann group when we mod out the action of $M$. This means that the last commutator above can be put to zero. We shall not be using the Galilean group in this thesis, but it has gained some popularity in the last few years because of its relevance in studies of condensed matter systems through string theory methods.\vspace{\baselineskip}

The Carroll algebra was first introduced in \mycite{LevyLeblond:1965tt} by performing a Wigner-\.In\"on\"u contraction on the Poincar\'e algebra. This contraction is different to the one used to obtain the Bargmann, and it is obtained by looking for the subalgebra of $\mathfrak{iso}(1,d)$ formed by $\{P_i,M_{ij},K_i,M,N\}$ modulo $N$, whose full commutator rules are
\[
\begin{tabular}{ll}
$\lbrack M_{ij},M_{kl}]=i(\eta_{ik}M_{jl}-\eta_{il}M_{jk}-\eta_{jk}M_{il}+\eta_{jl}M_{ik})\ ,\qquad$ &  $\lbrack M_{ij},P_k]=i(\eta_{ik}P_j-\eta_{jk}P_i)\ ,$\\
$\lbrack M_{ij},K_k]=i(\eta_{ik}K_j-\eta_{jk}K_i)\ ,$ & $\lbrack M_{ij},N]=0\ ,$\\
$\lbrack M_{ij},M]=0\ ,$ & $\lbrack P_i,P_j]=0\ ,$\\
$\lbrack P_i,N]=0\ ,$ & $\lbrack P_i,M]=0\ ,$ \\
$\lbrack K_i,K_j]=0\ ,$ & $\lbrack K_i,P_j]=i\eta_{ij}M\ ,$\\
$\lbrack K_i,N]=iK_i\ ,$ & $\lbrack K_i,M]=0\ ,$\\
$\lbrack N,N]=0\ ,$ & $\lbrack N,M]=-iM\ ,$\\
$\lbrack M,M]=0\ .$
\end{tabular}
\]
The \emph{Similitude} group $\Sim(d-1)$ is obtained as a subgroup of the Carroll group by keeping the subset $\{M_{ij},K_i,N\}$ of the Carollian algebra. Its algebra hence has $(d^2-d+2)/2$ generators and it can be seen to form a subalgebra of $\mathfrak{so}(1,d)$.
\[
\begin{tabular}{ll}
$\lbrack M_{ij},M_{kl}]=i(\eta_{ik}M_{jl}-\eta_{il}M_{jk}-\eta_{jk}M_{il}+\eta_{jl}M_{ik})\ ,\qquad$ & $\lbrack M_{ij},N]=0\ ,$\\
$\lbrack M_{ij},K_k]=i(\eta_{ik}K_j-\eta_{jk}K_i)\ ,$ & $\lbrack K_i,K_j]=0\ ,$\\
$\lbrack N,N]=0\ ,$ & $\lbrack K_i,N]=iK_i\ .$
\end{tabular}
\]
In fact, the group is isomorphic to the Euclidean group of $\mathbb{R}^{d-1}$, augmented by homotheties (similarities) parametrised by a scaling factor, and it is the maximal proper subgroup of the Lorentz group $SO(1,d)$.

If we add the translations operators $\{P_i,H,M\}$ the algebra obtained has $(d^2+d+4)/2$ elements, and it is a subalgebra of $\mathfrak{iso}(1,d)$. The group then formed is correspondingly labelled $\ISim(d-1)=\Sim(d-1) \ltimes \mathbb{R}^{1,d}$ \mycite{Kogut:1970xx}. For $d=3$, this is the symmetry group used in Very Special Relativity, a theory allowing a small Lorentz violation, consistent with the observed CP violation \mycite{Cohen:2006ky}.


\section{Holonomy}
\label{app:simhol}
Holonomy is a measure of how much parallel transportation along a closed loop on a smooth manifold fails to preserve a certain geometrical quantity. In a nutshell, when parallel-transporting a non-scalar object around a closed loop, it will only remain constant if the holonomy group is trivial; otherwise the new object will be the result of having acted upon the original one by a holonomic transformation. The set of all transformations for a given manifold $M$ and a connection $\nabla$ gives the holonomy group. In more formal language, this group is described as
\[\text{Hol}_x(\nabla)=\Big\{ P_\gamma \in GL(M_x), \text{ s.t. $\gamma$ is a closed loop based at x}\in M\Big\}\ ,\]
where $P_\gamma:M_x\rightarrow M_x$.\vspace{\baselineskip}

Since we deal with problems arising in the context of supergravity theories, we shall particularly deal with manifolds of special holonomy. These are defined as manifolds admitting the existence of parallel spinors (w.r.t.~Levi-Civit\`a), and thus they naturally appear in the context of (fake-) supersymmetric studies. The classification of Riemannian irreducible non-symmetric simply-connected holonomy manifolds was given by Berger \mycite{Berger:1955}. Moreover, Wang gave the dimensionalities of the spaces of (non-trivial) parallel spinors \mycite{Wang:1989xx}, thus establishing which manifolds have Ricci-flat metrics \mycite{Hitchin:1974,Friedrich:1979}.
\begin{table}[!h]
\hspace{\stretch{1}}
\begin{tabular}{|c|c|c|c|}
\hline
\textbf{Hol. group} & \textbf{dim($M$)} & \textbf{Associated manifold $(M,g)$} & \textbf{dim(\{$\epsilon$\}) s.t. $\nabla \epsilon=0$}\\
\hline
\hline
$SO(n)$ & $n\geq 2$ & generic orientable Riemannian&0\\
$U(n)$ & $2n$ &generic K\"ahler & 0\\
$SU(n)$	&$2n$& special K\"ahler (CY)&2 \\
$Sp(n)\cdot Sp(1)$ &$4n$& quaternionic-K\"ahler&0\\ 
$Sp(n)$ &$4n$& hyper-K\"ahler&$n+1$\\ 
$G_2$	&7&exceptional holonomy&1\\
$Spin(7)$&8 &exceptional holonomy&1\\
\hline
\end{tabular}
\hspace{\stretch{1}}
\label{Bergerslist}
\caption{Berger's classification of possible holonomy groups for irreducible non-symmetric simply-connected Riemannian manifolds. Notice how only the third, fifth, sixth and seventh cases are Ricci-flat, thus composing the list of special holonomy manifolds.}
\end{table}

In this thesis, however, we are interested in Lorentzian spaces, and holonomy is one of those features where Riemannian and pseudo-Riemannian geometry take on a very different form. The interested reader can consult \mycite{Berger:1955,Bryant:1996,Bohle:1998xx} for further information on this topic. We should also note that, since we consider connections that respect lengths, the maximal possible holonomy group is $SO(1,d)$. Given that $\Sim(d-1)$ is its maximal proper subgroup, this means that the minimal holonomy reduction that can occur in a Lorentzian spacetime $M^{1,d}$ is $\text{Hol}_M=\Sim(d-1)$. We now study the relation between having holonomy inside $\Sim(d-1)$ and recurrent vector fields.

\section{Recurrency}
\label{sec:recurr}
We say that a vector field $n^\mu$ is recurrent if
\begin{equation}
\label{eq:recurrency}
\nabla_\mu n^\nu=B_\mu n^\nu\ ,
\end{equation}
where $B_\mu$ is called the \emph{recurrence one-form} \mycite{Gibbons:2007zu}. Geometrically this means that $n^\mu$ does not change direction under parallel-transportation. When on top of that $n^\mu$ is null, \emph{i.e.}~$n_{\mu} n^{\mu}=0$, this implies that the connection has holonomy inside $\Sim(d-1)$. Let us see this.

Considering the musical isomorphism that takes $n^\mu \to^{\hspace{-0.7em}\flat}\: n_\nu=g_{\nu\mu}n^\mu$, one can show that
\[dn^\flat=B\wedge n^\flat\ ,\]
which implies
\[n^\flat=du \quad \rightarrow \quad dn^\flat=0 \quad\rightarrow \quad \nabla_\mu n_\nu=\xi n_\mu n_\nu\ ,\]
for some function $\xi$. This last equation says that $n^\mu$ generates a geodesic null congruence that is hypersurface orthogonal, non-expanding and shear-free. We parametrise this congruence by $v$, and hence $n=n^\mu \partial_\mu=\frac{\partial}{\partial v}$.\vspace{\baselineskip}

The components of $n$ and $n^\flat$ allow us to write explictly the metric in Walker form \mycite{Walker:1950}
\begin{equation}\label{eq:Walkerform}
ds^2=H(u,v,x)dudu+2du dv+2A_i(u,x)dudx^i + g_{ij}(u,x)dx^i dx^j\ ,
\end{equation}
where we have taken the function $u$, parameter $v$ and the $(d-1)$ transverse $x^i$ as coordinates. One can then use Ricci identity to obtain 
\begin{equation}
\label{eq:RecurRicciId}
{R_{\mu\nu\rho}}^\sigma n_\sigma= (dB)_{\mu\nu} n_\rho\ .
\end{equation}
If we interpret the curvature tensor ${R_{\mu\nu\rho}}^\sigma$ as a Lorentz algebra-valued two-form, eq.~(\ref{eq:RecurRicciId}) means that the rotation performed on $n$ for a given loop $\gamma=(X,Y)$ is related to the field strength of the recurrence one-form, \emph{i.e.}
\begin{equation}
(\mathcal{O}_{(X,Y)}\circ n)_\rho=F^{\phantom{|}(B)}(X,Y)n_\rho\ .
\end{equation}
It is an easy matter to see that $R_{\mu\nu+i}=0$. This implies that $(R_\gamma)_{+i}=0$ and hence that the local holonomy group is generated by $(d-1)$ generators less than those of $\text{SO}(1,d)$. In other words, the maximal holonomy group is $\Sim(d-1)$.

Furthermore, if $H=H(u,x)$, \emph{i.e.}~not depending on $v$, eq.~\ref{eq:Walkerform} becomes a Brinkmann wave \mycite{Brinkmann:1923xx}{Brinkmann:1923xy}{Brinkmann:1925xx}. This means that $\nabla_\mu  n^\nu=0$ and hence $n$ is a (null) Killing vector field. In this case the maximal holonomy group is not $\Sim(d-1)$ but rather a subgroup of it, the Euclidean group of $\mathbb{R}^{d-1}$, where one further generator is no longer needed, since now $R_{\mu\nu+-}=0$.

\cleardoublepage

\renewcommand{\leftmark}{\MakeUppercase{\chaptername\ \thechapter. The Lorentz and the Spin groups}}
\chapter{The Lorentz and the Spin groups}
\label{appsec:lorentz}
The group of diffeomorphisms, paramount to Einstein's theory of gravity, does not allow for half-spin representations\footnote{At least not ones that fall naturally within the framework of the gravity theories we shall be discussing in this thesis.}, and hence a different recipe is needed to include fermions in a formulation of gravity. This was first done by Weyl, who realised that the connection existent between the Lorentz and the Spin group could be exploited for this purpose \mycite{Weyl:1929fm}. By exponentiating the generators of the Lorentz algebra $\mathfrak{so}(1,d-1)$ in a spinorial representation, one obtains the simply-connected $\Spin(1,d-1)$ group. This is the famous 2-1 correspondence between $\Spin(1,d-1)$ and $SO(1,d-1)$, which we succinctly explain now.

The generators of the Lorentz algebra are those $M_{ab}$ with $M_{ab}=-M_{ba}$ (where $a,b=0,1,\ldots,d-1$) such that\footnote{Notice that the following rule coincides (modulo conventions) with the first equation in the system (\ref{eq:Poincarealgebraconm}). This is so because the Poincar\'e algebra is the Lorentz one plus translations.} 
\begin{equation}
\label{eq:lorentzalgebra}
[M_{ab},M_{cd}]\;=\;-\eta_{ac}\,M_{bd}-\eta_{bd}\,M_{ac}+\eta_{ad}\,M_{bc}+\eta_{bc}\,M_{ad}\ .
\end{equation}
This relation guarantees that the exponential map on the algebra gives the group of Lorentz-Fitzgerald transformations ${\Lambda^c}_d=e^{\sigma^{ab}{(M_{ab})^c}_d}$, which respects the metric $\eta=(+,-,...,-)$ in accordance with the special principle of relativity 
\begin{equation}
V^{'a}\eta_{ab}V^{'b}\;=\;V^{a}\eta_{ab}V^{b}\ ,
\end{equation}
where $V^{'a}={\Lambda^a}_b V^b$. This implies a generalised constraint for orthogonality
\begin{equation}
\Lambda\in SO(1,d-1)\quad\rightarrow\quad \eta_{ab}{\Lambda^b}_c\eta^{cd}\;=\;{{(\Lambda^{-1})}^d}_a\ ,\quad\ det(\Lambda)= 1\ ,
\end{equation}
where $\eta^{cd}$ is the inverse of $\eta$, and the flat metric can be used to raise/ lower Lorentz indices. This condition translates at the level of the Lie algebra into
\begin{equation}
(M_{ab})^T\;=\;-M_{ab}\quad\rightarrow\quad tr(M_{ab})\;=\;0\ ,
\end{equation} 
which implies that the generators are antisymmetric in their Lorentz-indices, \emph{i.e.} 
\begin{equation}
M_{ab}\;=\;-M_{ba}\ .
\end{equation}
The generators of the Lorentz algebra in the vector representation are thus given by
\begin{equation}
\Gamma_\text{v}{\left(M_{ab}\right)^c}_d\;=\;2\,{\eta_{[a}}^c\,\eta_{b]d}\ ,
\end{equation}
which obviously satisfy eq.~(\ref{eq:lorentzalgebra}).

The $\Spin(1,d-1)$ group, on the other hand, is generated by the product of an even number of elements of the Clifford algebra with an inverse, where
the defining relation for the Clifford algebra is given by 
\begin{equation}
\{\gamma_a,\gamma_b\}\equiv\gamma_a\gamma_b+\gamma_b\gamma_a\;=\;2\,\eta_{ab}\,\mathbbm{1}_{2^{\lfloor d/2 \rfloor} \times 2^{\lfloor d/2 \rfloor}}\ .
\end{equation}
The connection between Clifford and Lorentz algebras becomes obvious with the following choice for the spinorial representation of the Lorentz algebra
\begin{equation}
\Gamma_\text{s}{\left(M_{ab}\right)^\alpha}_\beta\;=\;\frac{1}{2}{(\gamma_{[a}\gamma_{b]})^\alpha}_\beta=\frac{1}{4}{(\gamma_a\gamma_b-\gamma_b\gamma_a)^\alpha}_\beta\ ,
\end{equation}
where the $\alpha$, $\beta$ indices run from 1 to $2^{\lfloor d/2 \rfloor}$, and label the components of a general complex spinor $\psi^\alpha$, which by definition is an irreducible representation of the Spin group. In other words, one can use the Clifford algebra for $\eta_{ab}$ to construct a spinorial representation for $\mathfrak{so}(1,d-1)$. And because spinors are irreps of $\Spin(1,d-1)$, which is generated from the Lorentz algebra, one has a way to speak of fermions locally, \emph{i.e.}~at every point on our curved manifold, on which one considers a Minkowskian tangent space with $SO(1,d-1)$ as the structure group for changes of frame.

The experienced reader will have noticed that the exponential map
\begin{equation}
e^{\sigma^{ab}M_{ab}}=\mathbbm{1}+\sigma^{ab}M_{ab}+\ldots
\end{equation}
will only give the elements of the Lorentz group connected to the identity, what is usually called the proper orthochronous, or restricted, Lorentz group $SO^+(1,d-1)$. To obtain the non-connected orbits\footnote{There are four orbits in total, relative to the possible orientations of space and time. Proper/ inproper ($det(\Lambda)=\pm 1$) determine whether the subgroup respects the orientation of space, and orthochronicity is a measure of the group respecting the direction of time (${\Lambda^1}_1>1$).}, one has to act with (discrete) parity $P=diag(1,-1,-1,-1)$ and time-reversal $T=diag(-1,1,1,1)$ transformations. However, as commonly done in the Physics literature, throughout this thesis -and with the exception of the following paragraph- we shall refer to the restricted Lorentz group simply as the Lorentz group, and we will generically label it by $SO(1,d-1)$, omitting the ($+$)-label. Likewise for the restricted $\Spin^+(1,d-1)$ group.
 
The generators of the Lorentz $\mathfrak{so}(1,d-1)$ algebra in the spinorial representation $\Gamma_{\text{s}}(M_{ab})$ will give the simply-connected\footnote{A group manifold $G$ is simply-connected, or 1-connected, if it is path-connected, and has trivial fundamental group, \emph{i.e.}~$\pi_1(G)=1$. The latter is defined as the set of loops defined on the space modulo continuous deformations (homotopies). This means that a simply-connected manifold presents no obstructions (holes) to deforming any loop homeomorphically into another loop with the same base point.} group, $\Spin^+(1,d-1)$, which is only locally isomorphic to $SO^+(1,d-1)$, \emph{i.e.}~the algebras are the same. At the level of the group, however, $\Spin^+(1,d-1)$ is covering $SO^+(1,d-1)$, and in particular for $d>2$ the former is 1-connected, so one says it is the universal covering group\footnote{This means that the Spin group will cover any other possible connected cover for the Lorentz group.} of the Lorentz group. Furthermore, this cover is a double cover, in that there are two elements of Spin mapping to each element of Lorentz.

{
\setlength{\textwidth}{360pt}
\hspace{1cm}\begin{minipage}{\textwidth}
The prototypical example of this is given by the homomorphism
\begin{equation}
\rho:SL(2,\mathbb{C})\simeq \Spin^+(1,3)\rightarrow SO^+(1,3)\ ,
\end{equation}
characterised by $\rho(P):X\rightarrow P X P^\dagger$, where $X$ is a generic Hermitian matrix with the extended Pauli matrices $\{\sigma^0=\mathbbm{1},\sigma^1,\sigma^2,\sigma^3\}$ as basis, which is identified with Minkowski space -where a generic vector $v^T=(t,x,y,z)$ lives- through the determinant function, \emph{i.e.}~$det(X)=t^2-x^2-y^2-z^2=v^T\eta v$, for 
\begin{equation}
X=\left(\begin{array}{cc}t+z&x-iy\\x+iy&t-z\end{array}\right)\ ,
\end{equation}
and $P\in SL(2,\mathbb{C})$. The endomorphism on the space of Hermitian matrices preserves the determinat, and hence its action on $M^{(1,3)}$ is isometric (vector length-preserving), which was of course expected since $\rho(P)$ is a Lorentz transformation. Furthermore, one can easily see that both $P=K$ and $P=-K$, for $K$ any element of $SL(2,\mathbb{C})$, give the same element of $M^{(1,3)}$, and hence the 2-1 mapping.
\end{minipage}}
\vspace{\baselineskip}

\noindent This is why it is common to hear that the latter is doubly-connected, namely that its fundamental group is isomorphic to $\mathbb{Z}_2$. Despite these differences, it is also quite common to hear them referred interchangeably.

\cleardoublepage

\renewcommand{\leftmark}{\MakeUppercase{Bibliography}}
\phantomsection
\addcontentsline{toc}{chapter}{References}
\bibliographystyle{thesisbibstyle}
\bibliography{bibdb}
\label{biblio}
\clearpage

\end{document}